\documentclass[a4paper,11pt]{article}
\pdfoutput=1
\usepackage{jheppub,booktabs}
\usepackage{amsmath,amssymb}
\usepackage{cancel}
\usepackage{subfigure}
\usepackage{mathrsfs}
\usepackage{graphicx}
\usepackage{hepunits}
\usepackage{wasysym}
\usepackage{feynmf}
\usepackage[utf8x]{inputenc}

\preprint{CERN-PH-TH-2014-206}
\author[a]{Adrián Carmona}
\author[a,b]{and Florian Goertz}
\affiliation[a]{
Institute for Theoretical Physics, \\
ETH Zurich, 8093 Zurich, Switzerland}
\affiliation[b]{Theory Division,\\
CERN, 1211 Geneva 23, Switzerland}
\emailAdd{carmona@itp.phys.ethz.ch}
\emailAdd{florian.goertz@cern.ch}
\title{A naturally light Higgs without light Top Partners}

\abstract{
We demonstrate that the inclusion of a realistic lepton sector can relax significantly the upper bound
 on top partner masses in minimal composite Higgs models, induced by the lightness of the Higgs boson. To
that extend, we present a comprehensive survey of the impact of different realizations of the fermion 
sectors on the Higgs potential, with a special emphasis on the role of the leptons.
The non-negligible compositeness of the $\tau_R$ in a general class of models that address the flavor 
structure of the lepton sector and the smallness of the corresponding FCNCs, can have a significant effect on 
the potential. We find that, with the $\tau_R$ in the symmetric representation of $SO(5)$,
an increase in the maximally allowed mass of the lightest top partner of $\gtrsim 1$\,TeV
 is possible for minimal quark setups like the MCHM$_{5,10}$, without increasing the tuning. 
A light Higgs boson $m_H\!\sim\!(100\!-\!200)\,$GeV is a natural prediction of such models,
which thus provide a new setup that can evade ultra-light top partners without
ad-hoc tuning in the Higgs mass.
Moreover, we advocate a more minimal realization of the lepton sector than generally used
in the literature, which still can avoid light partners due to its contributions 
to the Higgs mass in a different and very natural way, triggered by the seesaw mechanism.
This allows to construct the most economical $SO(5)/SO(4)$ composite Higgs models possible.
Using both a transparent 4D approach, as well as presenting numerical results
in the 5D holographic description, we demonstrate that, including leptons, 
minimality and naturalness do not imply light partners.
Leptonic effects, not considered before, could hence be crucial for the viability of composite models.}
\date{\today}
\begin{document}

\maketitle
\section{Introduction}
The LHC and its experiments have already delivered an outstanding contribution to our understanding of electroweak 
symmetry breaking (EWSB). With the discovery of the 125\,GeV scalar \cite{Aad:2012tfa,Chatrchyan:2012ufa}
and the first determination of  its 
properties, we are lead to the conclusion that a Higgs sector is responsible for EWSB. However, the question if this sector 
can be identified with the one appearing in the Standard Model (SM) of particle physics is still to be answered. There are 
various reasons to expect new particles beyond the SM (BSM) and then naturalness calls for a mechanism to avoid the 
sensitivity of the Higgs mass to large scales. Supersymmetric models or models of compositeness, like the most studied 
minimal composite Higgs models MCHM$_5$ and MCHM$_{10}$, provide an elegant incarnation of such a 
protection mechanism. Both ideas assume the presence of BSM physics not far above the electroweak scale $M_{\rm EW} 
\sim v$ in order to avoid the quadratic ultra-violet (UV) sensitivity of the Higgs mass but rather saturate it in the infra-red (IR). 
For the latter class of models, the separation between the Higgs mass and the BSM scale, where new resonances appear, 
can naively be larger since the Higgs is realized as a pseudo Goldstone boson of the coset $SO(5)/SO(4)$, providing an 
additional protection for its mass. 

However, in the MCHMs mentioned before, the fact that the Higgs boson is rather light, $m_H \approx 125\,$GeV, 
requires the presence of light partners of the top quark \cite{Contino:2006qr,Medina:2007hz,Csaki:2008zd,DeCurtis:2011yx,Matsedonskyi:2012ym,Marzocca:2012zn,
Pomarol:2012qf,Panico:2012uw,Archer:2014qga}. This tendency can be easily understood from the fact that generically 
the linear mixing terms between the top quark and the composite fermionic partners, needed to generate the top mass 
via the concept of partial compositeness, break the Goldstone symmetry and thus contribute to the Higgs mass. 
The large value of the top mass requires the masses of the top partners to be rather small in order to generate a 
large mixing with the composite sector without introducing too large coefficients of the linear mixing terms, that would 
make the Higgs too heavy. So in both classes of solutions to the naturalness problem one expects top partners at 
$\lesssim {\cal O}(1)$\, TeV. The non-observation of these particles at the LHC so far has already put both ideas 
under some pressure. 

For the MCHM$_{5,10}$, where the composite fermions are realized in fundamental and adjoint representations
of $SO(5)$, respectively, the presence of light partners significantly below the actual scale of these models has explicitly 
been demonstrated in  \cite{Contino:2006qr,Csaki:2008zd,DeCurtis:2011yx,Matsedonskyi:2012ym,Marzocca:2012zn,Pomarol:2012qf,Panico:2012uw}.\footnote{We will not consider the spinorial representation {\bf 4}, as in the fermion
sectors where it could have an impact on our analysis it is not viable due to a lack of protection 
for $Z \bar b_L b_L$ (or $Z \bar \tau_R \tau_R$) couplings.}
For a $125\,$GeV Higgs boson their masses have been shown to lie around $m_{\tilde t}^{(5)} \sim 600\,{\rm GeV}$ and
$m_{\tilde t}^{(10)}\sim 400\,{\rm GeV}$, given that the fundamental mass scale of the models resides at the TeV scale, as
suggested by naturalness. This is 
even in a region probed currently by experiments at the LHC and provides an option to discover signs of these models, 
but in the case of 
no observation also is a potential threat to the composite Higgs idea. Indeed, the MCHM$_{10}$ is already 
severely challenged by top partner searches, see also Section~\ref{sec:5D}.

The only viable way out of the necessity of such ultra-light states found so far, requires the embedding of
quarks in a symmetric representation {\bf 14} of $SO(5)$. These models however suffer generically from an ad-hoc tuning
\cite{Pomarol:2012qf,Panico:2012uw}. While in the MCHM$_{5,10}$ after EWSB the Higgs mass is automatically generated 
not too far from the experimental value, and light fermionic partners offer the option to arrive at 
$m_H=125\,$GeV, this is not true for this realization with a ${\bf 14}$. In contrast to 
the other models, one needs a sizable tuning of in general unrelated quantities to arrive at the correct Higgs mass, 
which is naturally predicted much too heavy, $m_H^{(14)} \sim 1\,$TeV. This is to be contrasted with the 
``double tuning'' in the MCHM$_{5,10}$, which is required to achieve a viable EWSB~\cite{Matsedonskyi:2012ym}, 
see Section~\ref{sec:Vht}.
Thus, while evading the necessary presence of problematic fermionic partners, the attractive prediction of a 
generically light Higgs boson is challenged and one needs to induce a different kind of {\it ad-hoc} cancellation, which is not linked to the necessity of a suitable EWSB, but rather to the particular experimental value of the Higgs mass, not favored by the model. 
While, besides that, there exist viable setups featuring a symmetric representation of $SO(5)$, one should also note that models like
the MCHM$_{14}$ suffer from a large modification of $\sigma(gg \to H)$ and are already disfavored from Higgs physics
at the LHC. Moreover, the setups considerably enlarge the particle content and parameter space with respect to the most minimal realizations without an obvious structural reason.

In this article, we will introduce a new class of options to lift the mass of the top partners. They can realize a light 
Higgs without an ad-hoc tuning and without large changes in Higgs production, thus adding new aspects to the question if 
the non-observation of fermionic partners below the TeV scale together with at most modest effects in Higgs production already 
put the $SO(5)/SO(4)$ composite Higgs framework under strong pressure. 
To that extend we analyze the influence of leptons on the Higgs 
potential. Naively, such contributions seem not to be relevant due to the small masses of the leptons, leading at 
first sight to a small mixing with the composite sector and thus a small Goldstone-symmetry breaking. However, in a general 
class of models that address the flavor structure of the lepton sector (e.g. via flavor symmetries) and in particular the 
smallness of leptonic flavor-changing neutral currents (FCNCs), there is a natural suppression of the Yukawa couplings in the composite sector as well as of the left-handed lepton compositeness (see \cite{delAguila:2010vg}). 
The reason for the former is that the Yukawa couplings control the size of the breaking of the flavor symmetries. The latter (more relevant) suppression is due to the fact that the left handed lepton couplings are in general not protected by custodial symmetry. Potentially dangerous corrections thus need to be suppressed by the elementary nature of the left-handed leptons.
In these flavor-protected models, the $\tau_R$ needs to mix stronger with the composite sector than naively expected, in order 
to generate its non negligible mass, which can lead to interesting effects in the Higgs potential.\footnote{Note that this is not 
the case in the bottom sector, where a relatively large degree
of compositeness of the doublet component $b_L$ is required due to the large top-quark mass, in turn not allowing for
 a sizable mixing of the $b_R$  with the composite sector (where the latter also in general features no custodial protection).
On the other hand, models of $\tau_R$ compositeness feature in general also enough protection of the $\tau_R$ couplings, 
not to be in conflict with precision tests \cite{Carmona:2013lva} and can have an interesting (possibly modest) impact on Higgs 
physics at the LHC \cite{Carmona:2013cq}.} 

Beyond that, we will point out a new motivation for charged lepton compositeness, built on the mere size of the neutrino 
masses. As it will turn out, the most minimal realization of the 
type-III seesaw mechanism in the composite Higgs framework projects the modest IR localization of right-handed 
neutrinos onto the charged leptons. The possible Majorana character of neutrinos thus leads to distinct new features in 
the lepton sector, compared to the light quarks.

It turns out that the contributions of the $\tau$ sector to the Higgs mass 
can interfere destructively with the top contribution and, for the symmetric $SO(5)$ representation, can lift the masses of 
the light fermionic resonances significantly above the region of $m_{\tilde t}\lesssim 1$\,TeV, currently 
tested at the LHC, even if the quark sector corresponds to the MCHM$_5$ or MCHM$_{10}$. 
Thus, such potential contributions should be taken into account when examining the viability of the composite Higgs idea. 
As we will show, lifting the top partner masses via leptonic contributions has several intriguing features.
First, one opens the possibility to avoid the large ad-hoc tuning appearing in equivalent realizations via the quark sector
in a way that ultra-light resonances are even disfavored from the point of view of the tuning.
Moreover, we will show that it is possible to evade light partners even without abandoning the concept of minimality of the setup. 

This article is organized as follows. In Section~\ref{sec:general} we provide a complete survey of possible realizations of the 
fermion sector of the composite setup for the chiral SM-fermions mixing with any one of the basic representations of $SO(5)$ 
up to a {\bf 14} and discuss the structure of the Higgs potential and its mass via a spurion analysis.
In particular, we review the emergence of light partners and demonstrate in a general way how they could be avoided via leptonic 
contributions, pointing out the virtues of this approach. While in this section we follow a generic, particular suited, 
4D approach that makes the 
important mechanisms more transparent, in Section~\ref{sec:5D} we will give explicit numerical results in the Gauge-Higgs 
unification (GHU) setup, which provides a weakly coupled dual description of the composite Higgs idea. Here, we will confirm the
general findings of the previous section and provide detailed results for the top partner masses and the tuning, in dependence
on the Higgs mass for all important incarnations of the fermion sector. In particular, we introduce a new realization of the lepton 
sector in GHU models, embedding both the charged and neutral leptons in a single ${\bf 5}_L+{\bf 14}_R$, thus working with 
less degrees of freedom than in the standard MCHM$_5$-like setup and a significantly reduced number of parameters. 
Employing a type-III seesaw mechanism, we make explicitly use of the additional $SU(2)_L$ triplets provided by the ${\bf 14}_R$,
presenting the most minimal composite model with such a mechanism - which even allows for an enhanced minimality in the 
quark sector with respect to known models. This avoids many additional colored states at the TeV scale and allows
for the least number of new degrees of freedom of all known viable setups of $SO(5)/SO(4)$.
We will show that this very minimal realization of leptons just belongs to the class of models that has the strongest impact on the 
Higgs mass and demonstrate its capabilities for lifting the light partners. We finally conclude in Section~\ref{sec:conclusions}.

\section{General Structure of the Higgs Potential in MCHMs and Light Partners}
\label{sec:general}
In this section, we review the structure of the Higgs potential in composite models and the emergence of ``anomalously'' 
light top partners in models with a naturally light Higgs boson. We will then illustrate how including the effects from a 
realistic lepton sector allows to construct models that evade the necessary presence of fermion partners with masses 
$\lesssim 1$\, TeV, without introducing a large (ad hoc) tuning. To make the important physics more transparent, while keeping the 
discussion as generic as possible, we will work with general 4D realizations of the composite Higgs framework 
\cite{Giudice:2007fh,Anastasiou:2009rv,Panico:2011pw,DeCurtis:2011yx,Azatov:2011qy,Matsedonskyi:2012ym} 
and later provide the connection to the dual 5D GHU setup \cite{Manton:1979kb, Hatanaka:1998yp, vonGersdorff:2002as, Csaki:2002ur, Contino:2003ve, Agashe:2004rs}, which adds explicit calculability to the strongly coupled 4D models.

\subsection{Generic (4D) Setup of the Models}

In composite Higgs models, the Higgs field is assumed to be a composite state of a new strong interaction. In 
consequence, corrections to the Higgs mass are cut off at the compositeness scale such that it is saturated in the IR.
Moreover, following the analogy with the pions in QCD, it is generically realized as a pseudo Nambu-Goldstone boson 
(pNGB) associated to the spontaneous breaking of a global symmetry \cite{Terazawa:1976xx,Terazawa:1979pj,Dimopoulos:1981xc,Kaplan:1983fs,Kaplan:1983sm,Georgi:1984ef,Banks:1984gj,Georgi:1984af,Dugan:1984hq},
see also \cite{Contino:2003ve, Agashe:2004rs}. 
This provides a natural reasoning for the fact that the Higgs is lighter than potential new resonances of 
the models. The minimal viable breaking pattern featuring a custodial symmetry for the $T$ parameter is $SO(5) \to SO(4)$, 
which leads to four Goldstone degrees of freedom. The pNGB Higgs can thus be described by the real scalar fields, 
$\Pi_{\hat a}$, $\hat a=1,..\,,4$\,, embedded in the $\Sigma$ field
	\begin{equation}
	\label{eq:NLS}
	\Sigma =  U\, \Sigma_0, \quad U= \mathrm{exp} \left(i \frac{\sqrt 2}{f_\pi} \Pi_{\hat a} T^{\hat a}\right)\,,
	\end{equation}
which transforms in the fundamental representation of $SO(5)$. Here, $\Sigma_0=(0,0,0,0,f_\pi)^T$ specifies the vacuum 
configuration, preserving $SO(4)$, $f_\pi$ is the pNGB-Higgs decay constant and $T^{\hat a}$ are the broken generators 
belonging to the coset $SO(5)/SO(4)$. These generators are defined in Appendix \ref{sec:gen}, together with the remaining 
$SO(4) \cong SU(2)_L \times SU(2)_R$ generators $T_L^a, T_R^a$.
 
Under $g \in SO(5)$, the Goldstone 
matrix $U$ appearing in the decomposition (\ref{eq:NLS}) transforms as \cite{Coleman:1969sm,Callan:1969sn}
	\begin{equation}
	\label{eq:Utraf}
	U \rightarrow g \cdot U \cdot \hat h^T(\Pi, g),\quad \hat h=\begin{pmatrix} \hat h_4 & 0 \\ 0 & 1 \end{pmatrix},\quad \hat h_4 \in SO(4),
	\end{equation}
such that $\Sigma \rightarrow g \cdot \Sigma$.
The above construction provides a non-linear realization of the $SO(5)$ symmetry on the $\Pi$ fields, which however 
transform in the fundamental representation of the unbroken $SO(4)$ ({\it i.e.}, as a bi-doublet under $SU(2)_L 
\times SU(2)_R$). Finally note that the $\Sigma$ field just corresponds to the last column of the Goldstone matrix $U$, 
$\Sigma= f_\pi \,U_{I5}$. The fact that the Higgs is realized as a Goldstone of $SO(5)/SO(4)$ leads to a vanishing 
potential at the tree level. Explicit $SO(5)$-breaking interactions then generate it at one loop, which induces a
Higgs vacuum expectation value (vev) $v$, taken along the scalar component $\Pi_{\hat 4} \equiv h = H+ v$, with $\langle H \rangle=0$,
mediating EWSB.

Beyond the pNGB Higgs, composite models generically contain fermionic and bosonic resonances with masses $m_{\Psi,\rho} 
\sim g_{\Psi,\rho}\,f_\pi \lesssim 4\pi f_\pi$, bound states of the new strong sector and transforming via $g$, in addition to the elementary 
SM-like fields. These bound states can be resolved only beyond a scale $\Lambda \sim 4 \pi f_\pi \gg m_H$, that 
defines the cutoff of the pNGB model. Since the effect of the gauge resonances is of minor importance for our study, we will
neglect them in the following, see below. Moreover, for our discussion of the Higgs potential only those fermionic 
resonances are important that appear in the breaking of the 
global SO(5) symmetry via large linear mixings to the SM, mediating the masses generated in the composite sector to the SM fields, 
as detailed below. They correspond to leading approximation just to the composite partners of the 
(up-type) third generation quarks $t_R$, $q_L$, as well as of the $\tau_R$. Note that these excitations contain in particular fields 
that are significantly lighter than the general scale of the new resonances, $m_{\rm cust} \lesssim f_\pi \ll m_{\Psi,\rho}$, 
dubbed light custodians, as for the models we consider these modes are present due to custodial symmetry
\cite{Agashe:2003zs,Agashe:2006at,Carena:2006bn,Contino:2006qr, Carena:2007ua, Carena:2007tn, Pomarol:2008bh, Panico:2010is,DeCurtis:2011yx}. 
These fields will be of special importance for the Higgs potential. 
In consequence of the above discussion, in this section we will consider 
an effective (low energy) realization of the composite setup, including only the resonances associated to the third generation top and 
$\tau$ sectors, to study the impact of different incarnations of the fermion sector on the Higgs potential, while the
other resonances are integrated out at zeroth order.\footnote{
We will comment on the effect of partners of lighter fermions in one case where they might become relevant numerically later.}
Finally, note that a generally subleading contribution to the potential still remains present inevitably from weakly gauging just the diagonal SM 
electroweak subgroup $G_{\rm EW}= SU(2)_L \times U(1)_Y$ of the global {\it composite} $SO(5) \times U(1)_X$ and {\it elementary} 
$SU(2)_L^0 \times U(1)_R^0$ groups, which also explicitly breaks the $SO(5)$ symmetry.\footnote{Switching off the SM gauge interactions 
and the linear fermion mixings, the Lagrangian is invariant~under separate global symmetries in the elementary 
and composite sectors. The additional $U(1)_X$\! factor is needed to arrive at the hypercharges of the SM-fermions,
via $Y\!=T_R^3 +\! X$, and we omit $SU(3)_{c}$.

Note that in a full two-site description, including composite gauge resonances, an 
additional $\sigma$-field $\Omega$ breaks the elementary $SO(5)^0$ (with $SU(2)_L^0 \times U(1)_R^0$ gauged) at the first site and 
the composite, completely gauged, $SO(5)$ at the second site to the diagonal subgroup $SO(5)_V$ \cite{DeCurtis:2011yx} 
(see also \cite{Panico:2011pw}).
One linear combination of the bosons corresponding to the $SU(2)_L^0 \times U(1)_R^0$ and $SU(2)_L \times U(1)_R$ 
subgroups remains massless, furnishing $G_{\rm EW}$, while the orthogonal combination and the coset 
$SO(5)/(SU(2)_L \times U(1)_R$) gauge fields acquire a mass at the scale $m_\rho$. 
The $\Sigma$ Goldstone bosons also contributes to the latter masses since $\Sigma$ breaks $SO(5) \to SO(4)$. A linear combination
of $\Omega$ and $\Sigma$ then actually provides the pseudo-Goldstone Higgs, which then delivers the longitudinal degrees of
freedom for the massless SM-like gauge fields in a second step. Since the gauge resonances have only a minor impact
on the Higgs potential (determined by the weak gauge coupling), here we study the limit of a large $\Omega$ 
decay constant for simplicity, which decouples all the heavy gauge resonances and leads to the Higgs sector just originating 
from $\Sigma$.}

Neglecting subleading effects due to the heavy resonances residing at the scale $m_{\rho,\Psi}$ and
not associated to the third generation fermions (and as such in particular irrelevant for our discussion of the Higgs potential),  our setup is thus described by the low-energy Lagrangian
	\begin{equation}
	\begin{split}
	{\cal L} =&\ {\cal L}^{\rm SM}[V^\mu,f]\, + {\cal L}_\Sigma  + {\cal L}_{\rm kin}^\Psi \\
	& + {\cal L}_{\rm mass}^\Psi - V(\Pi)\,.
	\end{split}
	\end{equation}
Here, the $\sigma$-model term ${\cal L}_\Sigma=\frac 1 2 \left(D_\mu \Sigma\right)^T D^\mu \Sigma = 
\frac {f_\pi^2}{2} \left(D_\mu U_{I5}\right)^T 
D^\mu U_{I5}$ contains the couplings of the composite Higgs to the SM gauge fields via the covariant derivative $D_\mu=
\partial_\mu -i g^\prime\, Y B_\mu -i g\, T^i W^i_\mu$, with $g$ and $g^\prime$ the SM gauge couplings.
 Note that the non-linearity of the Higgs sector, see  (\ref{eq:NLS}), induces a shift in the couplings of order $v^2/f_\pi^2$ 
with respect to the SM.
Moreover, ${\cal L}_{\rm kin}^\Psi = \sum_{f=T,t,{\cal T},\tau} {\rm Tr}[\bar \Psi^f \gamma^\mu D_\mu  \Psi^f]$ are 
the kinetic terms of the composite fermions, each associated to a chiral SM fermion, see below, and $ {\cal L}^{\rm SM}[V^\mu,f]$ encodes the spin 1 and spin 1/2 part of the SM Lagrangian containing only vector and/or fermion fields, {\it i.e.}, the 
field strength tensors and the covariant derivatives between SM-fermion bilinears (plus gauge fixing and ghost terms). 

The most relevant terms for our following discussion appear in the second row. Of particular importance is the fermion mass 
and mixing Lagrangian
	\begin{eqnarray}
	\label{eq:Lf1}
	 {\cal L}_{\rm mass}^\Psi & = &
	 {\rm Tr}\Big{\{}- \sum_{f,f^\prime=T,t}  m_\Psi^{f f^\prime}\, \bar \Psi_L^f \Psi_R^{f^\prime}  
	- f_\pi \sum_{f,f^\prime=T,t \atop i} Y^{f f^\prime}_i \, 
	\bar \Psi_L^f \cdot g^{f f^\prime}_i(\Sigma/f_\pi) \cdot \Psi_R^{f^\prime}\\
	& & \ \  -\, y_L^t f_\pi\  \bar q_L\, \Delta_L^t \Psi_R^T\, -\, y_R^t f_\pi\ \bar t_R\, \Delta_R^t 
	\Psi^t_L \Big{\}}+ {\rm h.c.}\, + (T \to {\cal T},\,t \to \tau,\,q \to \ell)\,, \nonumber
	\end{eqnarray}
where $m_\Psi^{ff^\prime}$ are the vector-like masses (and mass mixings) of the fermion resonances and the symbols
$\bar \Psi_L^f \cdot g^{f f^\prime}_i(\Sigma/f_\pi) \cdot \Psi_R^{f^\prime},
\,i=1,\dots,n$ denote all $SO(5)$ invariant combinations
that can be formed out of the bilinear $\bar \Psi_L^f\, \Psi_R^{f^\prime}$ and non-trivial functions of $\Sigma$ as well as, 
possibly, traces. The form of these Yukawa couplings in the strong sector, with coefficients $Y^{f f^\prime}_i$, depends 
on the particular representation chosen for the composite fermions.
Embedding for example all fermionic resonances in fundamental representations of $SO(5)$, as in the MCHM$_5$, we
obtain
	\begin{equation}
	\bar \Psi_L^f \cdot g^{f f^\prime}_1(\Sigma/f_\pi) \cdot \Psi_R^{f^\prime} \xrightarrow{\text{MCHM$_5$}}\
	\bar \Psi_L^f  \frac{\Sigma \Sigma^T}{f_\pi^2} \Psi_R^{f^\prime} \quad (n=1)\,,
	\end{equation}
and the global trace becomes trivial.

At this point, some comments are in order. The Lagrangian (\ref{eq:Lf1}) is constructed in the most general way that respects 
the global $SO(5)$ symmetry, up to linear mixing terms of the SM-like fermions with the composite resonances, where 
the former transform under $SU(2)_L^0 \times U(1)_R^0$ and not $SO(5)$. These elementary-composite mixings are parametrized 
by $y^{t,\tau}_{L,R}$, where the coupling matrices $\Delta^{t,\tau}_{L,R}$ are fixed by gauge invariance under 
$G_{\rm EW}$, {\it i.e.}, they couple appropriate linear combinations of the components of the composite operators
with definite charges under the SM gauge group to the corresponding SM fields, see below. After rotating to 
the mass basis only these bilinear terms induce masses for the SM-like fermions, which now feature composite components 
proportional to $y^{t,\tau}_{L,R}$, realizing the concept of partial compositeness \cite{Kaplan:1991dc,Contino:2003ve, Agashe:2004rs}. 
The chiral fermion masses are finally proportional to these mixings as well as to the Yukawa coupling in the strong sector,
mediating the transition between different components of the $SO(5)$ multiplets, $m_t \sim |y_L^t\, f_\pi/m_\Psi^{TT}\, Y_i^{Tt} v\, f_\pi/m_\Psi^{tt}\, y_R^t|$, and analogously for the lepton sector. 
Note that in the holographic 5D realization of this setup (see Section~\ref{sec:5D}), only the off-diagonal Yukawa couplings
$Y_i^{T t}$ are non zero, while the mass-mixing $m_\Psi^{tT}$ vanishes due to boundary conditions. 
The same is true in the corresponding deconstructed 2-site models, like studied here, if one requires finiteness of the Higgs potential~\cite{DeCurtis:2011yx,Carena:2014ria}.

As mentioned, the Lagrangian contains two vector-like resonances $\Psi^{T,t}\ (\Psi^{{\cal T},\tau})$ in each sector, associated to the two chiralities of SM fermions. In the (broken) conformal field theory picture, which turns out to be indeed dual to a 5D (GHU) picture \cite{Maldacena:1997re,Gubser:1998bc,Witten:1998qj,ArkaniHamed:2000ds,Contino:2003ve,Falkowski:2006vi}, 
the linear-mixing parameters $y^{t,\tau}_{L,R}$ just correspond to the anomalous dimensions of the composite operators that 
excite the resonances that the SM-like fermions couple to \cite{Contino:2004vy,Batell:2007jv,Batell:2007ez,Gherghetta:2010cj}.
Following this line of reasoning, in general each elementary chiral fermion 
couples to its own resonance (the Lagrangian described above can always be brought to such a basis).
Note that if the $q_L$ and the $t_R$ mix with composites that belong to different representations of $SO(5)$, the vector-masses
mixing those composites obviously vanish, $m_\Psi^{f f^\prime}=0$ for $f\neq f^\prime$. Moreover, we have neglected the allowed
terms $\sim |y_{L,R}^t| f_\pi\, \bar q_{L,R}\, \Delta_{L,R}^t\cdot  
g_i(\Sigma/f_\pi) \cdot \Psi_{R,L}^{T,t}$ in (\ref{eq:Lf1}), which will deliver no
new structure to the low energy theory we will be considering further below. They would however be needed in general if both chiral 
fermions would mix with the same single composite field.  Finally, the Lagrangian above 
can straightforwardly be generalized to the case of the SM fields mixing with more than one representation, each.

As we have explained, we do not consider the $b_R$, since it is expected to deliver a 
negligible contribution to the Higgs potential due to its small mixing with the composite sector.
Note in that context that large changes in the diagonal Higgs couplings of the (light) fermions would
generically also manifest themselves in large FCNCs \cite{Goertz:2014qia}. 
Note also that, in all models considered besides those mixing the right-handed sector with a {\bf 10} of $SO(5)$, 
the non-vanishing mass of the bottom quark would rely on an additional subleading mixing of $q_L$ with an appropriate 
multiplet $\Psi^B$ of the strong sector that features the correct $X$ charge to mix with the composite that couples 
with $b_R$. This term $\propto y^b_L$ has also been neglected, as it is again controlled by the rather small bottom mass 
 - its smallness also helps to protect the $Z b_L b_L$
coupling which would receive unprotected corrections from this second mixing of $q_L$.\footnote{Only in the {\bf 10} 
it is possible to mix both top and bottom sectors with composites with the same $U(1)_X$ charge, due to the presence 
of several $SU(2)_L$ singlets with different $T_R^3$, resulting in different $U(1)_Y$ charges. This allows to generate 
masses for both $t$ and $b$ from a composite sector with a single $U(1)_X$ charge, where the left handed mixing is 
parametrized by a common term $y_L$.}
A similar discussion holds for the neutrino sector, which in the case of Dirac neutrinos is completely analogous. For 
Majorana neutrinos, $y_R^\nu$ could be non negligible,  it's impact on the Higgs potential is however suppressed
by a very large scale. The $\tau_R$ on the other hand is expected to exhibit a sizable composite component, 
as explained before, which can have an impact on the Higgs potential and thus we include the $\tau$ sector.\footnote{
Remember that in our approximation we also neglect the masses and Higgs couplings of the lighter SM fermions. Those 
terms would analogously be induced via mixings with their heavy composite partners at $m_\Psi$, however 
these are in general strongly suppressed with the small fermion masses and thus lead to a negligible 
contribution to the Higgs potential. 
The full 4D Lagrangian would include the heavy fermionic resonances sharing 
the flavor quantum numbers of the first two fermion generations.
For a 4D setup including a comprehensive description of fermion and gauge resonances see e.g. \cite{Panico:2011pw,DeCurtis:2011yx}.}

To be more explicit, we now specify explicitly the representation of $SO(5)$ in which the composite fermions are embedded. In 
the following we will start with the fundamental representation {\bf 5} for concreteness, {\it i.e.}, the MCHM$_5$, however the 
discussion can easily be generalized to larger representations, which will be considered in the course of our study. Following 
the CCWZ approach \cite{Coleman:1969sm,Callan:1969sn},  
we can write the Lagrangian (\ref{eq:Lf1}) in terms of representations of the unbroken global $SO(4)$ symmetry, while keeping
it SO(5) symmetric. To that extend we decompose the {\bf 5} of $SO(5)$ into components $Q$ and $\tilde T$ that transform 
via $\hat h$ (see (\ref{eq:Utraf})) as a fourplet and a singlet under the unbroken $SO(4)$ symmetry. Writing 
\begin{equation}
	\begin{split}
	\Psi^T &= U (Q^T,\,
	\tilde T^T)^T,\,\Psi^t = U (Q^t,\,\tilde T^t )^T,\\[1.5mm]
	\Psi^{\cal T} &= U (L^{\cal T},\,\tilde {\cal T}^{\cal T})^T,\, 
	\Psi^\tau = U (L^\tau,\,\tilde {\cal T}^\tau )^T\,,
\end{split}
\end{equation} 
we arrive at
	\begin{eqnarray}
	\label{eq:L0mass}
	 {\cal L}_{\rm mass}^{\text{\tiny MCHM}_5\,\prime}  = & -&\! \sum_{f,f^\prime=T,t} 
	\left( m_{f f^\prime}\, \bar Q_L^f Q_R^{f^\prime} 
	+ \tilde m_{f  f^\prime}\, \bar{\tilde T}_L^f {\tilde T}_R^{f^\prime} \right)
	-\,y^t_L f_\pi\,  (\bar q_L \Delta^t_L)_I \left( U_{Ii}\, {Q_R^T}^i +  U_{I5}\, \tilde T_R^T \right) \nonumber \\
	& -&\,y^t_R f_\pi\,  (\bar t_R \Delta^t_R)_I \left( U_{Ii}\, {Q_L^t}^i + U_{I5}\, \tilde T_L^t \right)
	+{\rm h.c.}\\[1.8mm]
	& + & (t \to \tau,\,T \to {\cal T},\,q \to \ell,\,Q \to L) \,, \nonumber
	\end{eqnarray}
where $I=1,...,5,\, i= 1,...,4$, and the composite-Yukawa parameters $Y_1^{f f^\prime}$ have 
been absorbed in the vector-like mass of the composite singlets 
$ \tilde m_{f f^\prime} \equiv m_\Psi^{f f^\prime}+ Y_1^{f f^\prime} f_\pi$, and $ m_{f f^\prime}= m_\Psi^{f f^\prime}$.

We should note that while including both first layers of resonances, $\Psi^T$ and $\Psi^t$,
at the first place was necessary to see how the 4D fields can be identified with different bulk fields of AdS$_5$ (CFT) constructions of the composite Higgs idea, it is now convenient to remove the explicit appearance of one set of fields by going to a 
further effective low energy description, where now their effect is kept via new interactions.
By integrating out $Q^T$ and $\tilde T^t$ we arrive at a theory that can in particular be matched directly to the 5D theory
examined quantitatively in Section~\ref{sec:5D} (to leading approximation). The setup then describes one-to-one just those modes that are most relevant for the generation of the Higgs potential and the SM-like fermion masses.\footnote{In 
the 5D theory the SM-like fields $q_L, t_R$ will not mix with the resonances corresponding to $Q^T$ and $\tilde T^t$ at leading order, see Section~\ref{sec:5D}.} In this way, all the fields that we keep correspond in particular to light custodians. 

We thus finally arrive at the leading mass-mixing Lagrangian
	\begin{equation}
	\label{eq:mm}
	\begin{split}
	 {\cal L}_{\rm mass}^{\text{\tiny MCHM}_5} =  -\hat m_T \bar Q_L Q_R - \tilde m_T \bar{\tilde T}_L
	{\tilde T}_R &
	 -\,y^t_L f_\pi\,  (\bar q_L \Delta^t_L)_I \left( a_L^t U_{Ii} Q_R^i +  b_L^t U_{I5} \tilde T_R \right) \\
	&  -\,y^t_R f_\pi\,  (\bar t_R \Delta^t_R)_I \left( a_R^t U_{Ii} Q_L^i +  b_R^t U_{I5} \tilde T_L  \right)
	+{\rm h.c.}\\[1.5mm]
	& \hspace{-3.2cm}+(t \to \tau,\,T \to {\cal T},\,q \to \ell,\,Q \to L) \,,
	\end{split}
	\end{equation}
in agreement with \cite{Matsedonskyi:2012ym}, where we have removed the superscripts on 
$Q^t$ and $\tilde T^T$, while $\hat m_T\equiv m_{t t}, \tilde m_T \equiv \tilde m_{ T T}$ , and 
$a_L^t\equiv m_{T t}/m_{T T}\, b_L^t $ arises from integrating 
out $Q^T$ at zero momentum, which leads to a linear interaction between $q_L$ and $Q_R$ via the term $\sim m_{T t}$ in (\ref{eq:L0mass}). 
The ${\cal O}(1)$ coefficient $b_L^t$ ($a_R^t$) has been introduced for convenience \cite{Matsedonskyi:2012ym} by rescaling $y_L^t$ ($y_R^t$). Analogously, 
from integrating out $\tilde T^t$ one obtains $b_R^t\equiv (\tilde m_{T t}/\tilde m_{t t})^\ast\, a_R^t$.
Actually, in this discussion we neglect subleading terms in ratios like $m_{f f^\prime}/m_{f f}$, 
assuming implicitly that the mass-mixings
within the resonances are smaller than the diagonal mass terms. While this might be lifted in some regions of parameter space
(like for large brane masses in Section~\ref{sec:5D}), the qualitative picture will not be changed by neglecting such terms here for
simplicity. Note that in the following we will always work in a basis where the diagonal mass terms as well as the linear mixing parameters $y_{L,R}^t$ are real and positive.

While the Lagrangian (\ref{eq:mm}) can be directly mapped to a 5D theory, it in particular also provides a simple and viable 
complete 4D model itself, employing just one composite vector-like resonance
from the beginning \cite{Matsedonskyi:2012ym,Panico:2011pw}. Adding the extra term $\sim g_1(\Sigma/f_\pi) \sim \Sigma \Sigma^T$, mentioned above, to the linear mixing
in (\ref{eq:Lf1}) (as well as setting $T=t$ and removing the sum over the different composite excitations) one also obtains (\ref{eq:mm}) without two distinct 
vector-like composite fields, however the connection to the 5D setup is then not transparent anymore. As mentioned before, we neglect the additional four 
similar terms with $t\to b\,, T \to B\,, Q \to Q^b$, which are generically negligible for the Higgs potential.  
The Lagrangian (\ref{eq:mm}) will be the starting point for our analysis.

From invariance under $G_{\rm EW}$ one obtains for the fundamental representation of $SO(5)$ 
	\begin{equation}
	\label{eq:spu}
	\begin{split}
	\Delta^t_L=&\frac{1}{\sqrt 2}\begin{pmatrix} 0 & 0 & 1 & -i & 0 \\ 1 & i & 0 & 0 & 0\end{pmatrix}\,, \quad
	\Delta^f_R=- i \begin{pmatrix} 0 & 0 & 0 & 0 & 1 \end{pmatrix}\,, \\[1mm]
	\Delta^\tau_L=&\frac{1}{\sqrt 2}\begin{pmatrix} 1 & -i & 0 & 0 & 0 \\ 0 & 0 & 1 & i & 0\end{pmatrix}\,,
	\end{split}
	\end{equation}
$f=t,\tau$. The difference between the top-quark and $\tau$-lepton sectors arises since the corresponding SM-like 
doublets mix with different $T_R^3$ components of the $SU(2)_L \times SU(2)_R$ bi-doublets $Q^T,L^T$, respectively.
This is dictated by the quantum numbers of the SM fermions, which fix the $U(1)_X$ 
charges of the composites to $X_t=2/3$ and $X_\tau=-1$, for $\Psi^{T,t}$ and $\Psi^{{\cal T},\tau}$. 
Note that the setup includes a protection for the $Z \bar b_L b_L$ and $Z \bar \tau_R \tau_R$ couplings via a custodial
$P_{LR}$ symmetry exchanging $SU(2)_L \leftrightarrow SU(2)_R$ \cite{Agashe:2006at}.

To zeroth order in $v/f$, {\it i.e.}, neglecting EWSB, the Goldstone Matrix
	\begin{equation}
	U\sim {\bf 1}_{5\times 5} + {\cal O}(v/f_\pi, \Pi)
	\end{equation}
only mediates mixings of the SM doublet $q_L$ with $Q_R$ as well as of $t_R$ with $\tilde T_L$. 
Beyond that, the exact relative strength of the interactions of the SM fields with the fourplet and the singlet of $SO(4)$ 
are not fixed by symmetry so far but rather parametrized by the ratios of ${\cal O}(1)$ numbers $a_{L,R}^t/b_{L,R}^t$. 
In the same approximation, one obtains simple analytic formulas for the masses of the heavy resonances after diagonalizing the 
fermionic mass matrix. Employing, with some abuse of notation, the same names for the mass eigenstates as for 
the interaction eigenstates (which coincide in the limit $y_{L,R} \to 0$) one obtains for the masses of the physical
heavy resonances
	\begin{equation}
	m_T\simeq \sqrt{(\hat m_T)^2+|a_L^t|^2 y^{t\,2}_L\,  f_\pi^2}\, , \quad m_{\tilde T} \simeq \sqrt{(\tilde m_T )^2
	+|b_R^t|^2 y^{t\,2}_R\,  f_\pi^2}\,.
	\end{equation}

Expanding to the first non-trivial order in $v/f_\pi$ one arrives similarly at the top-quark mass
	\begin{equation}
	\label{eq:mt}
	m_t=\frac{|a_L^t a_R^t \tilde m_T - b_L^t b_R^t \hat m_T |}{2 \sqrt 2}\sin \varphi^t_L \sin \varphi^t_R\, 
	\sin \left(2 v/f_\pi\right)\,,
	\end{equation}
where the rotation angles
	\begin{equation}
	\sin \varphi^t_L=\frac{y^t_L f_\pi}{m_T} \,, \quad \sin \varphi^t_R=\frac{y^t_R f_\pi}{m_{\tilde T}}\,,
	\end{equation}
describe the admixture of the composite modes into the light SM-like top quark.
Inspecting these equations, one can thus see explicitly the mechanism of partial compositeness at work - the mass of a 
SM-like fermion is proportional to its mixings $\sin \varphi^t_{L,R}$ with the strong sector. Since after EWSB
this mixing involves fields with different quantum numbers under $G_{\rm SM}$, it is a source of tree-level FCNCs
via non-universal gauge interactions in the flavor basis.
This mechanism nicely matches with the experimental observation that FCNCs are more constrained for the light 
generations. Note that the same discussion as above holds for leptons with the aforementioned replacements of indices.
Finally, employing $a_L^t\sim a_R^t \sim b_L^t\sim b_R^t \sim 1$, we obtain to leading approximation
	\begin{equation}
	\label{eq:mtsim}
	m_t \sim \frac{y_L^t y_R^t f_\pi}{\sqrt 2 \min(m_T,m_{\tilde T})} v\,.
	\end{equation}

\subsection{The Higgs Potential and Light Partners}
\label{sec:Vht}
While the Higgs potential $ V(\Pi)$ is zero at tree level, it receives non-vanishing contributions at one loop from the 
explicit breaking of $SO(5)$, as discussed before. Since the gauge boson contributions are subleading (being proportional
to the weak gauge coupling) with no qualitative
effect on the mechanism of EWSB and in particular do not contribute significantly to the Higgs mass, we will neglect them for the discussion of this section. We will however take into account their full numerical impact, including the heavy resonances, 
in Section~\ref{sec:5D}.

Making use of symmetry arguments, we can already deduce the form of the potential directly from the mass-mixing
Lagrangian, {\it i.e.}, (\ref{eq:mm}) for the MCHM$_5$. To that extent, we restore the approximate SO(5) global 
symmetry formally by promoting the coupling matrices $\Delta^f_{L,R}$
to spurions $\hat \Delta^f_{L,R}$, transforming under the global $SO(5)$ of the strong sector in the same way the
corresponding resonances do. If the $\hat \Delta^f_{L,R}$ transform in addition appropriately under the elementary 
symmetry group $SU(2)_L^0 \times U(1)_R^0$ of the $q,t,l,\tau$ fermions, also the linear mixings are invariant under 
the full global symmetry of the rest of the Lagrangian.
As a consequence, also the Higgs potential needs to formally respect the $SO(5)$ symmetry (and the elementary 
symmetry), which should then be broken by the vevs of the spurions $\langle \hat \Delta^f_{L,R} \rangle = 
\Delta^f_{L,R}$ in order to generate a non-trivial potential. Thus the form of the Higgs potential can be constructed
by forming all possible invariants under the full global symmetry, containing at least one spurion $\Delta^f_{L,R}$,
set to its vev, and the Goldstone-Higgs matrix $U$. As the spurions are always accompanied by the linear mixing 
parameters $y^f_{L,R}$, taking the role of an expansion parameter, a series in powers of the spurions can be 
established. In order for the elementary $SU(2)_L^0 \times U(1)_R^0$ symmetry to be respected, the spurions can only 
enter in the combinations ${\Delta^f_L}^\dagger {\Delta^f_L}$ and ${\Delta^f_R}^\dagger {\Delta^f_R}$.
In the following we will first focus on the quark sector, whereas the impact of leptons will be analyzed in the next subsection.

\paragraph{MCHM$_5$}
For the discussed case of $q_L,t_R$ both mixing with a {\bf 5} of SO(5), the form of the potential at ${\cal O}(\Delta^2)$
is thus fixed to
	\begin{equation}
	\label{eq:V2}
	V_2^{(5)}(h)= \frac{N_c m_\Psi^4}{16 \pi^2}\left[\frac{y_L^{t\,2}}{g_\Psi^2}\, c_L^t\, v^{(5)}_L(h)
	+  \frac{y_R^{t\,2}}{g_\Psi^2}\, c_R^t\,  v^{(5)}_R(h) \right]\,,
	\end{equation}
where the prefactors follow from naive dimensional analysis and the fact that the considered fermions
enter in $N_c=3$ colors. The concrete values for the coefficients $c_{L,R}$, which are generically of ${\cal O}(1)$,  need to be determined from an explicit calculation and cannot be fixed by symmetries alone. 
Nevertheless, the $SO(5)$ symmetry already tells us that the Higgs field can only enter in two structures at this order
	\begin{equation}
	\label{eq:invs}
	\begin{split}
	v^{(5)}_L(h)&=\left(U^T {\Delta^t_L}^\dagger {\Delta^t_L} U\right)_{55}  = \frac 1 2 \sin^2(h/f_\pi)\, ,\\
	v^{(5)}_R(h)&=\left(U^T {\Delta^t_R}^\dagger {\Delta^t_R} U\right)_{55} =  \cos^2(h/f_\pi)=1-\sin^2(h/f_\pi)\, ,
	\end{split}
	\end{equation}
where we have employed (\ref{eq:spu}) and the explicit form of the Goldstone matrix as given in Appendix \ref{sec:gen}.
We inspect that, since the constant term in the second line can be neglected in the Higgs potential, only one functional
dependence on the Higgs field is present. The combinations of spurions exhibit a block-diagonal structure and do not mix
the fifth component of $U_{I5}=\Sigma/f_\pi$ with the other four components.
In consequence (dropping a constant), we get
	\begin{equation}
	\label{eq:V22}
	V_2^{(5)}(h) \cong   \frac{N_c m_\Psi^4}{16 \pi^2 g_\psi^2}\left[c_L^t \frac{y_L^{t\,2}}{2} - c_R^t\,
	y_R^{t\,2}\right] \sin^2(h/f_\pi)\,.
	\end{equation}

This leading contribution to the potential does however not yet lead to a viable phenomenology. Its minimum is 
realized for $h=0,f_\pi \frac \pi 2 ,...$\,, which means that we can not have a realistic symmetry breaking with $0<v<f_\pi$.
To fix this problem we need to take into account formally subleading contributions \cite{Matsedonskyi:2012ym,DeCurtis:2011yx,Pomarol:2012qf}. While no new independent $SO(5)$ invariant structures appear at 
${\cal O}(\Delta^4)$, one can have products of the structures (\ref{eq:invs}) which lead to a different trigonometric 
dependence on $h$,
	\begin{eqnarray}
	\label{eq:V45}
	V_4^{(5)}(h) & = &\, \frac{N_c m_\Psi^4}{16 \pi^2}\left[\frac{y_L^{t\,4}}{g_\Psi^4}\, c_{LL}^t\, 
	[v^{(5)}_L(h)]^2
	+  \frac{y_R^{t\,4}}{g_\Psi^4}\, c_{RR}^t\,  [v^{(5)}_R(h)]^2
	+ \frac{y_L^{t\,2} y_R^{t\,2}}{g_\Psi^4}\, c_{LR}^t\, v^{(5)}_L(h)\, v^{(5)}_R(h) \right]\nonumber \\
	&\cong&\, \frac{N_c m_\Psi^4}{16 \pi^2 g_\Psi^4}\left[\left(c_{LL}^t\frac{y_L^{t\,4}}{4}
	- c_{RR}^t y_R^{t\,4} \right) \sin^2(h/f_\pi)\right.\\
	& & \qquad \quad \ \left.-\left(c_{LL}^t\frac{y_L^{t\,4}}{4} + c_{RR}^ty_R^{t\,4}
	-c_{LR}^t\frac{y_L^{t\,2} y_R^{t\,2}}{2} \right)  \sin^2(h/f_\pi)  \cos^2(h/f_\pi) \right]\, \nonumber .
	\end{eqnarray}
However, since usually the elementary-composite mixings are significantly smaller than the strong coupling (see (\ref{eq:mt})),
$y_{L,R}^t \ll g_\Psi$,
the fact that $V_4^{(5)}(h)$ needs to have a comparable impact to $V_2^{(5)}(h)$ worsens the tuning of the model.\footnote{
Note that we indeed consider a strongly coupled new physics sector, $g_\Psi \gg 1$. Leaving this class of models, the considerations
regarding the tuning could be altered.} 

In particular, defining generally
	\begin{equation}
	\label{eq:Potc}
	V(h)=V_2(h)+V_4(h) \equiv \alpha  \sin^2(h/f_\pi) - \beta  \sin^2(h/f_\pi)\cos^2(h/f_\pi)\,,
	\end{equation}
we naturally obtain $\alpha \sim y^{t\,2}_{L,R}/g_\Psi^2 $ and $\beta \sim y^{t\,2}_{L,R} y^{t\,2}_{L,R}/g_\Psi^4$
in the MCHM$_5$. In order to allow for a viable EWSB, the leading contribution to $\alpha$, originating from $V_2^{(5)}(h)$, 
needs to feature a tuning within its contributions that brings it from its natural size of ${\cal O}(y^{t\,2}_{L,R}/g_\Psi^2)$
down to ${\cal O}(y^{t\,4}_{L,R}/g_\Psi^4)$.\footnote{\label{fn:aLaR5}
From (\ref{eq:V22}) we see that for the MCHM$_5$ realistic EWSB 
thus requires $c_L^t\, y_L^{t\,2} = 2c_R^t\, y_R^{t\,2} (1 + {\cal O}(\frac{y_{L,R}^2}{g_\Psi^2}))$.
In consequence we will denote $y_t \equiv y_R^t \sim y_L^t$. 
}
Explicitly, the (non-trivial) minimum of the potential (\ref{eq:Potc}) occurs at
	\begin{equation}
 	\sin^2(h/f_\pi)=\frac{\beta-\alpha}{2\beta}\,,
	\end{equation}
which requires $\alpha-\beta$ to be as small as 
\begin{equation}
\label{eq:EWSBcon}
\alpha-\beta=- 2 \beta \sin^2(v/f_\pi)
\end{equation}
in order to allow for the sought solution.
Comparing this required size to its natural size of $\alpha - \beta \sim y_t^2/g_\Psi^2$ , we obtain the famous 
``double tuning''  \cite{Matsedonskyi:2012ym}
	\begin{equation}
	\label{eq:DT5}
	\Delta_{(5)}^{-1}\sim  \frac{y_t^4/g_\Psi^4 \sin^2(v/f_\pi)}{y_t^2/g_\Psi^2}=\frac{y_t^2}{g_\Psi^2} \sin^2(v/f_\pi)\,,
	\end{equation}
{\it i.e.}, the coefficients entering $V(h)$ 
need not only to cancel to $\sim \sin^2(v/f_\pi) y_t^2/g_\Psi^2 \ll y_t^2/g_\Psi^2$ (the standard tuning due 
to $v\ll f$), but  in the MCHM$_5$ another tuning in the contributions to $V_2$ is required to make it also one 
order smaller in $y_t^2 / g_\Psi^2$.

Moreover we observe that
	\begin{equation}
	\label{eq:mh5}
	\begin{split}
 	m_H^2&\ =\ \frac{8}{f_\pi^2} \cos^2(v/f_\pi) \sin^2(v/f_\pi) \beta\\[1mm]
	&\hspace{-2mm} \overset{\text{\scriptsize MCHM$_5$}}{=}\ f_\pi^2 \frac{N_c}{2 \pi^2}
	\left( \frac{c^t_{LL}}{4} y_L^{t\,4} - \frac{c^t_{LR}}{2} y_L^{t\,2} y_R^{t\,2} + c^t_{RR} y_R^{t\,4}  \right) 
	\cos^2(v/f_\pi) \sin^2(v/f_\pi)\, ,
	\end{split}
	\end{equation}
and thus
	\begin{equation}
	\label{eq:mH1}
	m_H^{(5)} \sim \sqrt{\frac{3}{2 \pi^2}}\, y_t^2\, v \,.
	\end{equation}
Comparing this with (\ref{eq:mtsim}), one obtains the relation \cite{Matsedonskyi:2012ym}
	\begin{equation}
	\label{eq:mHmt}
	m_H^{(5)} \sim \frac{\sqrt 3}{\pi}\, \frac{\min(m_T,m_{\tilde T})}{f_\pi}\, m_t\, .
	\end{equation}
From this we can see explicitly that generically $m_H^{(5)} \gtrsim m_t$.
Finally, we also can read off directly that a light Higgs boson, as found at the LHC, requires light partners in the MCHM$_5$, 
$\min(m_T,m_{\tilde T}) \ll m_\Psi$. For example for $f_{\pi}=500\,$GeV as well as $m_H\lesssim 110\,$GeV 
(see Section~\ref{sec:mHfpi}) and $m_t \sim 150\,$GeV we obtain
\begin{equation}
\label{eq:mTEst}
\min(m_T,m_{\tilde T}) \lesssim 650\,{\rm GeV}\,,
\end{equation}
which is already in slight conflict with LHC searches \cite{Chatrchyan:2013uxa}. 
In contrast, the general scale of fermionic resonances would only be at $m_\Psi \sim 2$\, TeV (assuming $g_\Psi \sim 4$).
For $f_{\pi}=800\,$GeV this becomes still only $\min(m_T,m_{\tilde T}) \lesssim 1\,{\rm TeV}$, while the rest of the resonances clearly leave the LHC reach,  $m_\Psi > 3\,$TeV. These findings will be confirmed
for explicit 5D realizations of the composite Higgs framework in Section~\ref{sec:5D}.
Raising $f_{\pi}$ even further is problematic in the context of addressing the fine tuning problem, since the tuning increases quadratically with $f_\pi$.

In the following we will first review how the requirement for light partners and/or the double tuning could be changed
by enlarging the embedding of the quark sector of the composite models.
We will then show that, compared to modifications of the quark realizations, taking into account the effects from realistic 
embeddings of the lepton sector allows to significantly raise the masses of the top partners in a more natural way, 
{\it i.e.}, avoiding a significant ad-hoc increase in the complexity of the model with respect to
minimal realizations. We will also see that considering the lepton sector opens the possibility to avoid the ad-hoc tuning 
which generically emerges by modifying the quark embeddings in a way that avoids light partners.

\paragraph{MCHM$_{10}$} Let us start by mixing the fermions with {\bf 10}s of $SO(5)$, {\it i.e.}, the MCHM$_{10}$.
The decomposition into representations of $SO(4)\cong SU(2)_L \times SU(2)_R$ now follows ${\bf 10} = {\bf
(3,1) \oplus (1,3) \oplus (2,2)}$ and reads explicitly
\begin{equation}
	\begin{split}
	\Psi^T &= U\, [\,(T_{(3,1)}^T,T_{(1,3)}^T,Q^T)^A\, T^A]\, U^T,\quad
	\Psi^t = U\, [\,(T_{(3,1)}^t,T_{(1,3)}^t,Q^t)^A\, T^A]\, U^T \, ,\\[1.5mm]
	\Psi^{\cal T} &= U\, [\,(T_{(3,1)}^{\cal T},T_{(1,3)}^{\cal T},Q^{\cal T})^A\, T^A]\, U^T,\quad
	\Psi^\tau = U\, [\,(T_{(3,1)}^\tau,T_{(1,3)}^\tau,Q^\tau)^A\, T^A]\, U^T\, ,  
	\end{split}
\end{equation}
where the $SU(2)_L\ (SU(2)_R)$ triplet is represented by a three-component vector $T_{(3,1)}\ (T_{(1,3)})$,
while the bi-doublet $Q$ has four entries, such that $A=1,...,10$, and the generators $T^A$ are defined in Appendix \ref{sec:gen}.
The abelian charges read again $X_t=2/3$ and $X_\tau=-1$, for $\Psi^{T,t}$ and $\Psi^{{\cal T},\tau}$,
in order to protect the $Z \bar b_L b_L$ and $Z \bar \tau_R \tau_R$ vertices.
In complete analogy to as discussed for the MCHM$_5$, the Higgs potential can be studied to leading approximation by 
considering an effective theory containing only the composite resonances $Q^t, T_{(1,3)}^T, T_{(3,1)}^T$, 
while the other modes will be integrated out.\footnote{Again, they do not couple at leading order to the light SM-like modes in the 
5D picture and the resulting setup corresponds to a viable complete 4D model involving a single 10 
of $SO(5)$.}

Thus, in analogy to (\ref{eq:mm}), we finally obtain the mass-mixing Lagrangian 
\begin{eqnarray}
	 {\cal L}_{\rm mass}^{\text{\tiny MCHM}_{10}} &=&  -\hat m_T \bar Q_L Q_R - \tilde m_T 
	\left( \bar{T}_{(3,1)}{}_L\, T_{(3,1)}{}_R + \bar{T}_{(1,3)}{}_L\,  T_{(1,3)}{}_R\right) \nonumber \\[1.8mm]
	&&-\,y^t_L f_\pi\, {\rm Tr}\Big{\{} (\bar q_L {\bf \Delta}^t_L)\, U \left( a_L^t Q_R^{\hat a} T^{\hat a} 
	+  b_L^t {T_{(1,3)}^a}_R  T_R^a +  b_L^t {T_{(3,1)}^a}_R T_L^a\right) U^T \Big{\}}\nonumber \\
	&&-\,y^t_R f_\pi\,  {\rm Tr}\Big{\{}(\bar t_R {\bf \Delta}^t_R)\, U \left( a_R^t Q_L^{\hat a} T^{\hat a} 
	+  b_R^t  {T_{(1,3)}^a}_L T_R^a +  b_R^t {T_{(3,1)}^a}_L T_L^a  \right)U^T \Big{\}}
	+{\rm h.c.}\nonumber\\[1.9mm]
	&& +(t \to \tau,\,T \to {\cal T},\,q \to \ell,\,Q \to L) \,,
	\end{eqnarray}
where we removed the superscripts on $Q^t,
T_{(1,3)}^T, T_{(3,1)}^T$, while $\hat m_T\equiv m_\Psi^{tt} + Y_1^{t t}/2$, $\tilde m_T \equiv m_\Psi^{T T}$, and $\hat a=1,...,4,\, a=1,...,3$.
The coefficients $a_L^t\equiv (m_\Psi^{Tt} + Y_1^{T t}/2)/( m_\Psi^{TT} + Y_1^{T T}/2)\, b_L^t $ and $b_R^t\equiv 
(m_\Psi^{Tt}/ m_\Psi^{tt})^\ast\, a_R^t$ now arise from integrating out $Q^T, T_{(1,3)}^t, T_{(3,1)}^t$ at zero momentum,
 where $b_L^t$ ($a_R^t$) has again been introduced for convenience by rescaling $y_L^t$ ($y_R^t$).

The spurions that restore the $SO(5)$ symmetry now also transform in adjoint representations and take the vevs
\begin{eqnarray}
 ({\bf \Delta}^t_L)_\alpha & = &
\frac{1}{\sqrt 2}\begin{pmatrix}
0 & 0 & 0 & 0 & 0 & 0 & 0 & 0 & 1 & -i\nonumber \\
0 & 0 & 0 & 0 & 0 & 0 & 1 & i & 0 & 0  
\end{pmatrix}_{\alpha A}
T^A \, , \quad
({\bf \Delta}^f_R) =
\begin{pmatrix}
0 & 0 & 0 & 0 & 0 & 1 & 0 & 0 & 0 & 0
\end{pmatrix}_A
T^A \, ,\\[3mm]
 ({\bf \Delta}^\tau_L)_\alpha & = &
\frac{1}{\sqrt 2}\begin{pmatrix}
0 & 0 & 0 & 0 & 0 & 0  & 1 & -i & 0 & 0\\
0 & 0 & 0 & 0 & 0 & 0 & 0 & 0 & 1 & i 
\end{pmatrix}_{\alpha A}\,,
\end{eqnarray}
where $\alpha=1,2$ are $SU(2)_L$ indices.

In complete analogy to the case of the MCHM$_5$, the spurion analysis leads to the quark contribution
	\begin{equation}
	V_2^{(10)}(h)= \frac{N_c m_\Psi^4}{16 \pi^2}\left[\frac{y_L^{q\, 2}}{g_\Psi^2}\, c_L^q\, v^{(10)}_L(h)
	+  \frac{y_R^{t\,2}}{g_\Psi^2}\, c_R^t\,  v^{(10)}_R(h) \right]\,,
	\end{equation}
where now
	\begin{equation}
	\begin{split}
	v^{(10)}_L(h)&=\left(U^T ({{\bf \Delta}^q_L}^\dagger)^\alpha ({{\bf \Delta}^q_L})_\alpha U\right)_{55}  = \frac 1 2 - 
	\frac 3 8 \sin^2(h/f_\pi)\,,\\
	v^{(10)}_R(h)&=\left(U^T {{\bf \Delta}^t_R}^\dagger {{\bf \Delta}^t_R}\, U\right)_{55} = 	\frac 1 4 \sin^2(h/f_\pi)\, .
	\end{split}
	\end{equation}
With some abuse of notation we do not introduce new names for the other parameters, which can deviate from the case of the MCHM$_5$.
As in the MCHM$_5$, due to the block-diagonal form of both combinations of spurions, there is again only one trigonometric dependence on 
the Higgs field at second order in the spurions. We obtain
	\begin{equation}
	\label{eq:V210}
	V_2^{(10)}(h) \cong   \frac{N_c m_\Psi^4}{16 \pi^2 g_\psi^2}\left[- \frac 3 8 c_L^q y_L^{q\, 2} + \frac 1 4 c_R^t\,
	y_R^{t\, 2}\right] \sin^2(h/f_\pi)\,.
	\end{equation}

Once more, formally subleading contributions to the potential are essential for a viable electroweak symmetry breaking, 
while the above terms need to cancel at ${\cal O}(\frac{y_{L,R}^2}{g_\Psi^2})$.\footnote{
\label{fn:aLaR10} Specifically, one needs $\frac 
3 2 c_L^q\, y_L^{q\, 2} = c_R^t\,y_R^{t\,2} (1 + {\cal O}(\frac{y_{L,R}^2}{g_\Psi^2}))$ and thus again $ y_R^t \sim y_L^q$.}

At ${\cal O}({\bf \Delta}^4)$, new non-trivial $SO(5)$ invariant structures 
	\begin{equation}
	\begin{split}
	& v^{(10)}_{LL\,(k)} \sim (({\bf \Delta}_L)_{ij}^\alpha)^4\, (U_{m5})^{2,4}\\  
	& v^{(10)}_{LR\,(k)} \sim (({\bf \Delta}_L)_{ij}^\alpha)^2 (({\bf \Delta}_R)_{kl})^2\, (U_{m5})^{2,4} \\ 
	& v^{(10)}_{RR\,(k)} \sim (({\bf \Delta}_R)_{ij})^4 \, (U_{m5})^{2,4}
	\end{split}
	\end{equation}
can be formed in the MCHM$_{10}$, by contracting the matrix-indices of the spurions and Goldstone matrices in various 
different ways.\footnote{Note that  in $SO(5)$ invariant combinations the Goldstone matrices only enter as $U_{r5}$ and 
vanish to the identity as $U^T U = {\bf 1}$ in other possible contractions of indices.} 
In the end however, as for the MCHM$_5$, only the trigonometric functions $\sin^2(h/f_\pi)$ and $\sin^2(h/f_\pi)\cos^2(h/f_\pi)$ emerge,
	\begin{equation}
	\label{eq:V410}
	\begin{split}
	V_4^{(10)}(h) = &\, \frac{N_c m_\Psi^4}{16 \pi^2}\sum_k \left[\frac{y_L^{q\, 4}}{g_\Psi^4}\,  c_{LL}^{q\,(k)}\, 
	v^{(10)}_{LL\,(k)}
	+  \frac{y_R^{t\, 4}}{g_\Psi^4}\, c_{RR}^{t\,(k)}\, v^{(10)}_{RR\,(k)}
	+ \frac{y_L^{q\, 2} y_R^{t\, 2}}{g_\Psi^4}\,c_{LR}^{q\,(k)}\, v^{(10)}_{LR\,(k)}\right]\\
	\cong&\, \frac{N_c m_\Psi^4}{16 \pi^2 g_\Psi^4}\left[\left(c_{LL}^q y_L^{q\, 4}
	+ c_{RR}^t y_R^{t\, 4} + c_{LR}^q y_L^{q\, 2}y_R^{t\, 2}\right) \sin^2(h/f_\pi)\right.\\
	& \qquad \quad \ \left.+\left(c_{LL}^{\prime q} y_L^{q\, 4}
	+ c_{RR}^{\prime t} y_R^{t\, 4} + c_{LR}^{\prime q} y_L^{q\, 2}y_R^{t\, 2}\right)   \sin^2(h/f_\pi)  \cos^2(h/f_\pi) \right]\, ,
	\end{split}
	\end{equation}
where $c^{(\prime) f}_{LL,RR,LR}$ are generically ${\cal O}(1)$ coefficients emerging as linear combinations of the coefficients 
$c^{f\,(k)}_{LL,RR,LR}$ of the different $SO(5)$ invariant structures $v^{(10)}_{LL,RR,LR\,(k)}$. 
As a consequence of the above discussion, the tuning is clearly of the same order as in the MCHM$_5$ 
\begin{equation}
\Delta_{(10)}^{-1} \sim\Delta_{(5)}^{-1} \sim \frac{y_t^2}{g_\Psi^2} \sin^2(v/f_\pi)\,,
\end{equation}
and again the small Higgs mass suggests light partners.

\paragraph{MCHM$_{10-5}$, MCHM$_{5-10}$}
Mixing the left chirality with a {\bf 10} and the right with a {\bf 5} (denoted by the first and second subscript on MCHM, respectively),  like $\Psi^T\sim {\bf 10},\, \Psi^t\sim {\bf 5}$ in (\ref{eq:Lf1}), or vice versa, does not change the picture qualitatively. It just corresponds to replacing only one of the $v_A^{(5)}$
in (\ref{eq:V2}) with a $v_A^{(10)},\, A=L,R$. Since we found that both feature the same trigonometric dependence, this leads to 
the same conclusions, up to ${\cal O}(1)$ factors.
Moreover, it is evident from the discussion above that one can not mix the $t_R$ with a singlet of $SO(5)$ only -
not contributing to $V(h)$ - if one wants to achieve a realistic EWSB, given that the left handed top mixes with either a 
{\bf 5} or a {\bf 10}. We conclude that at least one chirality needs to mix with a more complex composite fermion 
realization than a single fundamental or adjoint if we want to see qualitatively new features in the quark sector.

\paragraph{MCHM$_{14}$}
If we embed the fermions in the symmetric representation, a {\bf 14} of $SO(5)$, new features emerge. 
We start with assigning both SM chiralities to a {\bf 14} of the strong sector, however as we will detail below, 
the generic features will remain if only one of them mixes with a symmetric representation.

The decomposition into $SU(2)_L \times SU(2)_R$ representations via ${\bf 14} = {\bf(3,3) \oplus (2,2) 
\oplus (1,1)}$ now reads
\begin{equation}
	\begin{split}
	\Psi^T &= U\, [\,(T_{(3,3)}^T,Q^T,{\tilde T}^T)^A\, \hat T^A]\, U^T,\quad
	\Psi^t = U\, [\,(T_{(3,3)}^t,Q^t,{\tilde T}^t)^A\, \hat T^A]\, U^T \, ,\\[1mm]
	\Psi^{\cal T} &= U\, [\,(T_{(3,3)}^{\cal T},Q^{\cal T},{\tilde T}^{\cal T})^A\, \hat T^A]\, U^T,\quad
	\Psi^\tau = U\, [\,(T_{(3,3)}^{\tau},Q^{\tau},{\tilde T}^{\tau})^A\, \hat T^A]\, U^T\, ,  
	\end{split}
\end{equation}
where the $SU(2)_L\times SU(2)_R$ bi-triplet is represented by the nine-component vector $T_{(3,3)}$,
while the bi-doublet is again denoted by $Q$ and the singlet by ${\tilde T}$, such that now $A=1,...,14$,
see Appendix \ref{sec:gen} for the definition of $\hat T^A$.
The abelian charges read once more $X_t=2/3$ and $X_\tau=-1$, for $\Psi^{T,t}$ and $\Psi^{{\cal T},\tau}$,
respectively, protecting still $Z \bar b_L b_L$ and $Z \bar \tau_R \tau_R$. 
Again, the Higgs potential can be studied conveniently to leading approximation after integrating 
out $Q^T, {\tilde T}^t, T_{(3,3)}^t$, leading again to a viable 4D model of effectively only one composite resonance.\footnote{
Also in the {\bf 14}, these modes do not couple at leading order to the light SM-like fields in the 5D picture.}

Thus, again in analogy to (\ref{eq:mm}), we obtain the final mass-mixing Lagrangian 
\begin{eqnarray}
	 {\cal L}_{\rm mass}^{\text{\tiny MCHM}_{14}} &=&  -\hat m_T \bar Q_L Q_R - \tilde m_T 
	\bar{\tilde T}_L {\tilde T}_R - m_{(3,3)}  \bar{T}_{(3,3)}{}_L T_{(3,3)}{}_R \nonumber \\[1.8mm]
	&&-\,y^t_L f_\pi\,  {\rm Tr}\Big{\{} (\bar q_L {\bf \Delta}^t_L)\, U \left( a_L^t Q_R^{\hat a} \hat T^{\hat a+9} 
	+  b_L^t {\tilde T}_R\,  \hat T^{14} +  b_L^t T{_{(3,3)}^{\beta}}_R \hat T^{\beta}\right) U^T \Big{\}}\nonumber \\
	&&-\,y^t_R f_\pi\,  {\rm Tr}\Big{\{} (\bar t_R {\bf \Delta}^t_R)\, U \left( a_R^t Q_L^{\hat a} \hat T^{\hat a+9} 
	+  b_R^t  {\tilde T}_L\,  \hat T^{14} +  c_R^t T{_{(3,3)}^{\beta}}_L \hat T^{\beta} \right)U^T\Big{\}}
	+{\rm h.c.}\nonumber \\[1.9mm]
	&& +(t \to \tau,\,T \to {\cal T},\,q \to \ell,\,Q \to L) \,,
	\end{eqnarray}
where we removed the superscripts on $Q^t, {\tilde T}^T, T_{(3,3)}^T$, while $\hat m_T\equiv m_\Psi^{tt} +
 Y_1^{t t}/2$, $\tilde m_T  \equiv m_\Psi^{T T} + 4Y_1^{T T}/5 + 4Y_2^{T T}/5$, $m_{(3,3)}
\equiv m_\Psi^{T T}$, and $\hat a=1,...,4,\, \beta=1,...,9$.
The coefficients $a_L^t\equiv (m_\Psi^{Tt} + Y_1^{T t}/2)/( m_\Psi^{TT} + Y_1^{T T}/2)\, b_L^t $, $b_R^t\equiv 
((m_\Psi^{Tt} + 4Y_1^{Tt}/5+ 4Y_2^{Tt}/5)/(m_\Psi^{tt}+ 4Y_1^{t t}/5+ 4Y_2^{t t}/5))^\ast\, a_R^t$ and
$c_R^t\equiv (m_\Psi^{Tt}/ m_\Psi^{tt})^\ast\, a_R^t $ now arise from integrating out $Q^T, {\tilde T}^t, T_{(3,3)}^t$ 
at zero momentum, where $b_L^t$ ($a_R^t$) has again been introduced for convenience by rescaling 
$y_L^t$ ($y_R^t$).

The spurions that restore the $SO(5)$ symmetry now transform in symmetric representations of $SO(5)$
and acquire the vevs
	\begin{equation}
	\begin{split}
	 ({\bf \it \Delta}^t_L)_\alpha & =
	\frac{1}{\sqrt 2}
	\left( 
	\begin{array}{cccccccccccccc}
	0 & 0 & 0 & 0 & 0 & 0 & 0 & 0 & 0 & 0 & 0 & 1 & -i & 0 \\
	0 & 0 & 0 & 0 & 0 & 0 & 0 & 0 & 0 & 1 & i & 0 & 0 & 0  
	\end{array}
	\right)_{\alpha A}
	\hat T^A \, , \\
	 ({\bf \it \Delta}^\tau_L)_\alpha & =
	\frac{1}{\sqrt 2}
	\left( 
	\begin{array}{cccccccccccccc}
	0 & 0 & 0 & 0 & 0 & 0 & 0 & 0 & 0 & 1 & -i & 0 & 0 & 0 \\
	0 & 0 & 0 & 0 & 0 & 0 & 0 & 0 & 0 &  0 & 0 & 1 & i & 0  
	\end{array}
	\right)_{\alpha A}
	\hat T^A \, , \\[2mm]
	 ({\bf \it \Delta}^f_R) & =
	\left( 
	\begin{array}{cccccccccccccc}
	0 & 0 & 0 & 0 & 0 & 0 & 0 & 0 & 0 & 0 & 0 & 0 & 0 & i
	\end{array}
	\right)_A
	\hat T^A \, ,
	\end{split}
	\end{equation}
where $\alpha=1,2$ are $SU(2)_L$ indices.

In contrast to the MCHM$_5$ and MCHM$_{10}$, the spurion analysis shows that now already at ${\cal O}({\bf \it \Delta}^2)$ 
two different trigonometric structures arise, which has interesting consequences for the quark \cite{Panico:2012uw} and in 
particular the lepton sector. Focusing first on the former, we find
	\begin{equation}
	V_2^{(14)}(h)= \frac{N_c m_\Psi^4}{16 \pi^2}\left[\frac{y_L^{t\, 2}}{g_\Psi^2}\, (c_L^{t\, (1)} \, v^{(14)}_{L\,(1)}
	+ c_L^{t\, (2)} \, v^{(14)}_{L\,(2)})
	+  \frac{y_R^{t\, 2}}{g_\Psi^2}\,  (c_R^{t\, (1)} \, v^{(14)}_{R\,(1)} + c_R^{t\, (2)} \, v^{(14)}_{R\,(2)}) \right]\,,
	\end{equation}
where now
	\begin{equation}
	\begin{split}
	v^{(14)}_{L\, (1)}&=\left(U^T ({{\bf \it \Delta}^t_L}^\dagger)^\alpha ({{\bf \it \Delta}^t_L})_\alpha U\right)_{55}  
	= 1- \frac 3 4 \sin^2(h/f_\pi)\\
	v^{(14)}_{L\, (2)}&=\left(U^T ({{\bf \it \Delta}^t_L}^\dagger)^\alpha\, U\right)_{55} 
	\Big(U^T({{\bf \it \Delta}^t_L})_\alpha U\Big)_{55}  = \cos^2(h/f_\pi) \sin^2(h/f_\pi)\\
	v^{(14)}_{R\, (1)}&=\left(U^T {{\bf \it \Delta}^t_R}^\dagger {{\bf \it \Delta}^t_R}\, U\right)_{55}  = \frac 4 5 - 
	\frac 3 4 \sin^2(h/f_\pi)\\
	v^{(14)}_{R\, (2)}&=\left(U^T {{\bf \it \Delta}^t_R}^\dagger\, U\right)_{55} 
	\Big(U^T{{\bf \it \Delta}^t_R}\, U\Big)_{55}  = \frac 4 5 - 
	\frac 1 4 (3+ 5 \cos^2(h/f_\pi)) \sin^2(h/f_\pi)\,.
	\end{split}
	\end{equation}
The symmetric representation has the crucial feature that the combinations of the type $\left(U^T {\bf \it \Delta}\,
U\right)_{55}$, which vanished before, now deliver a finite result. On the one hand, these allow to mix $U_{5i},i\!=\!1,\dots,4
\sim \sin(h/f_\pi)$ with $U_{55} \sim \cos(h/f_\pi)$ via a single insertion of $({{\bf \it \Delta}^t_L})_\alpha$, delivering 
directly the new $\left(\sin(h/f_\pi) \cos(h/f_\pi)\right)^2$ invariant, on the other hand they provide single trigonometric
functions to the {\it fourth} power via only two insertions of ${\bf \it \Delta}^t_R$, resulting in contributions to the 
same invariant.

As a consequence, we have
	\begin{equation}
	\label{eq:V2_14}
	\begin{split}
	V_2^{(14)}(h) \cong  \frac{N_c m_\Psi^4}{16 \pi^2 g_\psi^2} \left[- \frac 3 4 (c_L^{t\, (1)} 
	y_L^{t\,2} + (c_R^{t\, (1)}+c_R^{t\, (2)}) y_R^{t\,2} ) \right] \sin^2(h/f_\pi) \\
	 -\, \frac{N_c m_\Psi^4}{16 \pi^2 g_\psi^2} \left[- c_L^{t\, (2)} y_L^{t\, 2} 
	+ \frac 5 4 c_R^{t\, (2)} y_R^{\, t}  \right] \sin^2(h/f_\pi) \cos^2(h/f_\pi)
	\end{split}
	\end{equation}
and thus can accommodate a realistic EWSB just with $V_2^{(14)}(h)$ -  in principle without an additional tuning.
The coefficients $\alpha$ and $\beta$ (see (\ref{eq:Potc})) are both generated at ${\cal O}({\bf \it \Delta}^2)$ and so are generically 
of the same order. In consequence (\ref{eq:EWSBcon}) can be solved in a nontrivial way without relying on formally
subleading contributions and in particular for various hierarchies between $y_L^{t}$ and $y_R^{t}$, keeping still 
$c_{L,R}^{t\,(1),(2)} \sim {\cal O}(1)$.\footnote{
The latter fact will be interesting for constructing minimal models featuring a {\bf 14}, since essentially only one 
chirality is needed for a viable EWSB, see below.} Since we do not 
anymore need to artificially cancel one order in $y_{L,R}^{t\,2}/g_\Psi^2$ in $\alpha$, 
the formal tuning in EWSB is reduced to the minimal amount of
	\begin{equation}
	 \Delta_{(14)}^{-1} \sim \sin^2(v/f_\pi)\,.
	\end{equation}
In contrast to the MCHM$_{5,10}$, ${\cal O}(1)$ changes in the parameters on the left hand side of (\ref{eq:EWSBcon})
induce only moderately large changes in $v/f_\pi$, while the space of viable solutions is not directly left.

For the Higgs mass we obtain (see (\ref{eq:mh5}))
	\begin{equation}
	\label{eq:mh14_1}
 	m_H^2 \overset{\text{MCHM$_{14}$}}{=}\ m_\Psi^2 \frac{N_c}{2 \pi^2}
	\left( - c_L^{t\, (2)} y_L^{t\,2} 
	+ \frac 5 4 c_R^{t\, (2)} y_R^{t\,2} \right) 
	\cos^2(v/f_\pi) \sin^2(v/f_\pi)\, ,
	\end{equation}
and thus in general
	\begin{equation}
	\label{eq:mh14_2}
	m_H^{(14)} \sim \sqrt{\frac{3}{2 \pi^2}}\, \sqrt{\left|\frac 5 4 y_R^{t\, 2} - y_L^{t\, 2}\right|} 
	\frac{m_\Psi}{f_\pi} v\, \sim \frac{m_\Psi}{f_\pi} v\, ,
	\end{equation}
where the last similarity holds if at least one of the $y_{L,R}^t \sim {\cal O}(1)$, as expected due to the
large $m_t$, and no unnatural cancellation is happening. We observe that the Higgs boson is in general significantly 
heavier than the electroweak scale in this model.
In particular, it is heavier by a factor of $\sim m_\Psi/f_\pi = g_\Psi$ with respect
to the MCHM$_5$ and the MCHM$_{10}$, see (\ref{eq:mH1}).
Remember that $m_\Psi$ is the generic scale of the (heavy) fermionic resonances and in general not the one of the potentially
light partners. Finally, in the MCHM$_{14}$ light partners can not help to reduce $m_H$ up to the experimental value,
since even such partners can not allow $y^t_{L,R}$ to become very much smaller than one - then the top mass
could not be reproduced any longer. The general scale (\ref{eq:mh14_2}) is just too large \cite{Panico:2012uw}.

In the end, while in the MCHM$_{5,10}$  (in the MCHM$_{5,10}$ with light top partners) the Higgs is naturally 
expected to reside at (slightly below) the electroweak scale, $m_H \sim v\ (m_H < v)$, in the MCHM$_{14}$ as 
discussed above a significant {\it ad hoc} tuning of in general unrelated parameters in (\ref{eq:mh14_1}) is needed in order
to push the Higgs mass to $m_H \lesssim v$. Note in particular that while EWSB is a necessary condition for the universe being able
to host human beings, a heavier Higgs scenario could in principle be as viable as the light-Higgs one, which justifies to
consider the tuning in the MCHM$_{14}$ really as ad hoc, in comparison with the one in the MCHM$_{5,10}$. Although, due to the 
particular numerical value of the Higgs mass, in the end the total tuning in both classes of models 
turns out to be not too different (see Section~\ref{sec:5D}), for the MCHM$_{14}$ a light Higgs boson as found seems more~{\it 
unnatural}.\footnote{Note also that the tuning in the MCHM$_{14}$ has to be present {\it both} in $\beta$ and $\alpha$ and that
light partners can only reduce it marginally by allowing for somewhat smaller $y_{L,R}^t$.}

On the other hand, as we will see in particular in the numerical analysis, in the MCHM$_{14}$ one has indeed enough 
freedom to tune the parameters such as to accommodate a light Higgs mass {\it without} necessarily light top partners. 
This tuning also does not
need to spoil the power counting in $y^2/g_\Psi^2$.
In that context, remember that the other option of realizing a light Higgs with the help of light top partners in the MCHM$_{5,10}$ 
is already under pressure from LHC searches and could be excluded soon. As a consequence, taking only into account the 
quark sector, involving a {\bf 14} with a relatively "unnatural" light Higgs with respect to its actual scale, might be the 
last option for the composite Higgs to hide. We will now survey the most economical realizations of that idea.

\paragraph{MCHM$_{14-\rm X}$, MCHM$_{\rm X-14}$}

Already if only one SM-quark chirality mixes with a {\bf 14} of $SO(5)$ and the other with any of the representations,
the main qualitative features discussed above remain valid. The same remains true if the SM quarks mix with more than one representation at the same time, adding more $\Psi$ fields to  (\ref{eq:Lf1}), as long as one {\bf 14} is present (with a sizable
composite component). Only the numerical ${\cal O}(1)$ coefficients in front of the linear mixing parameters $y_{L,R}^t$ in
(\ref{eq:V2_14}) will change (with potentially the contribution of {\it one} chirality vanishing or additional mixings entering), 
while necessarily still both 
trigonometric structures will emerge at leading order in the spurions, thus avoiding the double tuning. In particular, the expression 
for the Higgs mass (see (\ref{eq:mh14_1}),(\ref{eq:mh14_2})) will remain as a dominant term, due to the unsuppressed 
contribution of the {\bf 14} to $\beta$ - so at least the requirement of some ad-hoc tuning remains. 
Mixings with other representations can have an impact on the final numerical value of the Higgs mass 
(after ad hoc tuning), however do not change the general picture.

In particular, the most minimal quark model avoiding the necessary presence of light top partners becomes evident, the MCHM$_{14-1}$, {\it i.e.}, the $t_L$ mixing with a {\bf 14} and the $t_R$ with a singlet \cite{Pomarol:2012qf,Panico:2012uw,Pappadopulo:2013vca}, see also Section~\ref{sec:5D}. 
Note that only if a {\bf 14}$_L$ is involved, guaranteeing EWSB by itself, it is possible to mix the $t_R$ with a singlet of 
$SO(5)$, thus not contributing to the breaking of $SO(5)$. Now, eqs. (\ref{eq:V2_14}) - (\ref{eq:mh14_2}) remain valid 
in the limit where the contributions $\sim y_R^t$ are set to zero.
Let us finish this subsection with adding up the degrees of freedom (dof) in the complete composite-fermion particle spectrum of 
this minimal model avoiding the presence of light fermion partners, in units of dirac fermions (neglecting color so far).
Taking into account the bottom sector in the most minimal way (${\bf 5}_{b_L}+{\bf 1}_{b_R}$) as well as the most 
minimalistic lepton embedding (${\bf 5}_{\tau_L}+{\bf 1}_{\tau_R} + {\bf 5}_{\nu_L}+{\bf 1}_{\nu_R}$), we 
arrive at a total number of $(14+1+5+1)_q+(5+1+5+1)_l = 21_q + 12_l = 33$ particles.\footnote{Note that in general the most 
minimal embedding of the quark sector, neglecting a potential impact of leptons on $V(h)$, is $5_{t_L}+5_{t_R}+5_{b_L}
+1_{b_R}$ or $10_{q_L}+5_{t_R}+1_{b_R}$ with 16 dof, compared to the 21 above.}

In the following subsection we will introduce a new class of models to avoid the presence of ultra-light fermionic partners, 
allowing on the one hand side to avoid the ad hoc tuning, while on the other hand they can even even increase 
the naturalness of the assumptions, the minimality, and in particular the predictivity with respect to this setup. 
We are lead to these models by considering for the first time the impact of a realistic lepton sector on the Higgs potential, 
making use of the formulae presented in this subsection.

\subsection{The Impact of the Leptonic Sector}
\label{sec:Vhl}

Since leptons are present in nature, they need to be included into the composite Higgs framework. 
Naively, due to their small masses, one might expect the impact of leptons on the Higgs potential to be small.
However, as explained before and as is well known, from the flavor structure in the lepton sector there is compelling 
motivation to assume the $\tau_R$ to be rather composite and thus coupled more significantly to the 
Higgs than suggested by the small mass of the $\tau$.
 
Beyond that, as we will show more explicitly in Section~\ref{sec:mMCHM3}, already the tiny neutrino masses 
provide a motivation for an enhanced compositeness in the whole right-handed lepton sector. The seesaw mechanism 
can be elegantly embedded in the composite framework, via a large Majorana mass in the elementary sector.
Since in such models the natural value for this mass is however the Planck scale \cite{Carena:2009yt,delAguila:2010vg,
Hagedorn:2011un,Hagedorn:2011pw,KerenZur:2012fr}, the right-handed neutrinos
should feature a non-negligible composite component that reduces their coupling to the Majorana-mass term by several 
orders of magnitude and thus should have a rather large $y_R^\nu$. As we will show, in very minimal setups for the lepton 
sector this is mirrored to a sizable $y_R^\ell$ for the charged leptons.

In particular this is the case for the model that we will advocate and examine in more detail in Section~\ref{sec:mMCHM3}, 
where the complete SM lepton sector will only mix with a single {\bf 5}$_L$+ {\bf 14}$_R$ (for each generation). While the
left-handed doublet will couple to a minimal {\bf 5}, right-handed neutrinos as well as charged leptons will
mix with the same {\bf 14}, realizing a type-III seesaw mechanism in the most minimal way that allows
for a protection of the $Z \tau_R \tau_R$ couplings. Moreover, this model features the least number of parameters 
possible for a realistic embedding of the lepton sector in general.

In the following we will give a comprehensive overview of the impact of possible realizations of the full fermion 
sector on the Higgs potential, the tuning, and the emergence of light partners, always assuming the $\tau_R$ to feature 
a non-negligible compositeness. We will again focus on the third generation for simplicity, while 
we will comment on the other generations relevant for the type-III seesaw model later.

\paragraph{MCHM$_{\rm X-Y}^{\rm Z-\{5,10,1\}}$}

We start with the option of mixing the right handed $\tau$ lepton with either a fundamental,
an adjoint, or a singlet representation of $SO(5)$, as denoted by ${\{5,10,1\}}$, where here and in the following 
the lepton representations are always given by superscripts. The left handed leptons as well as 
quarks are on the other hand allowed to mix with any of the representations considered.\footnote{Remember 
that both the right-handed bottom quark and neutrino are not important for the Higgs potential 
and will thus not be considered explicitly. We will nevertheless keep them in mind for the complete setup of the 
model, where we will include right handed neutrinos with the neutrino masses originating either from a (type I or III) 
seesaw mechanism or from a pure Dirac mass. The contribution of the left handed leptons will also always be negligible,
due to their small compositeness.} 
In that case the $\tau_R$ contributions to $\alpha$ and $\beta$ in (\ref{eq:Potc}), denoted as $\alpha_\tau, 
\beta_\tau$, arise at most at the same order in $y_{L,R}^{\tau, t}/g_\Psi$ as the top contribution. Now we need to note that 
the lepton contribution at a certain order is in general considerably smaller than the quark contribution.
It adds a similar term to $V_2(h)$ as given in (\ref{eq:V22}) or in (\ref{eq:V210}), with however
	\begin{equation}
	N_c  \to 1\,,\quad y_L^{t,q} \to 0\,,\quad t \to \tau
	\end{equation}
(as well as to $V_4(h)$ with the same replacements in (\ref{eq:V45}) and (\ref{eq:V410})).
First of all, this is not $N_c$ enhanced.
Beyond that, the $c_{R\,(RR)}^{\tau\,(\dots)}$, comprising of mass-related quantities 
$a_L^\tau, b_R^\tau$, etc. (see (\ref{eq:mt}), with $(t,T)\to (\tau,{\cal T})$), are in general somewhat 
smaller than $c_{R\,(RR)}^{t\,(\dots)}$ in viable lepton models, see Section~\ref{sec:5D}.
As a consequence the impact of the lepton sector on the Higgs mass via $\beta_\tau$ is directly negligible to good 
approximation.
Regarding the condition from EWSB (\ref{eq:EWSBcon}), the lepton contribution $\alpha_\tau$ is 
also significantly suppressed with respect to the quark contribution in general. Only after $\alpha_t$ has been tuned to be 
of the order of $\beta_t$ to guarantee EWSB, see Section~\ref{sec:Vht}, $\alpha_\tau$ might become relevant for 
the numerical minimization condition. It however does not change any of the qualitative conclusions in Section~\ref{sec:Vht}.
The same is true in general for the subleading contribution  $\alpha_W$ of the gauge bosons.

\paragraph{MCHM$_{\{5,10\} - \{5,10,1\}}^{\, \rm Z - 14}$}

We now move forward to the case of the $\tau_R$ mixing with a symmetric representation of $SO(5)$, where we 
will find interesting new features. The left handed SM-like quarks will first be restricted to mix with either a fundamental 
or an adjoint representation, while the right handed quarks could alternatively also mix with a singlet. The
relevant contributions to the Higgs potential now look like
	\begin{equation}
	\label{eq:Vlep}
	\begin{split}
	V(h)\cong \left[\frac{m_\Psi^4}{16 \pi^2 g_\psi^2} \left( N_c \left(\tilde c_L^t y_L^{t\,2} - \tilde c_R^t\,
	y_R^{t\, 2}\right) + \tilde c_R^\tau y_R^{\tau\,2} \right)\, +\, \alpha_t^{\!(4)}(y_{L,R}^{t\, 4}/g_\Psi^4)  \right] 
	\, \sin^2(h/f_\pi)\\
	- \left[  \frac{m_\Psi^4}{16 \pi^2 g_\psi^2}5/4\, c_R^{\tau\, (2)} y_R^{\tau\, 2}   
	\, +\, \beta_t (y_{L,R}^{t\, 4}/g_\Psi^4) \right] \,  \sin^2(h/f_\pi) \cos^2(h/f_\pi)\, .
	\end{split}
	\end{equation}
Here $\tilde c_L=\{c_L^t/2, -3/8\, c_L^q \}$ in the case of $t_L$ mixing with a \{{\bf 5, 10}\} of $SO(5)$, whereas 
$\tilde c_R=\{c_R^t,- 1/4\, c_R^t,0 \}$ for $t_R$ mixing with a \{{\bf 5, 10, 1}\}, respectively, see 
(\ref{eq:V22}) and (\ref{eq:V210}), while simply from (\ref{eq:V2_14}) we identify $\tilde c_R^\tau=-3/4 
(c_R^{\tau\, (1)}+c_R^{\tau\, (2)})$. The subleading quark terms, contributing to 
$\beta_t(y_{L,R}^{t\, 4}/g_\Psi^4) $ can be obtained easily from (\ref{eq:V45}) ((\ref{eq:V410})) for the case 
of $t_L$ and $t_R$ both mixing with a {\bf 5} ({\bf 10}) of $SO(5)$. The same is true
for the ${\cal O}(y_{L,R}^{t\, 4}/g_\Psi^4)$ contributions to $\alpha_t$, denoted as $\alpha_t^{(4)}$. In any case, a 
new trigonometric function beyond the $\sin^2(h/f_\pi)$ emerges from the quark sector only at ${\cal O}(y_{L,R}^{t\, 4}/g_\Psi^4)$. 

We inspect that the lepton 
sector can now deliver an essential contribution. First of all, it provides the formally leading term in the second row of
(\ref{eq:Vlep}), contributing to $\beta$ and thus to the Higgs mass at $y^2/g_\Psi^2$. Although, as discussed 
before, the lepton sector features a notable general suppression with respect to the top quark contributions to the
potential, this can be lifted by the smaller power suppression in $y^2/g_\Psi^2$ compared to $\beta_t$.
The $\tau_R$ contribution can thus help to allow for a light Higgs boson with $m_H \sim 125\,$GeV, without 
the need for light top partners.
The crucial point is that now we add two new mixing parameters $y_L^\tau$ and $y_R^\tau$ (as well as additional 
(${\cal O}(1)$) mass parameters) but only one new constraint, {\it i.e.}, the $\tau$ mass. Although only the latter 
mixing is relevant for the potential, this however now is basically a new free parameter. 

Without leptons, all mixing 
parameters were determined to good approximation by the top mass (\ref{eq:mtsim}) and the EWSB condition 
(\ref{eq:EWSBcon}), where the latter fixed $y_L^t  \sim y_R^t \equiv y_t$, as explained before.
Like this, $m_H$  was essentially fixed and generically too large, see e.g. (\ref{eq:mHmt}). The only 
additional freedom for a potential reduction in the Higgs mass was to lower the top-partner masses 
$m_T$ or $m_{\tilde T}$, to increase the mixing with the resonances which allows to lower $y_t$ for fixed 
$m_t$ and thus to lower $m_H$.
Now, the leptonic contribution offers an opportunity to break this pattern by providing the additional parameter 
$y_R^\tau$ entering $m_H$. Adding $\beta_\tau$ to (\ref{eq:mh5}) we arrive at  (setting $c_{LL,LR,RR}^t 
\to 1$)
	\begin{equation}
	\label{eq:mHlep}
	\begin{split}
	m_H & \sim \frac{1}{\sqrt 2 \pi} \sqrt{3 y_t^4\, +\, 5/4\, c_R^{\tau\, (2)}\, g_\Psi^2 y_R^{\tau\, 2} }\ v
	\\[1.5mm]
	& \sim \frac{1}{\pi} \sqrt{ 
	3\, \frac{\min(m_T,m_{\tilde T})^2}{f_\pi^2}\, m_t^2
	\, +\, 5/8\, c_R^{\tau\, (2)} g_\Psi^2 y_R^{\tau\, 2}\, v^2 }\,,
	\end{split}
	\end{equation}
and similarly for the other non-trivial quark embeddings mentioned before, while we will comment on the option of 
embedding the $t_R$ in a singlet further below.
Since the lepton sector delivers a non-negligible contribution to $m_H$, one can now use the additional freedom 
to arrive at the correct Higgs mass, without the need to tune the top partners light, only by moderately 
canceling the top contribution in (\ref{eq:mHlep}) above via the contribution of the $\tau_R$.
In Section~\ref{sec:5D} we will see explicitly that the effect of the latter is significant and can indeed interfere 
destructively with the top contribution, to allow for larger $\min(m_T,m_{\tilde T})$.

Moreover, in contrast to quarks in the {\bf 14}, no large ad-hoc tuning is needed, since in general $\beta_\tau$ 
is significantly smaller than $\beta_t$ when the fermions mix with a {\bf 14}. 
In the case of quarks, the ad hoc tuning was unavoidably generated due to the large top mass.
With the quarks now mixing with a fundamental or adjoint representation of $SO(5)$, their contribution to $\beta$ is 
suppressed one order further in $y_t^2/g_\Psi^2$ such that both the quark and the lepton contribution are 
already roughly as small as the electroweak scale ({\it c.f.}\ (\ref{eq:mh14_2})).
There is no need for light resonances whatsoever.
Finally, the impact of the $\tau_R$ on the EWSB condition (\ref{eq:EWSBcon}) is in general still similar to the 
case discussed before. It is subleading in the first place, while after the tuning in the quark sector, which
still more or less leads to $y_L^t \sim y_R^t$ (see first line of (\ref{eq:Vlep})), it can have a modest impact on the numerics.
We will see the whole mechanism at work explicitly in Section~\ref{sec:5D}.

\paragraph{MCHM$_{14 - \rm Y}^{\rm Z-W}$, MCHM$_{\rm  X - 14}^{\rm Z-W}$}

For completeness, we now discuss the case where the $t_L$ or the $t_R$ mixes with a {\bf 14} of $SO(5)$,
while the other fermions can mix with any one of the representations considered.
Now the leptonic effect is again subleading in general, since the top contributions
are always non-vanishing at the maximal possible order ($y^2/g_\Psi^2$), see Section~\ref{sec:Vht}.
Nevertheless the $\tau_R$ can have e.g. a numerical impact on the Higgs mass, 
if it mixes itself with a {\bf 14}, after the tuning of the large quark contribution
in order to reproduce the small $m_H$. 
After this reduction in the quark contribution, the leptons can also have again
a numerical influence on the EWSB condition for various representations.
However, they again deliver no qualitative change of the overall mechanism and findings with respect 
to the tuning and the absence of light partners if quarks are in the {\bf 14}.
\vspace{2mm}

In general, we have shown that if a composite $\tau_R$ mixes with a symmetric representation of $SO(5)$, 
interesting consequences can arise. In particular, light top partners are no longer needed, while still a large 
tuning in the Higgs mass can be avoided. We will see now that such models of the lepton sector can also 
have very interesting features regarding the minimality of the setup.
As explained, models of $\tau$ compositeness are well 
motivated from lepton-flavor physics, in particular from the absence of sizable FCNCs. However, the fact that leptons 
should mix with a {\bf 14} of $SO(5)$ is at this stage still rather ad hoc. In the end, although the setup offers 
an orthogonal approach and clearly has some virtues with respect to the {\bf 14} in the quark sector, 
so far it is has not been proven to be more motivated from a conceptual reasoning.

On the other hand, due to the mere fact that the neutrino masses are so tiny, a very attractive motivation
for having leptons in a {\bf 14} of SO(5) can be given, even without the need to rely on flavor protection to justify
right-handed lepton compositeness.
Indeed, the $SU(2)_L$ triplet present in the {\bf 14} is very welcome regarding neutrino masses. If it acquires 
a Majorana mass term, it provides heavy degrees of freedom that can induce the strongly suppressed dimension-5 
Weinberg operator, responsible for generating tiny neutrino masses via the well known (type-III) seesaw mechanism. 
As mentioned above, the {\bf 5}$_L$+ {\bf 14}$_R$ setup provides the most minimal composite Higgs model with a 
type-III seesaw for neutrino masses (with a protection of $Z \tau_R \tau_R$) and will thus be denoted the 
mMCHM$^{\rm III}$.  Beyond that minimality with respect to the particle content, since both the 
heavy lepton triplet containing the right-handed Majorana neutrino as well as the $\tau_R$ can mix with the 
same composite multiplet (containing both a $SU(2)_L$ triplet and singlet), the model features the least number 
of parameters possible, even in general.\footnote{Note that if the right-handed $\tau$ and the $N_R$ mix 
with the same multiplet of the strong sector, this suggests that they feature the same linear-mixing parameter 
(determined by the anomalous dimension of the composite operator) $y_R^\tau = y_R^{N_\tau}$, where the latter 
should be sizable, as explained before. The new fields that the {\bf 14} offers are thus used in an 
economical way. A further reduction in parameters (and fields) with respect to the MCHM$_{14}$ is reached since, 
with both right handed lepton multiplets being able to mix with the same composite representation with a single
$X$ charge, also the left handed fields can mix with a single representation (here a {\bf 5}). As a consequence,
the setup involves the minimal amount of composite SO(5) multiplets that is viable -  less than any other model  known before.}
Finally, note that in that context a similar type-I seesaw model would need at least an additional {\bf 1}$_{\nu_R}$ and in 
consequence another {\bf 5}$_L$, and would thus be much less minimal. Interestingly, the lack of light partners together with minimality points to a type-III seesaw model.

Moreover, an equivalently simple model with a {\bf 14} in the quark sector is not possible. It is a
peculiar feature of the lepton sector that the right-handed neutrinos can be part of a $SU(2)_L$ triplet,
while for quarks the right handed fields need both to mix with $SU(2)_L$ singlets which requires
a second composite multiplet (beyond the {\bf 14}) mixing with the right handed fermions, complicating 
the model.\footnote{While in the {\bf 10} it would in principle be possible to embed both 
$t_R$ and $b_R$, this is in any case disfavored, as it would introduce large corrections to $Z b_R \bar b_R$
couplings due to the large top mass.}

In that context, note that for the mMHCM$^{\rm III}$, considering the embeddings of the quarks, the very minimal and 
in principle obvious possibility of mixing the $SU(2)_L$ singlet $t_R$ with a singlet of $SO(5) \supset SU(2)_L$ arises naturally, 
while still the $t_L$ can mix with a {\bf 5} - the mMCHM$^{{\rm III}>}_{5-1}$.
This now becomes viable due to a potentially unsuppressed leptonic contribution in the  mMCHM$^{\rm III}$, 
which here can become 
larger than in the models of $\tau_R$ compositeness considered so far. First, the $N_c=3$ suppression can be lifted via
an $N_g=3$ enhancement, since now all right handed charged-lepton generations are expected to contribute 
(see Section~\ref{sec:mMCHM3}). Moreover, remember that the model does not need anymore to be based on a specific setup 
of flavor protection, coming usually with smaller $c_{R\,(RR)}^{\tau\,(\dots)}$ from a `Yukawa suppression' \cite{delAguila:2010vg}, to motivate a large right-handed lepton compositeness. The lifting of that suppression
is denoted by the superscript $>$. In this setup all terms $\sim y_R^t$ in (\ref{eq:Vlep}) vanish and it becomes possible 
that the now sizable $\tau_R$ contribution to $\alpha$ moderately cancels the $t_L$ one, such as to fulfill 
(\ref{eq:EWSBcon}).

In particular $y_L^t$ can also be somewhat smaller if the $t_R$ mixes with a singlet, since the top mass can always be 
matched via the now free $y_R^t$, which is not required to fulfill $y_R^t \sim y_L^t$ anymore from EWSB. This will just lead 
to a strongly composite $t_R$. In consequence of this freedom, the contribution from the quark sector to the Higgs mass
is also no longer fixed by $m_t$ and can become somewhat smaller, without light partners. 
Now generically $m_H$ becomes dominated by $\beta_\tau$
and the leptons do not need to interfere destructively with the quark contribution to $\beta$ anymore. Due to the generically  large 
$\beta_\tau$ the model will feature an ad-hoc tuning, similar to the model with a {\bf 14}$_{t_L}$
+ {\bf 1}$_{t_R}$, however the parametric ``double-tuning'' in EWSB can clearly be avoided, although the quark realization is 
minimal. Note that in such a scenario of a unsuppressed contribution of a leptonic ${\bf 14}_R$ to $\alpha$
and $\beta$, the lepton effects can help to avoid the ``double-tuning''  (and the emergence of light partners) also in other realizations of the quark sector than the ${\bf 5}_L+{\bf 1}_R$, breaking also the $y_L^t \sim y_R^t$ degeneracy in general. In all models, besides those featuring a {\bf 14} in the quark sector, the leptonic contributions, discussed in (\ref{eq:mHlep}), will then generically dominate $m_H$ before necessary 
cancellations take place. If quarks are however in a {\bf 14}, both sectors become comparable and the pure quark models 
are only modified in a sense that now all contributions become larger in general due to the additional lepton terms.

On the other hand, note that also in the mMCHM$^{\rm III}$ it is still possible to motivate a Yukawa suppression for leptons,
which will here be balanced by a slightly enhanced left-handed compositeness to keep the lepton masses fixed. This is in
particular feasible since, in contrast to the large top mass, one has $m_\tau \ll v$. This leads to the other limit
of the model, which results in rather similar predictions as for the models with a simple ${\bf 14}_{\tau_R}$, described before.
The most minimal viable version of this setup features a ${\bf 5}_{t_L}+{\bf 5}_{t_R}$ in the top-quark sector,
while the bottom sector (and light quarks) can be in a ${\bf 5}_{b_L}+{\bf 1}_{b_R}$, and will be denoted
the mMCHM$^{{\rm III}}_5$, while the corresponding optional version with large lepton Yukawas will again be called 
mMCHM$^{{\rm III}>}_5$.
 
Nevertheless, the most minimal complete model which avoids the presence of light top-partners belongs to the class MCHM$_{5 - 1}^{5 - 14}$, where the full embedding reads ${\bf 5}_{t_L} + {\bf 1}_{t_R} 
+ {\bf 5}_{b_L} + {\bf 1}_{b_R} + {\bf 5}_{\tau_L} + {\bf 14}_{\ell_R}$, equipped with a Majorana mass for 
the $SU(2)_L$ triplet and unsuppressed lepton Yukawas. It is just this model, 
featuring 31 dof, that we will think of as the mMCHM$^{{\rm III}>}_{5-1}$ in the following.
As discussed, the quarks can now mix with composites with only 12 
dof.\footnote{This corresponds to the minimal possible amount of fermions one might think of in general $SO(5)/SO(4)$ 
composite Higgs models, whereas the minimum for a sector that needs to trigger EWSB was $5_{t_L} + 1_{t_R} + 5_{b_L} 
+ 1_{b_R} = 16 = 5_{q_L} + 10_{t_R} + 1_{b_R}$.}
The model has thus less dof in the quark sector than any other composite model known, minimizing the
colored particle content, with further interesting consequences for phenomenology.

The most minimal model that avoids the presence of ultra-light partners by a modification of the quark sector 
is on the other hand defined as ${\bf 14}_{t_L} + {\bf 1}_{t_R} + {\bf 5}_{b_L} + {\bf 1}_{b_R} + {\bf 5}_{\tau_L} 
+ {\bf 1}_{\tau_R} 
+ {\bf 5}_{\nu_L} + {\bf 1}_{\nu_R}$. Beyond the fact that the {\bf 14} in such a model is conceptually 
less motivated from the pattern of the quark masses and seems ad hoc, it has 33 dof, and thus more than 
the  mMCHM$^{{\rm III}>}_{5-1}$  where the {\bf 14} is used to unify different SM multiplets. 
If nature should call for a {\bf 14} of $SO(5)$ via the non-discovery of light top partners, 
the lepton sector with a type-III seesaw seems to offer the most economical place to host this multiplet.

In summary, models with leptons in a symmetric representation of $SO(5)$, such as the  mMCHM$^{{\rm III}}_{5}$ offer 
the new possibility to create a {\it naturally} light Higgs without light partners in a well motivated setup following the 
principle of minimality in the lepton sector, while the total dof are in the ballpark of
minimal models (already the standard MCHM$_5$ with all fermions in a ${\bf 5}$ has more).
Moreover, they also invite to set up models with a maximal total amount of minimality, by
allowing the most minimal and natural quark embedding, with right handed singlets of $SO(5) \supset SU(2)_L$  
only, the  mMCHM$^{{\rm III}>}_{5-1}$. Counting color as a degree of freedom, such a model features $3\cdot12_q + 19_l = 
55$ fermionic dof for the third generation, 
compared to the most minimal viable model known before, with $3\cdot 16_q + 12_l = 60$ fermionic dof.\footnote{
The same counting goes trough to the case of three generations in the fully anarchic approach to flavor.}
It could thus be considered as the most minimal composite Higgs model in general and it does not predict
light partners. In consequence, minimality as an argument to expect light partners at the LHC might be 
questionable. In the following we will quantify the general findings of this section.

\section{Numerical Analysis in the GHU Approach - The Impact of Leptons}
\label{sec:5D}

In this section we are going to probe the general findings of last section numerically by studying explicit 
holographic realizations of the composite Higgs setup, {\it i.e.}, models of gauge-Higgs unification 
\cite{Manton:1979kb, Hatanaka:1998yp, vonGersdorff:2002as, Csaki:2002ur, Contino:2003ve, Agashe:2004rs}.
We will first present the 5D framework for the models under consideration and then discuss the calculation of
the Higgs potential. Finally we will present our numerical results and confront them with the general
predictions of Section~\ref{sec:general}.

\subsection{Setup of the (5D) GHU Models}
The 5D holographic realization of the MCHMs introduced in the previous section consists of a slice of AdS$_5$ with metric
\begin{eqnarray}
	\mathrm{d}s^2=a^2(z)\left(\eta_{\mu\nu}\mathrm{d}x^{\mu}x^{\nu}-\mathrm{d}z^2\right) \equiv \left(\frac{R}{z}\right)^2\left(\eta_{\mu\nu}\mathrm{d}x^{\mu}x^{\nu}-\mathrm{d}z^2\right),
\end{eqnarray}
where $z\in [R, R^{\prime}]$ is the coordinate of the additional spatial dimension and $R$ and $R^{\prime}$ are the positions of the UV and IR branes, respectively. We consider a bulk gauge symmetry $SO(5)\times U(1)_X$ broken by boundary conditions to the electroweak group $SU(2)_L\times U(1)_Y$ on the UV brane and to $SO(4)\times U(1)_X$ on the IR one. More explicitly, this setup correspond to gauge bosons with the following boundary conditions
\begin{eqnarray}
	L_{\mu}^{a}(+,+), \qquad R_{\mu}^{b}(-,+), \qquad B_{\mu}(+,+), \qquad Z_{\mu}^{\prime}(-,+), \qquad C_{\mu}^{\hat{a}}(-,-),
\end{eqnarray}
where $a=1,2,3$,  $b=1,2$, $\hat{a}=1,2,3,4$ and $-/+$ denote Dirichlet/Neumann boundary conditions at the corresponding brane.\footnote{The respective (4D) scalar components ($\mu \to 5$) have opposite boundary conditions, allowing for zero modes only in $C_5^{\hat{a}}$.}
In the above expression,  $L_{\mu}^{1,2,3}$ and $R_{\mu}^{1,2,3}$  are the 4D vector components of the 5D gauge bosons associated to $SU(2)_L$ and $SU(2)_R$, respectively, both subgroups of $SO(5)$. We have defined the linear combinations
\begin{eqnarray}
	B_{\mu}&=&s_{\phi}R_{\mu}^3+c_{\phi} X_{\mu},\qquad Z^{\prime}_{\mu}=c_{\phi} R_{\mu}^3-s_{\phi} X_{\mu},\nonumber\\
c_{\phi}&\equiv &\frac{g_5}{\sqrt{g_5^2+g_X^2}},\qquad \ \ \ \  s_{\phi}\equiv \frac{g_X}{\sqrt{g_5^2+g_X^2}},
\end{eqnarray}
with $g_5$ and $g_X$ the dimensionfull 5D gauge couplings of $SO(5)$ and $U(1)_X$, respectively, and $X_{\mu}$ the gauge boson associated with the additional $U(1)_X$. Finally, $C_{\mu}^{\hat{a}}$ are the gauge bosons corresponding to the broken coset space $SO(5)/SO(4)$, whose scalar counterparts zero-modes $C_{5,(0)}^{\hat{a}}(x,z)\equiv f_h^{\hat{a}}(z)h^{\hat{a}}(x)$ we will identify with the $SU(2)_L$ Higgs doublet. 

We typically fix $1/R\sim 10^{16}~ \TeV$ and, for each value of the warped down $1/R^{\prime}\sim \mathcal{O}(1)~\TeV$, 
addressing the hierarchy problem, we obtain $g_5$, $s_{\phi}$ and $\langle h^{\hat{a}}\rangle =v \delta_{\hat{a}4}$ in terms of $\alpha_{\rm QED}$, $M_W$ and $M_Z$. That means that, modulo the value of $R\sim M_{\rm Pl}^{-1}$, fixed by naturalness, in the 5D gauge sector the only free parameter is $R^{\prime}$, or equivalently, 
\begin{eqnarray}
	f_{\pi}\equiv \frac{2R^{1/2}}{g_5R^{\prime}}.
\end{eqnarray}
We will also consider the more general scenario where the dimensionless 5D gauge coupling $g_{\ast}\equiv g_5 R^{-1/2}$ (as well as $s_{\phi}$) can change for a fixed value of $f_{\pi}$. This can be done by allowing for $SU(2)_L$ and $U(1)_Y$ UV localized  brane kinetic terms, 
\begin{eqnarray}
	\mathcal{S}_{\rm UV}\supset\int\mathrm{d}^4x\left[-\frac{1}{4}\kappa^2 R \log \left(\frac{R^{\prime}}{R}\right) L^{a \mu\nu} L^a_{\mu\nu}-\frac{1}{4}\kappa^{\prime 2}R \log \left(\frac{R^{\prime}}{R}\right) B^{\mu\nu} B_{\mu\nu}\right]_{z=R},
	\label{bkgauge}
\end{eqnarray}
where $\kappa$ and $\kappa^{\prime}$ are dimensionless parameters. Then, for given values of $\{f_{\pi}, \kappa,\kappa^{\prime}\}$, we obtain with very good approximation
\begin{eqnarray}
\label{eq:gast}
	g_{\ast}\approx \frac{e}{\sin\theta_W}\sqrt{\log(R^{\prime}/R)}\sqrt{1+\kappa^2},\qquad  s_{\phi}\approx \tan\theta_W\sqrt{\frac{1+\kappa^{\prime 2}}{1+\kappa^2}}\,.
\end{eqnarray}
Moreover,
\begin{eqnarray}
	\sin(v/f_{\pi})\approx 2\frac{\sin\theta_W}{e}\frac{M_W }{f_{\pi}},
\end{eqnarray}
and $e=\sqrt{4\pi\alpha_{\rm QED}}$ is the electric charge while $\theta_W$ the Weinberg angle. In the dual CFT, changing $g_{\ast}$ corresponds to a change in the number of colors
\begin{eqnarray}
	\label{eq:NCFT}
	N_{\rm CFT}\equiv \frac{16\pi^2}{g_{\ast}^{2}}.
\end{eqnarray}
The value of $g_{\ast}$ can also be related to the free parameter $g_{\rho}$ introduced in Section~\ref{sec:general}, controlling the mass scale of the composite vector resonances $m_{\rho}\equiv g_{\rho}f_{\pi}$. Considering the first Kaluza-Klein (KK) resonance of a 5D gauge field with $(-,+)$ boundary conditions\footnote{In the holographic picture, these boundary conditions imply that this gauge field does not interact with the elementary sector.}, we obtain
\begin{eqnarray}
	g_{\rho}\approx 1.2024 \cdot g_{\ast}.
\end{eqnarray}

The fermion sector will depend on the $SO(5)$ representation in which the 5D fields transform,  $\mathbf{1}, \mathbf{4}, \mathbf{5}, \mathbf{10}$ or $\mathbf{14}$. 
Taking into account that $Y=T_R^3+X$ it is straightforward to work out all possible embeddings of the SM fermions. As mentioned in Section~\ref{sec:general}, henceforth we will just consider fermions with a sizable degree of compositeness  since they will be the only ones playing a non-negligible role in the generation of the Higgs potential and in the determination of the Higgs mass. That means in particular that we will neglect the first two quark generations, the right handed (RH) bottom and, with the exception of the mMCHM$^{\rm III}$ where all RH leptons will be composite, the first two lepton generations. Moreover, excepting again the mMCHM$^{\rm III}$,  possible RH neutrinos will also be neglected. In order to get the correct charged lepton masses we will still include left handed (LH) leptons when their RH counterparts are composite. 

Due to the absence of $P_{LR}$ symmetry protecting the $Z \bar{b}_L b_L$ coupling, the spinorial representation is usually not considered for quarks as it would lead to too large deviations from its measured value. For leptons with a moderate degree of compositeness, we would encounter a similar problem for the corresponding $Z$ couplings. Thus, we will not consider this case throughout this work, restricting ourselves to the other representations. Without trying to exhaust all possible combinations of fermion representations, which have been discussed qualitatively in Section~\ref{sec:general}, we will consider for the quark sector the cases where
\begin{itemize}
	\item both $q_L$ and $t_R$ are embedded in a $\mathbf{5}$, $\mathbf{10}$ or $\mathbf{14}$, each, denoted by MCHM$_5$, MCHM$_{10}$ and MCHM$_{14}$, respectively,
	\item  the quark doublet $q_L$ is living in a $\mathbf{14}$ whereas the $t_R$ is a full singlet of $SO(5)$, denoted by MCHM$_{14-1}$,
	\item the quark doublet $q_L$ is embedded in a $\mathbf{5}$ and the $t_R$ in a $\mathbf{14}$, denoted by MCHM$_{5-14}$.
\end{itemize}
Regarding leptons, the RH charged ones $e_R,\mu_R$ and $\tau_R$ will always be embedded in the $\mathbf{14}$, while the LH doublets will live either in the fundamental $\mathbf{5}$ or the symmetric representation~$\mathbf{14}$.

In order to fix the notation and illustrate further the different embeddings we will describe in some detail the cases cited above. For further details on the $SO(5)$ fermion representations we refer the reader to Appendix \ref{sec:ferm}.

\subsubsection{MCHM$_{5}$}
We consider two $\mathbf{5_{2/3}}$ of $SO(5)\times U(1)_X$ with the following boundary conditions
\begin{eqnarray}
\zeta_{1}&=&\left(\begin{array}{r}\tilde{\Lambda}_{1}[-,+]~  t_1[+,+]\\ \tilde{t}_{1}[-,+]~ b_1[+,+]\end{array}\right)\oplus t^{\prime}_{1}[-,+],\quad \zeta_{2}=\left(\begin{array}{r} \tilde{\Lambda}_{2}[+,-]~  t_2[+,-]\\ \tilde{t}_{2}[+,-]~ b_2[+,-]\end{array}\right) \oplus t^{\prime}_{2}[-,-],\qquad
\end{eqnarray}
where we have explicitly shown the decomposition under $SU(2)_L\times SU(2)_R$, with the bidoublet being represented by a $2\times 2$ matrix on which the $SU(2)_L$ rotation acts vertically and the $SU(2)_R$ one horizontally. More specifically, the left and right columns correspond to fields with $T_R^3=\pm 1/2$, whereas the upper and lower rows have $T_L^3=\pm 1/2$. The signs in square brackets denote the boundary conditions on the corresponding branes. A Dirichlet boundary condition for the RH chirality is denoted by $[+]$ while the opposite sign denotes the same boundary condition for the LH one. Therefore, before EWSB,  zero-modes with quantum numbers $\mathbf{2_{1/6}}$ and $\mathbf{1_{2/3}}$ under $SU(2)_L\times U(1)_Y$ live in $\zeta_1$ and $\zeta_2$, respectively.

The relevant part of the action reads
\begin{eqnarray}
	\mathcal{S}\supset\sum_{k=1,2}\int\mathrm{d}^4x\int_R^{R^{\prime}}\mathrm{d}z~a^4\left\{\bar{\zeta}_k\left[i\cancel{D}+\left(D_5+2\frac{a^{\prime}}{a}\right)\gamma^5-aM_k\right]\zeta_k\right\}+\mathcal{S}_{\rm UV}+\mathcal{S}_{\rm IR},
\end{eqnarray}
where\footnote{See Appendix \ref{sec:gen} for explicit expressions of the $SO(5)$ generators in the fundamental representation.}
\begin{eqnarray}
	D_{M}&=&\partial_M-ig_5T_L^a L_M^a-ig_5 T_R^b R_M^b-ig_Y Y B_M-i\frac{g_Y}{c_{\phi}s_{\phi}}Z_{M}^{\prime}\left(T_R^3-s_{\phi}^2Y\right)\nonumber\\
			   && -ig_5 T^{\hat{a}} C_M^{\hat{a}}, \mathrm{~~with~} M=\mu,5\quad \mathrm{and}\quad g_Y\equiv g_5g_X/\sqrt{g_5^2+g_X^2},
\end{eqnarray}
are the gauge covariant derivatives and $\mathcal{S}_{\rm UV}, \mathcal{S}_{\rm IR}$ are possible brane localized terms. We conventionally parametrize the bulk masses $M_k=c_k/R$ in terms of dimensionless bulk mass parameters $c_k$ and the fundamental scale $R$. 

In particular, the fifth component of the covariant derivative in the above action generates the Yukawa interactions
\begin{eqnarray}
	\mathcal{S}&\supset& -\sum_{k=1,2}ig_5\int\mathrm{d}^4x\int_R^{R^{\prime}}\mathrm{d}z~a^4\bar{\zeta}_k(x,z) \gamma^5 T^4 \zeta_k(x,z) C_5^4(x,z)\\
									   &=&-ig_5\left[\int_R^{R^{\prime}}\mathrm{d}za^{-1}\right]^{-1/2}\sum_{k=1,2}\int\mathrm{d}^4x\int_R^{R^{\prime}}\mathrm{d}z~a^3\bar{\zeta}_k(x,z) \gamma^5 T^4 \zeta_k(x,z) h(x)+\ldots,\quad \nonumber
\end{eqnarray}
where the dots stand for terms involving the non-physical KK excitations of the Higgs boson and we have used that the Higgs profile is given by
\begin{eqnarray}
	f_h^{4}(z)=a^{-1}\left[\int_R^{R^{\prime}}\mathrm{d}z~a^{-1}\right]^{-1/2}.
\end{eqnarray}
 
Looking at the specific form of the Yukawa interactions, we can see that in order to have a non-zero mass for the zero-modes after EWSB we need to have some IR brane terms splitting the zero-modes between the different multiplets. Therefore we consider the following brane action
\begin{eqnarray}
	\mathcal{S}_{\rm IR}=-\int\mathrm{d}^4x ~\left\{a(z)^4\left[M_{S}^q\overline{\zeta}^{(\mathbf{1},\mathbf{1})}_{1L} \zeta^{(\mathbf{1},\mathbf{1})}_{2R}+M_B^q\overline{\zeta}^{(\mathbf{2},\mathbf{2})}_{1L} \zeta^{(\mathbf{2},\mathbf{2})}_{2R}\right]\right\}_{z=R^{\prime}}+\mathrm{h.c.},
\end{eqnarray}
where we have used the decomposition $\zeta=\zeta^{(\mathbf{2},\mathbf{2})}+\zeta^{(\mathbf{1},\mathbf{1})}$ introduced explicitly in Appendix \ref{subsec:fun}. One could in principle add some UV brane masses between the exotic $SU(2)_L$ doublets of both bidoublets but their impact on the Higgs potential would be subleading, as they do not have zero-modes and the main contribution comes always from the elementary degrees of freedom. For the sake of simplicity, we will not consider the introduction of fermionic brane kinetic terms unless they are required to get the correct EWSB as, for instance, in the case of the MCHM$_{14-1}$,
see below.

One remarkable feature of these models is that we can remove the Higgs vev $\langle h^{\hat{a}}\rangle=v\delta_{\hat{a}4}$ from the bulk through the following 5D gauge transformation \cite{Hosotani:2005nz}
\begin{eqnarray}
	\zeta_k(x,z)&\to& \Omega(z)\zeta_k(x,z),\label{gtf}\\
	\mathbb{A}_M(x,z)&\to& \Omega(z) \mathbb{A}_M(x,z)\Omega(z)^{T}-(i/g_5)\left(\partial_M\Omega(z)\right) \Omega(z)^T\label{gtv},
\end{eqnarray}
where $\mathbb{A}_M=L_M^aT_L^a+R_M^aT_R^a+C_M^{\hat{a}}T^{\hat{a}}$ and 
\begin{eqnarray}
	\Omega(z)=\mathrm{exp}\left(-ig_5T^{\hat{a}} \langle h^{\hat{a}}(x)\rangle \int_R^z\mathrm{d}z^{\prime}~f_h^{\hat{a}}(z^{\prime})\right)=\mathrm{exp}\left(-ig_5T^{4}v\int_R^z\mathrm{d}z^{\prime}~f_h^4(z^{\prime})\right).
\end{eqnarray}
After such a transformation, we can solve the bulk equations of motion for fermions and gauge bosons neglecting the Higgs vev and impose the same UV boundary conditions, since $\Omega(R)=\mathbf{1}$. However, the IR boundary conditions of the transformed fields will mix through the Wilson lines $\Omega(R^{\prime})$ fields with the same electric charge.

Finally, similarly to the gauge sector,
we can obtain the 4D parameter $g_{\Psi}$ introduced in the previous section by identifying the mass scale of the fermionic resonances in the 4D picture, 
$m_{\Psi}\equiv g_{\Psi}f_{\pi}$, with the mass of the first KK mode of a 5D field with boundary conditions $[-,+]$ and bulk mass $c_{\Psi}=1/2$. This leads, in the absence of fermion brane kinetic terms, to
\begin{eqnarray}
	g_{\Psi}=g_{\rho}\approx 1.2024\cdot g_{\ast}.
\end{eqnarray}
Although in this case $g_{\Psi}$ and $g_{\rho}$ turn out to be equal, this degeneracy could be lifted by adding fermion kinetic terms. For completeness we also show  in Table~\ref{tab:cor} the rest of approximate identifications with the 4D setup, where the function $f_c$ 
\begin{eqnarray}
	f_c\equiv \left[\frac{1-2c}{1-\left(\frac{R}{R^{\prime}}\right)^{1-2c}}\right]^{\frac{1}{2}},
\end{eqnarray}
gives the wave function of the corresponding fermion zero-mode on the IR brane.
\begin{table}
\centering
\begin{tabular}{|c|c|}
\hline
5D & 4D \\
\hline
KK mass & $ \hat m_T, \tilde m_T$\\
\hline
$f_{c_1,-c_2}$ & $  y_{L,R}^tf_\pi/\ {\hat\ }{\!\!\! \tilde{\! m_T\,}} \approx \sin \varphi_{L,R}$\\
\hline
$M_B, M_S$ & $a_L, b_R$\\
\hline
fixed Higgs couplings $(C_5^{\hat a})$ & $b_L, a_R$\\
\hline
\end{tabular}
\caption{\label{tab:cor} Approximate correspondence between the parameters of the 5D model and those of the 4D model, introduced in Section~\ref{sec:general}.}
\end{table}
\subsubsection{MCHM$_{10}$}
In this case we consider two $\mathbf{10_{2/3}}$ multiplets,
\begin{eqnarray}
\zeta_{1}&=&\left(\begin{array}{r}\tilde{\Lambda}_{1}[-,+]~  t_1[+,+]\\ \tilde{t}_{1}[-,+]~ b_1[+,+]\end{array}\right)\oplus \left(\begin{array}{c}\hat{\Lambda}_1[-,+]\\\hat{t}_{1}[-,+]\\ \hat{b}_1[-,+]\end{array}\right)\oplus\left(\begin{array}{c}\Lambda^{\prime}_1[-,+]\\t^{\prime}_{1}[-,+]\\ b_1^{\prime}[-,+]\end{array}\right),\nonumber\\
\zeta_{2}&=&\left(\begin{array}{r} \tilde{\Lambda}_{2}[+,-]~  t_2[+,-]\\ \tilde{t}_{2}[+,-]~ b_2[+,-]\end{array}\right)\oplus \left(\begin{array}{c}\hat{\Lambda}_2[+,-]\\\hat{t}_{2}[+,-]\\ \hat{b}_2[+,-]\end{array}\right)\oplus\left(\begin{array}{c}\Lambda^{\prime}_2[+,-]\\t^{\prime}_{2}[-,-]\\ b_2^{\prime}[+,-]\end{array}\right),
\end{eqnarray}
where the first and second three-vectors in the $SO(4)$ decomposition correspond to the $(\mathbf{3},\mathbf{1})$ and $(\mathbf{1},\mathbf{3})$ of $SO(4)$, respectively. In the first case, the first, second and third components of the three-vector correspond to $T_L^{3}=+1,0,-1$, respectively, while the second case is analogous with the interchange $T_L^3\leftrightarrow T_R^3$. Similarly to the MCHM$_5$, we consider the following brane action
\begin{eqnarray}
	\mathcal{S}_{\rm IR}&=&-\int\mathrm{d}^4x ~\left\{a(z)^4\left[M_{T}^q\mathrm{Tr}\left(\overline{\zeta}^{(\mathbf{3},\mathbf{1})}_{1L} \zeta^{(\mathbf{3},\mathbf{1})}_{2R}+\overline{\zeta}^{(\mathbf{1},\mathbf{3})}_{1L} \zeta^{(\mathbf{1},\mathbf{3})}_{2R}\right)\right.\right.\nonumber\\
												&+&\left.\left.M_B^q\mathrm{Tr}\left(\overline{\zeta}^{(\mathbf{2},\mathbf{2})}_{1L} \zeta^{(\mathbf{2},\mathbf{2})}_{2R}\right)\right]\right\}_{z=R^{\prime}}+\mathrm{h.c.},
\end{eqnarray}
where the  decompositions $\zeta_{1,2}=\zeta_{1,2}^{(\mathbf{1},\mathbf{3})}+\zeta_{1,2}^{(\mathbf{3},\mathbf{1})}+\zeta_{1,2}^{(\mathbf{2},\mathbf{2})}$ are explicitly defined in Appendix~\ref{subsec:adj}.
\subsubsection{MCHM$_{14}$, MCHM$_{5-14}$, MCHM$_{14-1}$}
\label{sec:14s}
\paragraph{MCHM$_{14}$}
We will consider first the case of two $\mathbf{14_{2/3}}$
\begin{eqnarray}
	\zeta_{1}&=&t^{\prime}_{1}[-,+]\oplus \left(\begin{array}{r}\tilde{\Lambda}_{1}[-,+]~  t_1[+,+]\\ \tilde{t}_{1}[-,+]~ b_1[+,+]\end{array}\right)\oplus \zeta_1^{(\mathbf{3},\mathbf{3})}[-,+] ,\\
 \zeta_{2}&=&t^{\prime}_{2}[-,-]\oplus\left(\begin{array}{r} \tilde{\Lambda}_{2}[+,-]~  t_2[+,-]\\ \tilde{t}_{2}[+,-]~ b_2[+,-]\end{array}\right) \oplus \zeta_2^{(\mathbf{3},\mathbf{3})}[+,-],\qquad
\end{eqnarray}
where this time, for the sake of simplicity, we have left implicit the components of the two $SO(4)$ bi-triplets $\zeta_{1,2}^{(\mathbf{3},\mathbf{3})}$. The IR brane mass terms read
\begin{eqnarray}
	\mathcal{S}_{\rm IR}&=&-\int\mathrm{d}^4x ~\left\{a(z)^4\left[M_{S}^q\mathrm{Tr}\left(\overline{\zeta}^{(\mathbf{1},\mathbf{1})}_{1L} \zeta^{(\mathbf{1},\mathbf{1})}_{2R}\right)+M_B^q\mathrm{Tr}\left(\overline{\zeta}^{(\mathbf{2},\mathbf{2})}_{1L} \zeta^{(\mathbf{2},\mathbf{2})}_{2R}\right)\right.\right.\nonumber\\
																	 &&\left.\left.+M_T^q\mathrm{Tr}\left(\overline{\zeta}^{(\mathbf{3},\mathbf{3})}_{1L} \zeta^{(\mathbf{3},\mathbf{3})}_{2R}\right)\right]\right\}_{z=R^{\prime}}+\mathrm{h.c.},
	\label{ir14}
\end{eqnarray}
where the explicit form of the fields $\zeta_{1,2}^{(\mathbf{1},\mathbf{1})}$, $\zeta_{1,2}^{(\mathbf{2},\mathbf{2})}$, and $\zeta_{1,2}^{(\mathbf{3},\mathbf{3})}$ can be obtained from Appendix~\ref{subsec:sym}. 

Analogously to Ref.~\cite{Falkowski:2007hz} one can compute the contribution of the top quark and its tower of resonances to the $gg\to H$ loop induced process. This is particularly relevant for Higgs physics at the LHC and can impose some constraint on the composite parameters of the model, especially in this case where fields transform in $\mathbf{14}s$ of $SO(5)$ and some sensitivity to the particular composite sector is expected \cite{Azatov:2011qy, Carena:2014ria}. To this end, we have computed the infinite sum $\sum_{n}vy_{nn}^{\rm top}/m_{n}^{\rm top}$, where $y_{nn}^{\rm top}$ are the diagonal top-like Yukawa couplings and $m_n^{\rm top}$ the corresponding (KK) masses, via the UV value of the full 5D propagator evaluated at zero-momentum, obtaining
\begin{eqnarray}
	\label{eq:KKsum}
	\left.\sum_n \frac{vy_{nn}^{\rm top}}{m_n^{\rm top}}\right|_{\rm MCHM_{14}}&=&\frac{\theta_{\pi}}{\cos\theta_{\pi}\sin\theta_{\pi}}\\
																																																&\times&\frac{3(M_S^q-M_T^q) \cos 2\theta_{\pi}+(5 M_S^q-8 M_B^q+3M_T^q)\cos4\theta_{\pi}}{3(M_S^q-M_T^q)+(5 M_S^q-8 M_B^q+3M_T^q)\cos2\theta_{\pi}},\label{sumq}\nonumber\end{eqnarray}
	where for the sake of simplicity we have introduced the shorthand notation $\theta_{\pi}\equiv v/f_{\pi}$. The corresponding cross-section normalized to the SM one is then given by 
\begin{eqnarray}
	\frac{\sigma(gg\to H)_{\rm MCHM_{14}}}{\sigma(gg\to H)_{\rm SM}}\approx\frac{|\kappa_g^{\rm MCHM_{14}}|^2+|\kappa_{g,5}^{\rm MCHM_{14}}|^2}{\kappa_v^2},
	\label{eq:gglu}
\end{eqnarray}
where $\kappa_v=v/v_{\rm SM}$ is the ratio of the 5D Higgs vev over the SM one and we have defined
\begin{eqnarray}
	\kappa_g^{\rm MCHM_{14}}=\mathrm{Re}\left[\sum_n \frac{vy_{nn}^{\rm top}}{m_n^{\rm top}}\right]_{\rm MCHM_{14}}, \quad \kappa_{g,5}^{\rm MCHM_{14}}=\frac{3}{2}~\mathrm{Im}\left[\sum_n \frac{vy_{nn}^{\rm top}}{m_n^{\rm top}}\right]_{\rm MCHM_{14}}.
\end{eqnarray}
To write (\ref{eq:gglu}) we have implicitly assumed $4 m_{\rm top}^2/m_H^2\to \infty$ and neglected the subleading bottom contribution \cite{Carena:2014ria}.  

For the lepton case, we consider two $\mathbf{14_{-1}}$ under $SO(5)\times U(1)_X$,
\begin{eqnarray}
	\xi_{1}&=&\tau^{\prime}_{1}[-,+]\oplus \left(\begin{array}{r}\nu_{1}[+,+]~  \tilde{\tau}_1[-,+]\\ \tau_{1}[+,+]~ \tilde{Y}_1[-,+]\end{array}\right)\oplus \xi^{(\mathbf{3},\mathbf{3})}_1[-,+] ,\\
 \xi_{2}&=&\tau^{\prime}_{2}[-,-]\oplus\left(\begin{array}{r} \nu_{2}[+,-]~  \tilde{\tau}_2[+,-]\\ \tau_{2}[+,-]~ \tilde{Y}_2[+,-]\end{array}\right) \oplus \xi^{(\mathbf{3},\mathbf{3})}_2[+,-],\qquad
\end{eqnarray}
with an IR action analogous to the one of (\ref{ir14}) with brane masses $M_S^l, M_B^l$ and $M_T^l$. We can also compute the analogous to (\ref{eq:KKsum}) for the lepton case, which can be relevant for instance for $H\to \gamma \gamma$, obtaining the very same expression as in the quark case but with the change $\{M_S^q,M_B^q,M_T^q\}\to\{M_S^l,M_B^l,M_T^l\}$.

\paragraph{MCHM$_{5-14}$}
This case can be obtained easily from the previous, changing the first $\mathbf{14}$ for a $\mathbf{5}$, i.e., 
\begin{eqnarray}
\zeta_{1}&=&t^{\prime}_{1}[-,+]\oplus \left(\begin{array}{r}\tilde{\Lambda}_{1}[-,+]~  t_1[+,+]\\ \tilde{t}_{1}[-,+]~ b_1[+,+]\end{array}\right),\\
\zeta_{2}&=&t^{\prime}_{2}[-,-]\oplus\left(\begin{array}{r} \tilde{\Lambda}_{2}[+,-]~  t_2[+,-]\\ \tilde{t}_{2}[+,-]~ b_2[+,-]\end{array}\right) \oplus \zeta^{(\mathbf{3},\mathbf{3})}_2[+,-],\qquad
\end{eqnarray}
with the following IR action\footnote{For convenience, we have added prefactors $-\sqrt{5}/2$ and $-\sqrt{2}$ to the boundary masses $M_S^q$ and $M_B^q$, respectively, in order to compensate group factors that will otherwise appear along with the brane masses mixing the canonically normalized fermions (making for instance the brane mass mixing the bidoublets slightly smaller than the one mixing the singlets). \label{fot:pref}  }  
\begin{eqnarray}
	\mathcal{S}_{\rm IR}&=&\int\mathrm{d}^4x\int_R^{R^{\prime}}\mathrm{d} z\left\{a^4(z)\left[\frac{\sqrt{5}}{2}M_S^q \left(\bar{\zeta}_{1L}^{(\mathbf{1},\mathbf{1})}\zeta_{2R}^{(\mathbf{1},\mathbf{1})}\right)_{5}+\sqrt{2}M_B^q\left(\bar{\zeta}_{1L}^{(\mathbf{2},\mathbf{2})}\zeta_{2R}^{(\mathbf{2},\mathbf{2})}\right)_{5}\right]\right\}_{z=R^{\prime}}\nonumber\\
	&+&\mathrm{h.c.}.
	\label{ir14p5}
\end{eqnarray}

Similarly to the previous case, with two $\mathbf{14}s$, we can compute the top tower contribution to $gg\to H$, obtaining 
\begin{eqnarray}
   \label{14p5sum}
	&&\sqrt{2} \theta_{\pi} \left\{\left[\sqrt{10} \cot \theta_{\pi} \left((5 \cos 2\theta_{\pi}+3) M_S^{q\ast} \left(\cos 2\theta_{\pi} \left(5 \sqrt{2} M_S^q-4 \sqrt{5}
   M_B^q\right)-5 \sqrt{2} M_S^q\right)\right.\right.\right.\nonumber\\
   && \left.\left.\left.-40 \cos ^2\theta_{\pi} M_B^{q\ast} \left(\cos 2\theta_{\pi} \left(\sqrt{5} M_S^q-2 \sqrt{2}
   M_B^q\right)-\sqrt{5} M_S^q\right)\right)\right]\times\right.\nonumber\\
   &&\left.\left[(5 \cos 2\theta_{\pi}+3) M_S^{q\ast} \left(5 \cos 2\theta_{\pi} \left(\sqrt{10} M_S^q-4
   M_B^q\right)-20 M_B^q+3 \sqrt{10} M_S^q\right)\right.\right.\nonumber\\
   &&\left.\left.-40 \cos ^2\theta_{\pi} M_B^{q\ast} \left(M_S^q (5 \cos 2\theta_{\pi}+3)-4 \sqrt{10}
   M_B^q \cos ^2\theta_{\pi}\right)\right]^{-1}\right.\\
   &&\left.+\left[(5 \cos 2\theta_{\pi}+3) M_S^{q\ast} \left(8 M_S^q \cot \theta_{\pi}-\sin 2\theta_{\pi} \left(5 M_S^q-2 \sqrt{10}
   M_B^q\right)\right)\right.\right.\nonumber\\
   &&\left.\left.-4 \cos ^2\theta_{\pi} M_B^{q\ast} \left(40 M_B^q \sin \theta_{\pi} \cos \theta_{\pi}+\sqrt{10} M_S^q (5 \cos 2\theta_{\pi}+3) \cot
   \theta_{\pi}\right)\right]\times\right.\nonumber\\
   &&\left.\left[(5 \cos 2\theta_{\pi}+3) M_S^{q\ast} \left(\sqrt{2} M_S^q (5 \cos 2\theta_{\pi}+3)-8 \sqrt{5} M_B^q \cos ^2\theta_{\pi}\right)\right.\right.\nonumber\\
   &&  \left.\left.-8 \cos ^2\theta_{\pi}
   M_B^{q\ast} \left(\sqrt{5} M_S^q (5 \cos 2\theta_{\pi}+3)-20 \sqrt{2} M_B^q \cos ^2\theta_{\pi}\right)\right]\right\}=\left.\sum_n \frac{vy_{nn}^{\rm top}}{m_n^{\rm top}}\right|_{\rm MCHM_{5-14}}.\nonumber
\end{eqnarray}

It is again straightforward to write down the lepton case, where we consider the following $SO(5)$ multiplets, 
\begin{eqnarray}
\xi_{1}&=&\tau^{\prime}_{1}[-,+]\oplus \left(\begin{array}{r}\nu_{1}[+,+]~  \tilde{\tau}_1[-,+]\\ \tau_{1}[+,+]~ \tilde{Y}_1[-,+]\end{array}\right),\\
\xi_{2}&=&\tau^{\prime}_{2}[-,-]\oplus\left(\begin{array}{r} \nu_{2}[+,-]~  \tilde{\tau}_2[+,-]\\ \tau_{2}[+,-]~ \tilde{Y}_2[+,-]\end{array}\right) \oplus \xi^{(\mathbf{3},\mathbf{3})}_2[+,-].\qquad
\end{eqnarray}
Analogous expressions to (\ref{ir14p5}) and (\ref{14p5sum}) hold in this case with the corresponding changes $\{M_S^q,M_B^q\}\to \{M_S^l,M_B^l\}$.
\paragraph{MCHM$_{14-1}$}
In this case the RH top is embedded in a full singlet of $SO(5)\times U(1)_X$, $\sim \mathbf{1_{2/3}}$, while the $SU(2)_L$ doublet $q_L$ will live in $\mathbf{14_{2/3}}$,
\begin{eqnarray}
	\zeta_{1}&=&t^{\prime}_{1}[-,+]\oplus \left(\begin{array}{r}\tilde{\Lambda}_{1}[-,+]~  t_1[+,+]\\ \tilde{t}_{1}[-,+]~ b_1[+,+]\end{array}\right)\oplus \zeta^{(\mathbf{3},\mathbf{3})}_1[-,+] ,\\
\zeta_{2}&=&t^{\prime}_{2}[-,-].
\end{eqnarray}
As was pointed out in \cite{Pappadopulo:2013vca}, this setup has an accidental  $SU(9+4)=SU(13)$ global symmetry on the IR brane (coming from the fact that the $\mathbf{9_{2/3}}=\mathbf{(3,3)}_{\mathbf{2/3}}$ and the $\mathbf{4_{2/3}}=\mathbf{(2,2)}_{\mathbf{2/3}}$ $\subset \mathbf{14}$ can be rotated as a single multiplet), leading in particular to a vanishing contribution to the $\sin^2(h/f_\pi)$ coefficient of the Higgs effective potential. To break this symmetry, and be able to explore more general scenarios, we will introduce, also on the IR brane, fermion kinetic terms for both $SO(4)$ multiplets, $\mathbf{4_{2/3}}$ and  $\mathbf{9_{2/3}}$,
\begin{eqnarray}
	\mathcal{S}_{\rm IR}&=&\int\mathrm{d}^4x ~\left\{a(z)^4\left[M_{S}^q\overline{\zeta}^{(\mathbf{1},\mathbf{1})}_{1L} \zeta_{s2R}^{(\mathbf{1},\mathbf{1})}+\kappa_B^q R\mathrm{Tr}\left(\overline{\zeta}^{(\mathbf{2},\mathbf{2})}_{1L}i\cancel{\partial} \zeta^{(\mathbf{2},\mathbf{2})}_{1L}\right)\right.\right.\nonumber\\
																		   &&\left.\left.+\kappa_T^q R\mathrm{Tr}\left(\overline{\zeta}^{(\mathbf{3},\mathbf{3})}_{1L}i\cancel{\partial} \zeta^{(\mathbf{3},\mathbf{3})}_{1L}\right)\right]\right\}_{z=R^{\prime}}+\mathrm{h.c.}.
\end{eqnarray}

In this particular case, the sensitivity to the composite parameters in $gg\to H$ disappears, getting
\begin{eqnarray}
	\left.\sum_n \frac{vy_{nn}^{\rm top}}{m_n^{\rm top}}\right|_{\rm MCHM_{14-1}}=\theta_{\pi}\left(\cot\theta_{\pi}-\tan\theta_{\pi}\right).
\end{eqnarray}
\subsubsection{The mMCHM$^{\rm III}$: A New Minimal Model for Leptons}
\label{sec:mMCHM3}
Taking a closer look to the structure of the symmetric representation of $SO(5)$, $\mathbf{14}=(\mathbf{1},\mathbf{1})\oplus (\mathbf{2},\mathbf{2})\oplus (\mathbf{3},\mathbf{3}) $, one can readily see that it is the only one which can host at the same time a $P_{LR}$ protected $SU(2)_L \times U(1)_Y$ singlet and a triplet $\sim \mathbf{3}_0$. This feature implies in particular that, using this representation, we can build very minimal models in the lepton sector generating the neutrino masses through a type-III seesaw.  In the following, we will consider the most minimal of these scenarios, which we have called mMCHM$^{\rm III}$, realized with left- and right-handed leptons transforming as $\mathbf{5_{-1}}$ and $\mathbf{14_{-1}}$, respectively, under $SO(5)\times U(1)_X$ and the following boundary conditions 
\begin{eqnarray}
	\xi_{1\tau}&=&\tau^{\prime}_{1}[-,+]\oplus \left(\begin{array}{r}\nu_{1}^{\tau}[+,+]~  ~\tilde{\tau}_1[-,+]\\ \tau_{1}[+,+]~ \tilde{Y}_1^{\tau}[-,+]\end{array}\right),\\
 \xi_{2\tau}&=&\tau^{\prime}_{2}[-,-]\oplus\left(\begin{array}{r} \nu_{2}^{\tau}[+,-]~~  \tilde{\tau}_2[+,-]\\ \tau_{2}[+,-]~ \tilde{Y}_2^{\tau}[+,-]\end{array}\right) \oplus\left(\begin{array}{r} \hat{\lambda}^{\tau}_2[-,-]~~\nu_{2}^{\tau\prime\prime}[+,-]~  ~~\tau_2^{\prime\prime\prime}[+,-]\\ 
	\hat{\nu}_2^{\tau}[-,-]~~~\tau_{2}^{\prime\prime}[+,-]~Y_2^{\tau\prime\prime\prime}[+,-]\\ \hat{\tau}_2[-,-]~Y_2^{\tau\prime\prime}[+,-]~\Theta_2^{\tau\prime\prime\prime}[+,-] \end{array}\right).\qquad
\end{eqnarray}
Here, we have just shown the two multiplets for the third generation, the ones for the first two generations are completely analogous. At the level of zero-modes we have the $SU(2)_L\times U(1)_Y$ spectrum  $l_{\ell L}^{(0)}\sim \mathbf{2_{-1/2}}\subset \xi_{1\ell}$ and $\ell_R^{(0)}\sim\mathbf{1_{-1}},\Sigma_{\ell R}^{(0)}\sim \mathbf{3_0}\subset \xi_{2\ell }$, with $\ell =e,\mu,\tau$, which is the typical matter content for a 4D realization of the type-III seesaw \cite{Foot:1988aq}.  

In this case, we can write down the following UV Majorana mass term, 
\begin{eqnarray}
	\mathcal{S}_{\rm UV}=-\frac{1}{2}\sum_{\ell =e,\mu,\tau}\int\mathrm{d}^4x\int_{R}^{R^{\prime}}\mathrm{d}z\left\{a^4(z)M_{\Sigma}^{\ell}\mathrm{Tr}\left(\bar{\Sigma}_{\ell R} \Sigma_{\ell R}^c\right)\right\}_{z=R}+\mathrm{h.c.},
\end{eqnarray}
where 
\begin{eqnarray}
\Sigma_{\ell}=\begin{pmatrix}\hat{\nu}_2^\ell/\sqrt{2}&\hat{\lambda}_2^\ell\\ \ell_2& -\hat{\nu}_2^\ell/\sqrt{2}\end{pmatrix}, \qquad \ell=e,\mu,\tau,
\end{eqnarray}
are the 5D $SU(2)_L\times U(1)_Y$ triplets hosting the previously introduced $\Sigma_{\ell R}^{(0)}$ zero-modes. On the other hand, the IR brane masses read
\begin{eqnarray}
	\mathcal{S}_{\rm IR}&=&\sum_{\ell =e,\mu,\tau}\int\mathrm{d}^4x \left\{a(z)^4\left[\frac{\sqrt{5}}{2}M_{S}^{\ell}\left(\overline{\xi}^{(\mathbf{1},\mathbf{1})}_{1\ell L} \xi^{(\mathbf{1},\mathbf{1})}_{2\ell R}\right)_{5}+\sqrt{2}M_B^{\ell}\left(\overline{\xi}^{(\mathbf{2},\mathbf{2})}_{1\ell L} \xi^{(\mathbf{2},\mathbf{2})}_{2jR}\right)_{5}\right]\right\}_{z=R^{\prime}}+\mathrm{h.c.},\nonumber\\
	\label{ir14p5maj}
\end{eqnarray}
where, for the sake of simplicity, as we are just interested at the moment in the size of the contribution to the Higgs potential, we have assumed that all brane masses $M_{\Sigma}^{\ell},M_S^{\ell}$ and $M_B^{\ell}$ are flavor diagonal. 

The underlying reason for considering all three generations  in this case rather than just the third one is related to the size of the effective Majorana mass, i.e., the Majorana mass for the corresponding zero-modes
\begin{eqnarray}
	\mathcal{M}_{\rm M}^{\ell \ell^{\prime}}\approx \frac{f_{-c_2^{\ell}}^2}{R^{\prime}}\left(\frac{R^{\prime}}{R}\right)^{-2c_2^{\ell^{\prime}}}M_{\Sigma}^{\ell}\delta_{\ell \ell^{\prime}}, \qquad \ell ,\ell^{\prime}\in\{e,\mu,\tau\}.
\end{eqnarray}
This mass is typically too large unless the corresponding zero-mode profiles are pushed away from the UV brane. This leads to values of  $c_2^{\ell}\in (-1/2,0)$ and thus IR localized RH zero-modes. Therefore, we can see that in this scenario, just the quantum numbers of the lepton sector and the overall scale of the neutrino masses lead naturally to IR localized leptons for all three generations. As we will see below, this will allow us to compensate the relative color suppression of the lepton sector in the contribution to the Higgs potential,  making this setup particularly interesting for lifting the masses of the top partners.

\subsection{The One-Loop Higgs Mass}
\label{sec:MH}
We now move forward to derive the Higgs potential and in particular the Higgs mass in the explicit 5D models.
After some discussion on the matching to $m_H$, we review the calculation of the Coleman-Weinberg Potential in 
extra dimensional theories and finally present explicit results for the models studied in this work.

\subsubsection{The Higgs Mass at $f_\pi$}
\label{sec:mHfpi}

In the following, we want to estimate the value of the Higgs mass that should be matched to in the calculation of the 1-loop Higgs potential at the scale $f_\pi$ for the models under consideration.
Neglecting higher dimensional operators with $D\geq 6$, whose effect should be suppressed, the potential can be written as
\begin{equation}
V(h)=\alpha \sin^2(h/f_\pi) - \beta \sin^2(h/f_\pi) \cos^2(h/f_\pi) \simeq \mu^2 h^2 + \lambda h^4 + {\cal O}(h^6),
\end{equation}
where $\mu$ and $\lambda$ are functions of $\alpha$ and $\beta$.
Now these coefficients are fixed by the Higgs mass $m_H$ and the Higgs vev $v$. 
In particular,
\begin{equation}
\lambda = m_H^2/(8 v^2)\,,
\end{equation}
which means that we can use this standard relation to obtain the running of the Higgs mass from the running of the quartic as $m_H(\mu) =  \sqrt{8 \lambda(\mu)} v$.

Along the same lines, the running of $\lambda$ should be given to first approximation by the running of the $D\leq 4$ part of the Lagrangian, {\it i.e.}, the SM.\footnote{Although, approaching the scale $\mu = f_\pi$ the renormalization group evolution will receive modifications, most of the effect will be due to the SM, for $f_\pi \gg m_H$.} Setting $f_\pi \sim 1$\,TeV and employing the SM result (see e.g. \cite{Dev:2013ff})
\begin{equation}
\frac{\lambda(1\,{\rm TeV})}{\lambda(m_H=125\,{\rm GeV})} \approx 0.7,
\end{equation}
we finally arrive at a $\sim 20 \%$ correction
\begin{equation}
m_H(f_\pi \sim 1\,{\rm TeV}) \approx \sqrt{\frac{\lambda(1\,{\rm TeV})}{\lambda(m_H=125\, {\rm GeV})}}\, m_H\ \approx 105\,{\rm GeV}.
\end{equation}
Motivated by this result, we will employ $m_{H}(f_\pi)=105$\,GeV$\,(1 \pm 7.5 \%)$ in the following numerical analysis, accounting for the uncertainties in the running in a conservative way.
Thus, the top mass and the Higgs mass are treated on the same footing, evaluated at $f_\pi$.

\subsubsection{Coleman-Weinberg Potential in KK Theories}
The contribution to the one-loop Coleman-Weinberg potential of a particular KK tower can be written as follows
\begin{eqnarray}
	V(h)\supset \frac{N}{2}\sum_{n=1}^{\infty}\int \frac{\mathrm{d}^4 p}{(2\pi)^4}\log\left(p^2+m_n^2(h)\right),
\end{eqnarray}
where $N$  is the number of degrees of freedom of the corresponding type of resonance -- $N=+3$ for gauge bosons and $N=-4$ for fermions -- and $m_n(h)$, with $n\in\mathbb{N},$ are their masses in the presence of the Higgs background. It can be shown, see e.g. \cite{Oda:2004rm, Falkowski:2006vi},  that the previous infinite sum can be exchanged by an integral on the Minkowski space, leading to
\begin{eqnarray}
	V(h)=\sum_r \frac{N_r}{(4\pi)^2}\int_0^{\infty}\mathrm{d}p~p^3\log \rho_r(-p^2),
\end{eqnarray}
where $r$ sums over the different KK towers of the model and $\rho_r(w^2),~ w\in\mathbb{C},$ are some spectral functions, holomorphic in the $\mathrm{Re} (w)>0$ part of the complex plane and with roots in the real axis encoding the physical spectrum, i.e., 
\begin{eqnarray}
	\rho_r(m_{n;r}^2(h))=0, \quad n\in \mathbb{N}.
\end{eqnarray}
In models of GHU, where the Higgs background can be removed except for one brane through the transformations (\ref{gtf}) and (\ref{gtv}), it is actually possible to compute these spectral functions and therefore to have a calculable and finite Higgs potential, 
as these functions will go to zero fast enough for large values of $p^2$. We will focus on them in the following subsection.

\subsubsection{The Higgs Mass in the MCHMs}
We will consider first the contribution of the gauge boson sector, which is model independent for our choice of symmetry breaking $SO(5)\to SO(4)$. Then we will study the cases of the different fermion representations considered in this work. We would like to remark that the corresponding form factors will include obviously in all cases subleading effects of the  full mixing with the heavy resonances.
\paragraph{Gauge Bosons}
The form factors for gauge bosons  are well known in the literature \cite{Csaki:2008zd}. With our notation and the UV brane kinetic terms of (\ref{bkgauge}) they read 
\begin{eqnarray}
	\rho_{Z,W}(p^2)=1+f_{Z,W}(p^2)\sin^2\left(h/f_{\pi}\right),
\end{eqnarray}
with
\begin{eqnarray}
	f_Z(p^2)&=&\frac{p}{2}\left(\frac{R^{\prime}}{R}\right) \left[(1+s_{\phi}^2)C^{\prime}(p,R^{\prime})-p R\log(R^{\prime}/R)(\kappa^{\prime 2}+\kappa^2 s_{\phi}^2)S^{\prime}(p,R^{\prime})\right]\times \nonumber\\
				  &&\left[S(p,R^{\prime})(C^{\prime}(p,R^{\prime})-p R \log(R^{\prime}/R) \kappa^2  S^{\prime}(p,R^{\prime}))\right.\nonumber\\
		 &&\left.\qquad \qquad (C^{\prime}(p,R^{\prime})-p R \log(R^{\prime}/R) \kappa^{\prime 2}  S^{\prime}(p,R^{\prime}))\right]^{-1},\\
	f_W(p^2)&=&\frac{p}{2}\left(\frac{R^{\prime}}{R}\right)\frac{1}{S(p,R^{\prime})(C^{\prime}(p,R^{\prime})-p R \log(R^{\prime}/R) \kappa^2 S^{\prime}(p,R^{\prime}))}.
\end{eqnarray}
In the expressions above, $C(m,z)$ and $S(m,z)$ are functions satisfying the bulk equations of motion for gauge bosons and a vanishing Higgs vev, with boundary conditions $C(m,R)=1,~ C^{\prime}(m,R)=0,~S(m,R)=0,~ S^{\prime}(m,R)=m$, where the prime denotes a derivative with respect to the extra-dimensional coordinate. They are given by
\begin{eqnarray}
	C(m,z)&=&\frac{\pi}{2}mz\left[Y_0(mR)J_1(mz)-J_0(mR)Y_1(mz)\right],\\
	S(m,z)&=&\frac{\pi}{2}mz\left[Y_1(mz)J_1(mR)-J_1(mz)Y_1(mR)\right].
\end{eqnarray}
We obtain
\begin{eqnarray}
	V_{g}(h)&=&\frac{3}{(4\pi)^2}\int_0^{\infty}\mathrm{d}p~p^3\left[\log \rho_Z(-p^2)+2\log\rho_W(-p^2)\right]\nonumber\\
									&=&\frac{3}{32 \pi^2}\int_0^{\infty}\mathrm{d}t~t\left[\log \rho_Z(-t)+2\log\rho_W(-t)\right],
	\label{exgauge}
\end{eqnarray}
after performing the change of variables $t=p^2$. The above potential can be written to very good approximation as
\begin{eqnarray}
	V(h) &\approx &\alpha_g\sin^2(h/f_{\pi})-\beta_g\sin^2(h/f_{\pi})\cos^2(h/f_{\pi}),\nonumber
\end{eqnarray}
where we have defined
\begin{eqnarray}
	\alpha_{g}&=&\frac{3}{32 \pi^2}\int_{\mu_{\rm IR}}^{\infty}\mathrm{d}t~t\left\{\left[f_Z(-t)-\frac{1}{2}\left(f_Z(-t)\right)^2\right]+2\left[f_W(-t)-\frac{1}{2}\left(f_W(-t)\right)^2\right]\right\},	\quad \\
	\beta_{g}&=&-\frac{3}{64 \pi^2}\int_{\mu_{\rm IR}}^{\infty}\mathrm{d}t~t\left\{\left(f_Z(-t)\right)^2+2\left(f_W(-t)\right)^2\right\},	\quad 
\end{eqnarray}
and introduced an IR regulator $\mu_{\rm IR}$ in order to cure the spurious IR divergences arising from the former expansion \cite{Csaki:2008zd, Archer:2014qga}. Anyway, for the calculations perfomed in this work we just used the exact expression (\ref{exgauge}) as well as its analogues for the fermion case.

\paragraph{Fermions} Although the case of fermions is very model dependent, we can still perform some general discussion. Contrary to the gauge boson form factors, in the fermion case we will have in general further powers of $\sin^2(h/f_{\pi})$. In particular, for all the scenarios we consider in this work 
\begin{eqnarray}
	\rho_{f}(p^2)&=&1+f_2^f(p^2)\sin^2(h/f_{\pi})+f_4^f(p^2)\sin^4(h/f_{\pi})\nonumber\\
						  &+&f_6^f(p^2)\sin^6(h/f_{\pi})+f_8^f(p^2)\sin^8(h/f_{\pi}), \quad f\in\{t,e,\mu,\tau\},
\end{eqnarray}
where some of the functions $f_k^f(p^2)$, with $k\in\{2,4,6,8\}$, may be identically zero. We have computed all of them in a similar way to the previous case but their explicit expressions are too involved to justify the inclusion in this work. The corresponding contributions to the Higgs potential read
\begin{eqnarray}
	\hspace{-6mm} V_f(h)\!&=&\!-\frac{4}{32\pi^2}N_f\!\int_0^{\infty}\!\!\mathrm{d}t~t\log\rho_f(-t),
	\label{exferm}
\end{eqnarray}
where $N_f$ is a possible color factor, i.e., $N_t=3$ and $N_{e,\mu,\tau}=1$. Again, one could use with very good approximation
\begin{eqnarray}
	V(h)\approx \alpha_f\sin^2(h/f_{\pi})-\beta_f \sin^2(h/f_{\pi})\cos^2(h/f_{\pi}),
\end{eqnarray}
where we have neglected terms of $\mathcal{O}(\sin^6(h/f_{\pi}))$ and $\mathcal{O}(\sin^8(h/f_{\pi}))$ and defined
\begin{eqnarray}
	\alpha_f&=&-\frac{4}{32 \pi^2}N_f\int_{\mu_{\rm IR}}^{\infty}\mathrm{d}t~t\left[f_2^f(-t)+f_4^f(-t)-\frac{1}{2}(f_2^f(-t))^2\right],
	\quad\\
	\beta_f&=&-\frac{4}{32 \pi^2}N_f\int_{\mu_{\rm IR}}^{\infty}\mathrm{d}t~t \left[f_4^f(-t)-\frac{1}{2}(f_2^f(-t))^2\right]. 
\end{eqnarray}
However, as already mentioned, in practice we will just consider the exact expression in equation (\ref{exferm}).

\subsection{Lifting Light Partners with Leptons: Numerical Results and Discussion}
In order to be able to quantify the effect of leptons on the Higgs mass and the scale of the top partners we have performed a numerical scan focusing on two main scenarios considered in this work, namely, non-minimal models of composite leptons arising from a flavor protection mechanism featuring a composite $\tau_R$, and minimal models of type-III seesaw where brane masses can in principle be large and all three lepton generations have an impact. In both cases, we consider $R=10^{-16}$ TeV$^{-1}$ and fixed values of $f_{\pi}\lesssim 1$\,TeV, as well as $\kappa=\kappa^{\prime}=0$, which correspond to $g_{\ast}\approx 4.0$ and $s_{\phi}\approx \tan\theta_W$. We also scan uniformly $\kappa,\kappa^{\prime}\in[0,\sqrt{3}]$, which translates to  $g_{\ast}\sim[4.,8.]$ (or, equivalently, $N_{\rm CFT}\sim[2.5,9.8]$) and $s_{\phi}\sim [1/2\tan\theta_W,2\tan\theta_W]$. Regarding the fermion sector, we have assumed random complex brane masses 
\begin{eqnarray}
	| M_S^q|,|M_B^q|,|M_T^q|\le Y_\ast^q,\qquad | M_S^l|,|M_B^l|,|M_T^l|\le Y_\ast^l,\qquad M_S^{q,l},M_{B}^{q,l},M_{T}^{q,l}\in\mathbb{C},\quad
	\label{onemass}
\end{eqnarray}
in the first case, while we have taken real brane lepton masses in the second one for simplicity, 
\begin{eqnarray}
	|M_{\Sigma}^\ell|,| M_S^\ell|,|M_B^\ell|\le Y_\ast^l, \quad M_{\Sigma}^\ell,M_{S}^\ell,M_{B}^\ell\in \mathbb{R},  \quad \mathrm{with}\quad\ \ell=e,\mu,\tau,
	\label{twomass}
\end{eqnarray}
(with the quark brane masses analogous to (\ref{onemass})) as the introduction of the Majorana masses $M_{\Sigma}^\ell$ would otherwise double the size of the system of equations that we have to solve.\footnote{Which is already $10\times10$ in this case.} The numbers $Y_\ast^{q},Y_\ast^l\in\mathbb{R}^{+}$ are fixed to some benchmark values specified below. The quark bulk masses $c_1^q$ and $c_2^q$ were fixed in both cases requiring
\begin{eqnarray}
	\left.	\frac{\partial V(h)}{\partial h}\right|_{h=v}=0, \qquad m_{t}=m_t^{\rm ref},
\end{eqnarray}
with\footnote{Remember that $v$ was fixed from $\{\alpha_{\rm QED},M_W,M_Z\}$ together with $g_{\ast}$ and $s_{\phi}$. }
\begin{eqnarray}
V(h)=V_g(h)+\sum_f V_f(h), \qquad m_t^{\rm ref}\in [145, 155] ~\mathrm{GeV},
\end{eqnarray}
while the procedure of fixing the lepton bulk masses in each case differs. In the scenario of $\tau_R$ compositeness, where the neutrino sector is left unspecified and the only constraint is $m_{\tau}=1.7$ GeV, we randomly scan over $c_2^\tau \in[-0.5,0.5]$, fixing the other bulk mass via the $\tau$ mass. However, when considering minimal type-III seesaw models, we fix for each generation both bulk masses from the corresponding charged lepton and neutrino masses, imposing for the latter  
\begin{eqnarray}
	m_{\nu}^\ell=m_{\nu}^{\ell;\mathrm{ref}},\qquad \mathrm{with}\quad m_{\nu}^{\ell;\mathrm{ref}}=\varepsilon_\ell 10^{-p_\ell}~\mathrm{eV}\quad \mathrm{and}\quad  \varepsilon_\ell\in[0,1],~p_\ell\in[0,3].
\end{eqnarray}

We have ensured that all points in our random scan are in reasonable agreement with electroweak precision data by choosing $f_{\pi}=0.8$ and $1.0$ TeV, since
\begin{eqnarray}
	T=0,\qquad S\approx \frac{3}{2}\pi v^2R^{\prime 2}=6\pi \left(\frac{v}{f_{\pi}}\right)^2g_{\ast}^{-2}, \qquad U=0,
\end{eqnarray}
in these models at tree level \cite{Agashe:2004rs, Csaki:2008zd, Archer:2014qga} and $f_{\pi}=0.8~(1.0)~\mathrm{TeV}$ implies in particular 
\begin{eqnarray}
 S\in[0.028~(0.018), 0.112~(0.071)].
\end{eqnarray}
In general, although $\left.S\right|_{U=0}=0.05\pm 0.09$ at $95\%$ confidence level \cite{Baak:2012kk}, the correlation with the predicted value of the $T$ parameter is large,  $\rho_{\rm corr.}=0.91$, which would in principle disfavor a large portion of the 
values above. We should however keep in mind the limitations of this analysis, as we are neglecting non-oblique effects as well as radiative fermion corrections which can give an non-negligible positive contribution to the $T$ parameter \cite{Anastasiou:2009rv}. In addition to this, we have also checked that $|\delta g_{Z \bar{\ell} \ell}/g_{Z \bar{\ell}\ell}|\leq 2\permil$, with $\ell=e,\mu,\tau,$ for all points on the scan.

Finally, before going to the specific results, let us briefly comment on how we measured the fine-tuning of the models at hand. Although a full examination of the tuning of these models including all possible brane kinetic terms is well  beyond the scope of this paper, see e.g  \cite{Archer:2014qga} for the case of the MCHM$_5$, we still wanted to have a qualitative idea of the amount of tuning required by these setups, as well as being able to compare them with other scenarios lifting the masses of the top partners. To this end, we have computed the Barbieri-Giudice measure of the tuning \cite{Barbieri:1987fn}, defined as
\begin{eqnarray}
	\Delta_{\rm BG}=\mathrm{max}_{i,j}\left|\frac{\partial \log \mathcal{O}_i}{\partial X_j}\right|,
	\end{eqnarray}
	where $\mathcal{O}_i$ are the observables considered and $X_j$ the different inputs of the model. In particular, for every point of our numerical scan, we have computed the derivatives of the logarithm of both the $W$ and the Higgs mass\footnote{Note that the condition $\left.\partial V(h)/\partial h\right|_{h=v}=0$ defines  $v$ and thus the W mass as a function of all other variables.} imposing EWSB ({\it i.e.}, $\left.\partial V(h)/\partial h\right|_{h=v}=0$) with respect of all variables of the model and taken the maximum.  

	Before checking explicitly the impact of leptons in both scenarios, let us first review the behavior of the different quark representations that we are considering.

	\subsubsection{Minimal Quark Setups}

\begin{figure}[t!]
\begin{center} 
	\subfigure[ $Y_\ast^q=0.7$  (left plot) and $Y_\ast^q=1.4$ (right plot) in the MCHM$_5$ with $f_{\pi}=0.8$\,TeV and $\kappa=\kappa^{\prime}=0.$]{%
		   \includegraphics[width=0.43\textwidth]{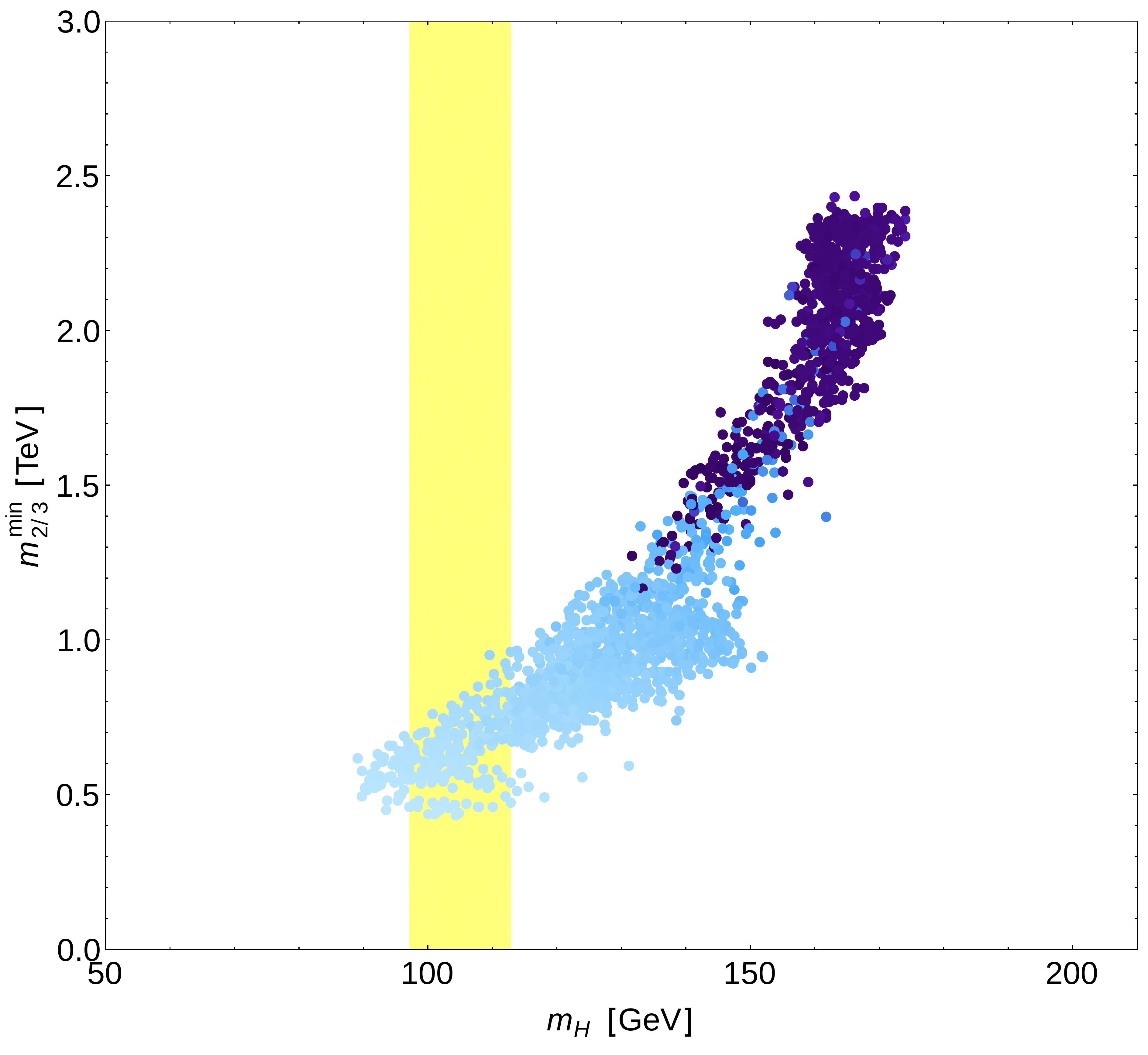}
		   \includegraphics[width=0.0865\textwidth]{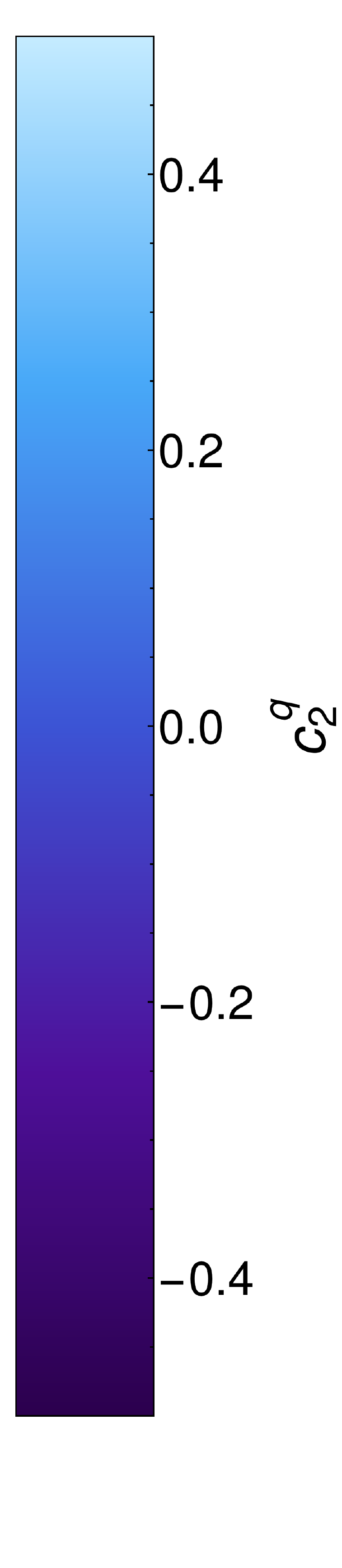}
			\includegraphics[width=0.43\textwidth]{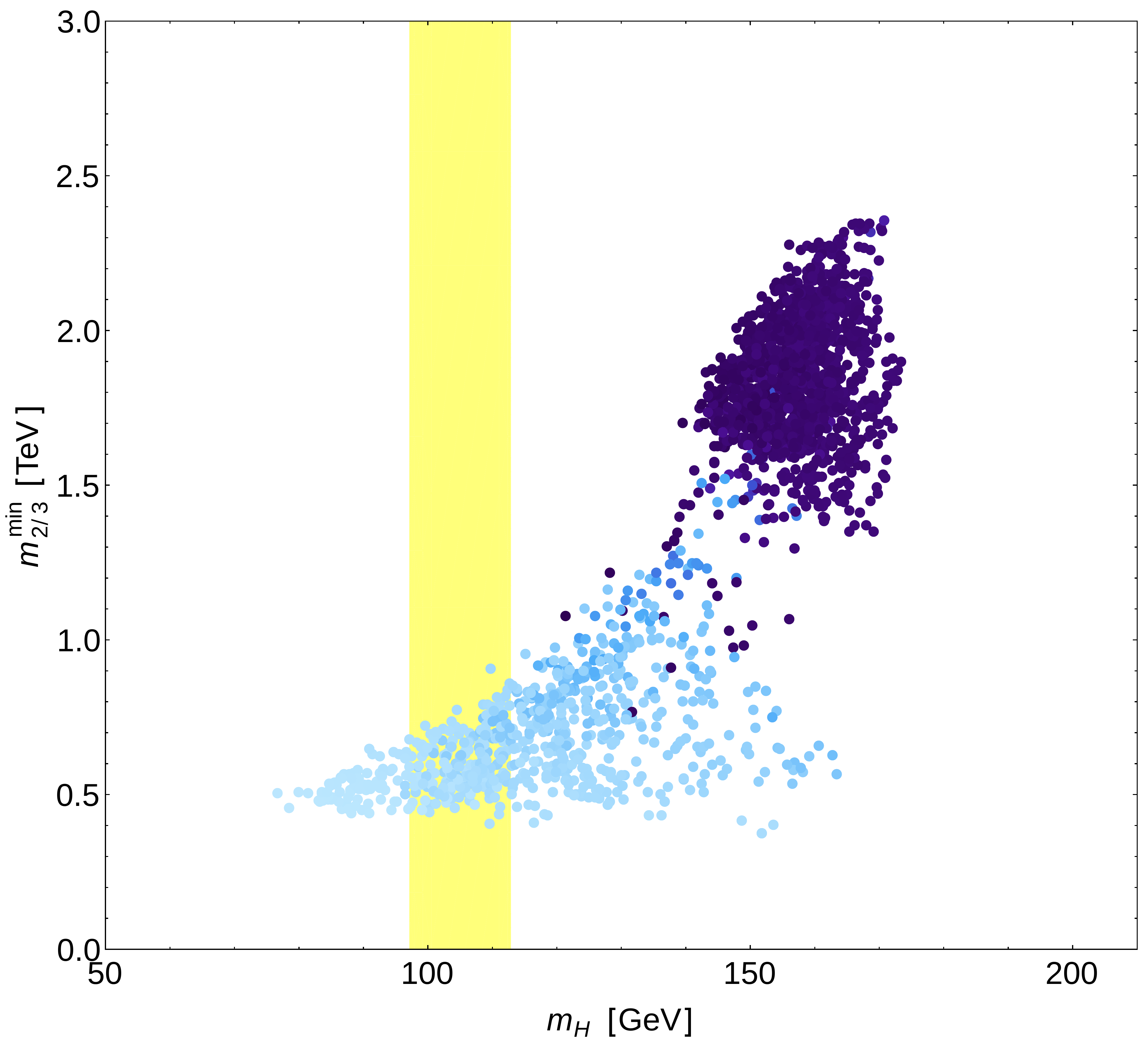}
			 \includegraphics[width=0.0865\textwidth]{figures/bar_cq2.pdf}
			 \label{fig:mchm5mcusvsmh}
		}
\\
\subfigure[$Y_\ast^q=0.7$  (left plot) and $Y_\ast^q=1.4$ (right plot) in the MCHM$_{10}$ with $f_{\pi}=0.8$\,TeV and $\kappa=\kappa^{\prime}=0.$ ]{%
		   \includegraphics[width=0.43\textwidth]{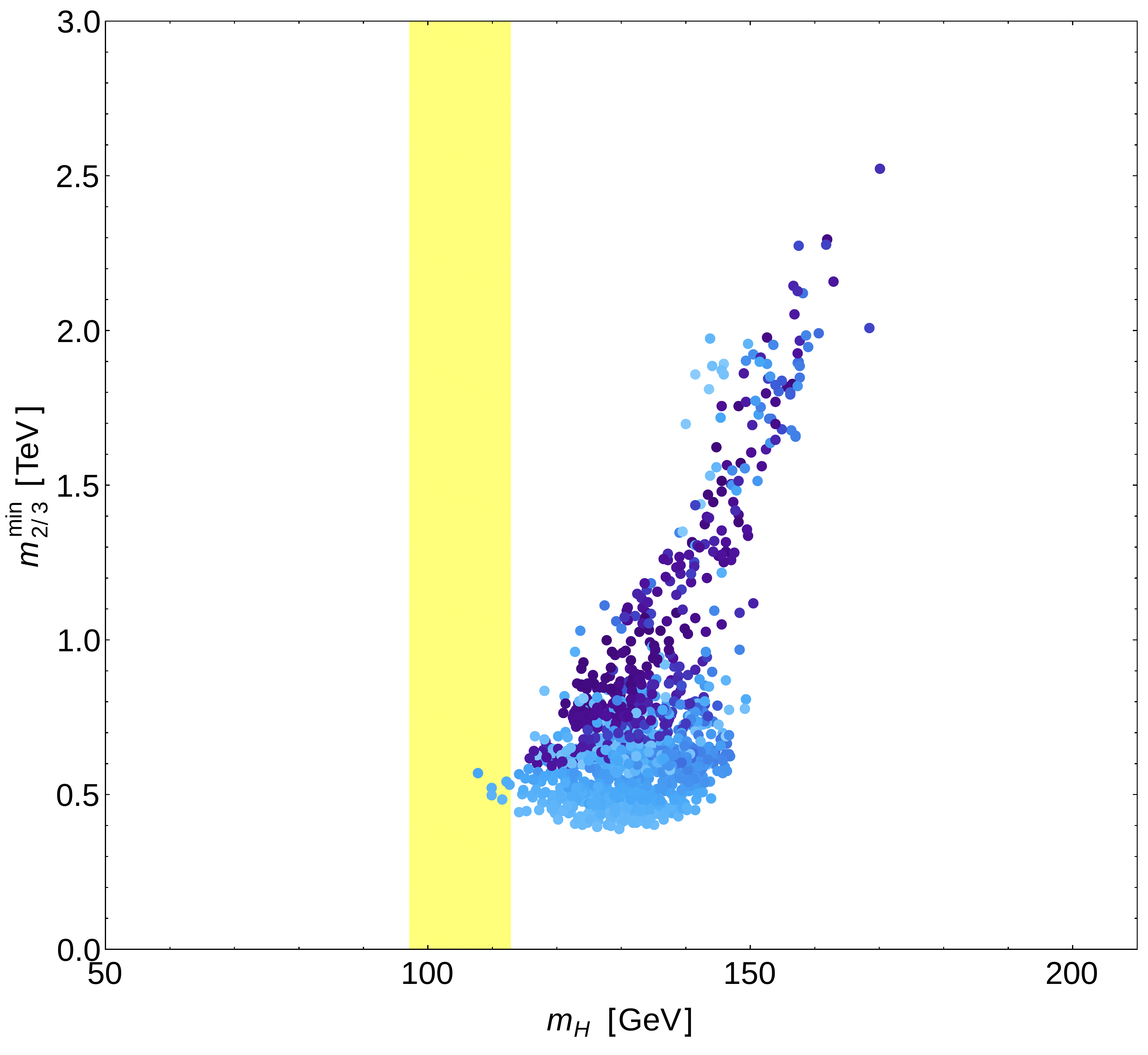}
		   \includegraphics[width=0.0865\textwidth]{figures/bar_cq2.pdf}
			\includegraphics[width=0.43\textwidth]{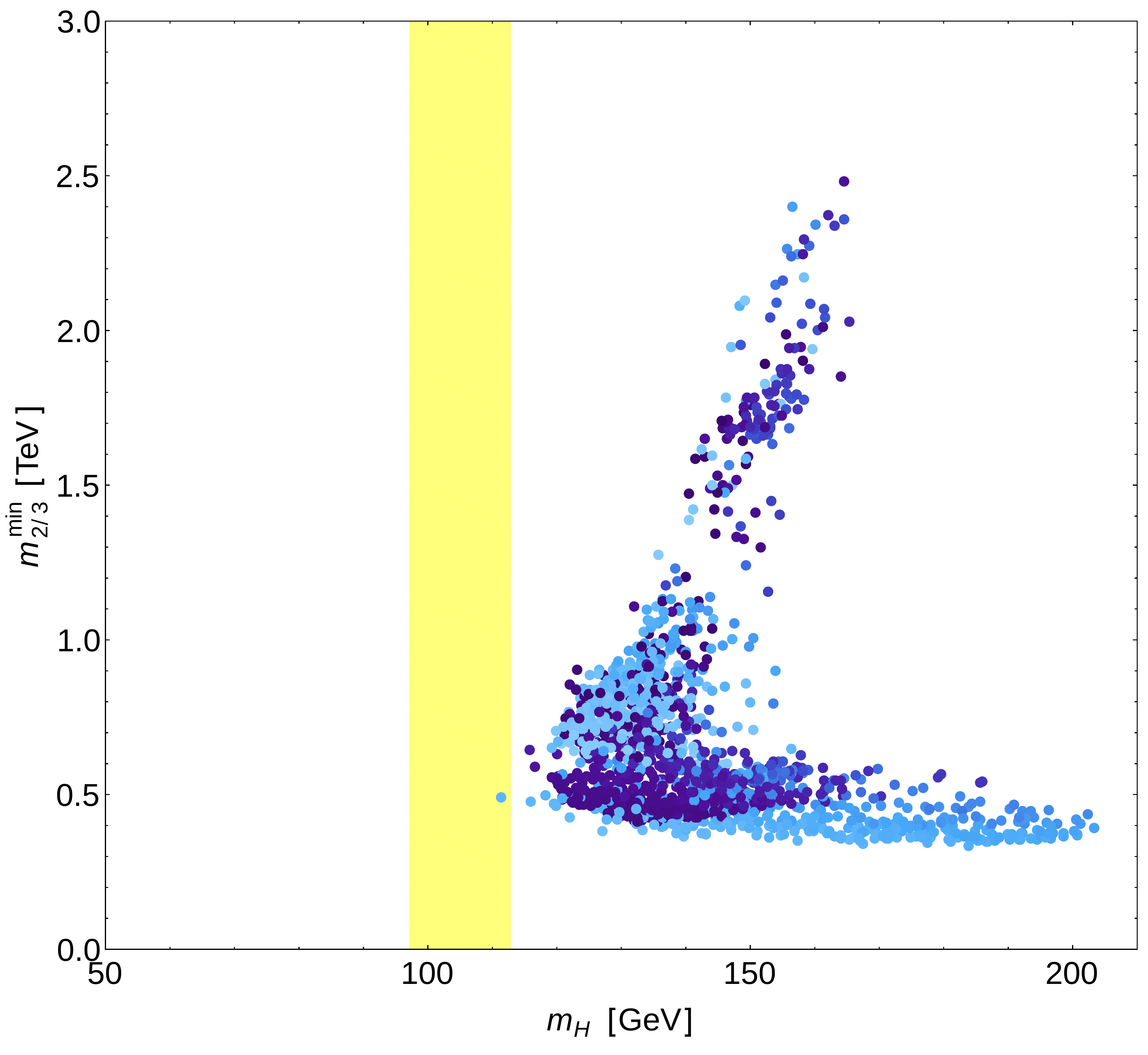}
			 \includegraphics[width=0.0865\textwidth]{figures/bar_cq2.pdf}
			 \label{fig:mchm10mcusvsmh}
		}
		\caption{Mass of the first top partner as a function of the Higgs mass in the MCHM$_5$ and MCHM$_{10}$, figures~\ref{fig:mchm5mcusvsmh} and \ref{fig:mchm10mcusvsmh}. In both cases, we have fixed $f_{\pi}=0.8~\TeV$ and $\kappa=\kappa^{\prime}=0$, corresponding roughly to $g_{\ast}\sim 4$ or $N_{\rm CFT}\sim 10$, and $s_\phi\approx \tan \theta_W$. Lighter points correspond to more IR localized $t_R$. The yellow band corresponds to a $\pm 7.5\%$ variation around $m_H(f_\pi)=105$\,GeV.}
\label{fig:mcusvsmh}
\end{center}
\end{figure}

	We start by studying the MCHM$_{5}$ and MCHM$_{10}$ as two representatives of models mixing the top quarks with
fundamental and adjoint representations of $SO(5)$.
Here, only the fact that both $t_L$ and $t_R$ mix with a {\bf 5} or a {\bf 10} is important, while the embedding of the composites 
that mix with the lighter quarks is irrelevant to excellent approximation. Also, as discussed in Section~\ref{sec:Vht}, setups of this class that mix the left handed top with another representations than the right handed top do not show a qualitatively 
new behavior.  In consequence, these benchmark models 
exploit the full range of predictions of setups mixing each top chirality with a representation smaller than the symmetric one.
For the time being, the contribution of the leptons will be rendered irrelevant by embedding them in single fundamental or adjoint representations and/or assuming a negligible lepton compositeness. 

\begin{figure}[t!]
\begin{center} 
			\includegraphics[width=0.493\textwidth]{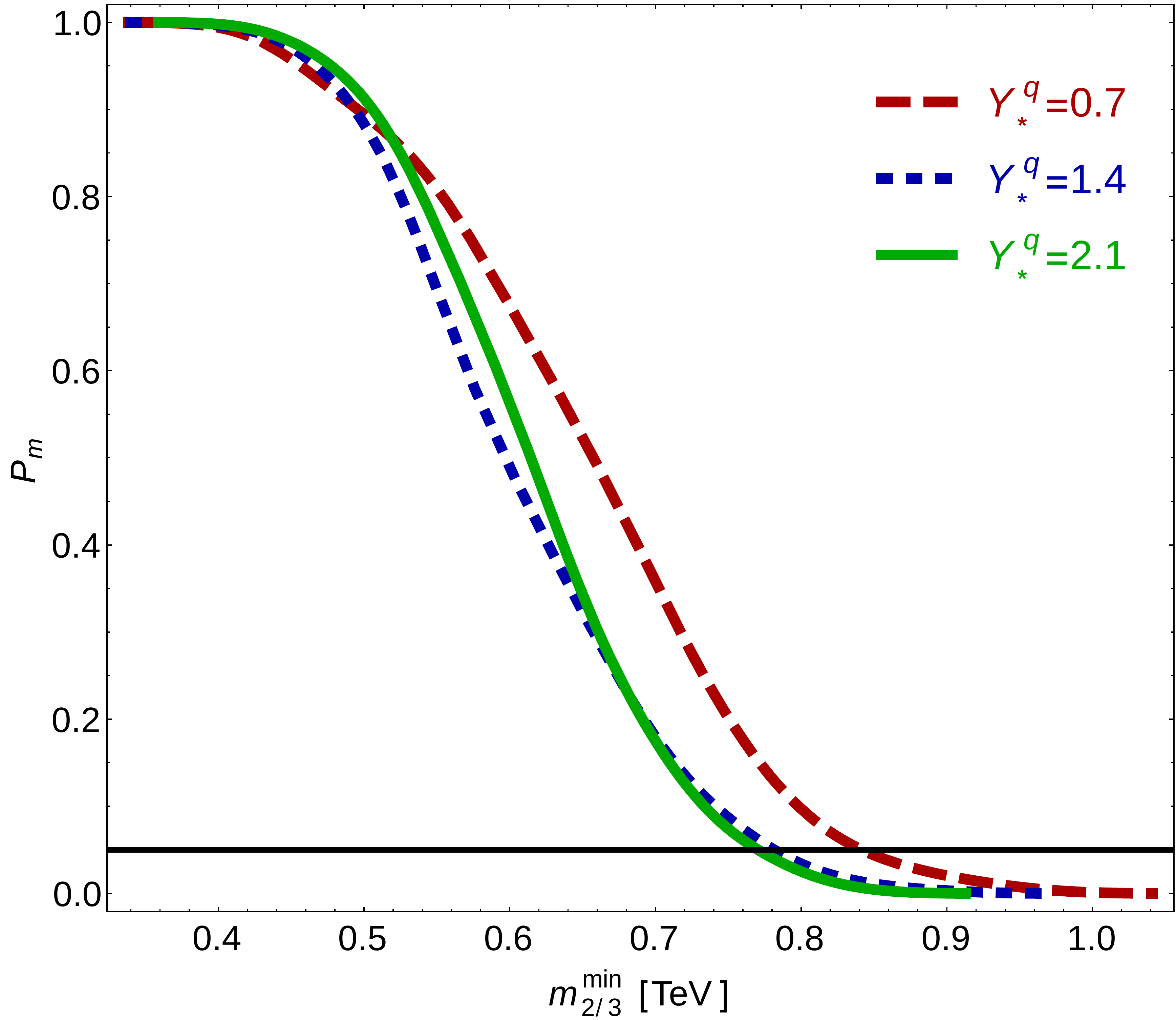}
			\caption{Survival function ${\cal P}_m$ of the first top partner mass in the MCHM$_5$, for $f_{\pi}=0.8$\,TeV, $\kappa=\kappa^{\prime}=0$, and $Y_\ast^q\in\{0.7,1.4,2.1\}$, assuming $m_{H}(f_\pi)=105$\,GeV$\,(1 \pm 7.5 \%)$.}
\label{fig:pdfmcus}
\end{center}
\end{figure}

In Figure~\ref{fig:mchm5mcusvsmh} we show the mass of the lightest $Q=2/3$ resonance, $m^{\rm min}_{2/3}$, as a function 
of the Higgs mass evaluated at $f_\pi$ for the MCHM$_5$, considering two values for the maximally allowed brane masses, $Y_\ast^q=0.7$ and $Y_\ast^q=1.4$.\footnote{Here, $m^{\rm min}_{2/3}$ corresponds to $\min(m_T,m_{\tilde T})$ in (\ref{eq:mHmt}).
Moreover, since every $SU(2)_L$ representation present in the setups we consider features a $Q=2/3$ entry, 
$m^{\rm min}_{2/3}$ also corresponds to good approximation to the mass of the lightest mode of the whole spectrum of
resonances.} We also give the dependence on the bulk mass of the $SO(5)$ multiplet hosting mostly the right-handed top, $c_2^q$
(where $f_{-c_2}$ corresponds to $y_R^tf_\pi/\tilde m_T \approx \sin \varphi_R^t $ in the 4D setup, see Table~\ref{tab:cor}). The darker the hue of the blue points, the smaller $c_2^q$. Finally, note that we set
$f_{\pi}=0.8~\TeV$ and assume no brane localized kinetic terms for the main part of the analysis. We will comment on their impact
at the end of this subsection.
The plots confirm the discussion of Section~\ref{sec:Vht}. First, the general scale of the predicted Higgs mass
is as low as $m_H \lesssim v$, in agreement with the fact that $\beta_t$ is only generated at subleading order in 
the spurions, see (\ref{eq:mH1}). Moreover, we see clearly that
the smaller the mass of the Higgs boson, the lighter the top partners need to be, confirming (\ref{eq:mHmt}). 
In particular, the light Higgs with mass  $m_{H}(f_\pi)=105$\,GeV$\,(1 \pm 7.5 \%)$, 
evaluated at $f_\pi$ (see Section~\ref{sec:mHfpi}) and 
depicted by the yellow band, requires top partners below a TeV. 
We can also see that, as expected from Table~\ref{tab:cor}, the lighter the partners, the bigger in general the values of 
$c_2^q$, which in the 5D setup corresponds to a larger mixing with the composite sector due to the stronger IR localization. 
On the other hand, there exist gradients in the $m_{H}- m^{\rm min}_{2/3}$ plane with constant $c_2^q$, where
the larger Goldstone-symmetry breaking due to a larger $y_R^t$ is coming with an increased partner mass such as to mostly
cancel the effect on $c_2^q$.
The analogous plots for the MCHM$_{10}$ are shown in Figure~\ref{fig:mchm10mcusvsmh}. While the qualitative behavior is similar,
quantitatively the situation is more severe in this model where also the $b_L$ contributes to the potential, since the top partners need to be even lighter to arrive at the correct Higgs mass.
The required $m^{\rm min}_{2/3} \lesssim (500-600)$\,GeV is already in conflict with LHC searches. 
Moreover, the experimental Higgs mass is at the boundary of the viable parameter space of the model.
For the MCHM$_5$, this is quantified further in Figure~\ref{fig:pdfmcus}, where we show the fraction of parameter points $P_m$ out of all points 
that reproduce the correct Higgs mass  $m_{H}(f_\pi)=105$\,GeV$\,(1 \pm 7.5 \%)$, for which the lightest top partner is heavier than 
$m^{\rm min}_{2/3}$ , versus $m^{\rm min}_{2/3}$, assuming $f_\pi=0.8$\,TeV. The curve has been obtained by smoothening the corresponding histograms and the black line depicts the $95\%$ quantile. 
For all brane masses considered, we find that $\gtrsim 90\%$ of the points
feature a top partner below 800\,GeV, in good agreement with (\ref{eq:mTEst}). 
The experimental bounds already cut significantly into the parameter space of the model \cite{Chatrchyan:2013uxa}.
As in the MCHM$_{10}$ the viable $m_H$ region is hardly reachable, we do not show the corresponding plot for
that model. It is evident that the Higgs mass (in particular together with the absence of light partners) already practically rules
out this model. 
If the LHC excludes $Q=2/3$ partners up to a TeV, which is possible with less than $300\,{\rm fb}^{-1}$ of data at the LHC14
\cite{Gershtein:2013iqa,Agashe:2013hma}, also the MCHM$_5$ would be excluded for practical purposes as a full solutions to 
the hierarchy problem.

\begin{figure}[t!]
\begin{center} 
	\subfigure[ $Y_\ast^q=0.7$  (left plot) and $Y_\ast^q=1.4$ (right plot) in the MCHM$_5$ with $f_{\pi}=0.8$\,TeV and $\kappa=\kappa^{\prime}=0$.]{%
		   \includegraphics[width=0.43\textwidth]{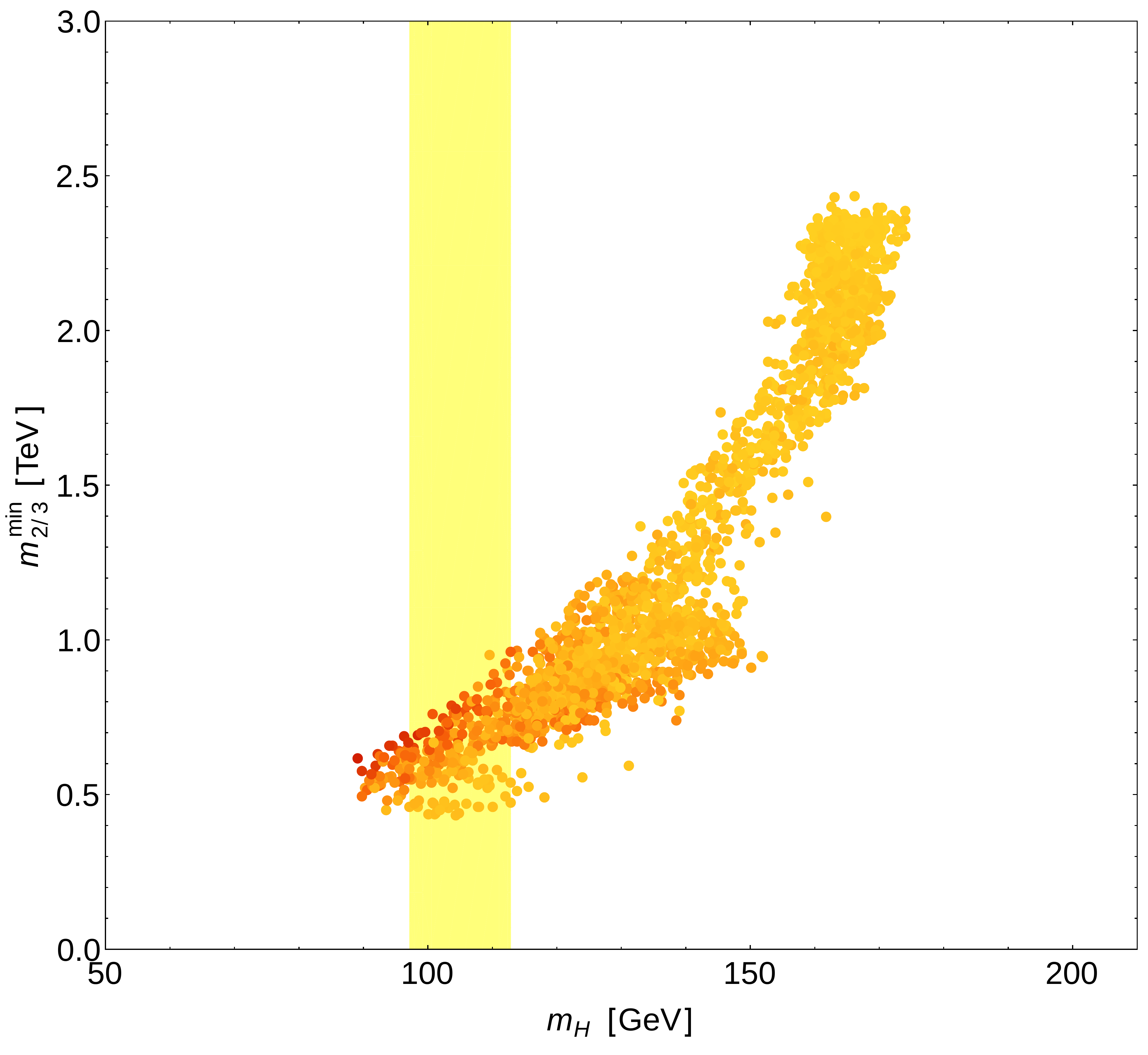}
		   \includegraphics[width=0.0865\textwidth]{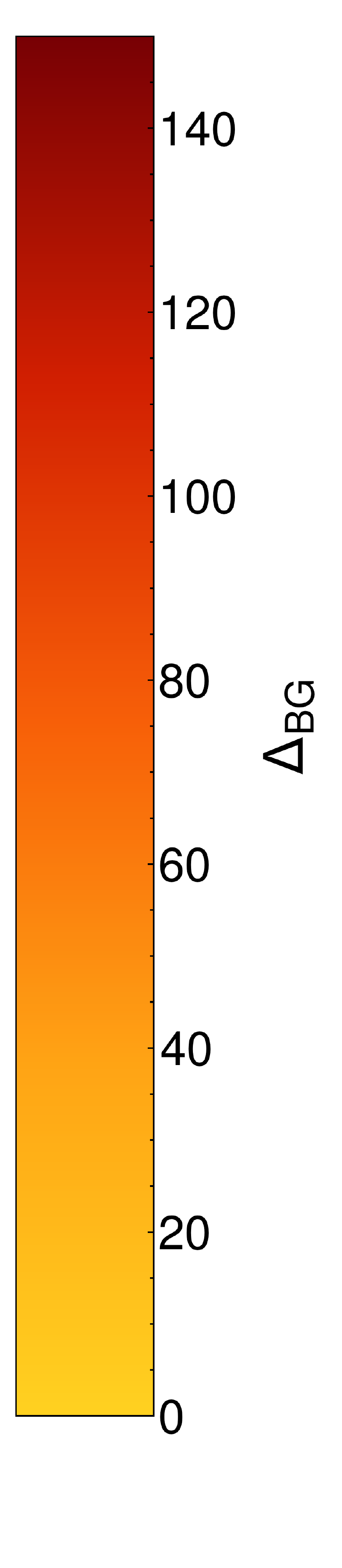}
			\includegraphics[width=0.43\textwidth]{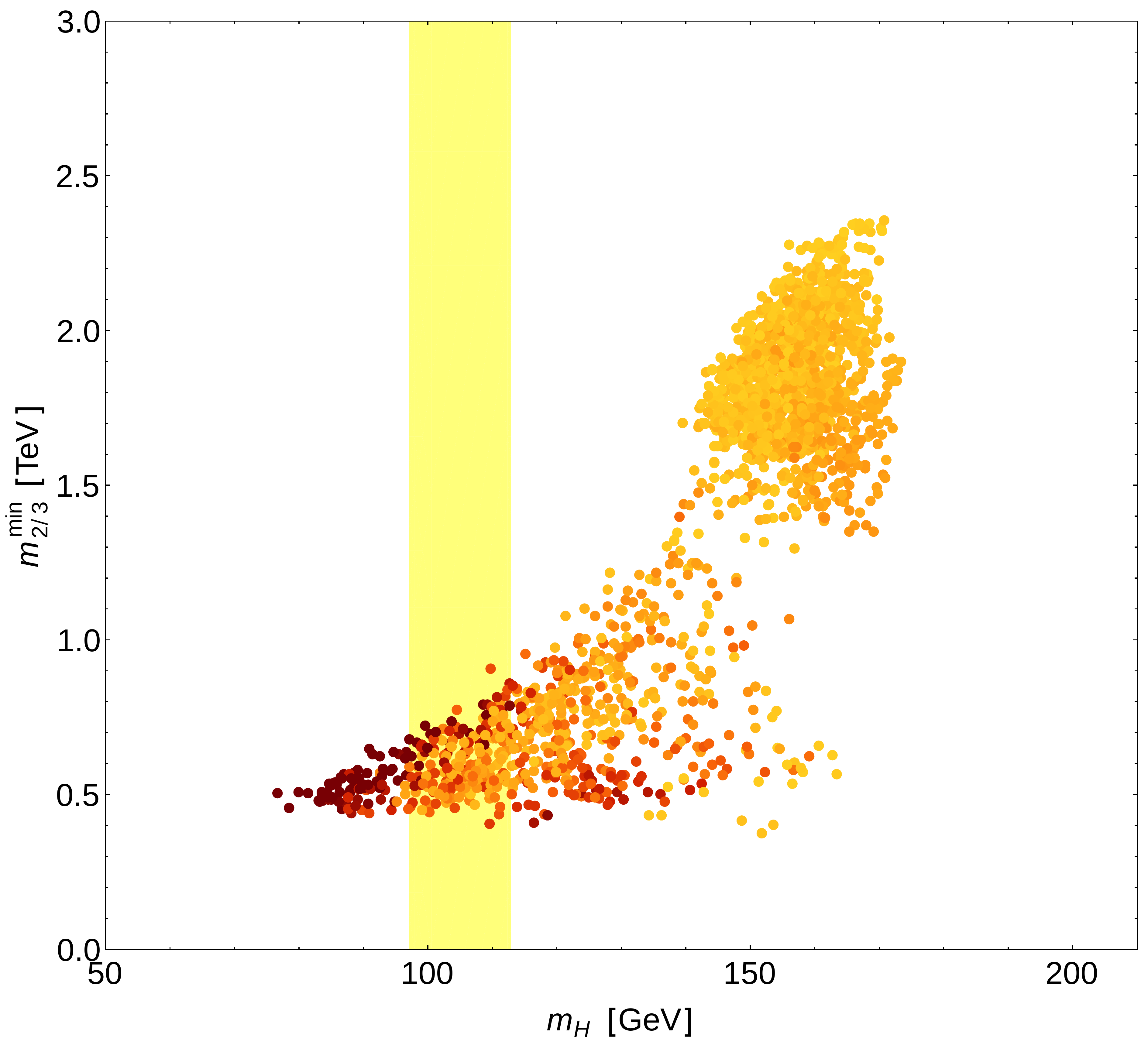}
			 \includegraphics[width=0.0865\textwidth]{figures/bar_ft.pdf}
			 \label{fig:mchm5mcusvsmhft}
		}
\\
\subfigure[$Y_\ast^q=0.7$  (left plot) and $Y_\ast^q=1.4$ (right plot) in the MCHM$_{10}$ with $f_{\pi}= 0.8$\,TeV and  $\kappa=\kappa^{\prime}=0$.  ]{%
		   \includegraphics[width=0.43\textwidth]{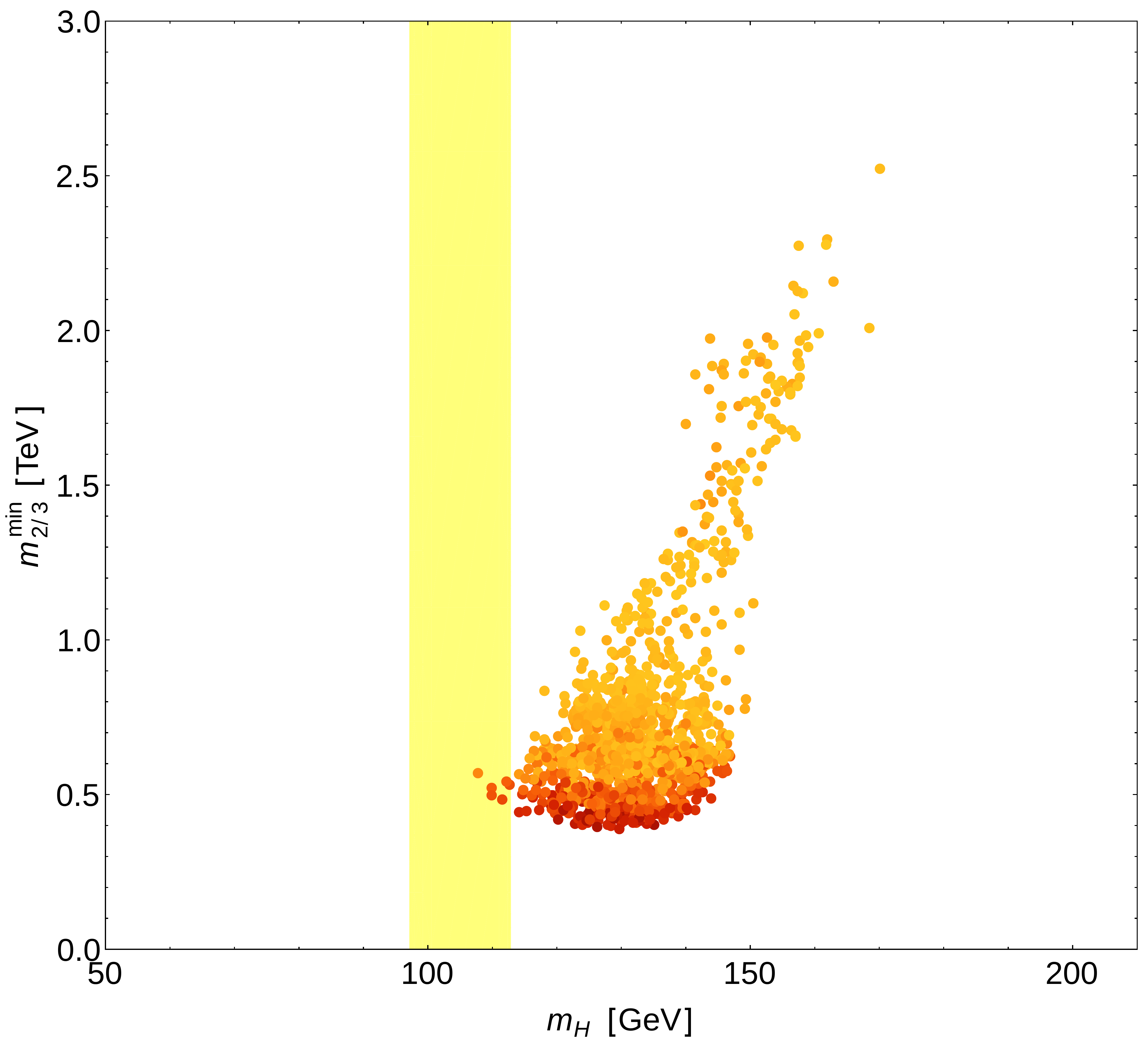}
		   \includegraphics[width=0.0865\textwidth]{figures/bar_ft.pdf}
			\includegraphics[width=0.43\textwidth]{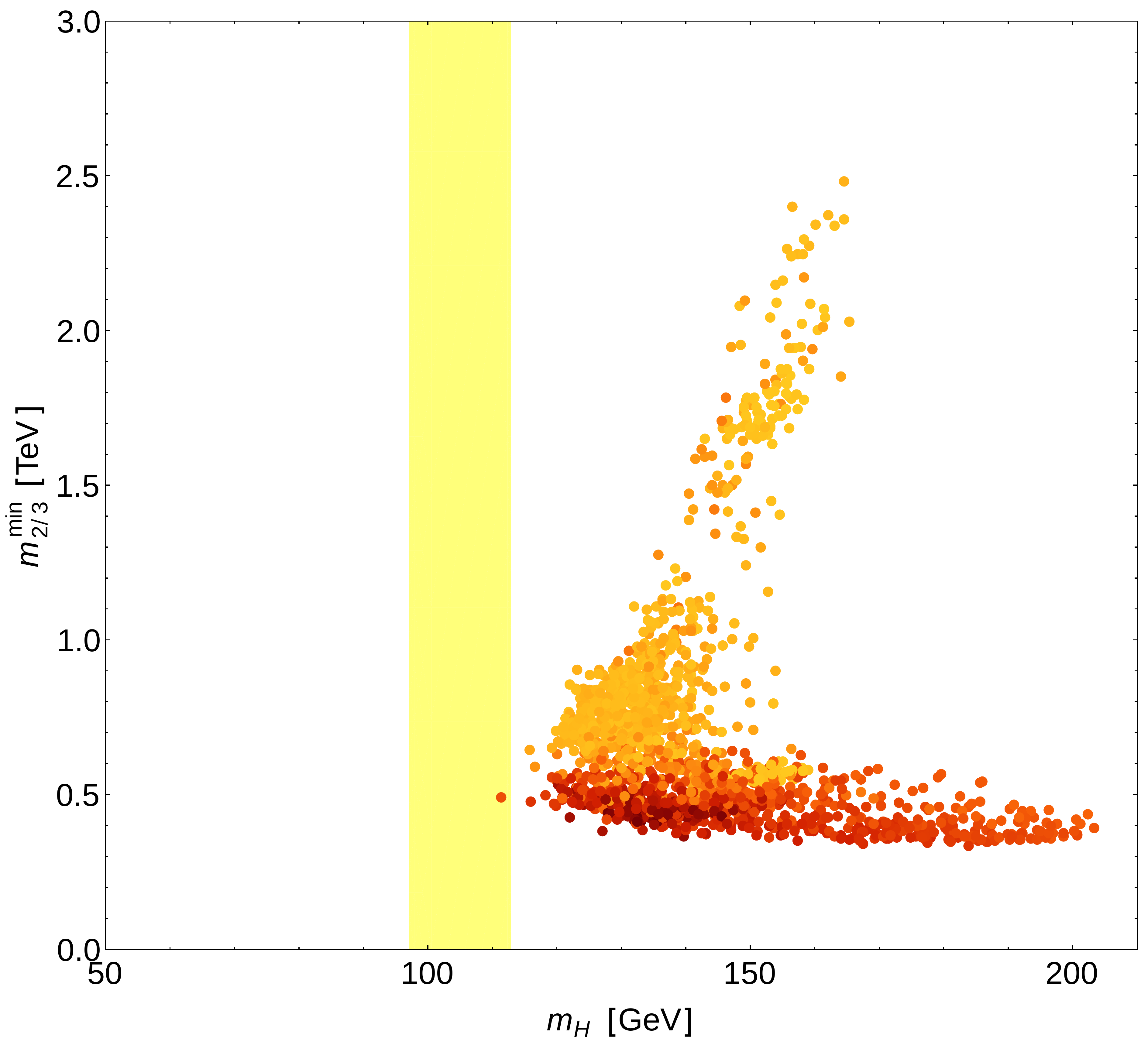}
			 \includegraphics[width=0.0865\textwidth]{figures/bar_ft.pdf}
			 \label{fig:mchm10mcusvsmhft}
		}
		\caption{Mass of the first top partner as a function of the Higgs mass in the MCHM$_5$ and the MCHM$_{10}$, figures \ref{fig:mchm5mcusvsmhft} and \ref{fig:mchm10mcusvsmhft}, respectively, for  $\kappa=\kappa^{\prime}=0$ and $f_{\pi}=0.8$\,TeV. Lighter points correspond to smaller values of $\Delta_{\rm BG}$ and therefore to less tuned points. }
\label{fig:mcusvsmhft}
\end{center}
\end{figure}
\begin{figure}[t!]
\begin{center} 
	\subfigure[ $Y_\ast^q=0.7$  (left plot) and $Y_\ast^q=1.4$ (right plot) in the MCHM$_5$ with $f_{\pi}=1$\,TeV and $\kappa=\kappa^{\prime}=0$.]{%
		   \includegraphics[width=0.43\textwidth]{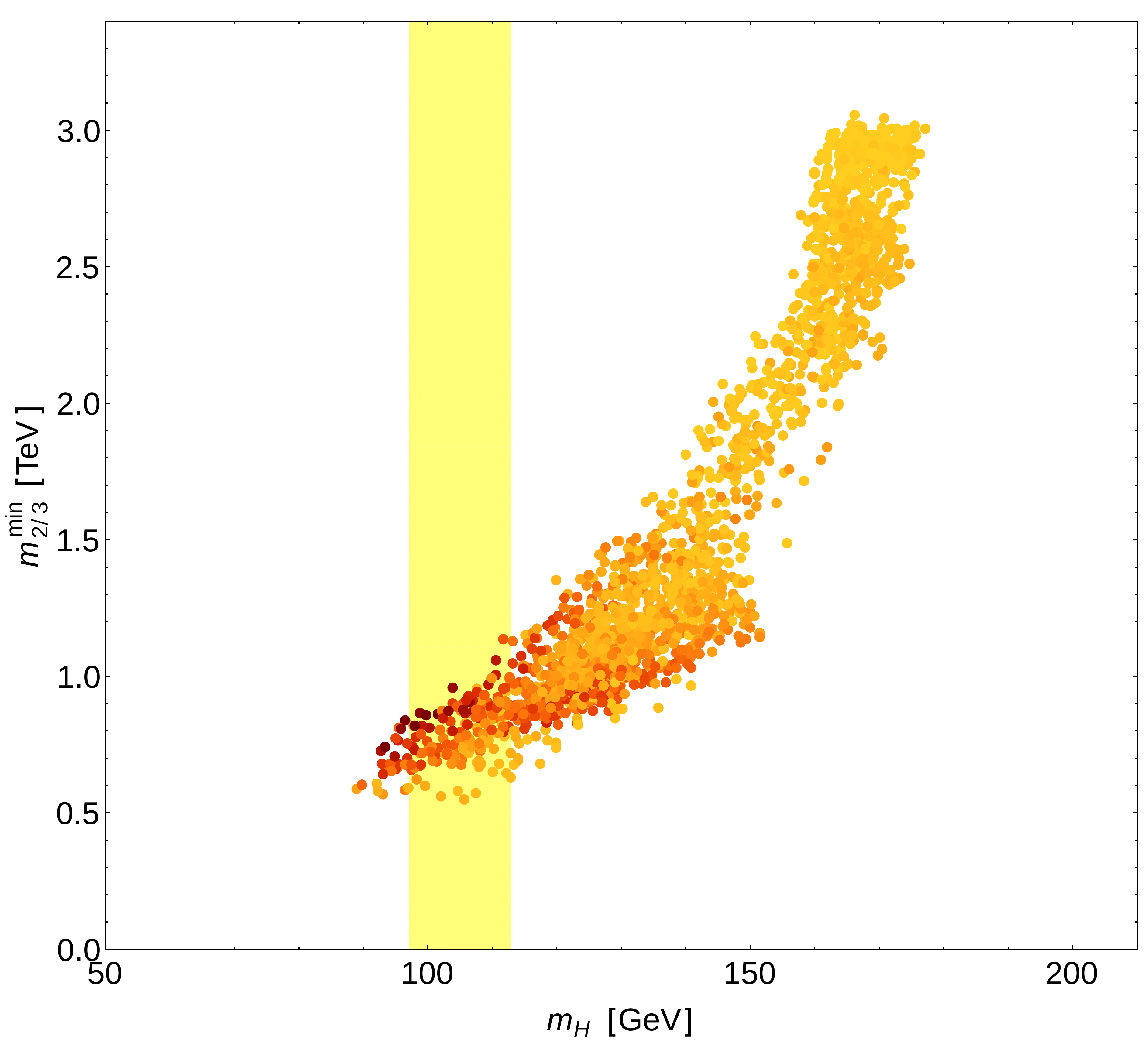}
		   \includegraphics[width=0.0865\textwidth]{figures/bar_ft.pdf}
			\includegraphics[width=0.43\textwidth]{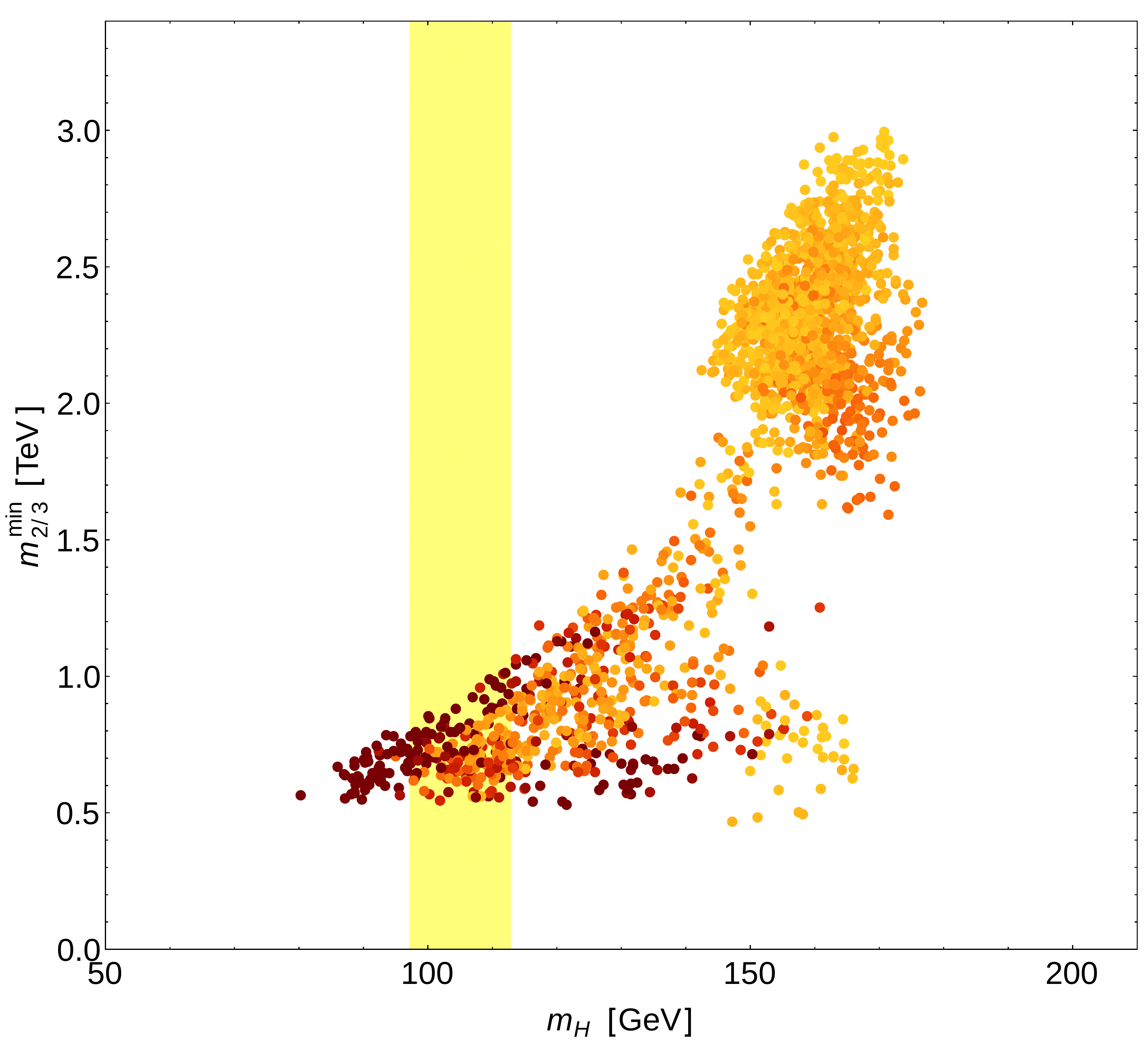}
			 \includegraphics[width=0.0865\textwidth]{figures/bar_ft.pdf}
			 \label{fig:mchm5mcusvsmhftfpi1}
		}
\\
\subfigure[$Y_\ast^q=0.7$  (left plot) and $Y_\ast^q=1.4$ (right plot) in the MCHM$_{10}$ with $f_{\pi}= 1$\,TeV and  $\kappa=\kappa^{\prime}=0$.  ]{%
		   \includegraphics[width=0.43\textwidth]{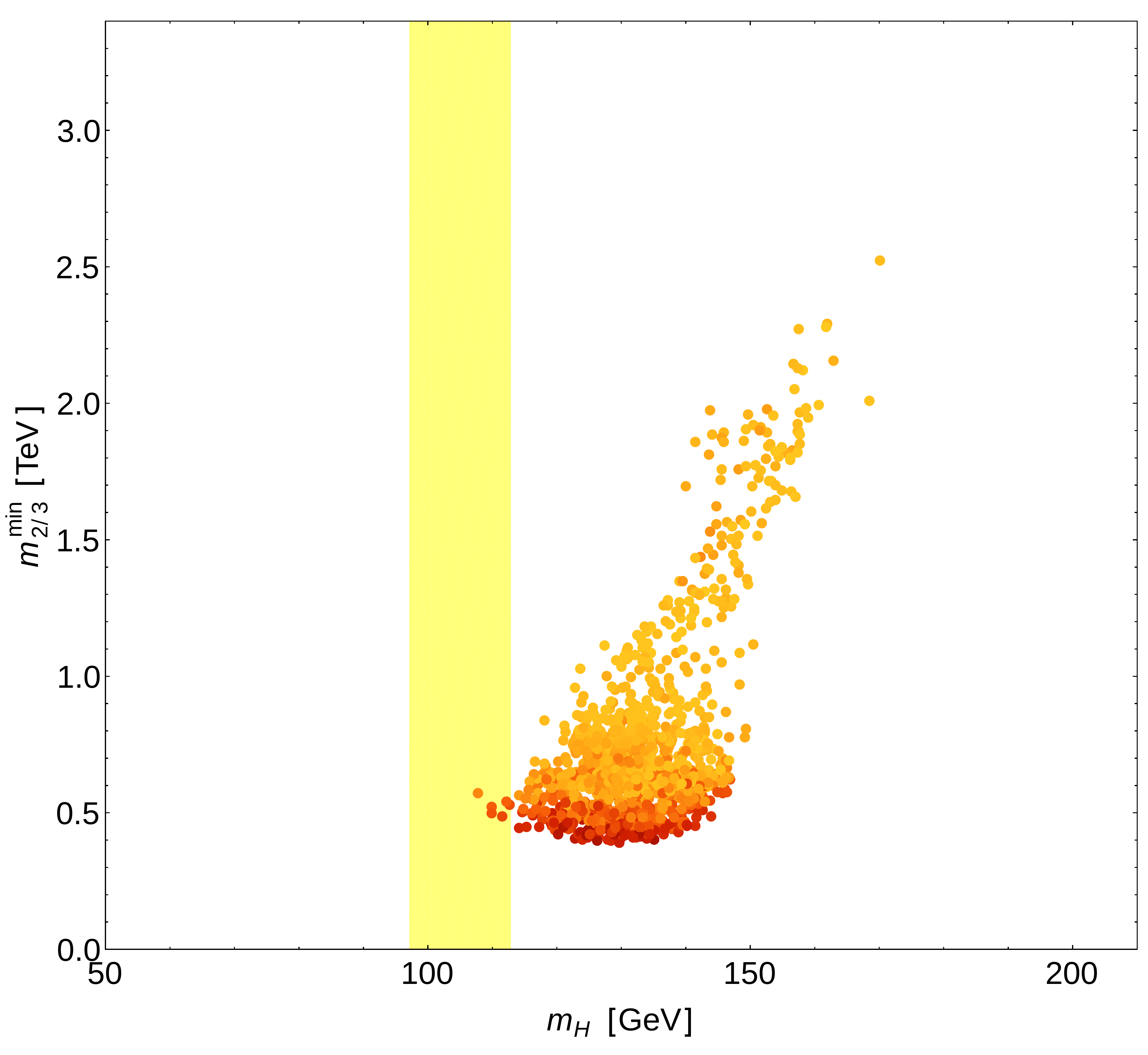}
		   \includegraphics[width=0.0865\textwidth]{figures/bar_ft.pdf}
			\includegraphics[width=0.43\textwidth]{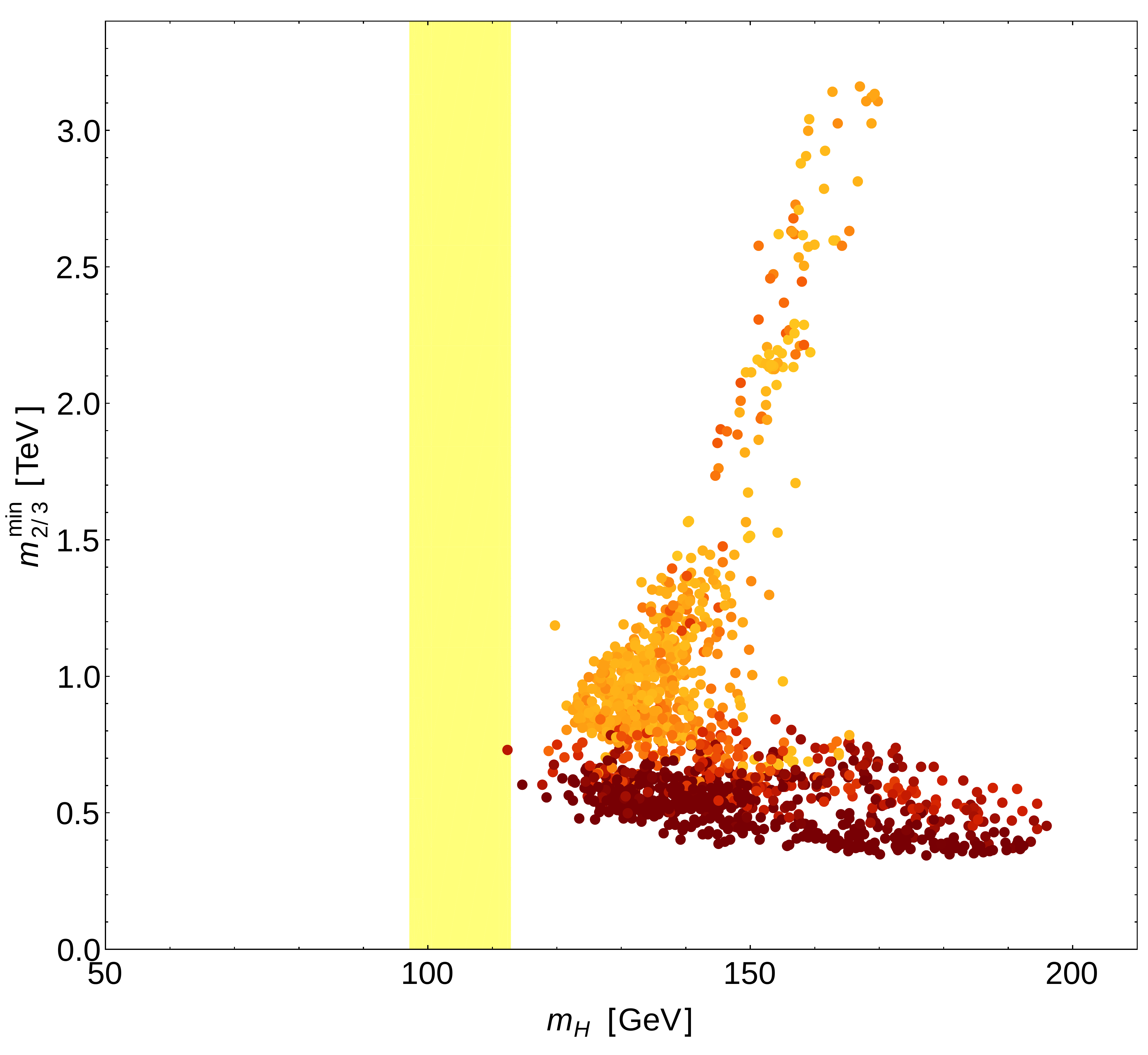}
			 \includegraphics[width=0.0865\textwidth]{figures/bar_ft.pdf}
			 \label{fig:mchm10mcusvsmhftfpi1}
		}
		\caption{Mass of the first top partner as a function of the Higgs mass in the MCHM$_5$ and the MCHM$_{10}$, figures \ref{fig:mchm5mcusvsmhftfpi1} and \ref{fig:mchm10mcusvsmhftfpi1}, respectively, for  $\kappa=\kappa^{\prime}=0$ and $f_{\pi}=1$\,TeV. Lighter points correspond to smaller values of $\Delta_{\rm BG}$ and therefore to less tuned points. }
\label{fig:mcusvsmhftfpi1}
\end{center}
\end{figure}
Now we turn to the tuning present in the models, which we show in Figures~\ref{fig:mcusvsmhft} and~\ref{fig:mcusvsmhftfpi1} 
for $f_\pi=0.8\,$TeV and $f_\pi=1\,$TeV, respectively.
As explained before, we describe the tuning by the Barbieri-Giudice measure, which we visualize by the color of every point in 
the $m_H-m^{\rm min}_{2/3}$ plane, where light yellow corresponds to a negligible tuning $\Delta_{\rm BG} \sim 0$, 
while dark red depicts a large tuning $\Delta_{\rm BG} \gtrsim 100$. The measure includes both the tuning to get the correct 
EWSB (the tuning entering the Higgs vev) as well as a potential ad-hoc tuning in the Higgs mass.
In the upper panels of each figure we show the results for the MCHM$_5$ employing $Y_\ast^q=0.7$ and $Y_\ast^q=
1.4$, while in the lower panels we provide the same plots for the MCHM$_{10}$. 
As a general trend, one can clearly see that lighter Higgs masses ({\it i.e.}, lighter partner masses) increase the tuning. This can easily be understood for example from (\ref{eq:mh5}) and the tuning condition (\ref{eq:EWSBcon}),(\ref{eq:DT5}), which leads to an increased tuning for smaller $\beta$.
In the left panel of Figure~\ref{fig:mchm5mcusvsmhft}, we inspect that for the correct Higgs mass, depicted again by the yellow 
band, the fine-tuning in the MCHM$_5$ is of the order of $\Delta_{\rm BG}\sim (10-80)$ for $Y_\ast^q=0.7$. For a larger $Y_\ast^q=1.4$ the tuning is generically increased since, keeping the dependence on the brane masses, it is determined by the combination $Y_\ast^q y^t$, which still grows $\sim \sqrt{Y_\ast^q}$ for constant $m_t$, see
(\ref{eq:mt}) (and Table~\ref{tab:cor}). For that case we see that $\Delta_{\rm BG} \sim (30-150)$.
The qualitative behavior in the MCHM$_{10}$ is similar, however as the physical $m_H$ can hardly be reached, it is difficult to
draw quantitative conclusions on the tuning.\footnote{In this particular realization, the $b_L$ contribution to the potential becomes relevant, as explained before. As it turns out, especially for the case of large maximal brane masses, this enlarges the parameter region, allowing e.g. $y^q_L$ ($y^t_R$) to be significantly reduced while the bottom contribution keeps $m_H$ fixed as well as potentially saves the EWSB condition (\ref{eq:EWSBcon}). This new region is characterized by light partner masses, needed (together with sizable brane masses) to arrive at the large $m_t$ in the presence of a moderate linear mixing, as well as a large tuning, since the required cancellations in the fermion sector to arrive at a viable EWSB, are now even more challenging.}
Increasing the Higgs-decay constant to $f_\pi=1\,$TeV should allow the top partners to be slightly heavier for fixed $m_H$,
as can be seen from (\ref{eq:mHmt}) (and similarly for the MCHM$_{10}$). This behavior is confirmed in our numerical analysis,
presented in Figure~\ref{fig:mcusvsmhftfpi1}, which shows the same plots as the previous figure, now employing
the larger Higgs decay constant. This comes however at the price of a considerably larger tuning, due to a smaller ratio $v/f_\pi$,
as is obvious from (\ref{eq:DT5}). In particular for those points that feature the maximal $m^{\rm min}_{2/3}$ of $\sim (900-1000)$\,~GeV 
the tuning becomes in general at least $\Delta_{\rm BG}\sim{\cal O}(100)$.

\begin{figure}[t!]
\begin{center} 
		   \includegraphics[width=0.403\textwidth]{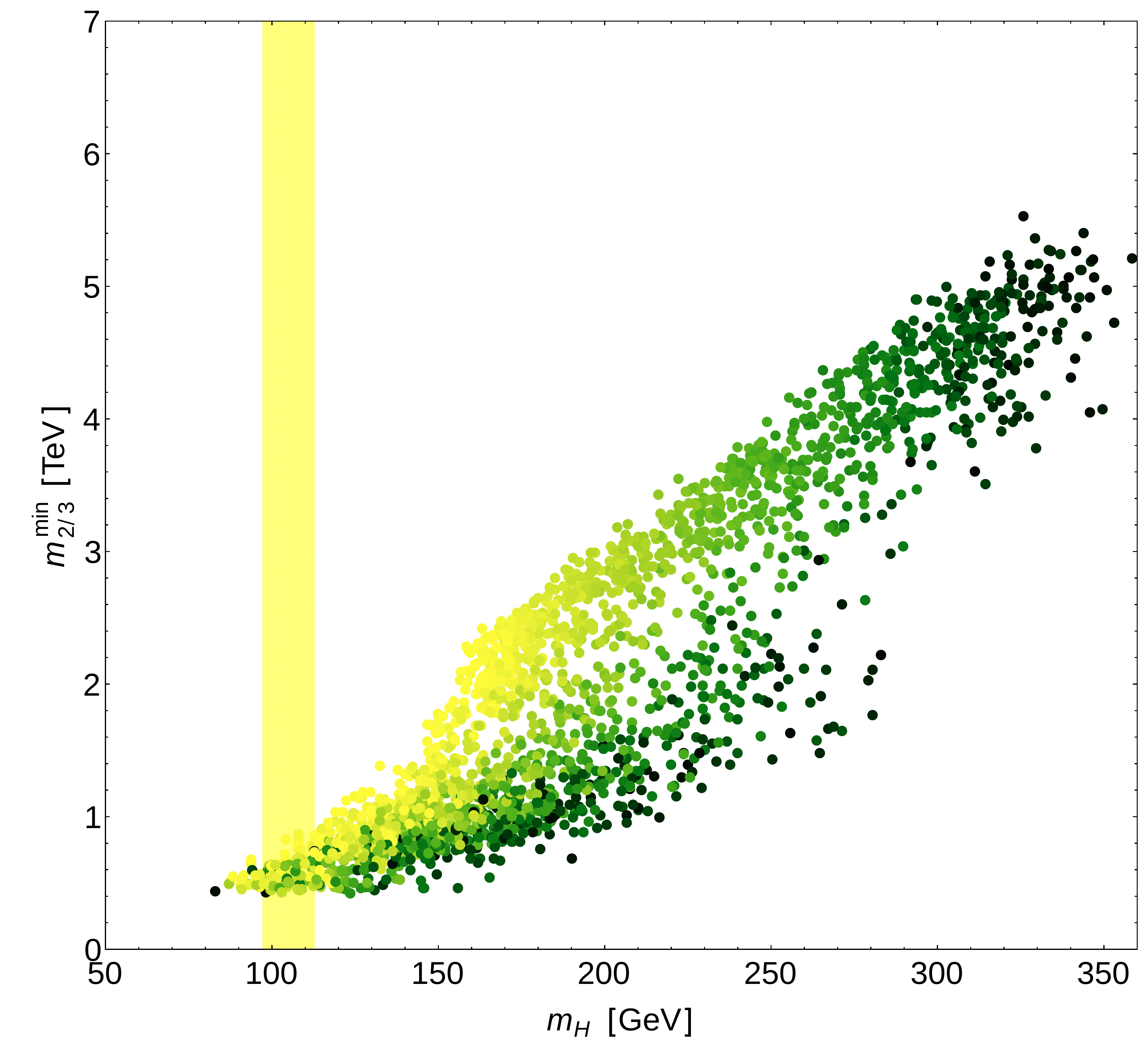}
		   \includegraphics[width=0.081\textwidth]{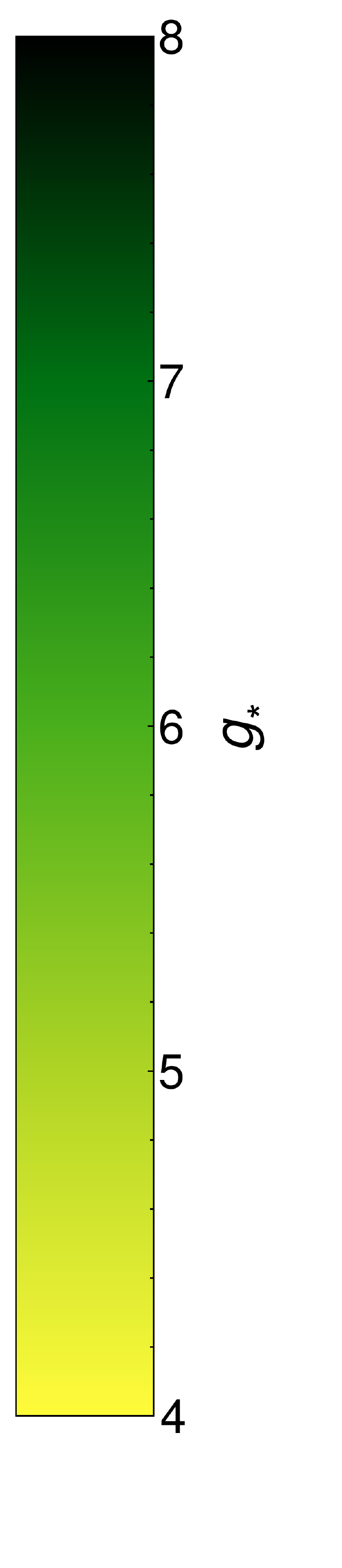}
		  \includegraphics[width=0.403\textwidth]{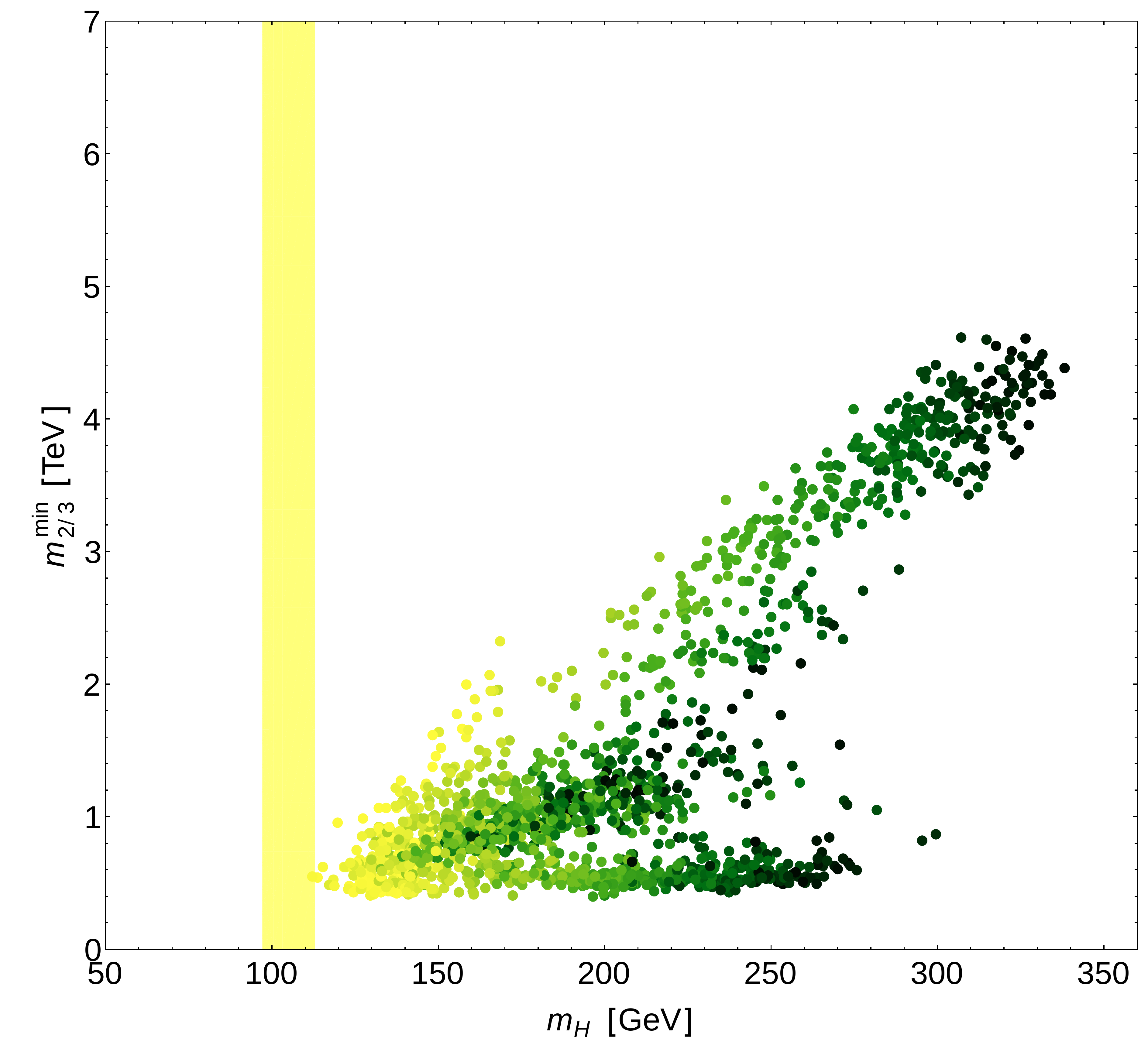}
		   \includegraphics[width=0.081\textwidth]{figures/bar_g5.pdf}
		\caption{Mass of the first top partner as a function of the Higgs mass in the MCHM$_5$ (left plot) and MCHM$_{10}$
(right plot). 
In both cases we have taken $\kappa,\kappa^{\prime}\in [0,\sqrt 3]$, which corresponds roughly to $g_\ast \in [4, 8]$, and have chosen the benchmark value $Y_\ast^q=0.7$ as well as $f_{\pi}=0.8$\,TeV.}
\label{fig:mcusvsmhncft}
\end{center}
\end{figure}
Finally, we study in addition the dependence on the 5D gauge coupling $g_\ast$, the ``inverse'' of the number of colors in the strong sector $N_{\rm CFT}$ and determined by the coefficient $\kappa$ of the brane localized kinetic terms of the $SU(2)_L$ gauge bosons, see (\ref{eq:gast}) and (\ref{eq:NCFT}). In Figure~\ref{fig:mcusvsmhncft} we show again the 
mass of the lightest top partner in dependence on the Higgs mass, for $f_\pi=0.8$\,TeV and $Y_\ast^q=0.7$, 
indicating $g_\ast$ by the hue of the green color.
We see that if the 5D gauge coupling is increased with respect to the default value $g_\ast \sim 4$
(by turning on the brane kinetics), which corresponds to a darker green, the Higgs boson gets generally heavier for a 
given $m^{\rm min}_{2/3}$. This can be understood easily as the increased 5D coupling enhances all contributions to the potential.
While in the MCHM$_5$, shown in the left plot, 
it is still possible to arrive at the correct Higgs mass, in the MCHM$_{10}$ with non-negligible brane kinetics, the Higgs 
becomes unavoidably much too heavy. The fact that the Higgs mass generically increases for a given mass
of the lightest top partner directly means that also changing $N_{\rm CFT}$ can not raise the mass of 
the light partners, induced by $m_H$.

In summary, our numerical analysis confirms the qualitative features described in Section~\ref{sec:Vht}.
Models mixing the different SM chiralities with a fundamental or an adjoint representation require light partners below the TeV scale.
Moreover, they feature a larger tuning than the naive estimate of $v^2/f_\pi^2 \sim 10 \%$. We will now see quantitatively how
a composite lepton sector can naturally break the correlation between $m^{\rm min}_{2/3}$ and $m_H$, 
without the need to change the colored particle content.

\subsubsection{The Impact of Leptons}

\begin{figure}[t!]
\begin{center} 
	\subfigure[ $Y_\ast^q=0.7$  (left plot) and $Y_\ast^q=1.4$ (right plot) in the MCHM$_5^{5-14}$ with $f_{\pi}=0.8$\,TeV and $\kappa=\kappa^{\prime}=0$. ]{%
		   \includegraphics[width=0.43\textwidth]{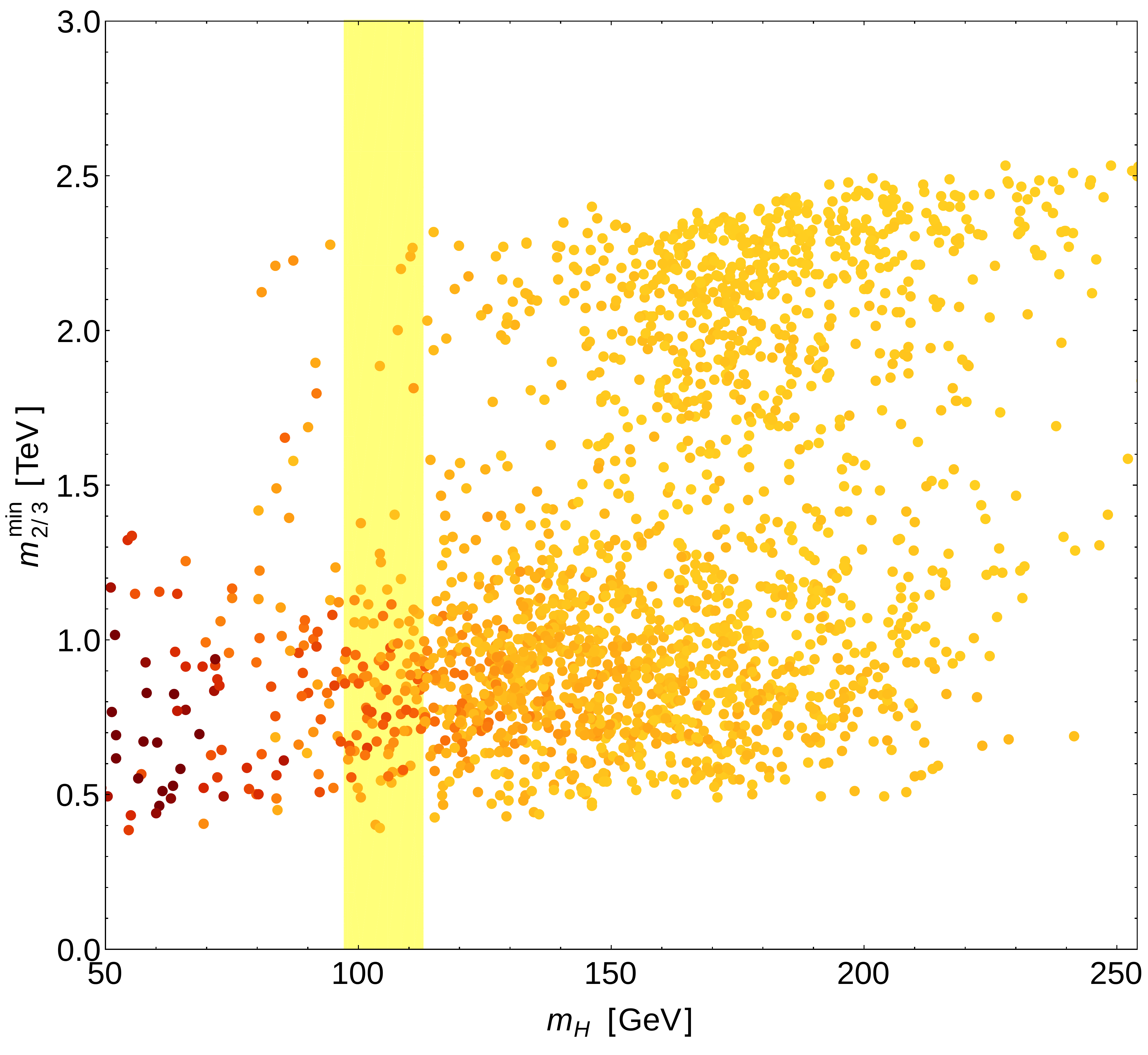}
		   \includegraphics[width=0.0865\textwidth]{figures/bar_ft.pdf}
			\includegraphics[width=0.43\textwidth]{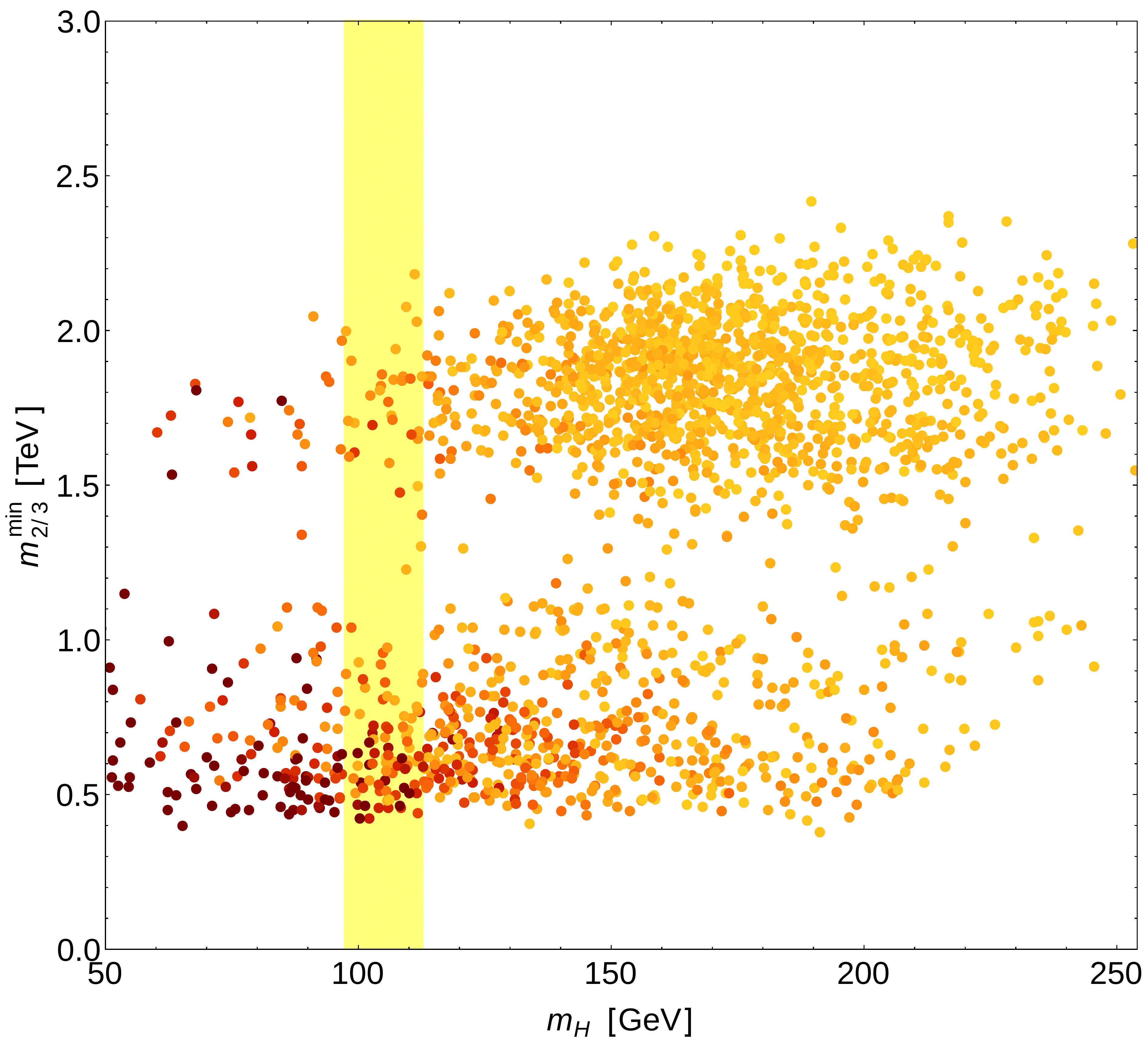}
			 \includegraphics[width=0.0865\textwidth]{figures/bar_ft.pdf}
			 \label{fig:14p5mcusvsmhft1}
		}
\\
\subfigure[$Y_\ast^q=0.7$  (left plot) and $Y_\ast^q=1.4$ (right plot) in the MCHM$_{5}^{5-14}$ with $f_{\pi}= 1$\,TeV and $\kappa=\kappa^{\prime}=0$.  ]{%
		   \includegraphics[width=0.43\textwidth]{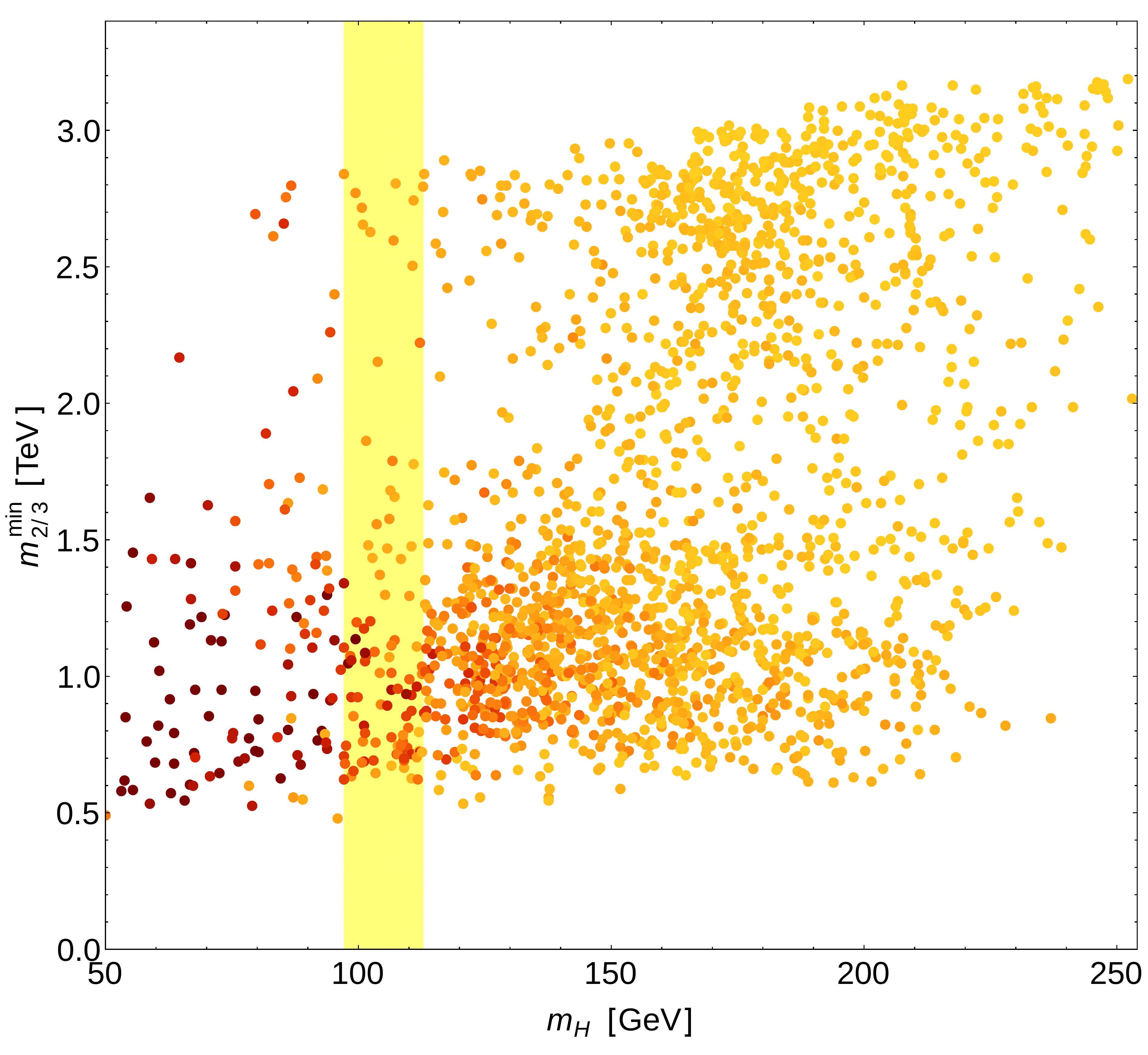}
		   \includegraphics[width=0.0865\textwidth]{figures/bar_ft.pdf}
			\includegraphics[width=0.43\textwidth]{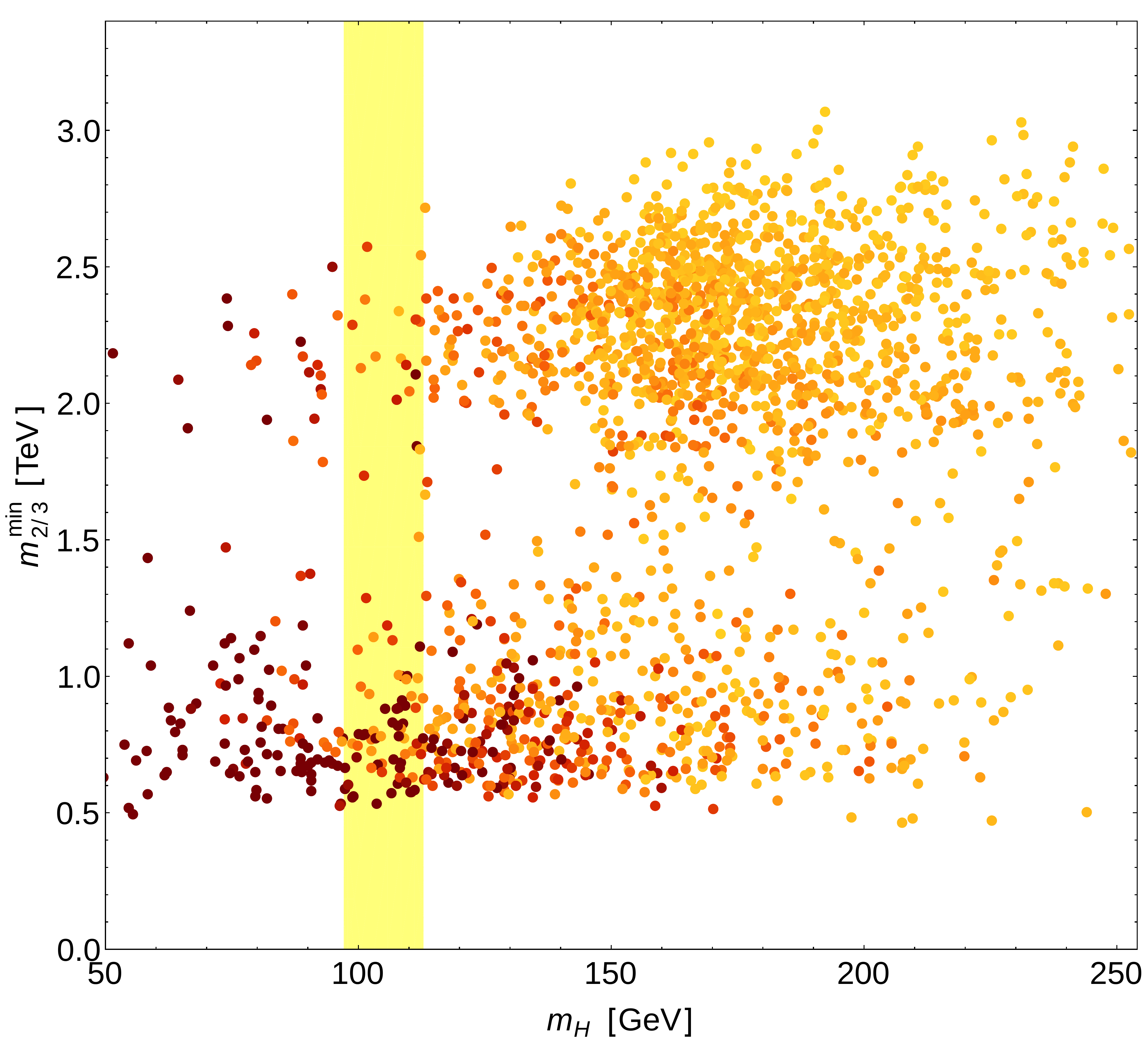}
			 \includegraphics[width=0.0865\textwidth]{figures/bar_ft.pdf}
			 \label{fig:14p5mcusvsmhft2}
		}
		\caption{Mass of the first top partner as a function of the Higgs mass in the MCHM$_5^{5-14}$ for  $f_{\pi}=0.8$\,TeV and $f_{\pi}=1$\,TeV, figures \ref{fig:14p5mcusvsmhft1} and \ref{fig:14p5mcusvsmhft2}, respectively. In both cases we have fixed $\kappa=\kappa^{\prime}=0$, while $Y_\ast^l=0.35$. Lighter points correspond to smaller values of $\Delta_{\rm BG}$ and therefore to less tuned points. }
\label{fig:14p5mcusvsmhft}
\end{center}
\end{figure}
We start with the standard models of $\tau_R$ compositeness due to flavor protection, without a unification 
of right-handed lepton and neutrino multiplets, {\it i.e.}, the ordinary MCHM$_5^{5-14}$.
In Figure~\ref{fig:14p5mcusvsmhft1}, we show again the mass of the lightest $Q=2/3$ resonance, as a function 
of the Higgs mass for $f_{\pi}=0.8~\TeV$, assuming once more $Y_\ast^q=0.7$ and $Y_\ast^q=1.4$
for the maximally allowed brane masses in the quark sector. For the leptons, we employ $Y_\ast^l=0.35$,
accounting for the `Yukawa suppression' generally present in these models, and here and in the following
again assume no brane localized kinetic terms. 
The plots confirm nicely our discussion of Section~\ref{sec:Vhl}. The basically unsuppressed lepton contributions to $\beta$, that
arise for a non-negligible mixing with a {\bf 14}, allow for a destructive interference with the quark contributions such as to raise
the masses of the lightest top partners well beyond the 1 TeV scale. 
In particular, for both values of the brane masses considered, lightest partners as heavy as 2 TeV are possible,
removing considerably the pressure coming from current LHC searches.
Due to the same mechanism, the Higgs mass could now be even below $100$\, GeV and its measured 
value lies not anymore at the boundary of the possible values. Because of the modest leptonic contribution to $\beta$,
the models also feature no ad-hoc tuning. As before, the general tuning $\Delta_{\rm BG}$
is visualized by the hue of the points in the $m_H-m^{\rm min}_{2/3}$ plane.
One can see that the lepton sector adds no additional relevant tuning to the setup and the dependence on $m_H$ 
follows the same pattern as in the MCHM$_5$, dominated by the large top contribution to $\alpha$ which needs to be tuned
to become of the order of $\beta$. Also the increased tuning for larger $Y_\ast^q$ can be explained along the same lines as before.
An interesting feature however becomes apparent, if one studies the tuning versus $m^{\rm min}_{2/3}$.
The tuning is decreased for {\it heavier} top partners.
So, not only are light partners not necessary anymore in the setup, they are even less natural and the expectation
would correspond to top partners $\gtrsim 1.5\,$TeV. This can be understood since points with a larger $m^{\rm min}_{2/3}$
feature usually a larger $y_t$ and this makes the ``double tuning''  (\ref{eq:DT5}) less severe. 
All these qualitative features remain after increasing the Higgs decay constant to $f_\pi = 1\,$TeV,
which is shown in \ref{fig:14p5mcusvsmhft2}. The slightly higher partner masses and the increase in the tuning can 
be explained as before in the MCHM$_{5,10}$.
	
\begin{figure}[t!]
\begin{center} 
		   \includegraphics[width=0.403\textwidth]{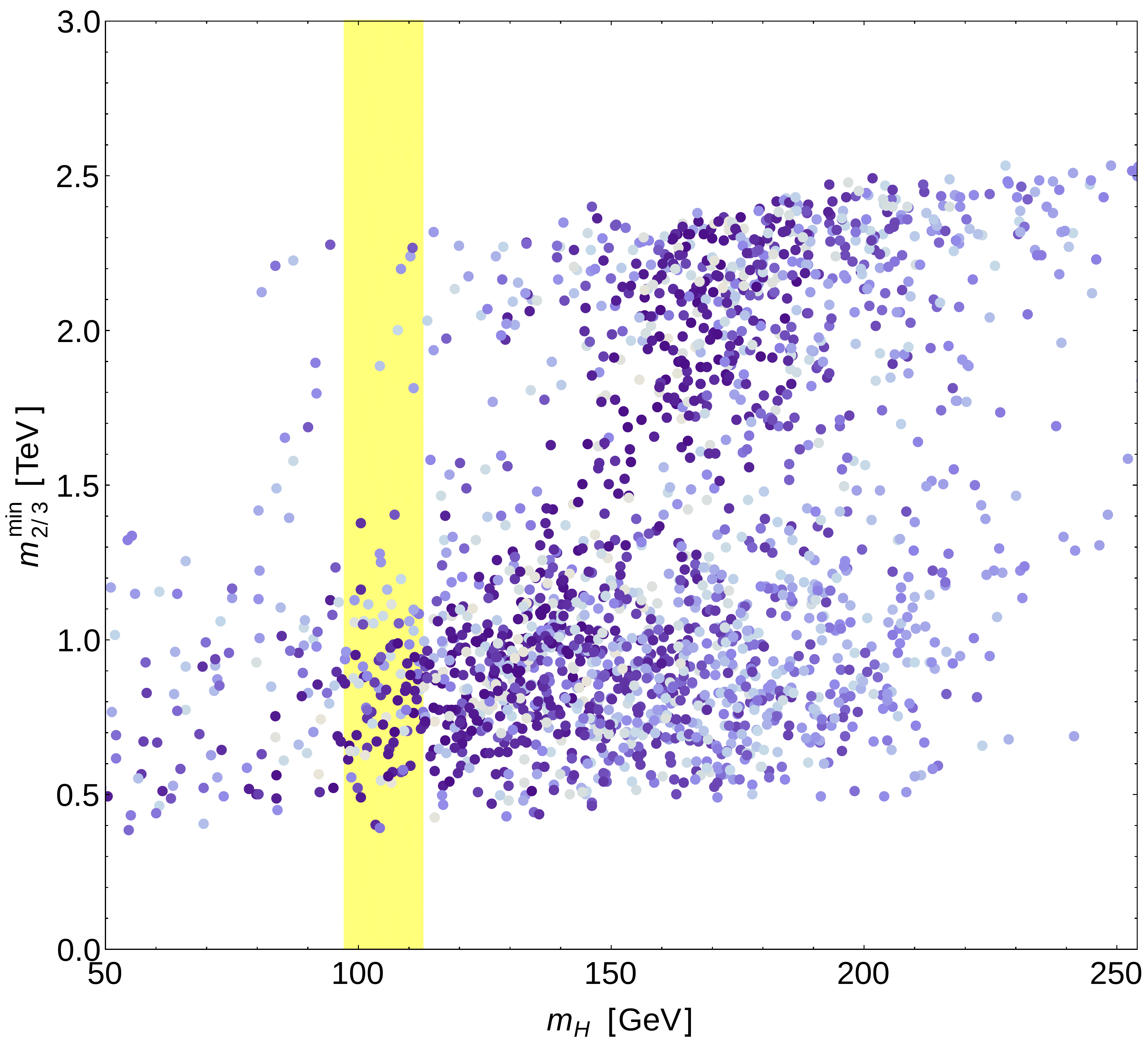}
		   \includegraphics[width=0.081\textwidth]{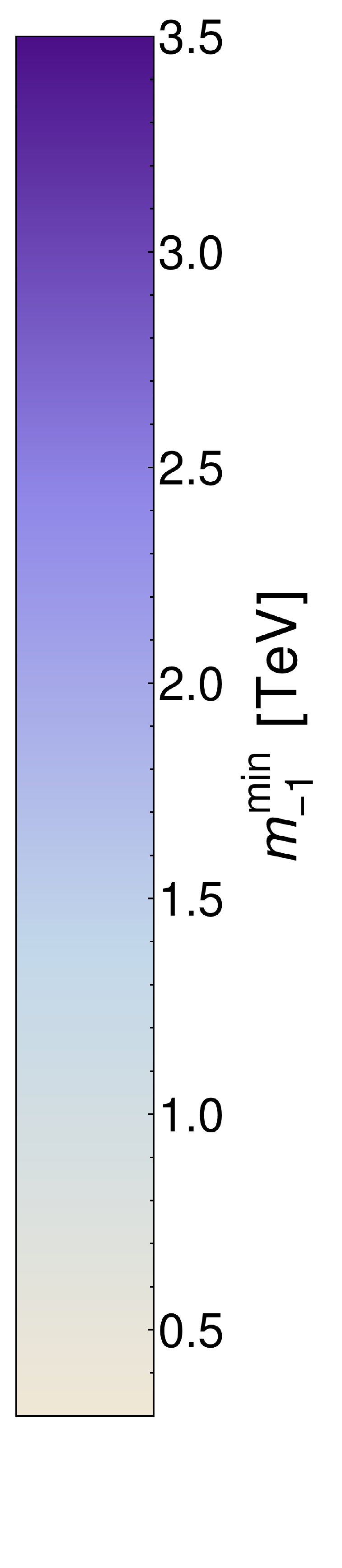}
			\includegraphics[width=0.403\textwidth]{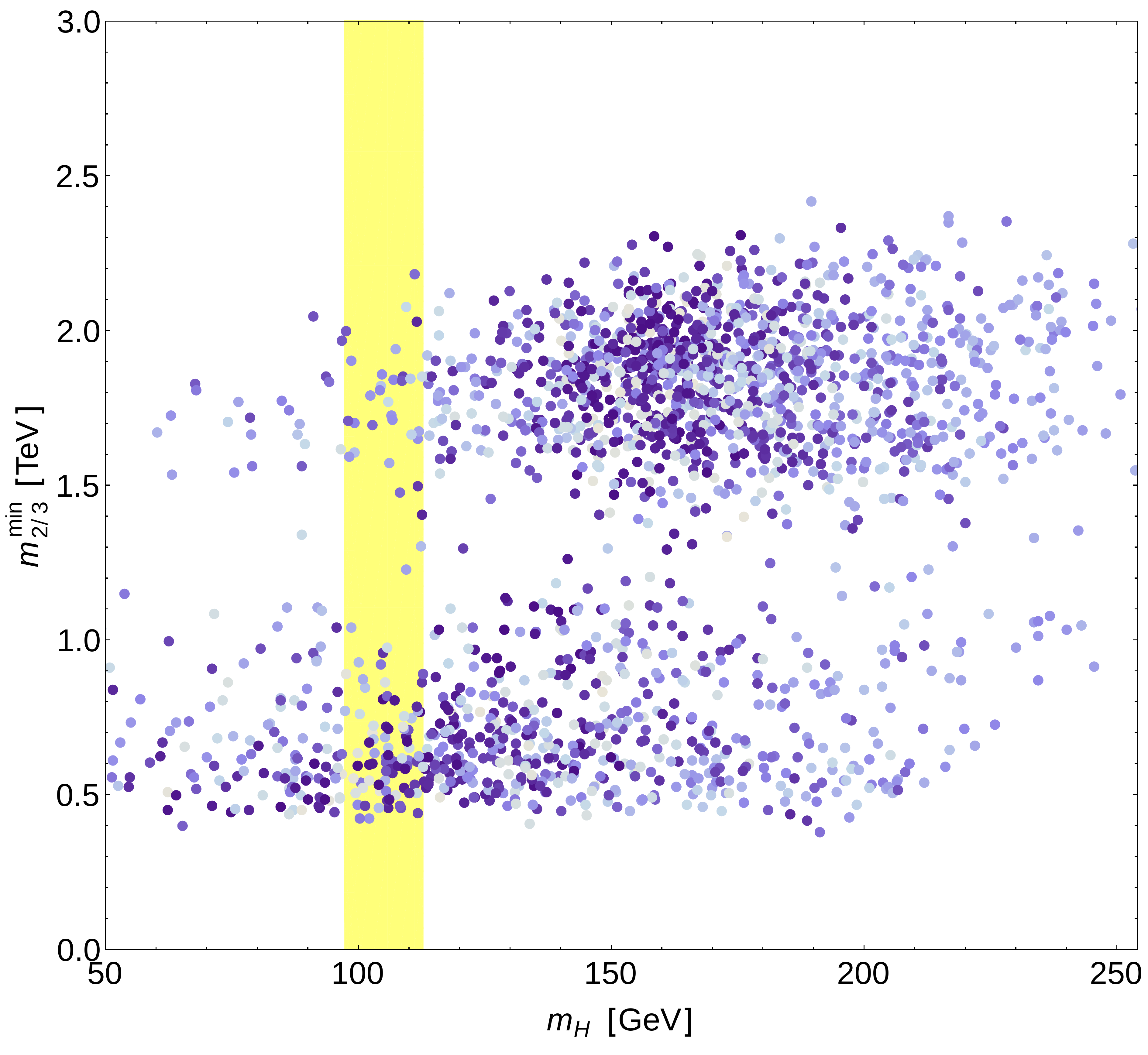}
			 \includegraphics[width=0.081\textwidth]{figures/bar_ml.pdf}
		\caption{Mass of the first top partner as a function of the Higgs mass in the MCHM$_5^{5-14}$ for $Y_\ast^q=0.7$  (left plot) and $Y_\ast^q=1.4$ (right plot), where always $f_{\pi}=0.8$\,TeV, $\kappa=\kappa^{\prime}=0$, and $Y_\ast^l=0.35$. Lighter points correspond to smaller values of $m_{-1}^{\rm min}$. }
\label{fig:14p5mcusvsmhml}
\end{center}
\end{figure}
To show that our solution does not introduce ultra-light $\tau$ partners in conflict with observation, we show in Figure~\ref{fig:14p5mcusvsmhml} 
the same plots as before (with $f_\pi=0.8$\,TeV), now indicating the mass of the lightest $Q=-1$ resonance, $m^{\rm min}_{-1}$,
by the hue of the blue color. We inspect that the $\tau$ partners are in general heavier than a TeV.
In particular, there are always points with both $m^{\rm min}_{2/3}>2\,$TeV, $m^{\rm min}_{-1}>2\,$TeV.

\begin{figure}[t!]
\begin{center} 
			\includegraphics[width=0.493\textwidth]{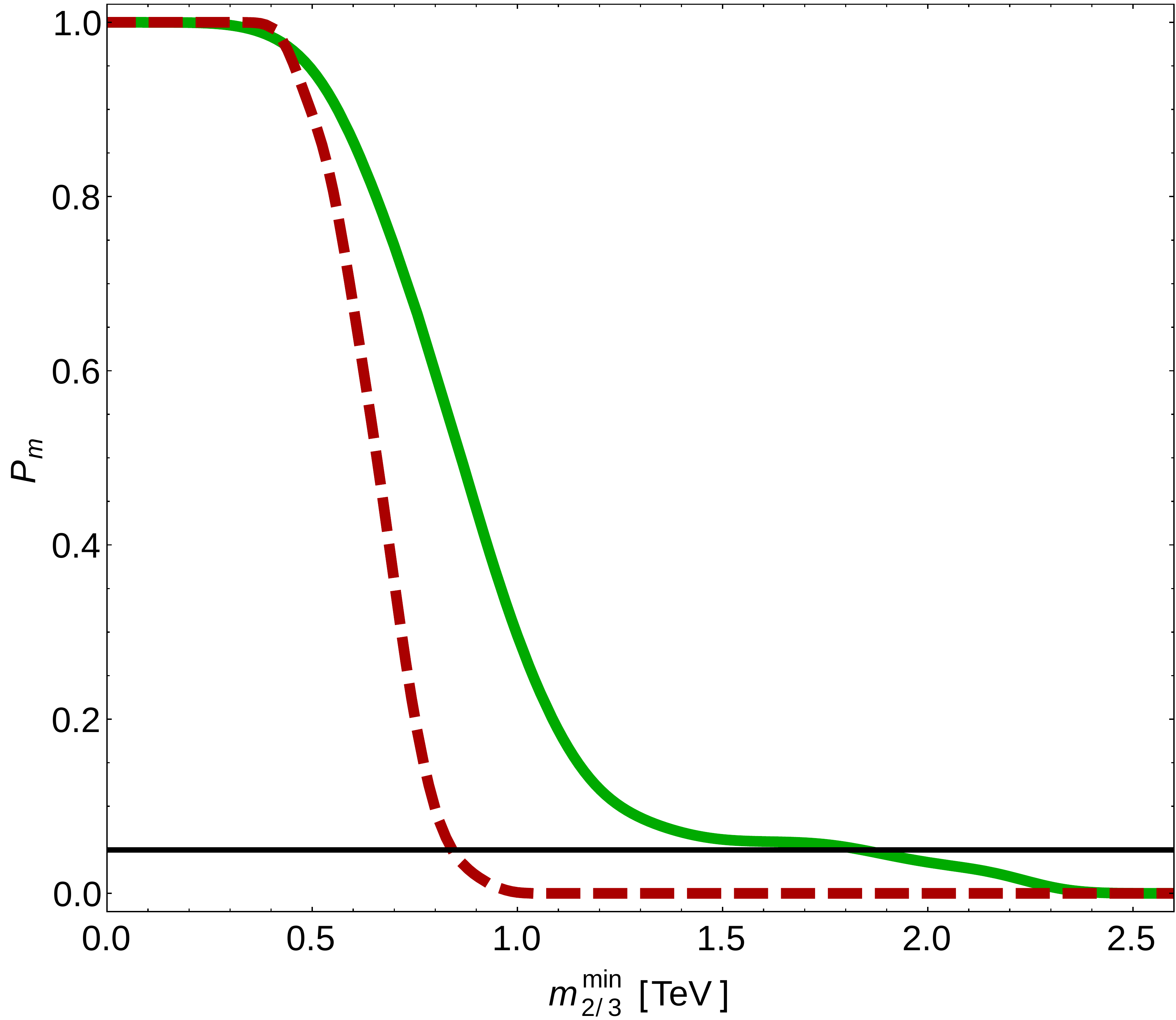}
			\includegraphics[width=0.493\textwidth]{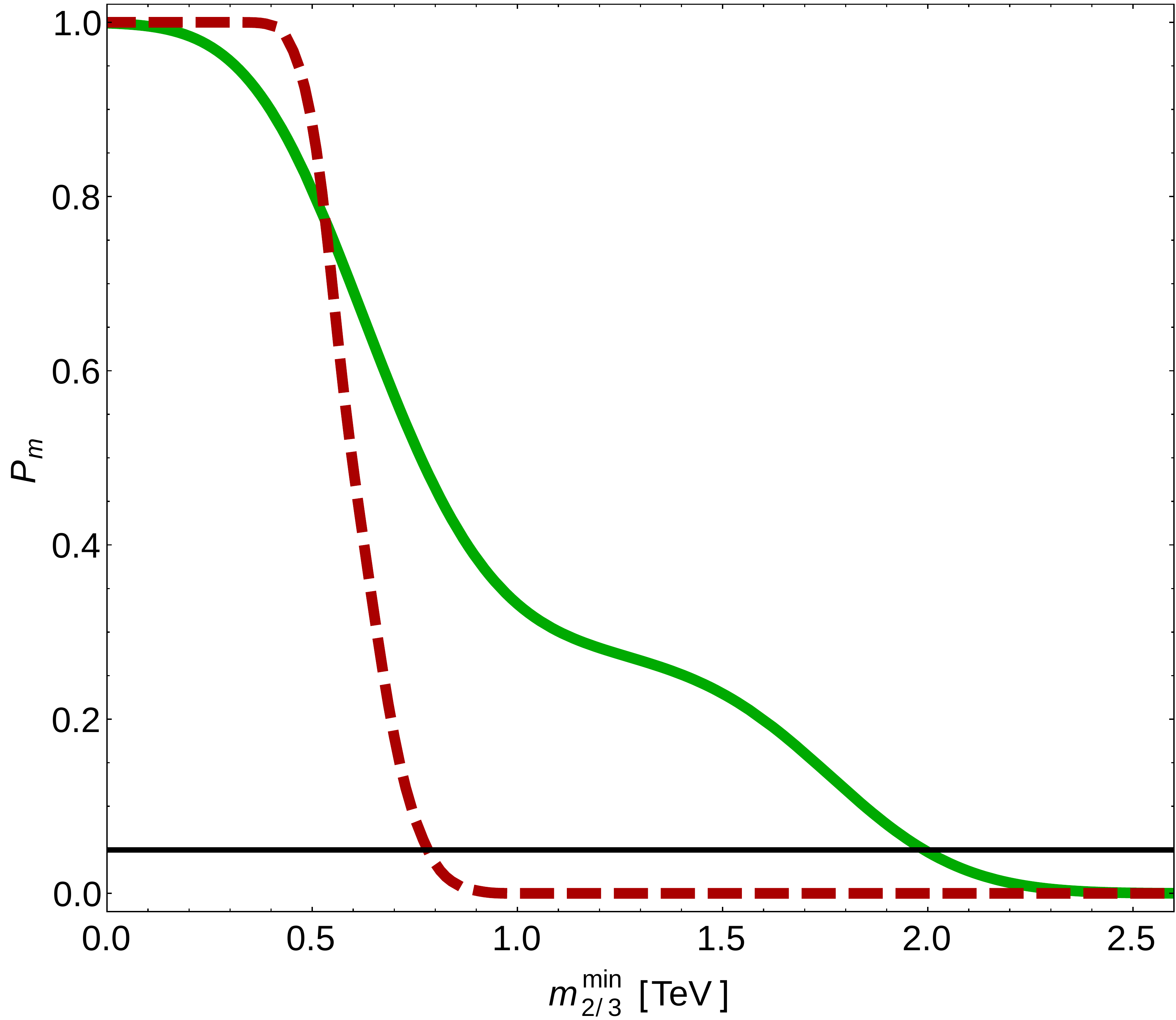}
			\caption{Survival function $\mathcal{P}_m$ of the first top partner mass for $Y_\ast^q=0.7$ (left plot) and $Y_\ast^q=1.4$ (right plot). In both cases we have plotted the MCHM$_5^{5-14}$ (solid green) with $Y_\ast^l=0.35$ vs. the MCHM$_5$ (dashed red) for $\kappa=\kappa^{\prime}=0$ and $f_{\pi}=0.8$\,TeV, assuming  $m_{H}(f_\pi)=105$\,GeV$\,(1 \pm 7.5 \%)$.}
\label{fig:14p5sfmcus}
\end{center}
\end{figure}
We now move on to compare more quantitatively the predictions for $m^{\rm min}_{2/3}$ in the MCHM$_5^{5-14}$
with the minimal quark setups, where we take the MCHM$_5$ as a reference. Checking the corresponding survival function 
for the MCHM$_5^{5-14}$ 
(green solid) against the one of the MCHM$_5$ (red-dashed) in the left panel of Figure~\ref{fig:14p5sfmcus}, where  
$Y_\ast^q=0.7$ and always $f_{\pi}=0.8$\,TeV, one can see that while $95\%$ ($99\%$) of the parameter points of the MCHM$_5$ feature light partners 
below 800\,GeV (950 \,GeV), in the MCHM$_5^{5-14}$ still $60\%$ ($35\%$) of the points have 
$m^{\rm min}_{2/3}$ bigger than these values. In particular the $95\%$ quantile is reached only 
for $m^{\rm min}_{2/3}= 1.8$\,TeV.
The difference becomes in general even more dramatic for larger brane masses, $Y_\ast^q=1.4$, shown in the right panel
of Figure~\ref{fig:14p5sfmcus}. 
While the values for the MCHM$_5$ are very similar, for the MCHM$_5^{5-14}$
now even $25\%$ of the points feature $m^{\rm min}_{2/3}>1.5$\,TeV.
This can be understood from the fact that in that case the large top mass allows more easily larger values of the partner 
masses, see (\ref{eq:mt}), while the correct Higgs mass can still be obtained via the lepton contributions.

\begin{figure}[t!]
\begin{center} 
		   \includegraphics[width=0.403\textwidth]{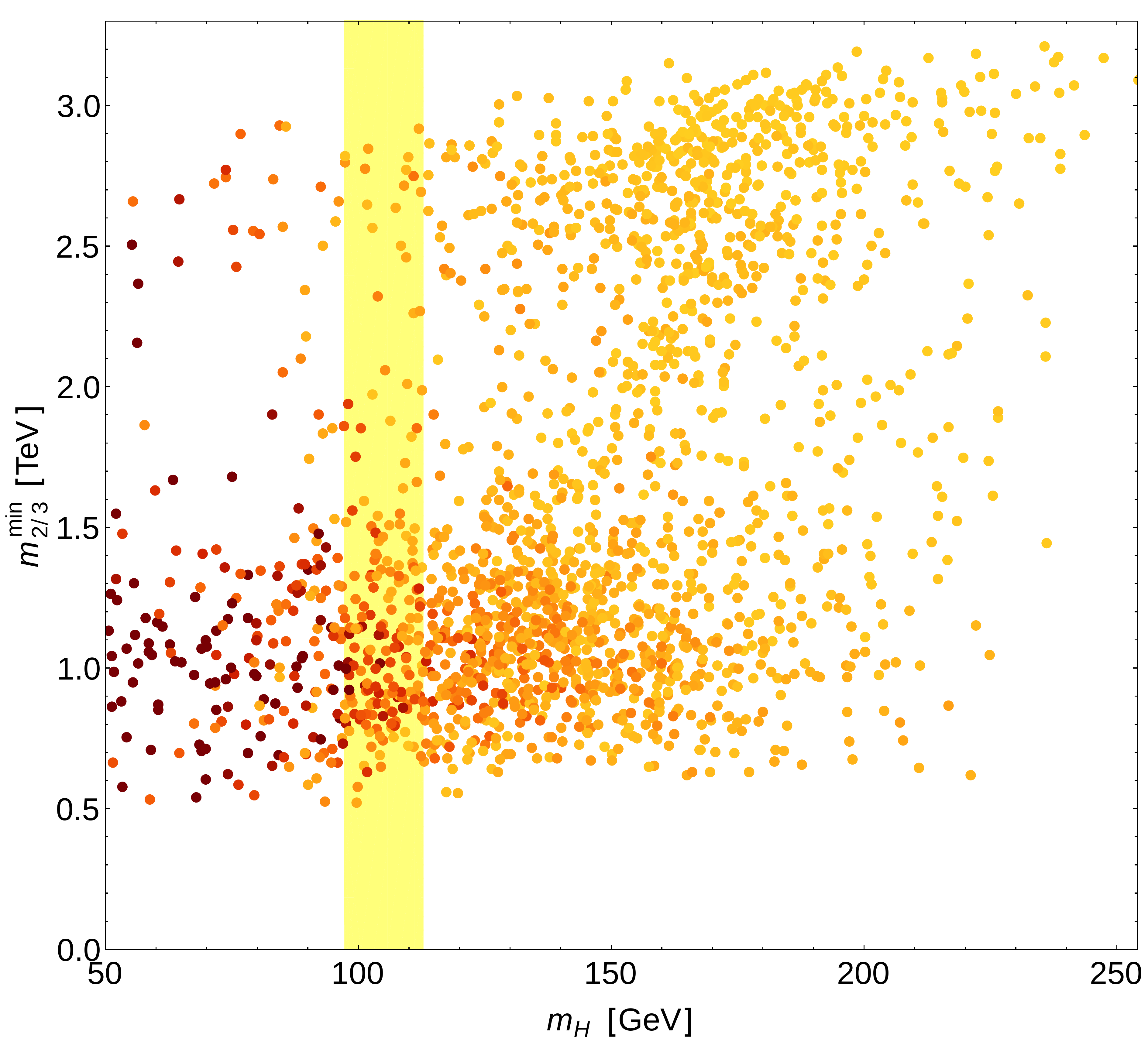}
		   \includegraphics[width=0.081\textwidth]{figures/bar_ft.pdf}
			\includegraphics[width=0.403\textwidth]{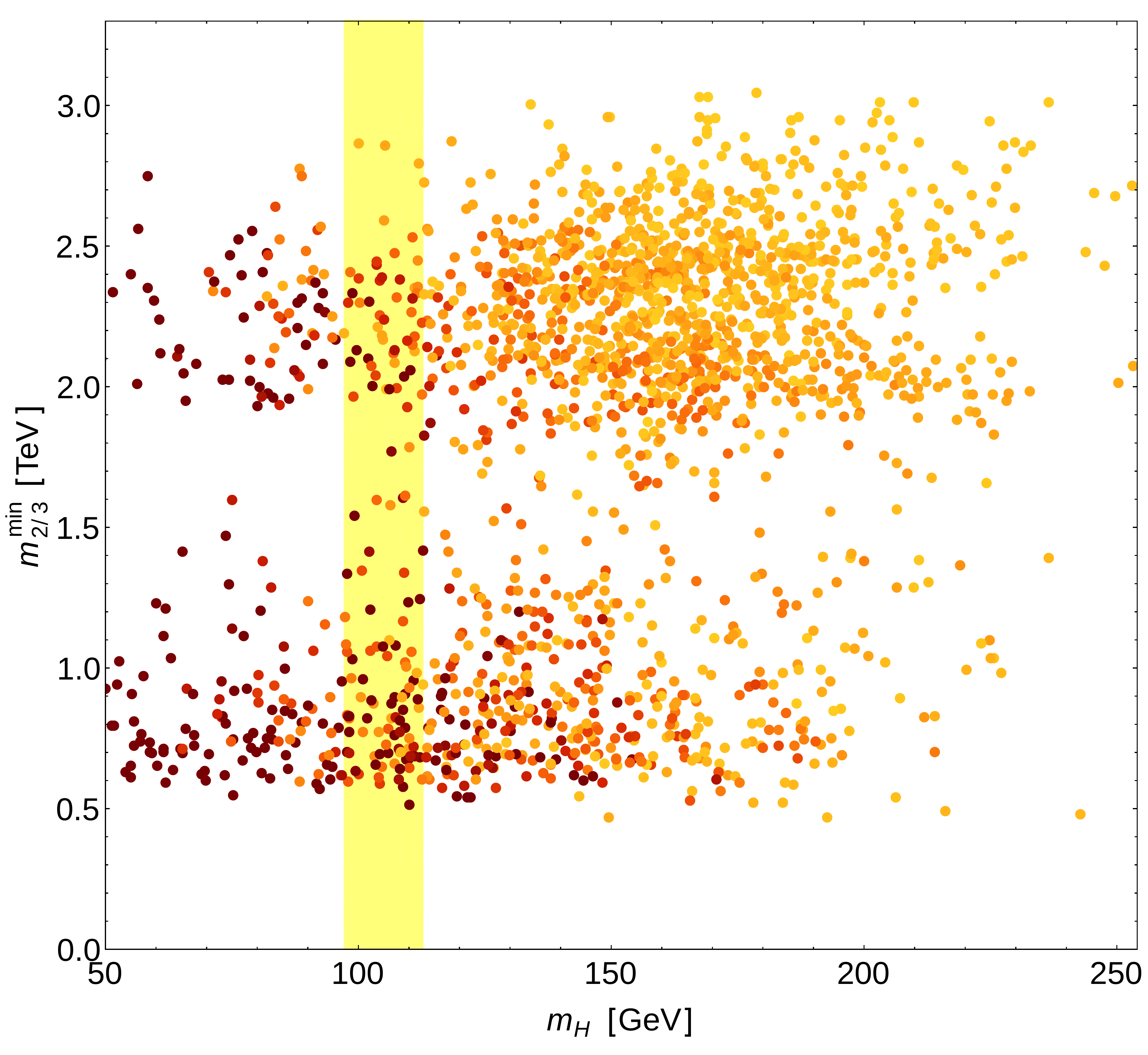}
			 \includegraphics[width=0.081\textwidth]{figures/bar_ft.pdf}
		\caption{Mass of the first top partner as a function of the Higgs mass in the MCHM$_5^{14}$ for $Y_\ast^q=0.7$ (left plot) and $Y_\ast^q=1.4$ (right plot), where always  $Y_\ast^l=0.35$,  $f_{\pi}=1$\,TeV, and $\kappa=\kappa^{\prime}=0$. Lighter points correspond to smaller values of $\Delta_{\rm BG}$ and therefore to less tuned points. }
\label{fig:mchm14mcusvsmhft}
\end{center}
\end{figure}

\begin{figure}[t!]
\begin{center} 
		   \includegraphics[width=0.403\textwidth]{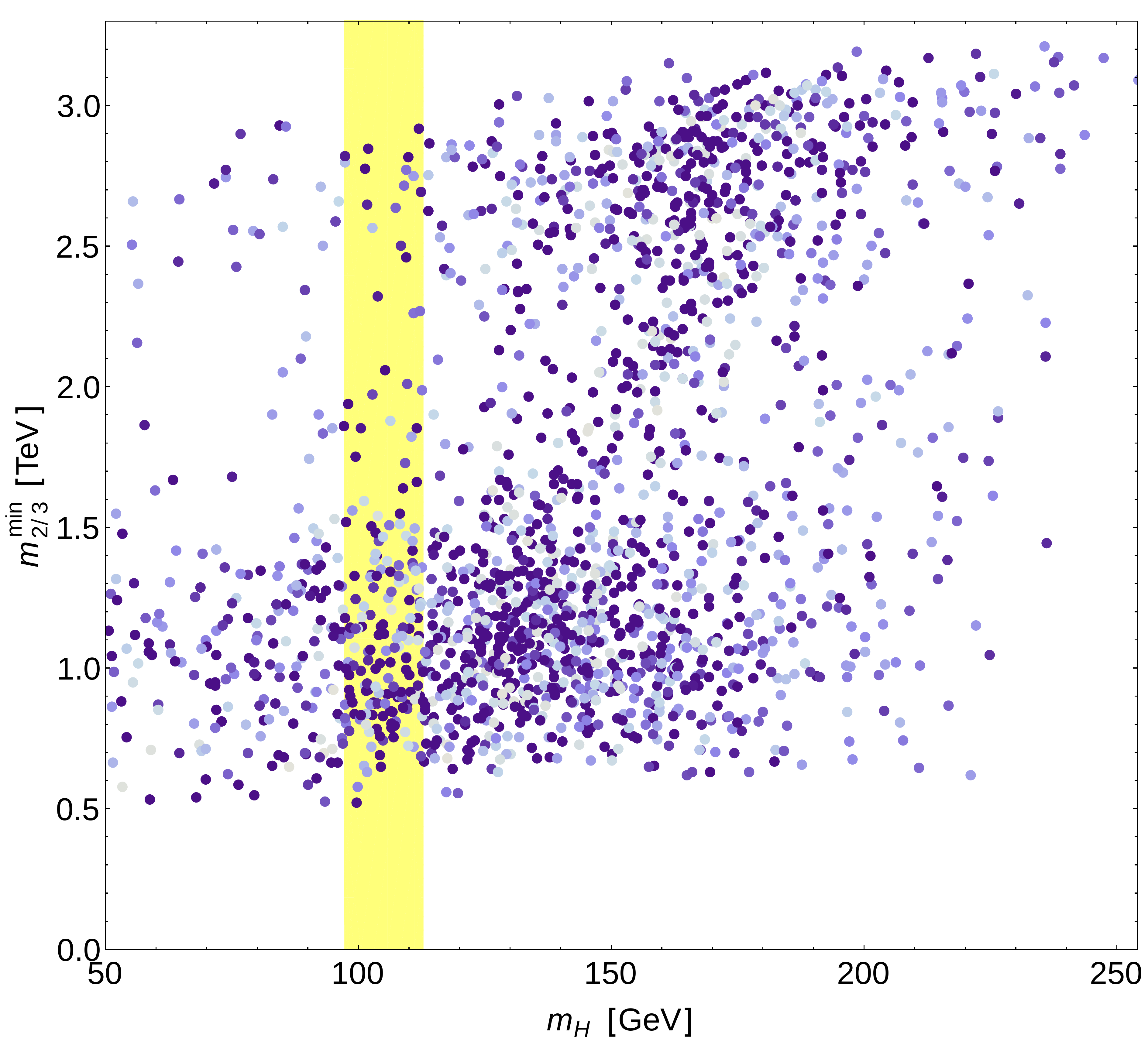}
		   \includegraphics[width=0.081\textwidth]{figures/bar_ml.pdf}
			\includegraphics[width=0.403\textwidth]{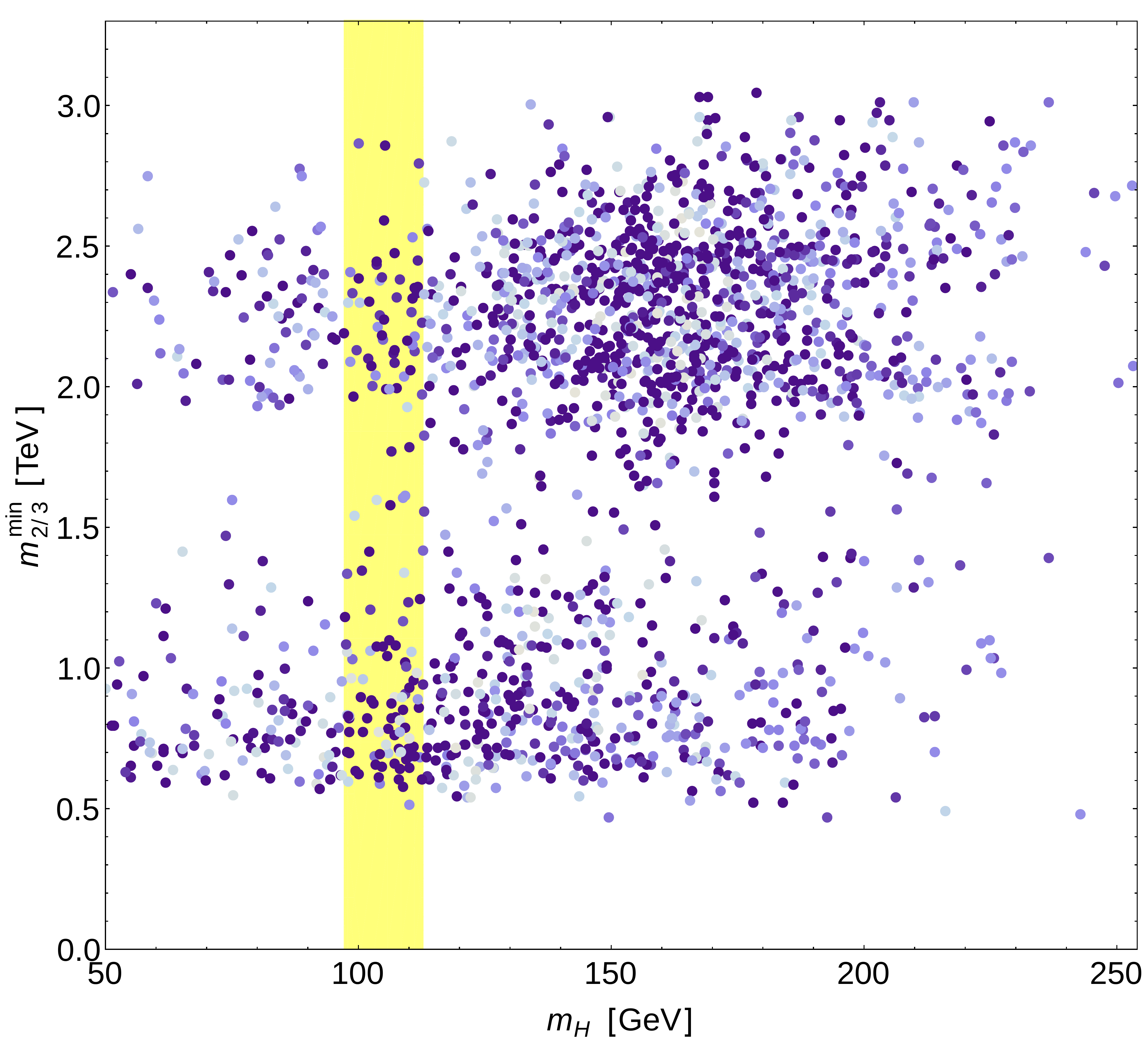}
			 \includegraphics[width=0.081\textwidth]{figures/bar_ml.pdf}
			 \caption{Mass of the first top partner as a function of the Higgs mass in the MCHM$_5^{14}$ 
 for $Y_\ast^q=0.7$ (left plot) and $Y_\ast^q=1.4$ (right plot), where always $Y_\ast^l=0.35$, $f_{\pi}=1$\,TeV, and $\kappa=\kappa^{\prime}=0$. Lighter points correspond to smaller values of $m_{-1}^{\rm min}$.}
\label{fig:mchm14mcusvsmhm1}
\end{center}
\end{figure}
\begin{figure}[t!]
\begin{center} 
			\includegraphics[width=0.493\textwidth]{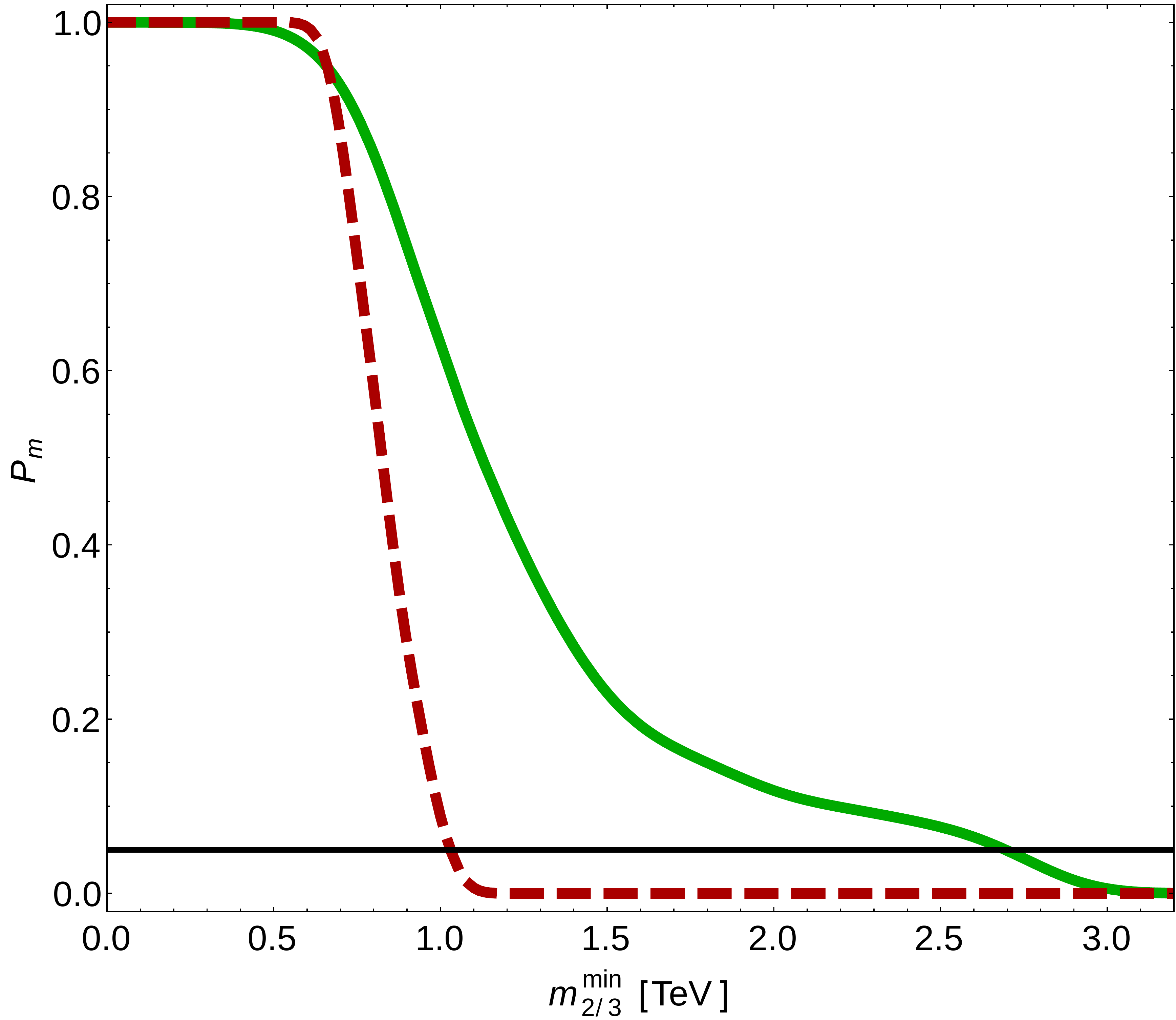}
			\includegraphics[width=0.493\textwidth]{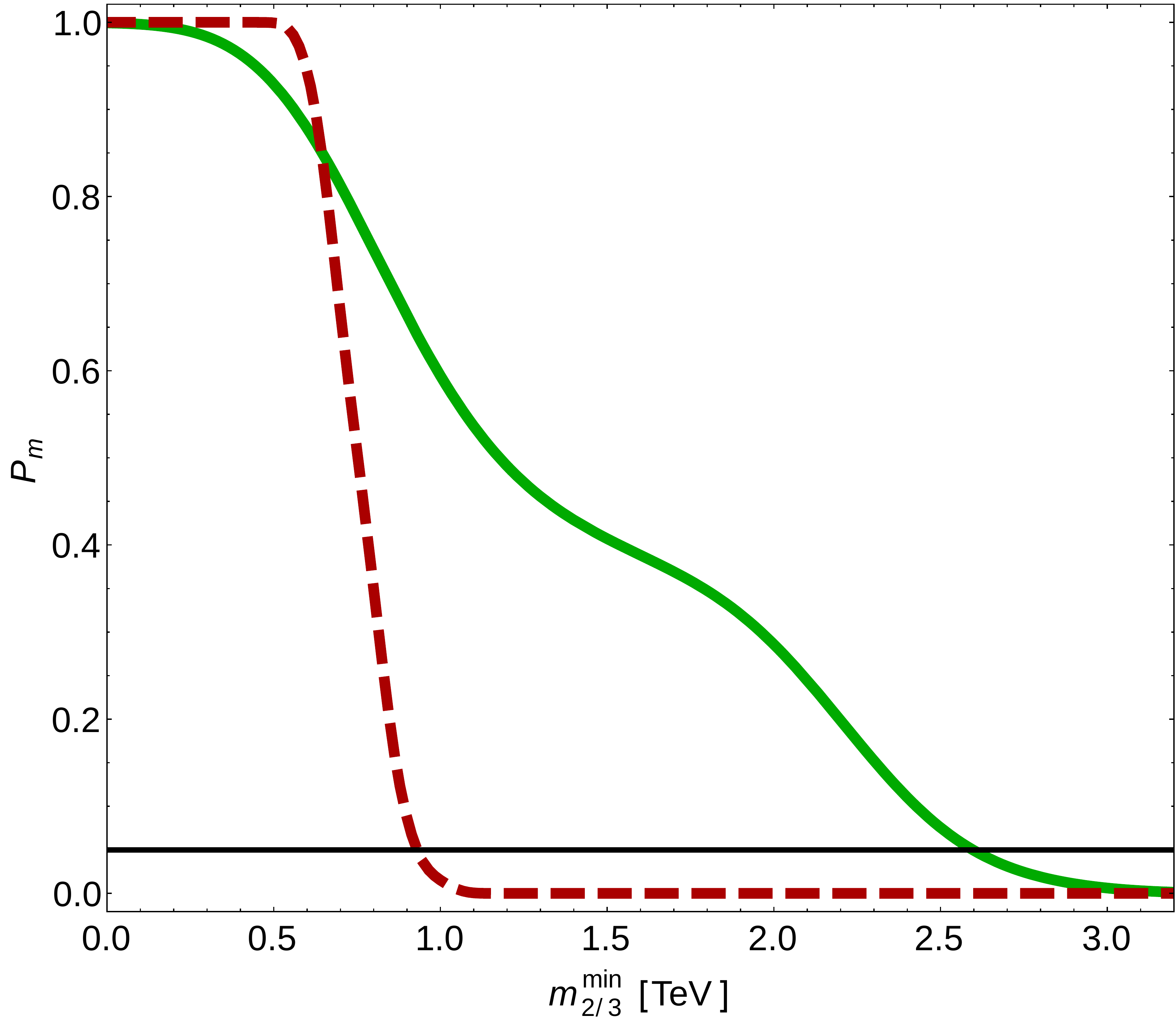}
			\caption{Survival function $\mathcal{P}_m$ of the first top partner mass for $Y_\ast^q=0.7$ (left plot) and $Y_\ast^q=1.4$ (right plot). In both cases we have plotted the MCHM$_5^{14}$ (solid green) with  $Y_\ast^l=0.35$ vs. the MCHM$_5$ (dashed red), for $\kappa=\kappa^{\prime}=0$ and $f_{\pi}=1$\,TeV, assuming  $m_{H}(f_\pi)=105$\,GeV$\,(1 \pm 7.5 \%)$.}
\label{fig:mchm14sfmcus}
\end{center}
\end{figure}
For completeness we also present the corresponding predictions for the MCHM$_5^{14}$, where both the left-handed and
right-handed leptons are embedded in symmetric representations of $SO(5)$, in Figures~\ref{fig:mchm14mcusvsmhft}-\ref{fig:mchm14sfmcus}. Note that here we show only the results for $f_\pi=1$\,TeV, since the model
generically features larger corrections to the Higgs couplings, compared to the MCHM$_5^{5-14}$-like setup, which has a more efficient protection, as can be seen from the results of Section~\ref{sec:14s} and taking into account that the infinite KK sum
is saturated to a large extend by the first mode.
As expected, although some numerical details change, the qualitative 
behavior is not modified by mixing the (elementary) left handed leptons with another representation. The fine tuning, depicted 
in Figure~\ref{fig:mchm14mcusvsmhft}, is similar to the case of $f_\pi=1$\,TeV examined before, and the same is true for the 
masses of the $\tau$ partners, which are still not required to be light, see Figure~\ref{fig:mchm14mcusvsmhm1}. In the end, 
in the MCHM$_5^{14}$ with $f_\pi=1$\,TeV the top partners can be lifted to high values of $m^{\rm min}_{2/3} 
\gtrsim 2.5$\, TeV without going to extreme corners of the parameter space, see Figure~\ref{fig:mchm14sfmcus}.

\begin{figure}[t!]
\begin{center} 
{%
		   \includegraphics[width=0.406\textwidth]{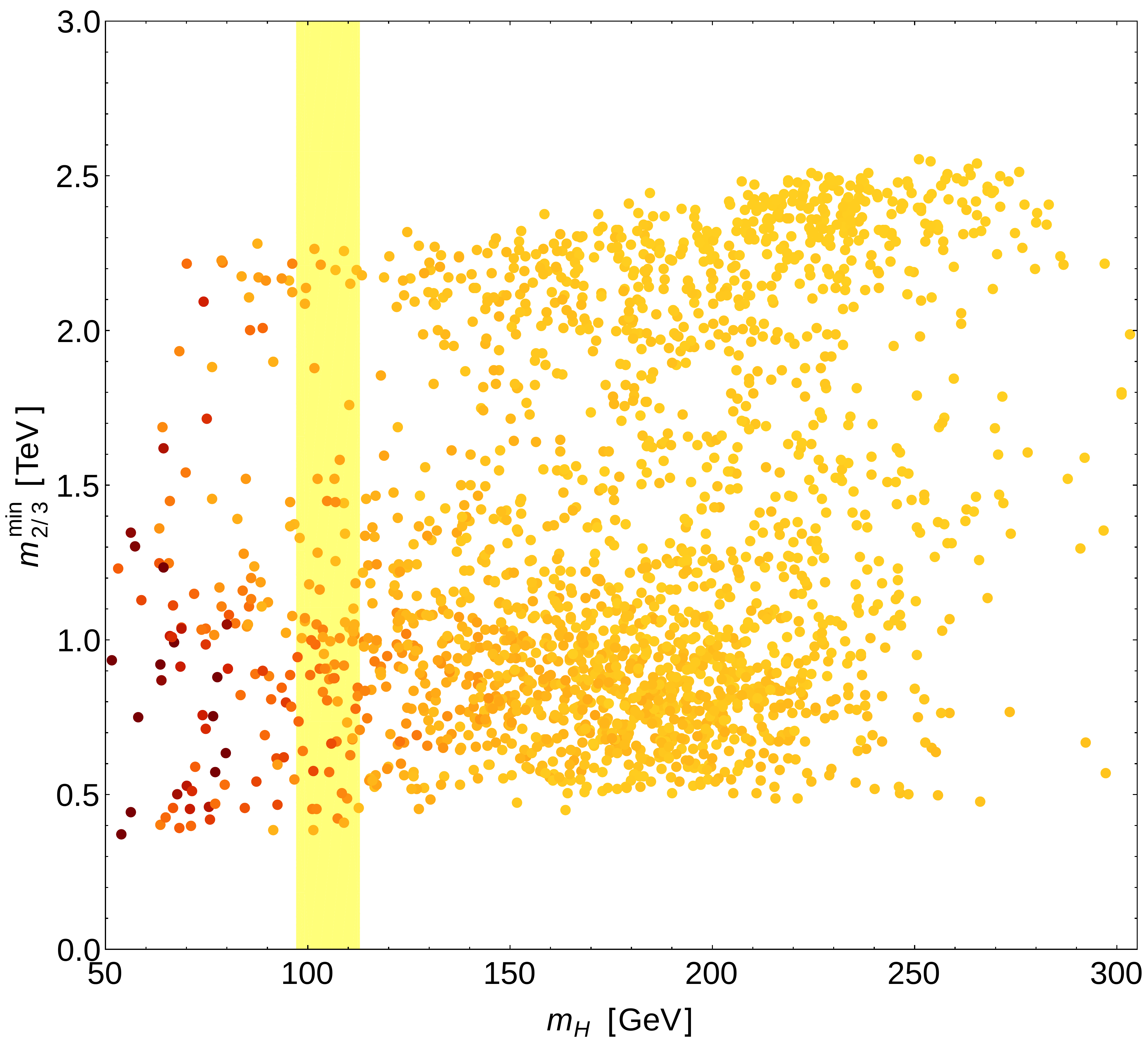}
		   \includegraphics[width=0.0816\textwidth]{figures/bar_ft.pdf}
			\includegraphics[width=0.406\textwidth]{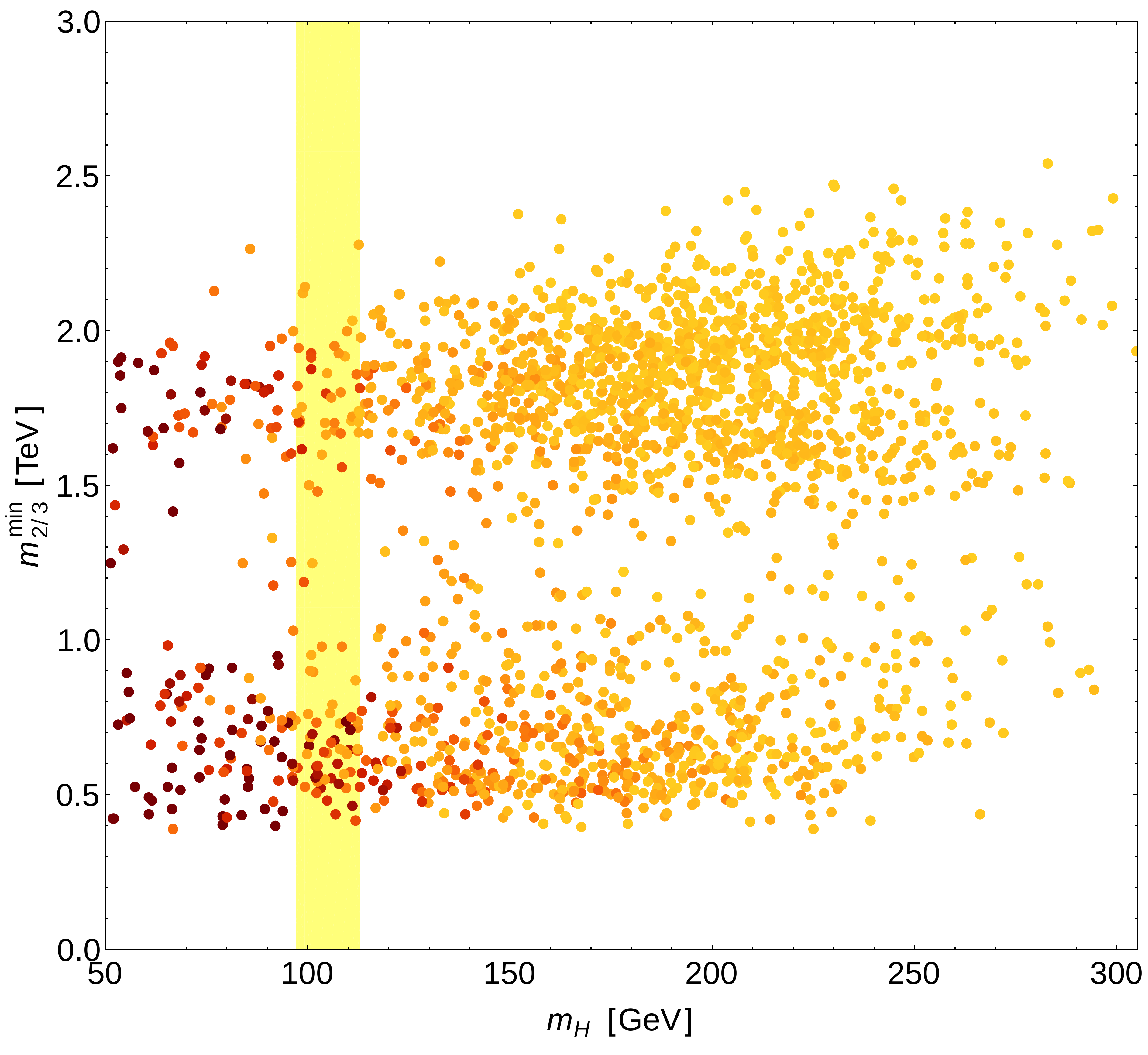}
			 \includegraphics[width=0.0816\textwidth]{figures/bar_ft.pdf}
		}
		\caption{Mass of the first top partner as a function of the Higgs mass in the mMCHM$^{\rm III}_{5}$ for 
$Y_\ast^q=0.7$ (left plot) and $Y_\ast^q=1.4$ (right plot). In both cases we have fixed
 $Y_\ast^l=0.35$, $\kappa=\kappa^{\prime}=0$, 
as well as $f_{\pi}=0.8$\,TeV. Lighter points correspond to smaller values of $\Delta_{\rm BG}$ and therefore to less tuned points. }
\label{fig:14p5majmcusvsmhft}
\end{center}
\end{figure}

Let us continue with more minimal models of the lepton sector. We start with the minimal type-III seesaw model  
mMCHM$^{\rm III}_{5}$, introduced before - assuming still a Yukawa suppression due to flavor symmetries or the like, 
{\it i.e.}, $Y_\ast^l=0.35$. We study again the mass of the 
lightest $Q=2/3$ resonance, as a function of the Higgs mass, which we display in Figure~\ref{fig:14p5majmcusvsmhft}, assuming 
once more $Y_\ast^q=0.7$ and $Y_\ast^q=1.4$, as well as $f_{\pi}=0.8~\TeV$. As before, the tuning $\Delta_{\rm BG}$ is
visualized by the hue of the points in the $m_H-m^{\rm min}_{2/3}$ plane.
It is evident from the plots that this model, as expected, still allows to lift the masses of the lightest top partner well beyond 
a TeV, without increasing the tuning. On the contrary, just as for the MCHM$_5^{5-14}$, heavier top partners are more natural
in this model (which is not true for models with negligible impact of the lepton sector).
Compared to the  MCHM$_5^{5-14}$, due to the $N_g$-enhanced leptonic contribution in the mMCHM$^{\rm III}_{5}$, 
the fraction of the parameter space featuring $m^{\rm min}_{2/3}\sim (1-2.5)$\, TeV for a viable $m_H$
(displayed again by the yellow band) is however significantly enhanced.
Since a considerable portion of these points features a tuning of $\Delta_{\rm BG} \lesssim (10-20)$ the model even allows to
reconcile the absence of light partners with the minimization of the fine tuning.\footnote{Still the model features no
``ad-hoc'' tuning and the observed $m_H$ remains in a rather central region of the parameter space.} Moreover, the masses
of the lepton partners are in general at the high scale $m_\Psi$ in the type-III seesaw models, as the IR localization of the 
right-handed leptons is rather modest (and $m_{\ell}\ll m_t$).

\begin{figure}[t!]
\begin{center} 
		   \includegraphics[width=0.403\textwidth]{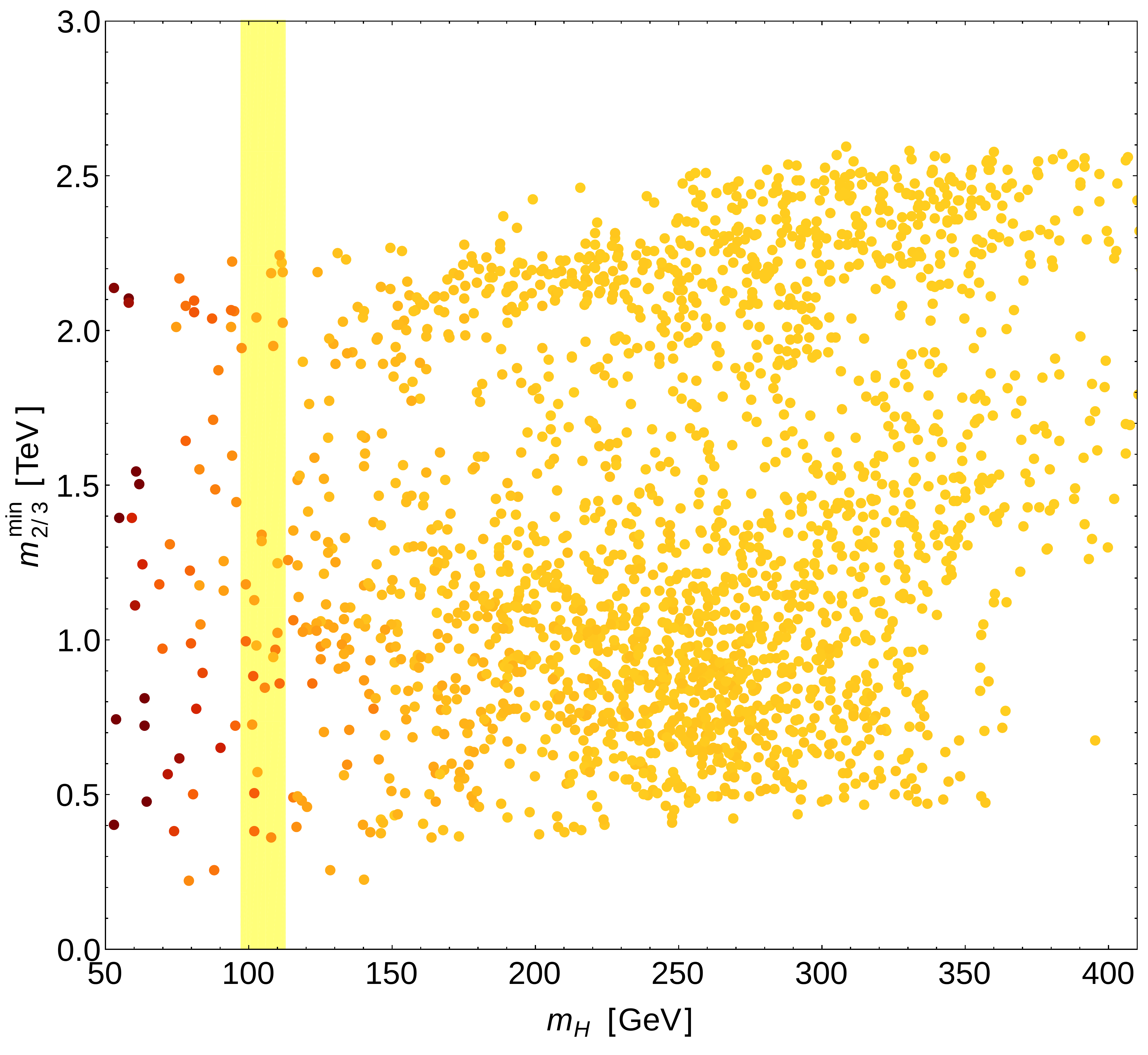}
		   \includegraphics[width=0.081\textwidth]{figures/bar_ft.pdf}
			\includegraphics[width=0.403\textwidth]{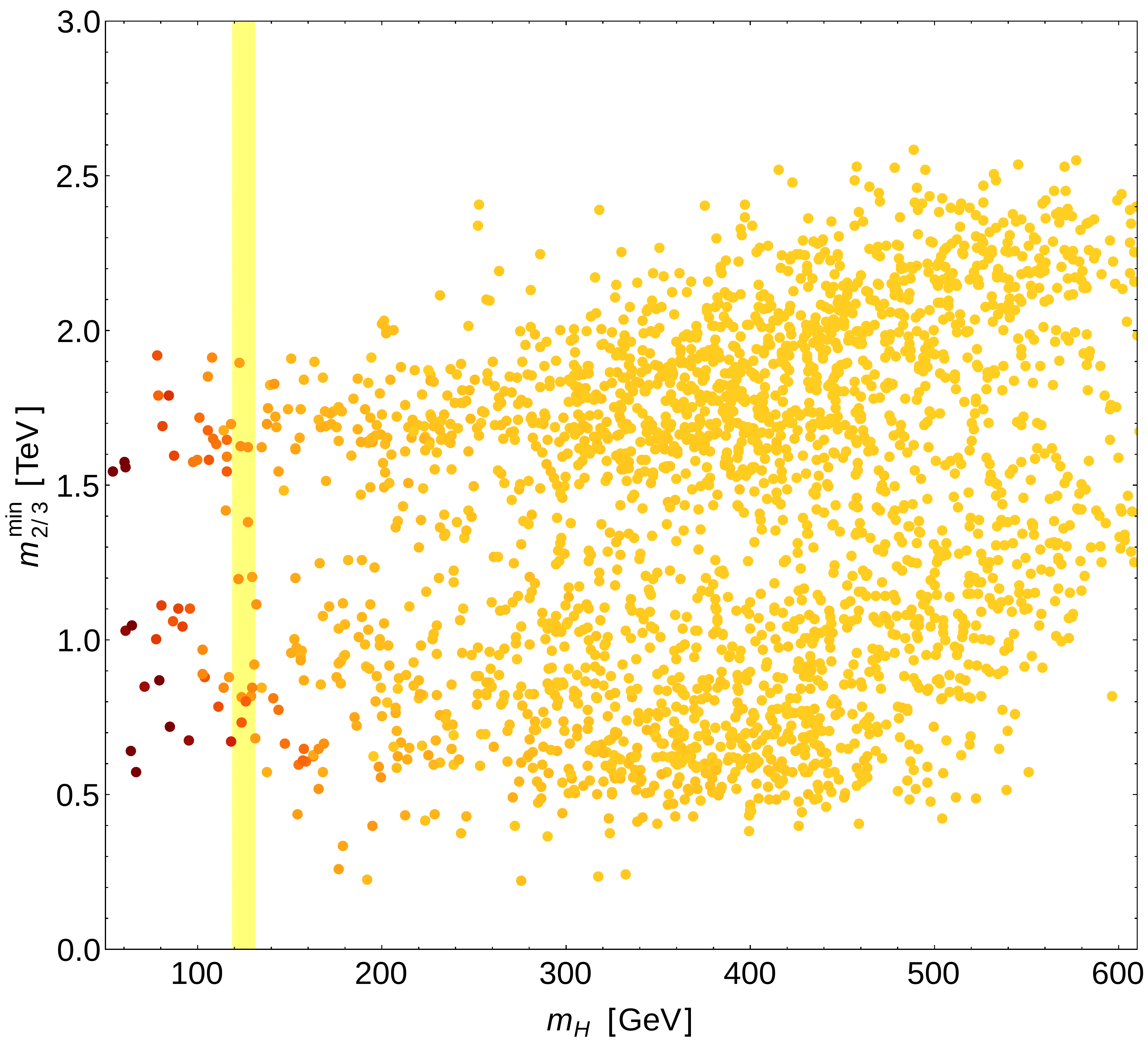}
			 \includegraphics[width=0.081\textwidth]{figures/bar_ft.pdf}
			 \caption{Mass of the first top partner as a function of the Higgs mass in the mMCHM$^{\rm III>}_{5}$ for $Y_\ast^l=Y_\ast^q=0.7$ (left plot) and $Y_\ast^l=Y_\ast^q=1.4$ (right plot). In both cases we have fixed $\kappa=\kappa^{\prime}=0$ and $f_{\pi}=0.8$\,TeV. Lighter points correspond to smaller values of $\Delta_{\rm BG}$ and therefore to less~tuned~points. }
\label{fig:14p5majmcusvsmhftnh}
\end{center}
\end{figure}

For completeness we also examine the mMCHM$^{\rm III>}_{5}$, {\it i.e.}, we lift the `Yukawa suppression' so that we now study,
in Figure~\ref{fig:14p5majmcusvsmhftnh}, the cases of $Y_\ast^l=Y_\ast^q=0.7$ (left plot) and $Y_\ast^l=Y_\ast^q
=1.4$ (right plot), while the other parameters remain as given before. While the qualitative situation remains similar, now the 
more pronounced leptonic contribution leads to a further increase of the average $m^{\rm min}_{2/3}$ in the viable 
$m_H$ region, which however now features slightly less points. The increase is due to the fact that in this class of models the localization of the right handed leptons is mostly fixed by the Majorana masses and remains constant under changes of $Y_\ast^l$. 

\begin{figure}[t!]
\begin{center} 

			\includegraphics[width=0.493\textwidth]{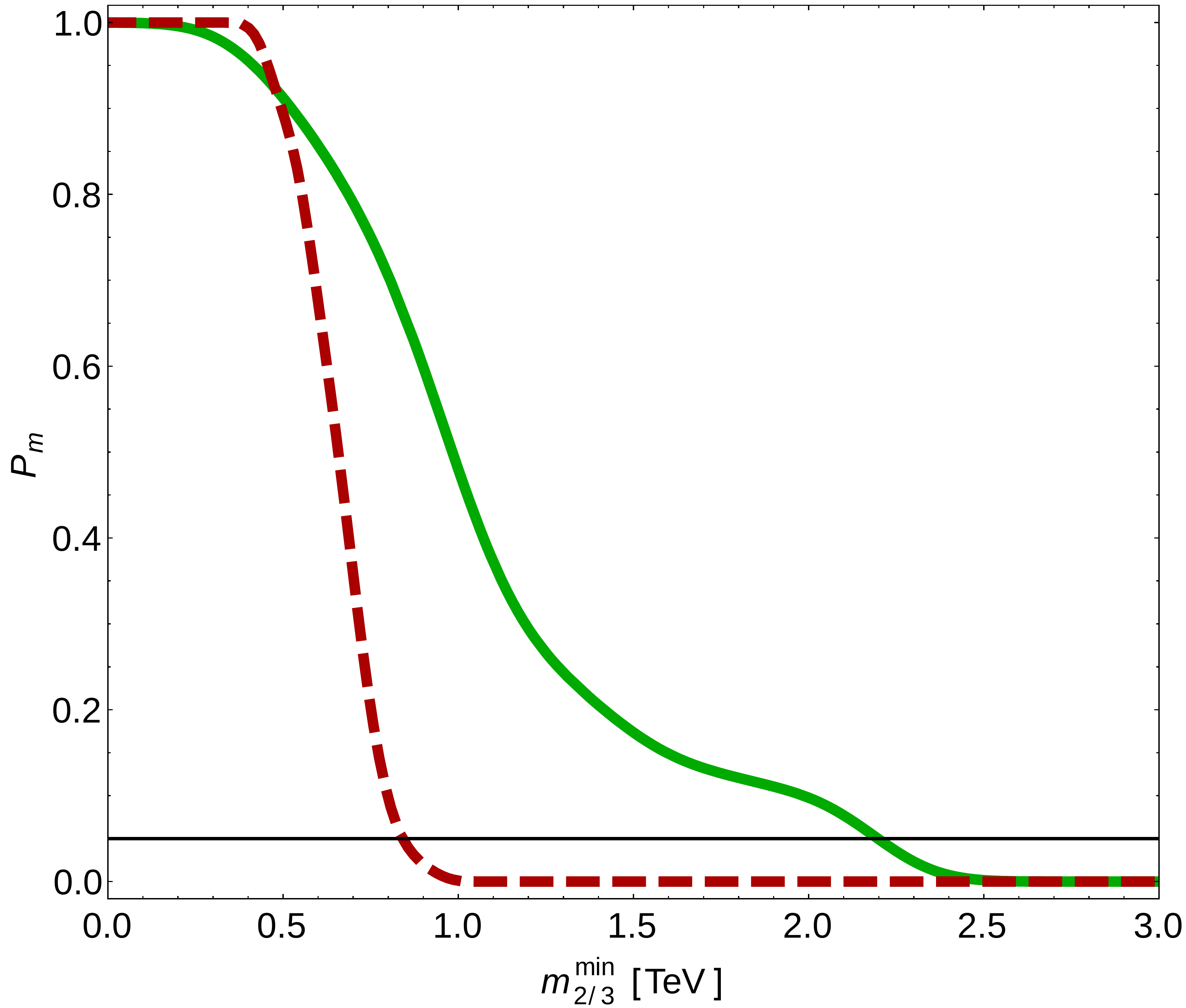}
			\includegraphics[width=0.493\textwidth]{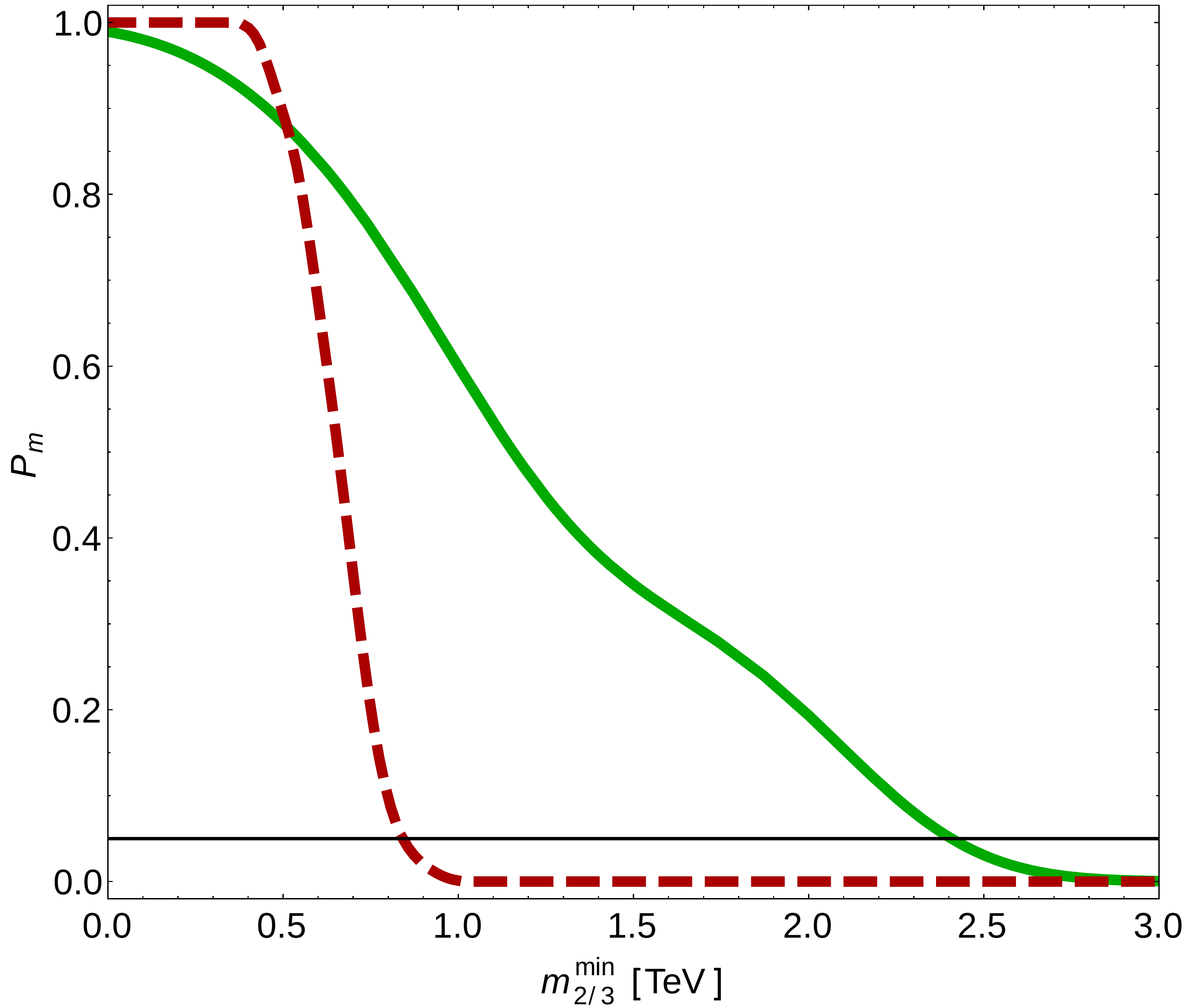}
			\caption{Survival function $\mathcal{P}_m$ of the first top partner mass 
in the mMCHM$^{\rm III}_5$, with $Y_\ast^l=0.35$ (solid green, left plot) and in the mMCHM$^{\rm III>}_{5}$, with
$Y_\ast^l=0.7$ (solid green, right plot), plotted in each case vs. the MCHM$_5$ (dashed red). We always assume $Y_\ast^q=0.7$, $\kappa=\kappa^{\prime}=0$, $f_{\pi}=0.8$\,TeV, as well as  $m_{H}(f_\pi)=105$\,GeV$\,(1 \pm 7.5 \%)$.}
\label{fig:14p5majsfmcus}
\end{center}
\end{figure}

The distribution of the partner masses in the minimal lepton models is examined in more detail in Figure~\ref{fig:14p5majsfmcus}
where we display the survival 
function for the first top partner mass in the mMCHM$^{\rm III}_{5}$ (green solid, left panel) as well as in the 
mMCHM$^{\rm III>}_{5}$ (green solid, right panel), employing $Y_\ast^l=0.35$ and $Y_\ast^l=0.7$, respectively, 
always against the one in the MCHM$_5$ (red-dashed), with $Y_\ast^q=0.7$, and always $f_\pi=0.8$\,TeV.
Although the size of the brane masses is still rather modest, we can see a large relaxation of the need for light partners.
For the case of the mMCHM$^{\rm III}_{5}$, still $\gtrsim10\%$ of the parameter space (in agreement with the Higgs mass) features partner masses of even $m^{\rm min}_{2/3}\gtrsim 2 $\, TeV while for the mMCHM$^{\rm III>}_{5}$ the same holds true for $\sim 20\%$ of the viable points.

\begin{figure}[!t]
	\begin{center}
			\includegraphics[width=0.403\textwidth]{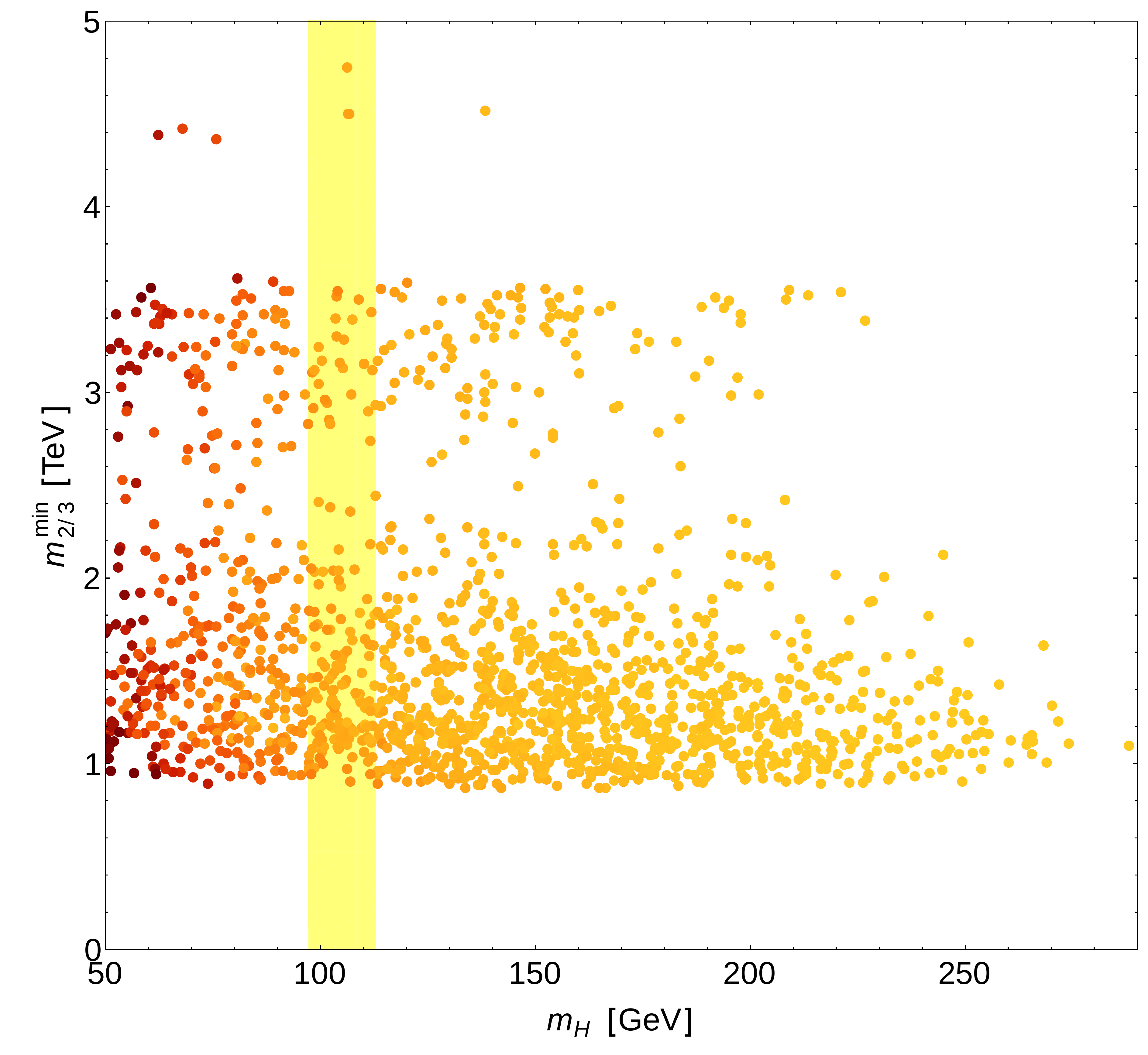}
			 \includegraphics[width=0.081\textwidth]{figures/bar_ft.pdf}
			\includegraphics[width=0.403\textwidth]{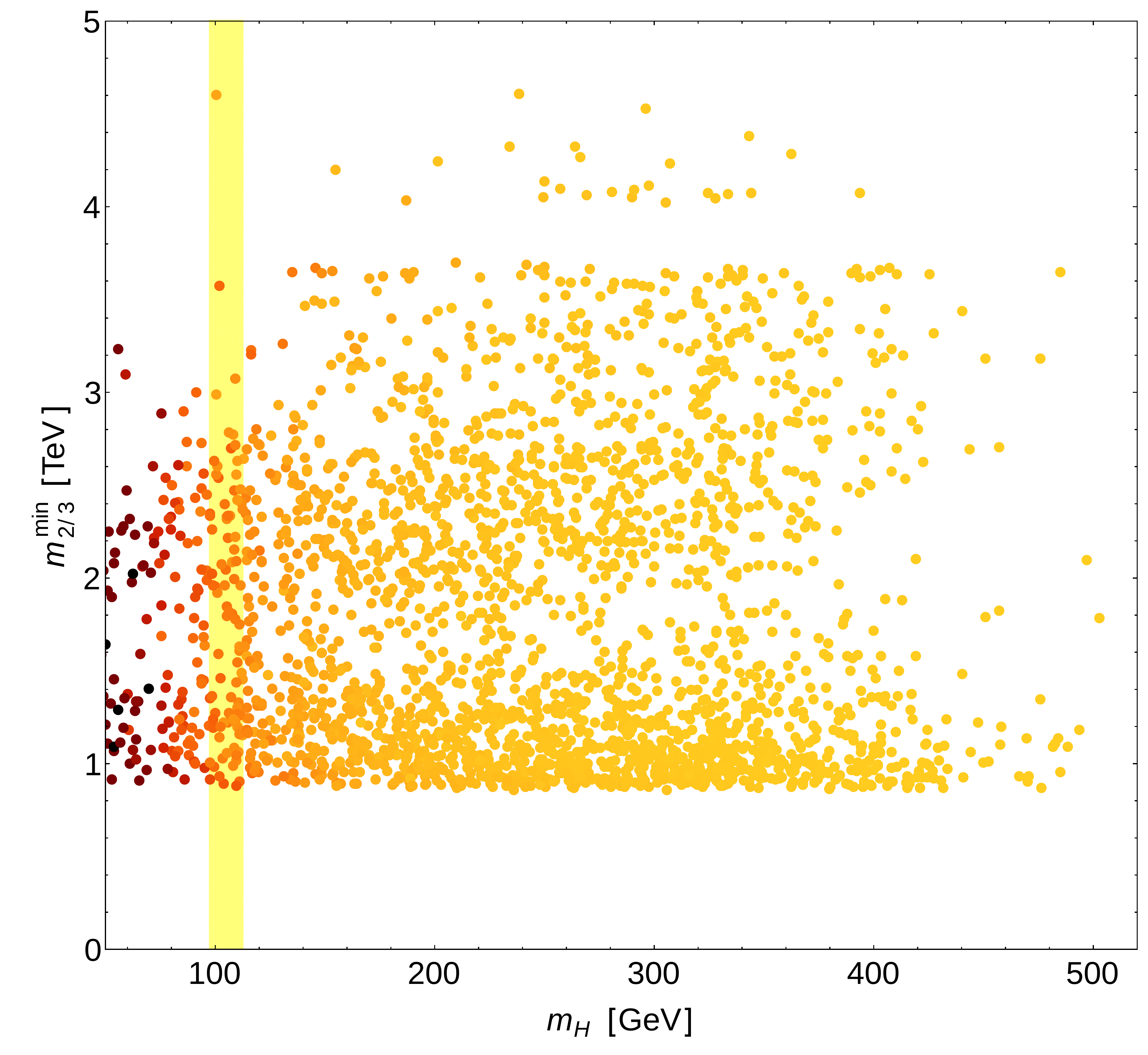}
		   	\includegraphics[width=0.081\textwidth]{figures/bar_ft.pdf}
			
			 \caption{Mass of the first top partner as a function of the Higgs mass in the mMCHM$^{\rm III >}_{5-1}$ with $Y_\ast^l=Y_\ast^q=0.7$ (left plot) and $Y_\ast^l=Y_\ast^q=1.4$  (right plot). In both cases we have again taken $\kappa=\kappa^{\prime}=0$ and $f_{\pi}=0.8$\,TeV. Lighter points correspond to smaller values of $\Delta_{\rm BG}$ and therefore to less tuned points.
\label{fig:14p5majmcusvsmhftnh5p1}}
		 \end{center}
\end{figure}

\begin{figure}[t!]
\begin{center} 

			\includegraphics[width=0.493\textwidth]{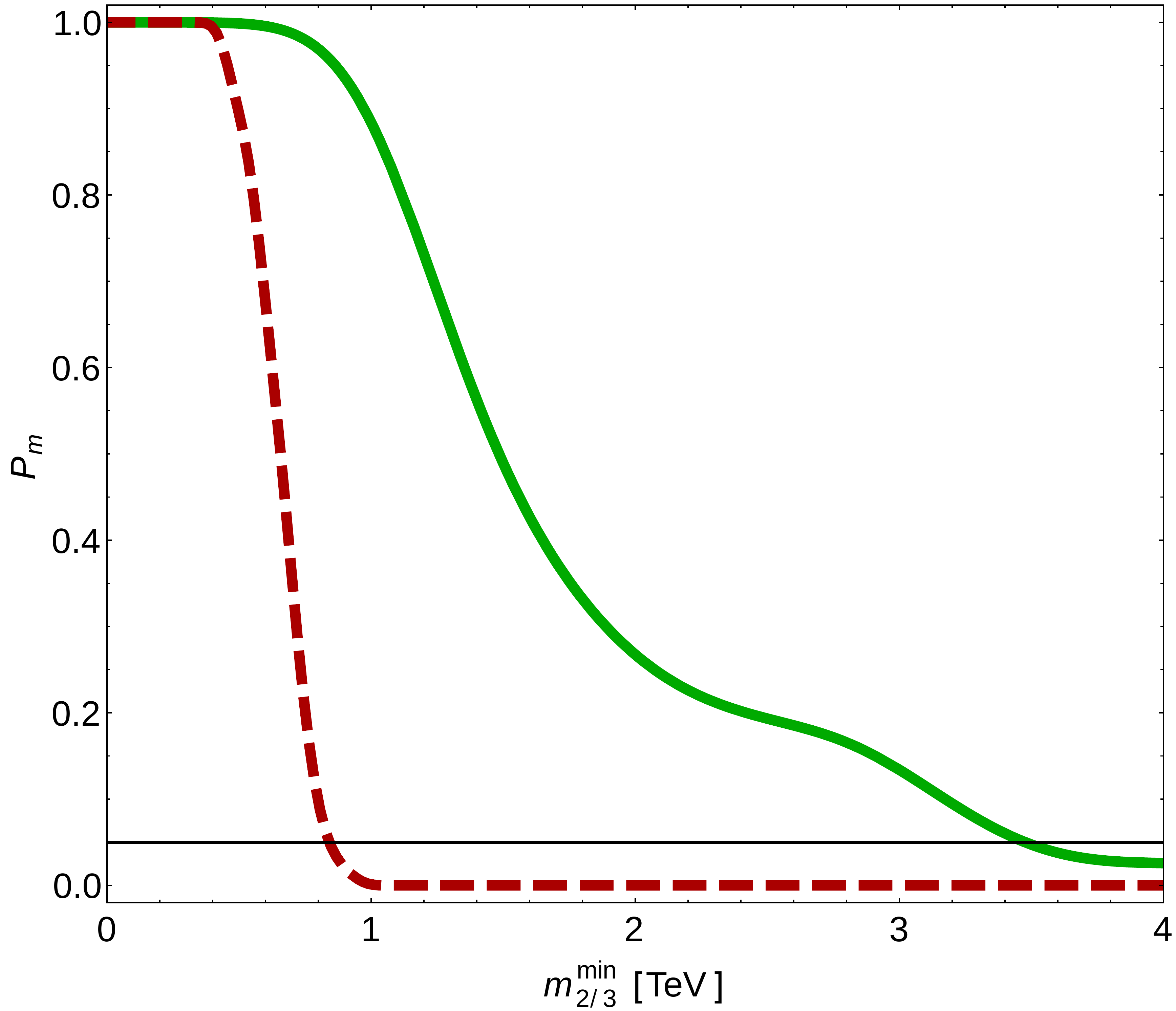}
			\includegraphics[width=0.493\textwidth]{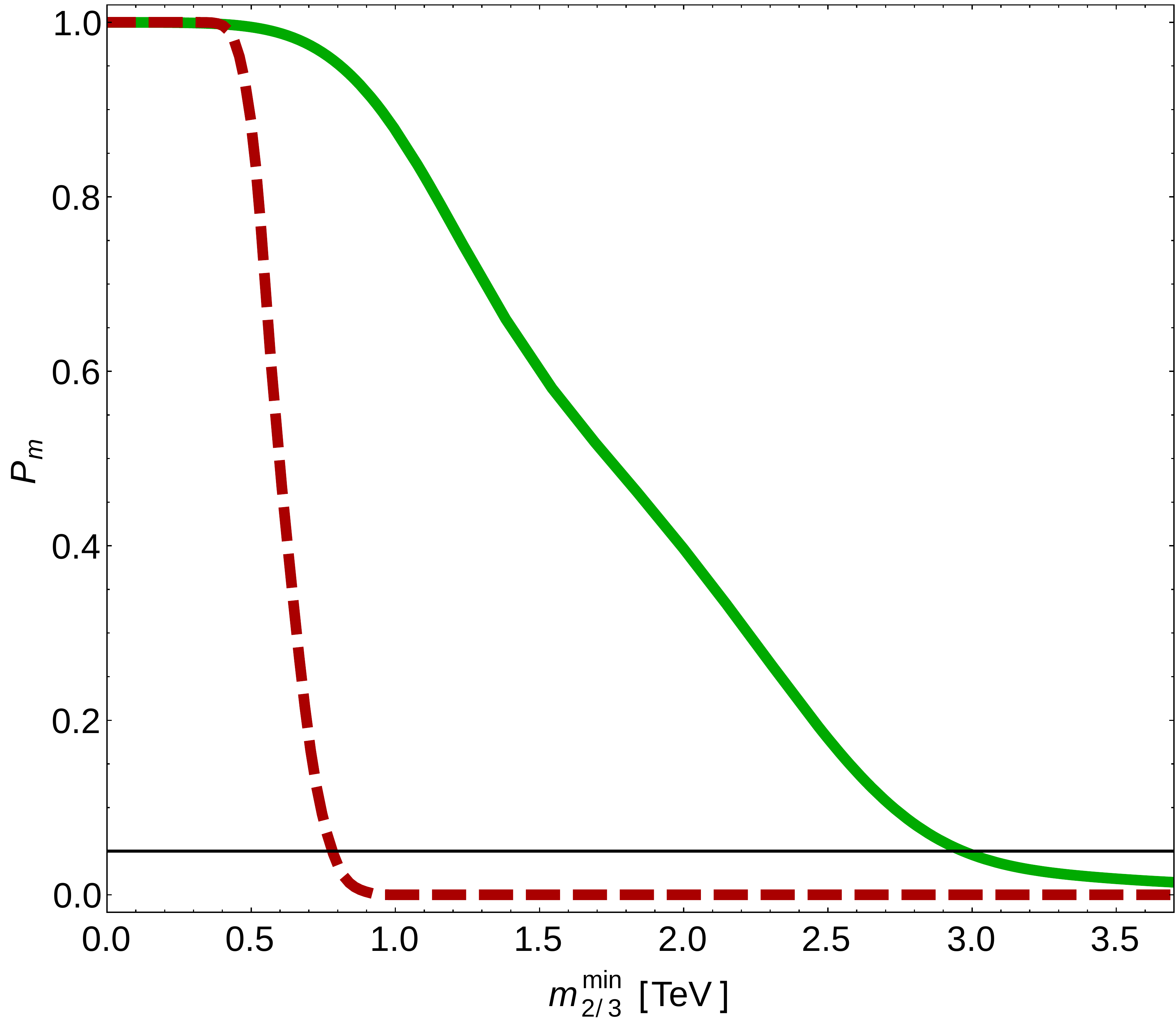}
			\caption{Survival function $\mathcal{P}_m$ of the first top partner mass 
in the mMCHM$^{\rm III>}_{5-1}$ with $Y_\ast^l=Y_{\ast}^q=0.7$ (solid green, left plot) and $Y_\ast^l=Y_{\ast}^q=1.4$ (solid green, right plot), plotted in each case vs. the MCHM$_5$ (dashed red) with $Y_\ast^q=0.7$ and $Y_{\ast}^q=1.4$, respectively. In both cases we assume $\kappa=\kappa^{\prime}=0$, $f_{\pi}=0.8$\,TeV, as well as $m_{H}(f_\pi)=105$\,GeV$\,(1 \pm 7.5 \%)$.
\label{fig:14p5majsfnh5p1}}
\end{center}
\end{figure}

Finally, we explore the most minimal of all models of the composite framework studied in this work, the mMCHM$^{\rm III >}_{5-1}$.
With this model, we expect to leave finally the class of setups getting along without any ad-hoc tuning.
The contribution $\alpha_l$ due to the leptons in the symmetric representation of $SO(5)$ is now required to cancel to a significant extend the sizable contribution of the top quark to allow for EWSB and, as discussed in Section~$\ref{sec:general}$, the now 
also large $\beta_l$ will generically lead to a relatively large $m_H$.\footnote{While already in the intermediate case of the
mMCHM$^{\rm III>}_{5}$ the leptonic contribution could be large, the model still allowed for more freedom, compared to the mMCHM$^{\rm III >}_{5-1}$, which has less parameters and where a sizable leptonic contribution to $\beta$ is unavoidable due to the large top mass. However, compared to the mMCHM$^{\rm III>}_{5}$, the even more natural mMCHM$^{\rm III}_{5}$ seems to be the preferred way out of the light partner issue with a light Higgs.}
Indeed, we inspect in Figure~\ref{fig:14p5majmcusvsmhftnh5p1},
which shows again $\Delta_{\rm BG}$ in the $m_H-m^{\rm min}_{2/3}$ plane, assuming $f_{\pi}=0.8~\TeV$,
that for very light Higgs masses the tuning raises strongly, already for the lower brane masses. 
However, although the model belongs to the category of ad-hoc tuning, in the end the situation is still not so bad 
and the correct Higgs mass can be reached with a modest tuning of $\Delta_{\rm BG} \sim (30-40) $ in the case of $Y_\ast^q=Y_\ast^l=0.7$ (left panel), while even for $Y_\ast^q=Y_\ast^l=1.4$ (right panel) the corresponding tuning is 
rather moderate, about $\Delta_{\rm BG} \sim (40-80)$.
As explained before, the increase is not excessive in particular due to the option of a relatively fully composite 
$t_R$, not contributing to $V(h)$ in this setup, which in turn allows for a less IR localized $t_L$ and thus a reduced 
quark contribution to the Higgs potential.
As is clearly visible from the plot, the model does not even feature ultra light partners below a TeV anywhere in its parameter space
- the discovery of such states could exclude the mMCHM$^{\rm III >}_{5-1}$.
The corresponding survival functions, depicted by the solid green lines in Figure~\ref{fig:14p5majsfnh5p1}, do not drop under 
$5 \%$ even until $m^{\rm min}_{2/3}\gtrsim 3$\, TeV both for $Y_\ast^l=Y_{\ast}^q=0.7$ (left plot) and $Y_\ast^l=Y_{\ast}^q=1.4$ (right plot).

\begin{figure}[!t]
\begin{center} 
		   \includegraphics[width=0.403\textwidth]{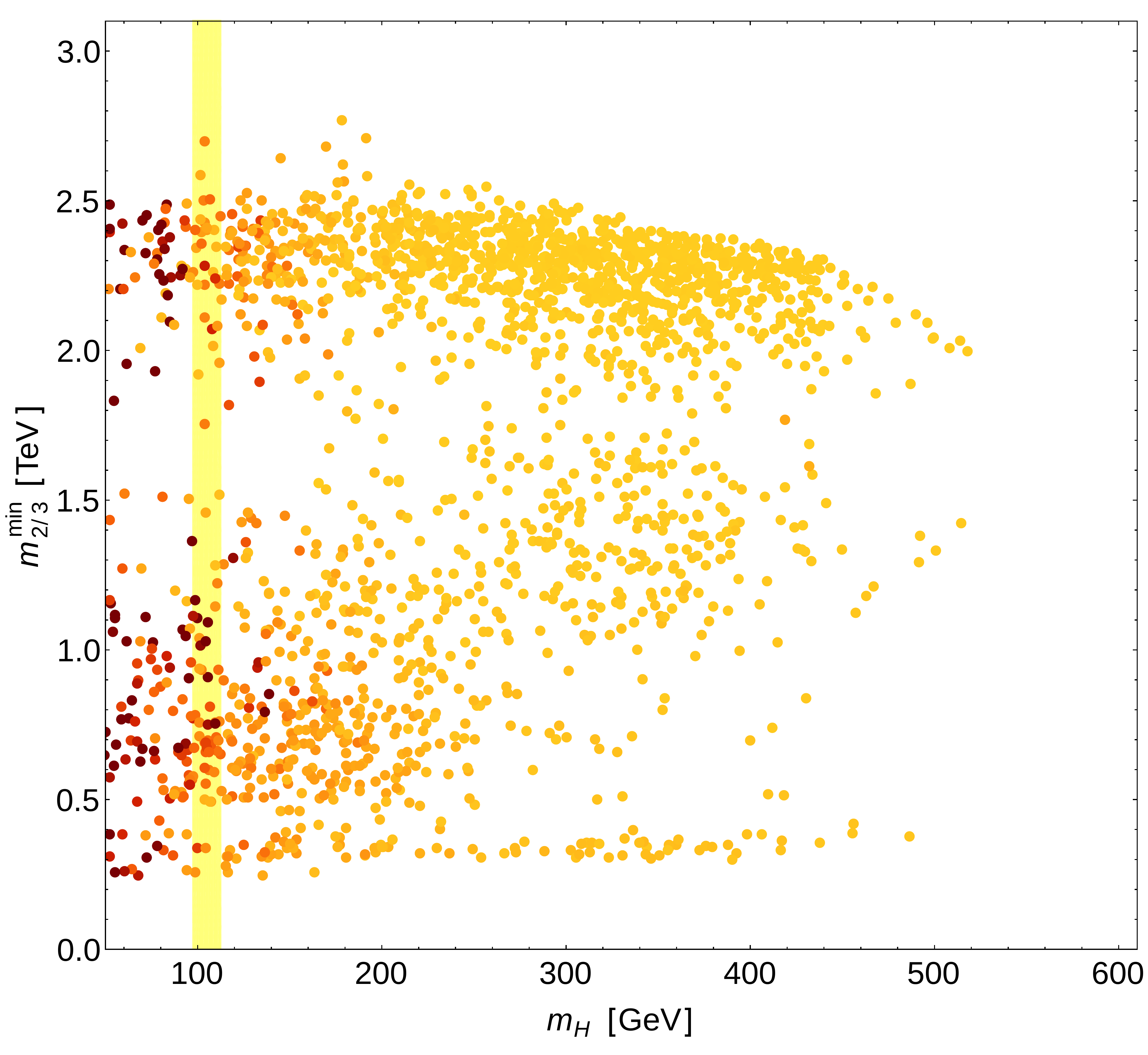}
		   \includegraphics[width=0.081\textwidth]{figures/bar_ft.pdf}
			\includegraphics[width=0.403\textwidth]{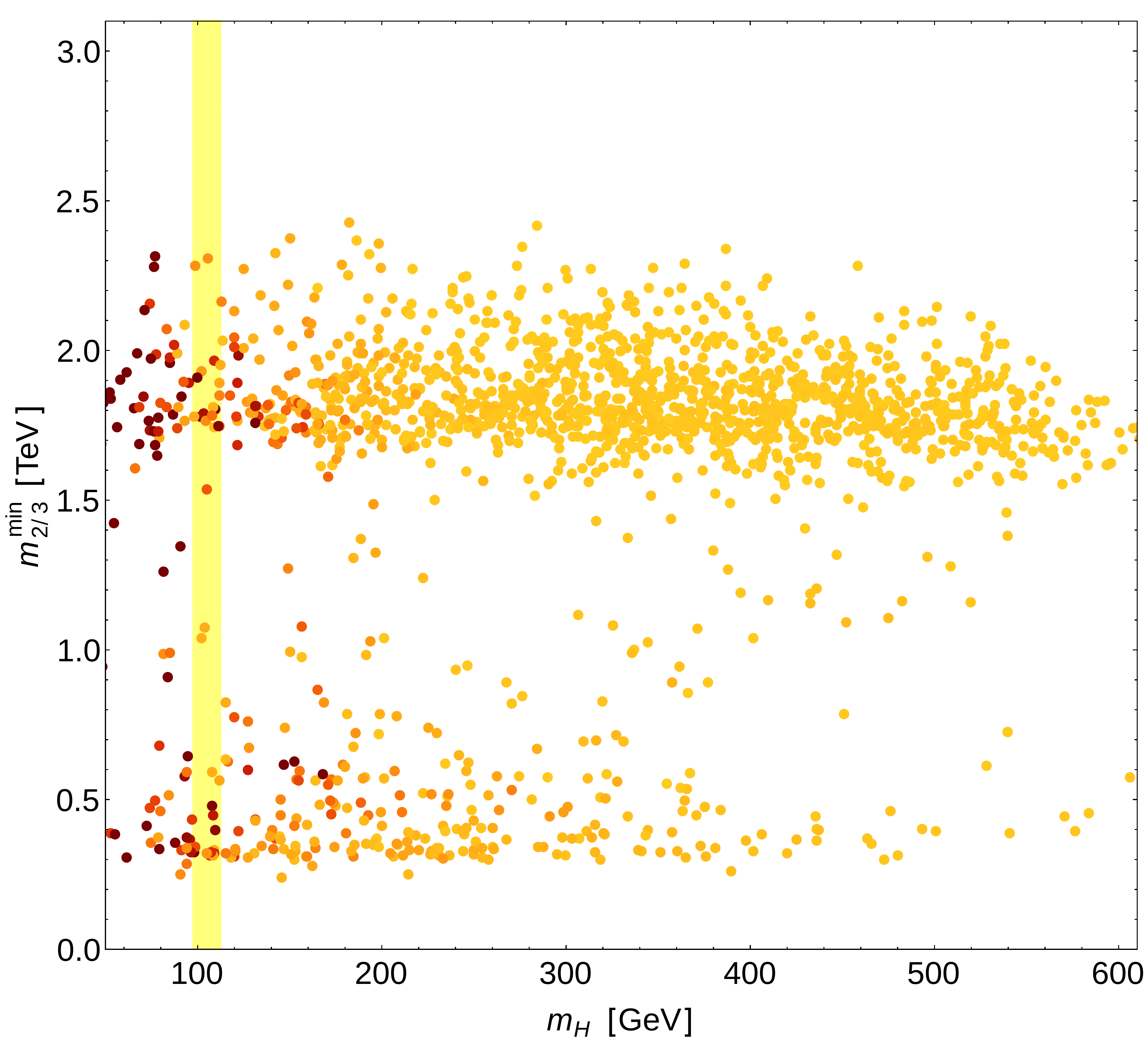}
			 \includegraphics[width=0.081\textwidth]{figures/bar_ft.pdf}
			 \caption{Mass of the first top partner as a function of the Higgs mass in the MCHM$_{14}$ for $Y_{q}^{\ast}=0.7$ (left) and $Y_q^{\ast}=1.4$ (right). In both cases we have assumed $f_{\pi}=0.8$\,TeV and $\kappa=\kappa^{\prime}=0$. Lighter points correspond to smaller values of $\Delta_{\rm BG}$ and therefore to less tuned points. }
\label{fig:MCHM14}
\end{center}
\end{figure}

\begin{figure}[!t]
\begin{center} 
		   \includegraphics[width=0.48\textwidth]{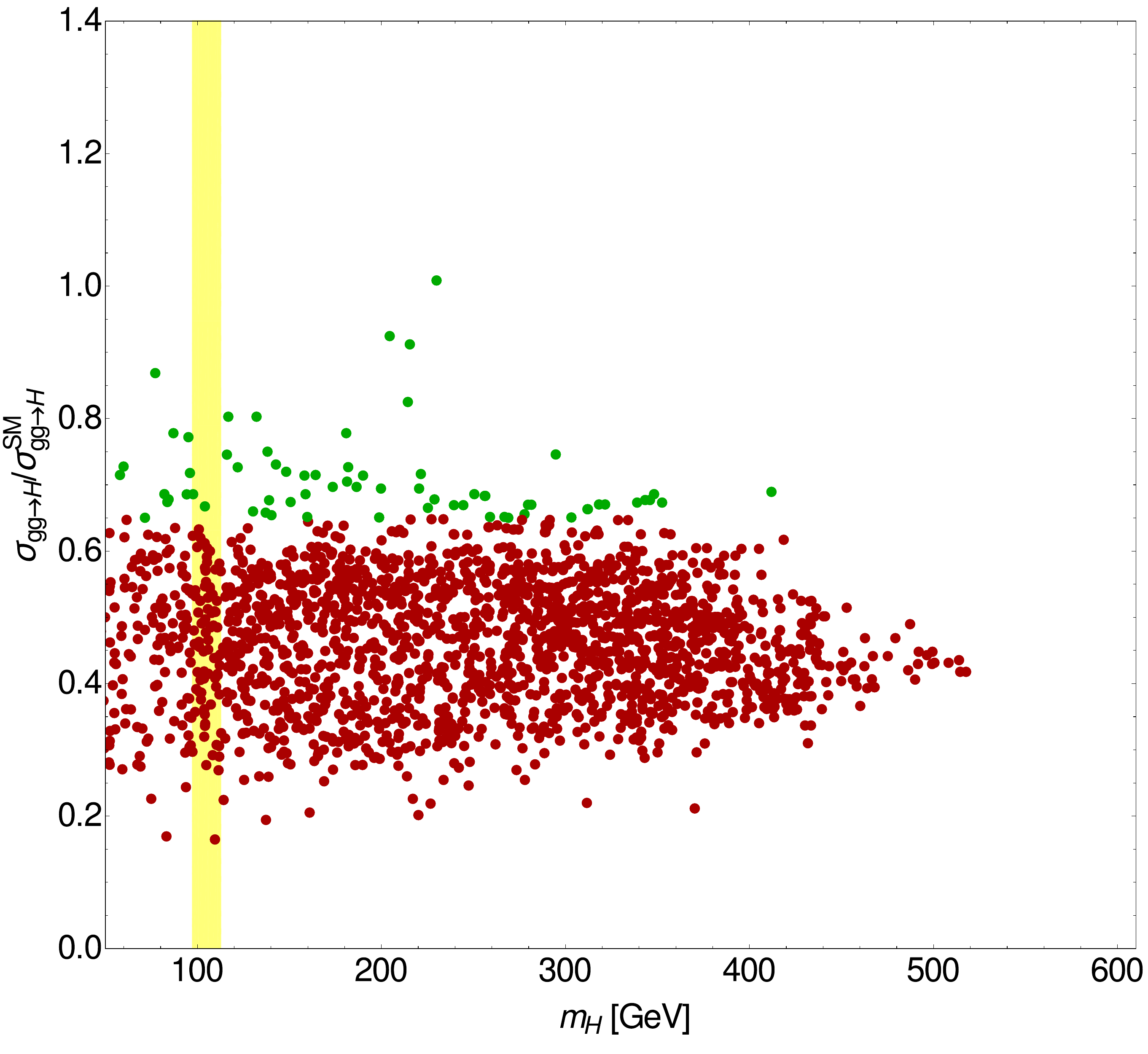}
		   \hspace{0.1cm}
			\includegraphics[width=0.48\textwidth]{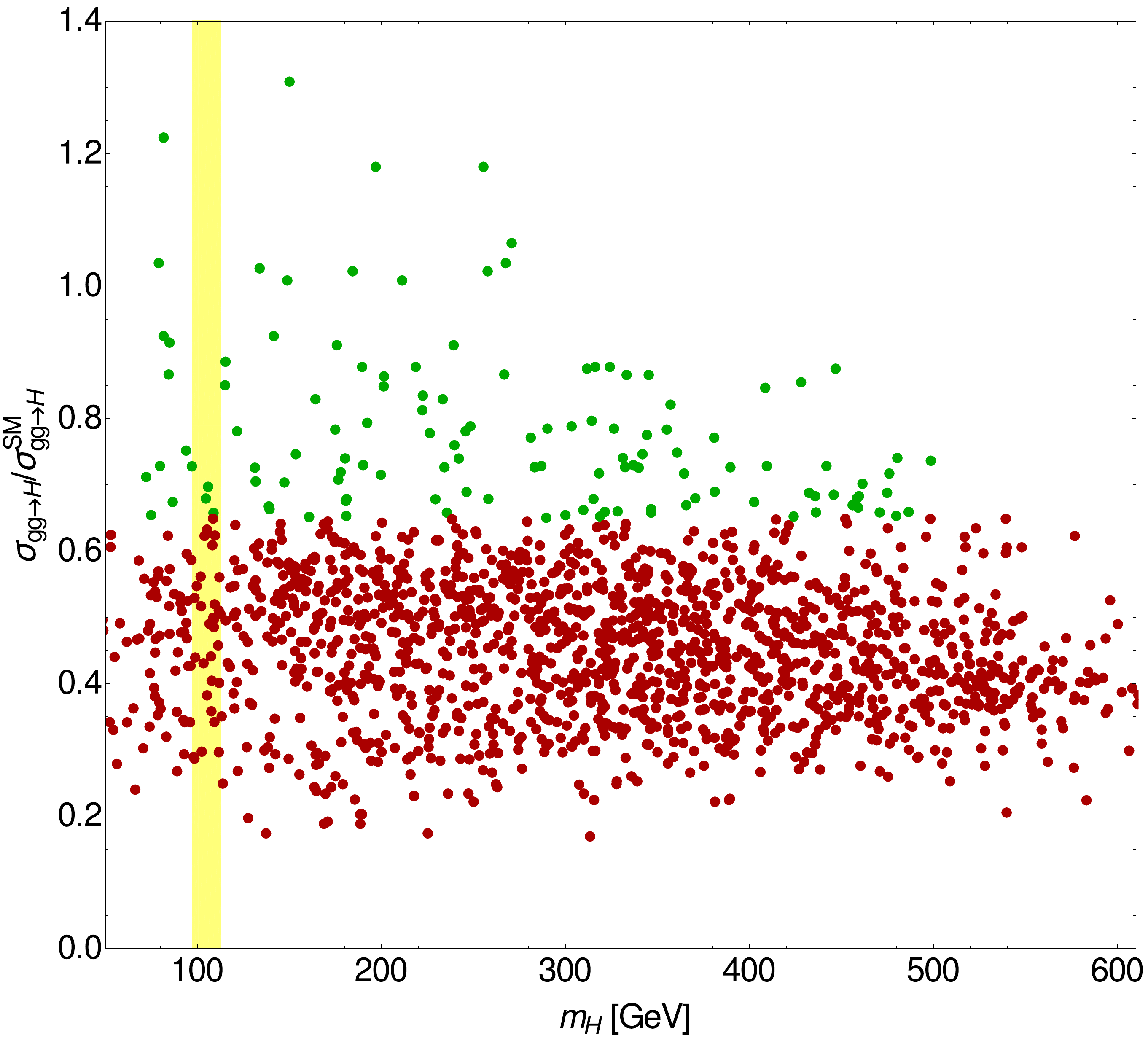}
			 \caption{$gg\to H$ cross section normalized to the SM value as a function of the Higgs mass in the MCHM$_{14}$ for $Y_{q}^{\ast}=0.7$ (left) and $Y_\ast^q=1.4$ (right). In both cases we have assumed $f_{\pi}=0.8$\,TeV and $\kappa=\kappa^{\prime}=0$. The red color indicates points that deviate by more than $35 \%$ from the SM prediction. \label{fig:MCHM14ggf}}
\end{center}
\end{figure}

After presenting the numerical results for the models that allow to lift the top-partner masses via significant 
leptonic contributions, we now want to compare them quantitatively to models that employ a {\bf 14} in the quark sector to
lift the partner masses. For these models we expect in general a sizable ad-hoc tuning, as explained in Section~\ref{sec:general}.
We start with the MCHM$_{14}$, where both top chiralities mix with a {\bf 14} of $SO(5)$.
Looking at the $m_H-m^{\rm min}_{2/3}$ plane presented in Figure~\ref{fig:MCHM14}, we can on the one hand see explicitly 
the option to lift the lightest top partners beyond a TeV in part of the parameter space. 
On the other hand, the significant ad-hoc tuning in this model also becomes apparent.
Not only is the generic scale of the Higgs mass raised to $m_H \sim 300$\,GeV and beyond ({\it c.f.} (\ref{eq:mh14_2})),
in particular a Higgs around $m_H \sim 100$\, GeV leads in general to a large tuning of $\Delta_{\rm BG}\sim{\cal O}(50-100)$
in the case of $Y_{q}^{\ast}=0.7$ (left panel), while $Y_\ast^q=1.4$ (right panel) leads generically to $\Delta_{\rm BG}\sim{\cal O}(50-200)$, where we again employed $f_{\pi}=0.8$\,TeV. 

Moreover, remember that this model induces large corrections to the SM Higgs couplings, in particular the $g g \to H$ cross section
receives sizable changes, as shown in Section~\ref{sec:14s}.
 The numerical predictions for $\sigma_{gg \to H}$, normalized to the SM value, are given in Figure~\ref{fig:MCHM14ggf} 
with analogous parameters as before. 
We can see that most of the parameter space features large corrections
of more than $35 \%$, as indicated by the red color. Thus, this solution to the light-partner issue seems indeed 
already to be disfavored from Higgs physics at the LHC.

\begin{figure}[t!]
\begin{center} 
	\subfigure[$m^{\rm min}_{2/3}$ vs.\ $m_H$  in the MCHM$_{14-1}$, with 
$Y_\ast^q=0.7$  (left), $Y_\ast^q=1.4$ (right), while
$0\leq \kappa_B^q,\kappa_T^q \leq Y_\ast^q$.\newline Lighter points correspond to smaller values of $\Delta_{\rm BG}$ and therefore to less tuned points.]{%
		   \includegraphics[width=0.43\textwidth]{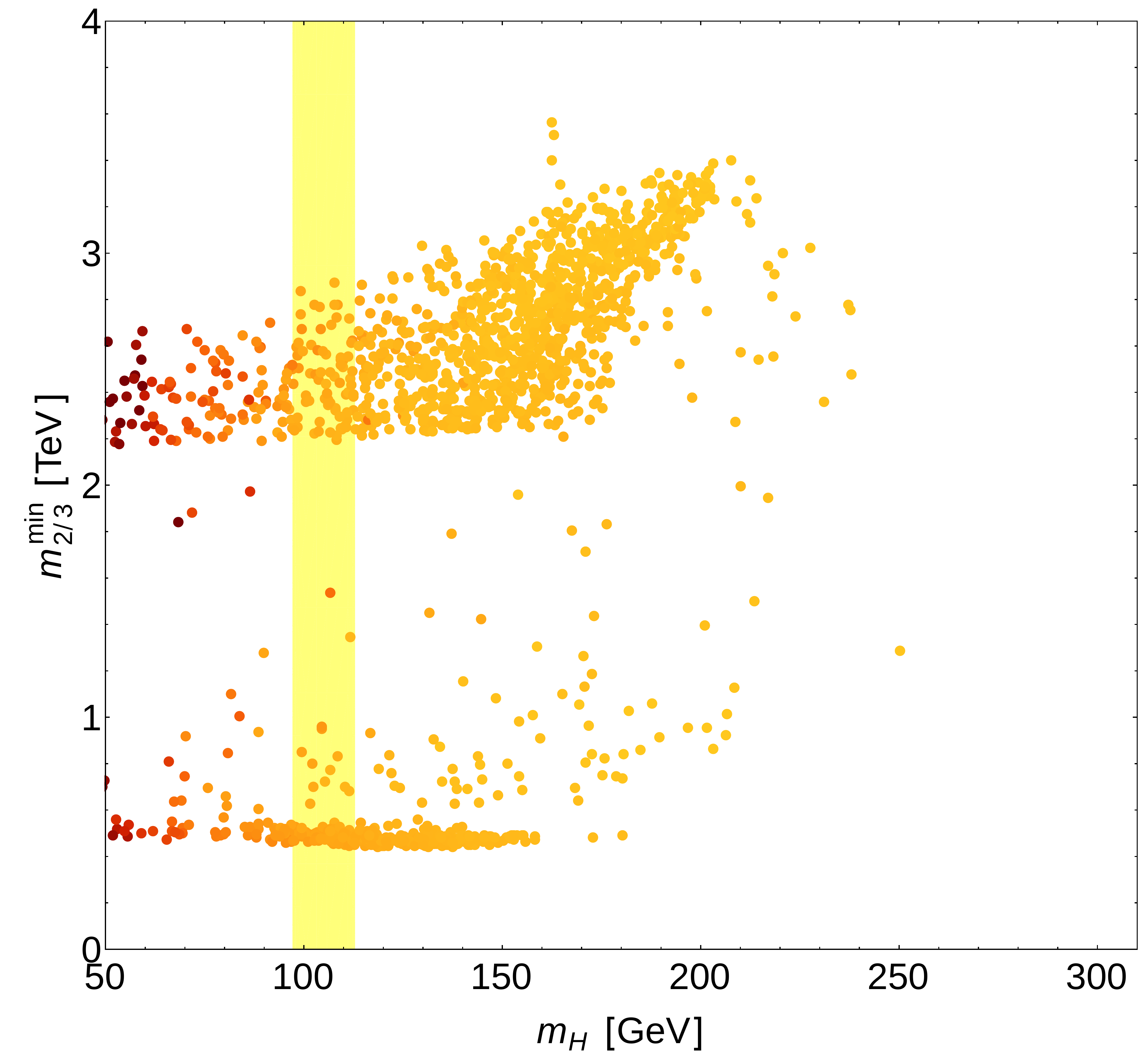}
		   \includegraphics[width=0.0865\textwidth]{figures/bar_ft.pdf}
			\includegraphics[width=0.43\textwidth]{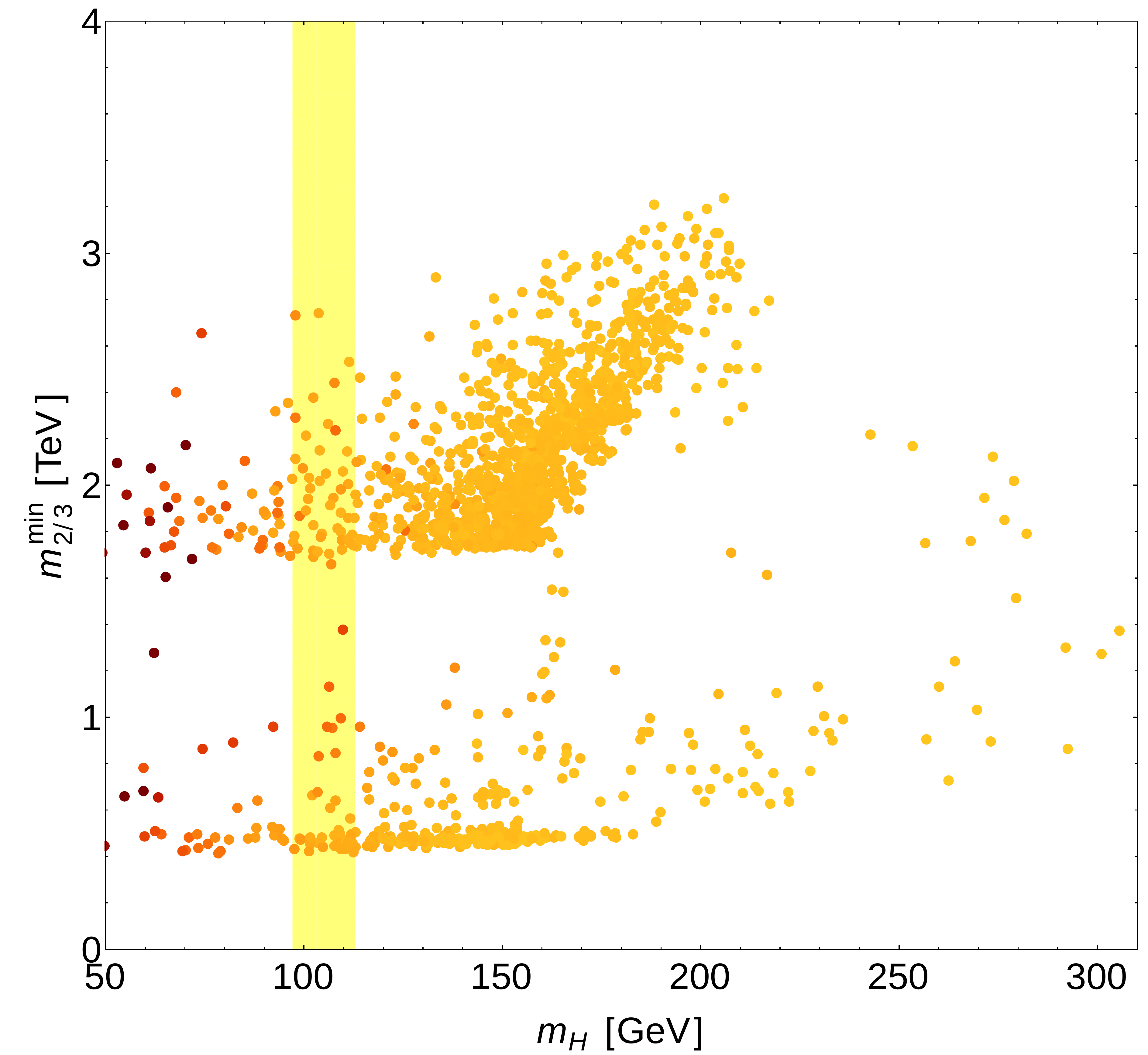}
			 \includegraphics[width=0.0865\textwidth]{figures/bar_ft.pdf}
			 \label{fig:14p1vssmin}
		}

\subfigure[Survival function $\mathcal{P}_\Delta$ for the fine tuning $\Delta_{\rm BG}$ 
in the MCHM$_{14-1}$ (dotted blue),  mMCHM$^{\rm III>}_{5-1}$ with $Y_\ast^l=0.7$ (dashed red), and mMCHM$^{\rm III}_5$ with $Y_\ast^l=0.35$  (solid green), always employing $Y_{\ast}^q=0.7$.~In~the left panel we show the full parameter range in agreement with $m_H$, while in the right we require $m^{\rm min}_{2/3}>1\,$TeV.]{%
\includegraphics[width=0.493\textwidth]{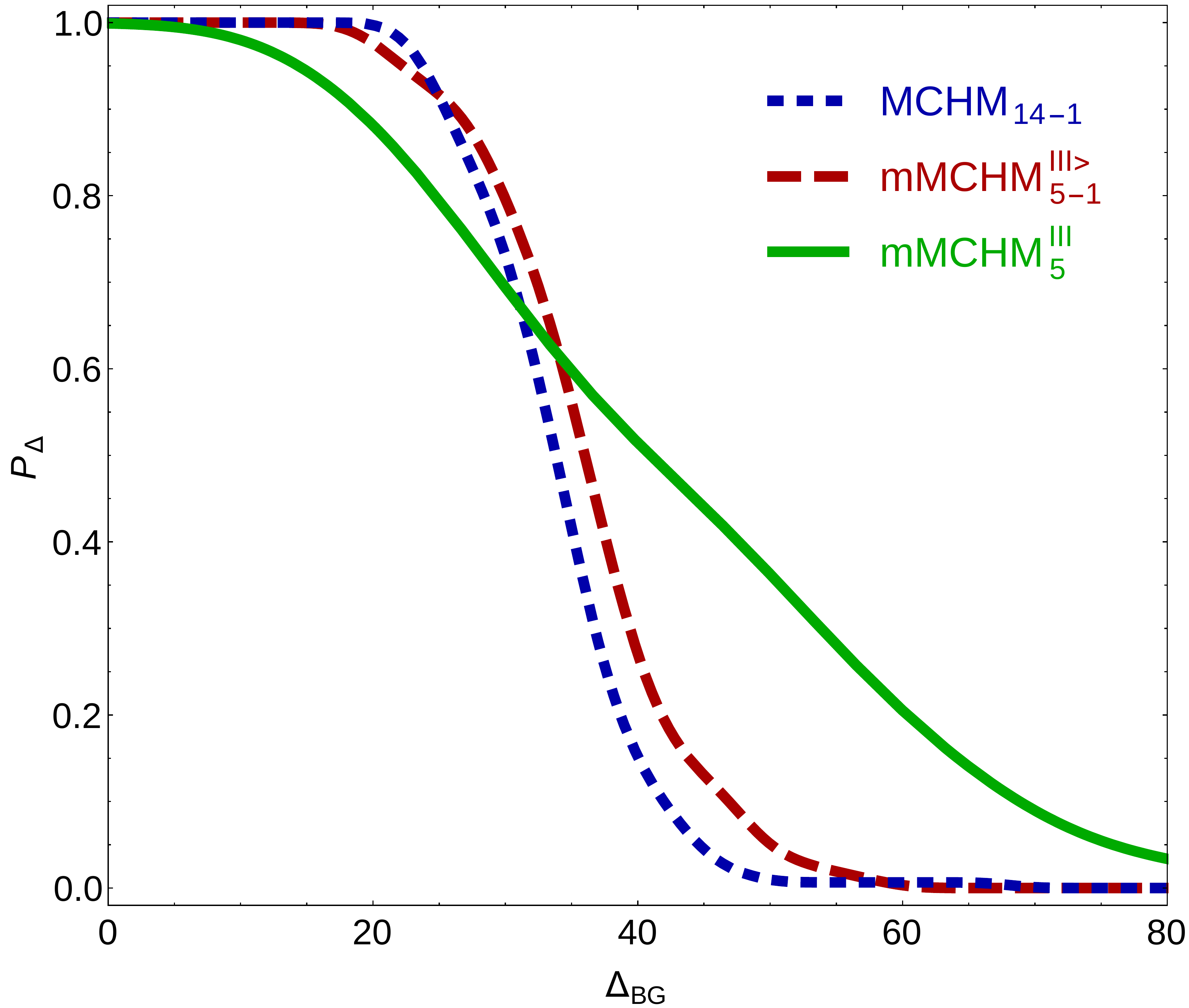}
			\includegraphics[width=0.493\textwidth]{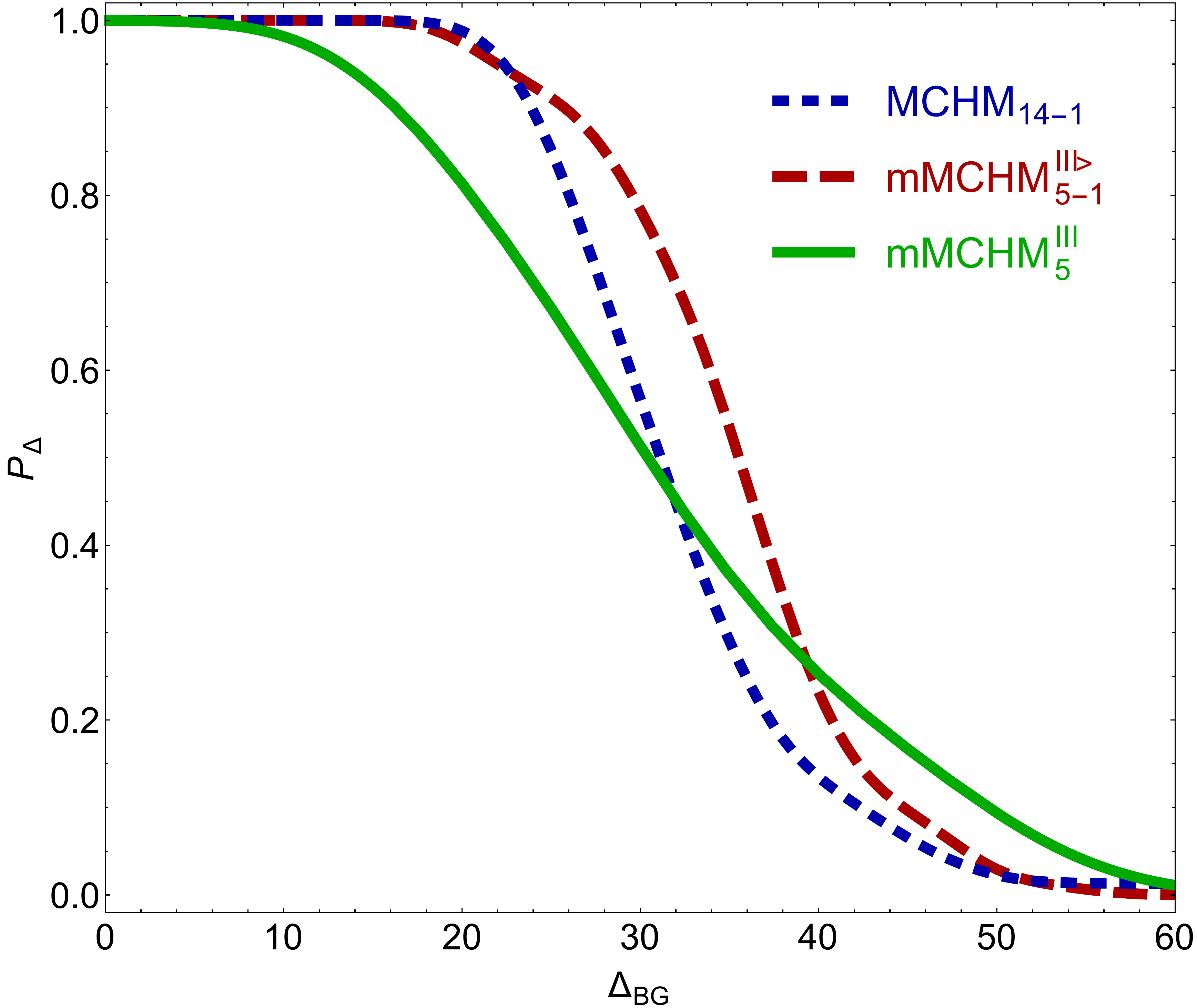}
		 \label{fig:FTcomp}
		}
		\caption{Fine tuning $\Delta_{\rm BG}$ in the MCHM$_{14-1}$ (Figure~\ref{fig:14p1vssmin}) as well as compared to the minimal models of lepton compositeness (Figure~\ref{fig:FTcomp}). We always employ $\kappa=\kappa^{\prime}=0$ and \mbox{$f_{\pi}=0.8$\,TeV}}
\end{center}
\end{figure}

We now move forward to the more economical model of mixing the $t_L$ with a {\bf 14}, while the $t_R$ 
mixes with a singlet, {\it i.e.} the MCHM$_{14-1}$. 
In that case, it turns out that one needs brane-localized fermion kinetic terms $\kappa_B^q,\kappa_T^q>0$ to allow for a non-vanishing $\alpha_t$, see Section~\ref{sec:14s}.
Once more we however also expect to be able to lift the top-partner masses beyond the TeV scale, again at the price of a non-negligible ad-hoc tuning via the sizable quark contribution to $\beta$, which due to the ${\bf 1}_R$ should however again be more moderate than before, comparable to the case of the mMCHM$^{\rm III >}_{5-1}$.
Indeed, this picture is supported from the $m^{\rm min}_{2/3}$ vs. $m_H$ plots presented in Figure~\ref{fig:14p1vssmin}, where we see that, while still light partners are possible, in a large part of the viable parameter space they are lifted to the $\sim 2\,$TeV scale, both for $Y_\ast^q=0.7$  (left plot) and $Y_\ast^q=1.4$ (right plot), where always $f_{\pi}=0.8$\,TeV. 
Although the density of points in our scan gets slightly more diluted approaching the experimental Higgs mass, compared to the MCHM$_5$, the yellow band can be reached relatively easily with a moderate tuning of $\Delta_{\rm BG} \sim (30-40) $, 
comparable to the case of the  mMCHM$^{\rm III >}_{5-1}$.
Nevertheless, the numbers show that this ad-hoc tuning is in general somewhat more severe than the ``double tuning'' of the MCHM$_5$ or the mMCHM$^{\rm III}_5$. 

We examine this behavior more quantitatively in Figure~\ref{fig:FTcomp}, confronting
the fine tuning of the MCHM$_{14-1}$ and the mMCHM$^{\rm III >}_{5-1}$ with that of the mMCHM$^{\rm III}_5$,
which can completely avoid the ad-hoc tuning, employing $Y_\ast^q=0.7$, $f_\pi=0.8$, as well as lepton brane masses as before. In the left panel we display the survival function $P_\Delta$, describing the fraction of points with a given fine-tuning larger or equal to the value of $\Delta_{\rm BG}$ on the x-axis, for all points in the viable Higgs-mass band. This confirms clearly that, while the general tuning in both the mMCHM$_{14-1}$ and the mMCHM$^{\rm III >}_{5-1}$ is moderate - $P_\Delta$ starts to drop considerably below 1 around  $\Delta_{\rm BG} \gtrsim 25$ - 
the drop in the mMCHM$^{\rm III}_5$ already starts at $\Delta_{\rm BG} \gtrsim 10$, improving the tuning by a factor
of 2 to 3. The difference becomes even more apparent once we use the constraint $m^{\rm min}_{2/3}>1$\,TeV, 
employed in the right panel of Figure~\ref{fig:FTcomp}, which ensures the absence of light partners. As explained before, this selects the points with the smallest tuning within the mMCHM$^{\rm III}_5$, 
opening for the first time the parameter space to allow for a minimal tuning of even less than $10\%$ while at the same time not
predicting ultra-light partners.
While already this model provides a motivation for the appearance of a symmetric representation of $SO(5)$ and does not 
introduce many new particles, a big virtue of the mMCHM$^{\rm III >}_{5-1}$ on the other hand is its highest degree of 
minimality and naturalness.
This is true in the quark sector, embedding each SM fermion in the most minimal $SO(5)$ multiplet one can think of 
(respecting custodial protection of the $Z$ couplings), as well as in the lepton sector, where it provides the most minimal realization 
of the type-III seesaw, leading to the least number of dof in the full fermion sector for viable models. Obviously, raising top partners via this model is possible in a much more minimal way than in the MCHM$_{14-1}$, which adds many colored dof at the TeV scale, while the amount of tuning remains the same.

\section{Conclusions}
\label{sec:conclusions}

We have presented a comprehensive survey of the impact of different realizations of the fermion sector on the potential of the composite $SO(5)/SO(4)$ Higgs. Putting a special emphasis on the lepton sector, mostly neglected in other studies, we presented new models that allow to lift the masses of the lightest top partners well above the region currently probed by the LHC in a way orthogonal to previous studies, {\it i.e.} without a significant enlargement of the colored fermion sector and, as such, not sacrificing
the prediction of a fully natural light Higgs.
In particular, we pointed out the possible minimality of this solution by presenting the mMCHM$^{\rm III}$-type models, 
which, although featuring a symmetric representation of $SO(5)$, allow for less new particles than e.g. the standard MCHM$_5$,
by unifying both left-handed and right-handed SM-leptons in a single multiplet, respectively. 
We also demonstrated how, in contrast to the quark sector, the lepton sector provides a compelling motivation for the emergence of a symmetric representation via the seesaw mechanism and how the  mMCHM$^{\rm III}_5$ allows to reconcile the absence of light partners with a minimal tuning. It is indeed the particular type-III seesaw, explored for the first time, that allows the unification of the right-handed lepton fields - for quarks the $SU(2)$ breaking masses and/or the quantum numbers 
do not allow for a similar mechanism. We thus provided the first example of a composite model that can describe
a full fermion sector with only two composite fields.

Indeed, in the case of no discovery of light top partners rather soon after the restart of the LHC (with ${\cal O}(100\,{\rm fb}^{-1}$)), this class of models with non-negligible leptonic effects could furnish the only economical composite models 
that are still compatible with a minimal tuning of not more than $10 \%$.
Although here the sizable contribution of the lepton sector to the Higgs mass has a quite different origin from the sizable top quark contribution ({\it i.e.}, it is not caused by a large $m_f$), the setup still intertwines the flavor and/or neutrino-mass structure with the Higgs potential. In particular, models with pronounced leptonic effects do not suffer from large corrections in Higgs production,
which, as we demonstrated explicitly, already challenge the MCHM$_{14}$.
To exclude the bulk of the parameter space of the models featuring a symmetric representation in the lepton  sector directly via top partner searches would be challenging, even at the HL-LHC
with $3000\,{\rm fb}^{-1}$, which has a reach of roughly 1.7\,TeV \cite{Gershtein:2013iqa,Agashe:2013hma}. A 100\,TeV collider on the other hand is expected to probe the full parameter space of natural models as discussed here.
Indirect detection via the modification of the Higgs couplings, e.g. in channels involving leptons, or the search for color octet resonances could be alternative options to unveil signs of the models.
The impact on lepton flavor, which might require some additional model building, will be investigated elsewhere \cite{next}.
However, let us already note that all setups considered here feature a first layer of flavor protection since the right-handed leptons 
come with a custodial protection due to the $P_{LR}$ symmetry, while the left-handed ones are rather strongly UV localized, {\it i.e.}, they feature a very small compositeness.
Since the lepton sector, which needs to be included in a fully realistic composite model, allows to construct minimal models
with minimal tuning and a naturally light Higgs, that do not predict light top partners, the absence of corresponding LHC signals
should not threaten the idea of the minimal composite Higgs.

\paragraph{Acknowledgements}
We are grateful to Roberto Contino, Giuliano Panico, and José Santiago for useful discussions. The research of the authors was supported by the Swiss National Foundation under contract SNF 200021-143781. The research of FG was also supported by a Marie Curie Intra European Fellowship within the 7th European Community Framework Programme (grant no. PIEF-GA-2013-628224).

\begin{appendix}

\section{Generators and Explicit Form of Goldstone Matrix}
\label{sec:gen}

The generators of $SO(5)$ in the fundamental representation read
\begin{eqnarray}
\label{genso5}
T_{L,ij}^a&=&-\frac{i}{2}\left[\frac{1}{2}\epsilon^{abc}\left(\delta^b_i
\delta^c_j-\delta^b_j\delta_i^c\right)
+\left(\delta_i^a\delta_j^4-\delta^a_j\delta_i^4\right)\right],\qquad
a=1,2,3, \nonumber \\ 
T_{R,ij}^a&=&-\frac{i}{2}\left[\frac{1}{2}\epsilon^{abc}\left(\delta^b_i
  \delta_j^c-\delta_j^b\delta^c_i\right)-\left(\delta^a_i\delta_j^4
-\delta_j^a\delta_i^4\right)\right],\qquad
a=1,2,3, \\ 
T_{ij}^{\hat{a}}&=&
-\frac{i}{\sqrt{2}}\left[\delta^{\hat{a}}_i\delta^5_j
-\delta^{\hat{a}}_j\delta^5_i\right],\qquad
\hat{a}=1,2,3,4, \nonumber
\end{eqnarray}
which have been chosen orthonormal with respect to the Cartan-Killing inner product, i.e. $\textrm{Tr}(T^{\alpha}\cdot T^{\beta})=\delta^{\alpha\beta}$, for all $T^{\alpha},T^{\beta}\in\{T_L^{a},T^{a}_R,T^{\hat{a}}\}.$ As $[T_L^3,T_R^3]=0$ we can go to a new basis where both generators are diagonal at the same time. This will make the explicit form of the fermion representations much simpler. To this end, we perform the rotation 
\begin{eqnarray}
	T\to \mathbf{A}\cdot T\cdot \mathbf{A}^{\dagger},\quad \mathrm{with}\quad  T\in\{T_L^a,T_R^a,T^{\hat{a}}\},
	\label{rot}
\end{eqnarray}
where $\mathbf{A}$ is the unitary matrix
\begin{eqnarray}
	\mathbf{A}=\frac{1}{\sqrt{2}}\begin{pmatrix}1&-i&0&0&0\\ 0&0&-i&1&0\\ 0&0&i&1&0 \\ -1 &-i &0&0&0 \\ 0&0&0&0&\sqrt{2}\end{pmatrix}.
\end{eqnarray}

The previous set of generators, together with the identity, can be extended to span a basis of the linear space of square matrices of order $n$ over the complex field $\mathbb{C}$, $FL(n,\mathbb{C})$. In particular, in the basis of (\ref{genso5}),  this can be done by adding the following set of matrices
	\begin{eqnarray}
		\tilde{T}^{\hat{a}}_{ij}&=&\frac{1}{\sqrt{2}}\left[\delta^{\hat{a}}_i\delta^{5}_j+\delta_j^{\hat{a}}\delta_i^5\right], \qquad\qquad \qquad  ~ \hat{a}=1,2,3,4,\\
		\tilde{T}^{\hat{a}\hat{b}}_{ij}&=&\frac{1}{\sqrt{2}}\left[\delta_i^{\hat{a}}\delta_j^{\hat{b}}+\delta_j^{\hat{a}}\delta_i^{\hat{b}}\right],\qquad \qquad \qquad~ \hat{a}<\hat{b}\in\{1,2,3,4\},\\
		\tilde{T}^{aa}_{ij}&=&\delta^{\hat{a}}_i\delta^{\hat{a}}_j+\delta^4_i\delta_j^4-\frac{1}{2}\delta_{ij}+\frac{1}{2}\delta^5_i\delta_j^5,\qquad ~a=1,2,3,\phantom{4}\\
		\tilde{T}^s_{ij}&=&\frac{1}{2\sqrt{5}}\mathrm{diag}(1,1,1,1,-4), 
	\end{eqnarray}
	that can be collectively denoted by $\hat{T}^{A}\in\{\tilde{T}^{\hat{a}\hat{b}},\tilde{T}^{aa},\tilde{T}^{\hat{a}},\tilde{T}^s\}$.

\section{Fermion Representations}
\label{sec:ferm}
\subsection{Fundamental}
\label{subsec:fun}
The fundamental representation of $SO(5)$ decompose under $SO(4)\cong SU(2)_L\times SU(2)_R$ as $\mathbf{5}=(\mathbf{2},\mathbf{2})\oplus(\mathbf{1},\mathbf{1})$. More explictly, in the basis given by the transformation (\ref{rot}), 
a field $\chi_{\mathbf{5}}$ transforming under this representation will read 
\begin{eqnarray}
	\chi_{\mathbf{5}}=\chi^{(\mathbf{2},\mathbf{2})}+\chi^{(\mathbf{1},\mathbf{1})},
	\end{eqnarray}
	with
	\begin{eqnarray}
		\chi^{(\mathbf{2},\mathbf{2})}=\begin{pmatrix}\chi^{(\mathbf{2},\mathbf{2})}_{++}\\ \chi^{(\mathbf{2},\mathbf{2})}_{-+}\\ \chi^{(\mathbf{2},\mathbf{2})}_{+-}\\ \chi^{(\mathbf{2},\mathbf{2})}_{--}\\ 0\end{pmatrix},\qquad 	\chi^{(\mathbf{1},\mathbf{1})}=\begin{pmatrix}0\\ 0\\ 0\\ 0\\ \chi^{(\mathbf{1},\mathbf{1})}_s\end{pmatrix}.
\end{eqnarray}
In the above equation the first and second subscripts $\pm$ on the $\chi^{(\mathbf{2},\mathbf{2})}$ fields correspond to $T_L^3=\pm 1/2$ and $T_R^3=\pm 1/2$, respectively.
\subsection{Adjoint}
\label{subsec:adj}
Similarly, the adjoint representation of $SO(5)$ decomposes as $\mathbf{10}=(\mathbf{3},\mathbf{1})\oplus (\mathbf{1},\mathbf{3})\oplus(\mathbf{2},\mathbf{2})$. Still in the same basis, a field $\chi_{\mathbf{10}}$ transforming under this representation  will read
\begin{eqnarray}
	\chi_{\mathbf{10}}=\chi^{(\mathbf{3},\mathbf{1})}+\chi^{(\mathbf{1},\mathbf{3})}+\chi^{(\mathbf{2},\mathbf{2})},
\end{eqnarray}
	where
\begin{eqnarray}
	\chi^{(\mathbf{3},\mathbf{1})}&=&\begin{pmatrix}\frac{1}{2}\chi^{(\mathbf{3},\mathbf{1})}_{0}& -\frac{i}{\sqrt{2}} \chi^{(\mathbf{3},\mathbf{1})}_{+} & 0&0& 0\\ \frac{i}{\sqrt{2}}\chi^{(\mathbf{3},\mathbf{1})}_{-} &-\frac{1}{2}\chi^{(\mathbf{3},\mathbf{1})}_{0}&0&0& 0\\   0&0&\frac{1}{2}\chi^{(\mathbf{3},\mathbf{1})}_{0}&  -\frac{i}{\sqrt{2}}\chi^{(\mathbf{3},\mathbf{1})}_{+}&0\\ 0&0& \frac{i}{\sqrt{2}}\chi_{-}^{(\mathbf{3},\mathbf{1})} &-\frac{1}{2}\chi_{0}^{(\mathbf{3},\mathbf{1})}&0 \\   0 &0 &0 &0 &0\end{pmatrix},\\
	\chi^{(\mathbf{1},\mathbf{3})}&=&\begin{pmatrix}\frac{1}{2}\chi^{(\mathbf{1},\mathbf{3})}_{0}& 0&\frac{i}{\sqrt{2}} \chi^{(\mathbf{1},\mathbf{3})}_{+} & 0& 0\\0&\frac{1}{2}\chi^{(\mathbf{1},\mathbf{3})}_{0}&0& \frac{i}{\sqrt{2}}\chi^{(\mathbf{1},\mathbf{3})}_{+} & 0\\ -\frac{i}{\sqrt{2}}\chi^{(\mathbf{1},\mathbf{3})}_{-} &0&-\frac{1}{2}\chi^{(\mathbf{1},\mathbf{3})}_{0}&  0&0\\ 0& -\frac{i}{\sqrt{2}}\chi_{-}^{(\mathbf{1},\mathbf{3})} &0&-\frac{1}{2}\chi_{0}^{(\mathbf{1},\mathbf{3})}&0 \\   0 &0 &0 &0 &0\end{pmatrix},\\
	\chi^{(\mathbf{2},\mathbf{2})}&=&\frac{1}{\sqrt{2}}\begin{pmatrix}0& 0&0&0&-i\chi_{++}^{(\mathbf{2},\mathbf{2})}\\  0&0&0&0& -\chi_{-+}^{(\mathbf{2},\mathbf{2})}\\ 0& 0&0&0&\chi_{+-}^{(\mathbf{2},\mathbf{2})} \\ 0 &0 &0 &0& i\chi^{(\mathbf{2},\mathbf{2})}_{--}\\ i\chi_{--}^{(\mathbf{2},\mathbf{2})} & -\chi_{+-}^{(\mathbf{2},\mathbf{2})} & \chi_{-+}^{(\mathbf{2},\mathbf{2})}&-i\chi^{(\mathbf{2},\mathbf{2})}_{++}&0\end{pmatrix},
\end{eqnarray}
and the subscripts $\pm,0$ on $\chi^{(\mathbf{3},\mathbf{1})}$ ($\chi^{(\mathbf{1},\mathbf{3})}$) correspond to $T_L^3=\pm 1,0$ ($T_R^3=\pm1, 0$), respectively. 
\subsection{Symmetric}
\label{subsec:sym}
Finally, we consider the symmetric representation of $SO(5)$, $\mathbf{14}$, decomposing under $SO(4)$ as $\mathbf{14}=(\mathbf{1},\mathbf{1})\oplus (\mathbf{2},\mathbf{2})\oplus (\mathbf{3},\mathbf{3})$. A field $\chi_{\mathbf{14}}$ transforming under this representation can be expressed in the basis given by (\ref{rot}) as
\begin{eqnarray}
	\chi_{\mathbf{14}}=\chi^{(\mathbf{0},\mathbf{0})}+\chi^{(\mathbf{2},\mathbf{2})}+\chi^{(\mathbf{3},\mathbf{3})},
\end{eqnarray}
with
\begin{eqnarray}
	\chi^{(\mathbf{0},\mathbf{0})}&=&	\chi^{(\mathbf{0},\mathbf{0})}_{s}\frac{1}{2\sqrt{5}}\mathrm{diag}(1,1,1,1,-4),\\
	\chi^{(\mathbf{2},\mathbf{2})}&=&\frac{1}{\sqrt{2}}\begin{pmatrix}0& 0&0&0&\chi_{++}^{(\mathbf{2},\mathbf{2})}\\  0&0&0&0& \chi_{-+}^{(\mathbf{2},\mathbf{2})}\\ 0& 0&0&0&-\chi_{+-}^{(\mathbf{2},\mathbf{2})} \\ 0 &0 &0 &0& -\chi^{(\mathbf{2},\mathbf{2})}_{--}\\ \chi_{--}^{(\mathbf{2},\mathbf{2})} & -\chi_{+-}^{(\mathbf{2},\mathbf{2})} & \chi_{-+}^{(\mathbf{2},\mathbf{2})}&-\chi^{(\mathbf{2},\mathbf{2})}_{++}&0\end{pmatrix},
\end{eqnarray}
and
\begin{eqnarray}
	\chi^{(\mathbf{3},\mathbf{3})}&=&\begin{pmatrix}-\frac{1}{2}\chi^{(\mathbf{3},\mathbf{3})}_{00}& \frac{1}{\sqrt{2}} \chi^{(\mathbf{3},\mathbf{3})}_{+0} & \frac{1}{\sqrt{2}}\chi_{0+}^{(\mathbf{3},\mathbf{3})} & -\chi_{++}^{(\mathbf{3},\mathbf{3})}& 0\\ -\frac{1}{\sqrt{2}}\chi^{(\mathbf{3},\mathbf{3})}_{-0} &\frac{1}{2}\chi^{(\mathbf{3},\mathbf{3})}_{00}&\chi_{-+}^{(\mathbf{3},\mathbf{3})}& -\frac{1}{\sqrt{2}}\chi_{0+}^{(\mathbf{3},\mathbf{3})}& 0\\   -\frac{1}{\sqrt{2}} \chi_{0-}^{(\mathbf{3},\mathbf{3})} & -\chi_{+-}^{(\mathbf{3},\mathbf{3})}&\frac{1}{2}\chi^{(\mathbf{3},\mathbf{3})}_{00}&  -\frac{1}{\sqrt{2}}\chi^{(\mathbf{3},\mathbf{3})}_{+0}&0\\ \chi_{--}^{(\mathbf{3},\mathbf{3})} & \frac{1}{\sqrt{2}}\chi_{0-}^{(\mathbf{3},\mathbf{3})}& \frac{1}{\sqrt{2}}\chi_{-0}^{(\mathbf{3},\mathbf{3})} &-\frac{1}{2}\chi_{00}^{(\mathbf{3},\mathbf{3})}&0 \\   0 &0 &0 &0 &0\end{pmatrix}.
\end{eqnarray}
Analagously, the first and second subscripts $\pm,0$ in $\chi^{(\mathbf{3},\mathbf{3})}$ correspond to $T_L^3=\pm1,0$ and $T_R^3=\pm1,0$, respectively.
\end{appendix}

\bibliographystyle{JHEP}
	\bibliography{Hpot}{}

\providecommand{\href}[2]{#2}\begingroup\raggedright\begin{thebibliography}{10}

\bibitem{Aad:2012tfa}
{\bf ATLAS Collaboration} Collaboration, G.~Aad et~al., {\it {Observation of a
  new particle in the search for the Standard Model Higgs boson with the ATLAS
  detector at the LHC}},  {\em Phys.Lett.} {\bf B716} (2012) 1--29,
  [\href{http://xxx.lanl.gov/abs/1207.7214}{{\tt arXiv:1207.7214}}].

\bibitem{Chatrchyan:2012ufa}
{\bf CMS Collaboration} Collaboration, S.~Chatrchyan et~al., {\it {Observation
  of a new boson at a mass of 125 GeV with the CMS experiment at the LHC}},
  {\em Phys.Lett.} {\bf B716} (2012) 30--61,
  [\href{http://xxx.lanl.gov/abs/1207.7235}{{\tt arXiv:1207.7235}}].

\bibitem{Contino:2006qr}
R.~Contino, L.~Da~Rold, and A.~Pomarol, {\it {Light custodians in natural
  composite Higgs models}},  {\em Phys.Rev.} {\bf D75} (2007) 055014,
  [\href{http://xxx.lanl.gov/abs/hep-ph/0612048}{{\tt hep-ph/0612048}}].

\bibitem{Medina:2007hz}
A.~D. Medina, N.~R. Shah, and C.~E. Wagner, {\it {Gauge-Higgs Unification and
  Radiative Electroweak Symmetry Breaking in Warped Extra Dimensions}},  {\em
  Phys.Rev.} {\bf D76} (2007) 095010,
  [\href{http://xxx.lanl.gov/abs/0706.1281}{{\tt arXiv:0706.1281}}].

\bibitem{Csaki:2008zd}
C.~Csaki, A.~Falkowski, and A.~Weiler, {\it {The Flavor of the Composite
  Pseudo-Goldstone Higgs}},  {\em JHEP} {\bf 0809} (2008) 008,
  [\href{http://xxx.lanl.gov/abs/0804.1954}{{\tt arXiv:0804.1954}}].

\bibitem{DeCurtis:2011yx}
S.~De~Curtis, M.~Redi, and A.~Tesi, {\it {The 4D Composite Higgs}},  {\em JHEP}
  {\bf 1204} (2012) 042, [\href{http://xxx.lanl.gov/abs/1110.1613}{{\tt
  arXiv:1110.1613}}].

\bibitem{Matsedonskyi:2012ym}
O.~Matsedonskyi, G.~Panico, and A.~Wulzer, {\it {Light Top Partners for a Light
  Composite Higgs}},  {\em JHEP} {\bf 1301} (2013) 164,
  [\href{http://xxx.lanl.gov/abs/1204.6333}{{\tt arXiv:1204.6333}}].

\bibitem{Marzocca:2012zn}
D.~Marzocca, M.~Serone, and J.~Shu, {\it {General Composite Higgs Models}},
  {\em JHEP} {\bf 1208} (2012) 013,
  [\href{http://xxx.lanl.gov/abs/1205.0770}{{\tt arXiv:1205.0770}}].

\bibitem{Pomarol:2012qf}
A.~Pomarol and F.~Riva, {\it {The Composite Higgs and Light Resonance
  Connection}},  {\em JHEP} {\bf 1208} (2012) 135,
  [\href{http://xxx.lanl.gov/abs/1205.6434}{{\tt arXiv:1205.6434}}].

\bibitem{Panico:2012uw}
G.~Panico, M.~Redi, A.~Tesi, and A.~Wulzer, {\it {On the Tuning and the Mass of
  the Composite Higgs}},  {\em JHEP} {\bf 1303} (2013) 051,
  [\href{http://xxx.lanl.gov/abs/1210.7114}{{\tt arXiv:1210.7114}}].

\bibitem{Archer:2014qga}
P.~R. Archer, {\it {Fine Tuning in the Holographic Minimal Composite Higgs
  Model}},  \href{http://xxx.lanl.gov/abs/1403.8048}{{\tt arXiv:1403.8048}}.

\bibitem{delAguila:2010vg}
F.~del Aguila, A.~Carmona, and J.~Santiago, {\it {Neutrino Masses from an A4
  Symmetry in Holographic Composite Higgs Models}},  {\em JHEP} {\bf 1008}
  (2010) 127, [\href{http://xxx.lanl.gov/abs/1001.5151}{{\tt
  arXiv:1001.5151}}].

\bibitem{Carmona:2013lva}
A.~Carmona and F.~Goertz, {\it {Composite Taus and Higgs Decays}},  {\em PoS}
  {\bf EPS-HEP2013} (2013) 267, [\href{http://xxx.lanl.gov/abs/1310.3825}{{\tt
  arXiv:1310.3825}}].

\bibitem{Carmona:2013cq}
A.~Carmona and F.~Goertz, {\it {Custodial Leptons and Higgs Decays}},  {\em
  JHEP} {\bf 1304} (2013) 163, [\href{http://xxx.lanl.gov/abs/1301.5856}{{\tt
  arXiv:1301.5856}}].

\bibitem{Giudice:2007fh}
G.~Giudice, C.~Grojean, A.~Pomarol, and R.~Rattazzi, {\it {The
  Strongly-Interacting Light Higgs}},  {\em JHEP} {\bf 0706} (2007) 045,
  [\href{http://xxx.lanl.gov/abs/hep-ph/0703164}{{\tt hep-ph/0703164}}].

\bibitem{Anastasiou:2009rv}
C.~Anastasiou, E.~Furlan, and J.~Santiago, {\it {Realistic Composite Higgs
  Models}},  {\em Phys.Rev.} {\bf D79} (2009) 075003,
  [\href{http://xxx.lanl.gov/abs/0901.2117}{{\tt arXiv:0901.2117}}].

\bibitem{Panico:2011pw}
G.~Panico and A.~Wulzer, {\it {The Discrete Composite Higgs Model}},  {\em
  JHEP} {\bf 1109} (2011) 135, [\href{http://xxx.lanl.gov/abs/1106.2719}{{\tt
  arXiv:1106.2719}}].

\bibitem{Azatov:2011qy}
A.~Azatov and J.~Galloway, {\it {Light Custodians and Higgs Physics in
  Composite Models}},  {\em Phys.Rev.} {\bf D85} (2012) 055013,
  [\href{http://xxx.lanl.gov/abs/1110.5646}{{\tt arXiv:1110.5646}}].

\bibitem{Manton:1979kb}
N.~Manton, {\it {A New Six-Dimensional Approach to the Weinberg-Salam Model}},
  {\em Nucl.Phys.} {\bf B158} (1979) 141.

\bibitem{Hatanaka:1998yp}
H.~Hatanaka, T.~Inami, and C.~Lim, {\it {The Gauge hierarchy problem and higher
  dimensional gauge theories}},  {\em Mod.Phys.Lett.} {\bf A13} (1998)
  2601--2612, [\href{http://xxx.lanl.gov/abs/hep-th/9805067}{{\tt
  hep-th/9805067}}].

\bibitem{vonGersdorff:2002as}
G.~von Gersdorff, N.~Irges, and M.~Quiros, {\it {Bulk and brane radiative
  effects in gauge theories on orbifolds}},  {\em Nucl.Phys.} {\bf B635} (2002)
  127--157, [\href{http://xxx.lanl.gov/abs/hep-th/0204223}{{\tt
  hep-th/0204223}}].

\bibitem{Csaki:2002ur}
C.~Csaki, C.~Grojean, and H.~Murayama, {\it {Standard model Higgs from higher
  dimensional gauge fields}},  {\em Phys.Rev.} {\bf D67} (2003) 085012,
  [\href{http://xxx.lanl.gov/abs/hep-ph/0210133}{{\tt hep-ph/0210133}}].

\bibitem{Contino:2003ve}
R.~Contino, Y.~Nomura, and A.~Pomarol, {\it {Higgs as a holographic
  pseudoGoldstone boson}},  {\em Nucl.Phys.} {\bf B671} (2003) 148--174,
  [\href{http://xxx.lanl.gov/abs/hep-ph/0306259}{{\tt hep-ph/0306259}}].

\bibitem{Agashe:2004rs}
K.~Agashe, R.~Contino, and A.~Pomarol, {\it {The Minimal composite Higgs
  model}},  {\em Nucl.Phys.} {\bf B719} (2005) 165--187,
  [\href{http://xxx.lanl.gov/abs/hep-ph/0412089}{{\tt hep-ph/0412089}}].

\bibitem{Terazawa:1976xx}
H.~Terazawa, K.~Akama, and Y.~Chikashige, {\it {Unified Model of the
  Nambu-Jona-Lasinio Type for All Elementary Particle Forces}},  {\em
  Phys.Rev.} {\bf D15} (1977) 480.

\bibitem{Terazawa:1979pj}
H.~Terazawa, {\it {Subquark Model of Leptons and Quarks}},  {\em Phys.Rev.}
  {\bf D22} (1980) 184.

\bibitem{Dimopoulos:1981xc}
S.~Dimopoulos and J.~Preskill, {\it {Massless Composites With Massive
  Constituents}},  {\em Nucl.Phys.} {\bf B199} (1982) 206.

\bibitem{Kaplan:1983fs}
D.~B. Kaplan and H.~Georgi, {\it {SU(2) x U(1) Breaking by Vacuum
  Misalignment}},  {\em Phys.Lett.} {\bf B136} (1984) 183.

\bibitem{Kaplan:1983sm}
D.~B. Kaplan, H.~Georgi, and S.~Dimopoulos, {\it {Composite Higgs Scalars}},
  {\em Phys.Lett.} {\bf B136} (1984) 187.

\bibitem{Georgi:1984ef}
H.~Georgi, D.~B. Kaplan, and P.~Galison, {\it {Calculation of the Composite
  Higgs Mass}},  {\em Phys.Lett.} {\bf B143} (1984) 152.

\bibitem{Banks:1984gj}
T.~Banks, {\it {Constraints on SU(2) x U(1) Breaking by Vacuum Misalignment}},
  {\em Nucl.Phys.} {\bf B243} (1984) 125.

\bibitem{Georgi:1984af}
H.~Georgi and D.~B. Kaplan, {\it {Composite Higgs and Custodial SU(2)}},  {\em
  Phys.Lett.} {\bf B145} (1984) 216.

\bibitem{Dugan:1984hq}
M.~J. Dugan, H.~Georgi, and D.~B. Kaplan, {\it {Anatomy of a Composite Higgs
  Model}},  {\em Nucl.Phys.} {\bf B254} (1985) 299.

\bibitem{Coleman:1969sm}
S.~R. Coleman, J.~Wess, and B.~Zumino, {\it {Structure of phenomenological
  Lagrangians. 1.}},  {\em Phys.Rev.} {\bf 177} (1969) 2239--2247.

\bibitem{Callan:1969sn}
J.~Callan, Curtis~G., S.~R. Coleman, J.~Wess, and B.~Zumino, {\it {Structure of
  phenomenological Lagrangians. 2.}},  {\em Phys.Rev.} {\bf 177} (1969)
  2247--2250.

\bibitem{Agashe:2003zs}
K.~Agashe, A.~Delgado, M.~J. May, and R.~Sundrum, {\it {RS1, custodial isospin
  and precision tests}},  {\em JHEP} {\bf 0308} (2003) 050,
  [\href{http://xxx.lanl.gov/abs/hep-ph/0308036}{{\tt hep-ph/0308036}}].

\bibitem{Agashe:2006at}
K.~Agashe, R.~Contino, L.~Da~Rold, and A.~Pomarol, {\it {A Custodial symmetry
  for Zb anti-b}},  {\em Phys.Lett.} {\bf B641} (2006) 62--66,
  [\href{http://xxx.lanl.gov/abs/hep-ph/0605341}{{\tt hep-ph/0605341}}].

\bibitem{Carena:2006bn}
M.~S. Carena, E.~Ponton, J.~Santiago, and C.~E. Wagner, {\it {Light Kaluza
  Klein States in Randall-Sundrum Models with Custodial SU(2)}},  {\em
  Nucl.Phys.} {\bf B759} (2006) 202--227,
  [\href{http://xxx.lanl.gov/abs/hep-ph/0607106}{{\tt hep-ph/0607106}}].

\bibitem{Carena:2007ua}
M.~S. Carena, E.~Ponton, J.~Santiago, and C.~Wagner, {\it {Electroweak
  constraints on warped models with custodial symmetry}},  {\em Phys.Rev.} {\bf
  D76} (2007) 035006, [\href{http://xxx.lanl.gov/abs/hep-ph/0701055}{{\tt
  hep-ph/0701055}}].

\bibitem{Carena:2007tn}
M.~Carena, A.~D. Medina, B.~Panes, N.~R. Shah, and C.~E. Wagner, {\it {Collider
  phenomenology of gauge-Higgs unification scenarios in warped extra
  dimensions}},  {\em Phys.Rev.} {\bf D77} (2008) 076003,
  [\href{http://xxx.lanl.gov/abs/0712.0095}{{\tt arXiv:0712.0095}}].

\bibitem{Pomarol:2008bh}
A.~Pomarol and J.~Serra, {\it {Top Quark Compositeness: Feasibility and
  Implications}},  {\em Phys.Rev.} {\bf D78} (2008) 074026,
  [\href{http://xxx.lanl.gov/abs/0806.3247}{{\tt arXiv:0806.3247}}].

\bibitem{Panico:2010is}
G.~Panico, M.~Safari, and M.~Serone, {\it {Simple and Realistic Composite Higgs
  Models in Flat Extra Dimensions}},  {\em JHEP} {\bf 1102} (2011) 103,
  [\href{http://xxx.lanl.gov/abs/1012.2875}{{\tt arXiv:1012.2875}}].

\bibitem{Kaplan:1991dc}
D.~B. Kaplan, {\it {Flavor at SSC energies: A New mechanism for dynamically
  generated fermion masses}},  {\em Nucl.Phys.} {\bf B365} (1991) 259--278.

\bibitem{Carena:2014ria}
M.~Carena, L.~Da~Rold, and E.~Pontón, {\it {Minimal Composite Higgs Models at
  the LHC}},  {\em JHEP} {\bf 1406} (2014) 159,
  [\href{http://xxx.lanl.gov/abs/1402.2987}{{\tt arXiv:1402.2987}}].

\bibitem{Maldacena:1997re}
J.~M. Maldacena, {\it {The Large N limit of superconformal field theories and
  supergravity}},  {\em Int.J.Theor.Phys.} {\bf 38} (1999) 1113--1133,
  [\href{http://xxx.lanl.gov/abs/hep-th/9711200}{{\tt hep-th/9711200}}].

\bibitem{Gubser:1998bc}
S.~Gubser, I.~R. Klebanov, and A.~M. Polyakov, {\it {Gauge theory correlators
  from noncritical string theory}},  {\em Phys.Lett.} {\bf B428} (1998)
  105--114, [\href{http://xxx.lanl.gov/abs/hep-th/9802109}{{\tt
  hep-th/9802109}}].

\bibitem{Witten:1998qj}
E.~Witten, {\it {Anti-de Sitter space and holography}},  {\em
  Adv.Theor.Math.Phys.} {\bf 2} (1998) 253--291,
  [\href{http://xxx.lanl.gov/abs/hep-th/9802150}{{\tt hep-th/9802150}}].

\bibitem{ArkaniHamed:2000ds}
N.~Arkani-Hamed, M.~Porrati, and L.~Randall, {\it {Holography and
  phenomenology}},  {\em JHEP} {\bf 0108} (2001) 017,
  [\href{http://xxx.lanl.gov/abs/hep-th/0012148}{{\tt hep-th/0012148}}].

\bibitem{Falkowski:2006vi}
A.~Falkowski, {\it {About the holographic pseudo-Goldstone boson}},  {\em
  Phys.Rev.} {\bf D75} (2007) 025017,
  [\href{http://xxx.lanl.gov/abs/hep-ph/0610336}{{\tt hep-ph/0610336}}].

\bibitem{Contino:2004vy}
R.~Contino and A.~Pomarol, {\it {Holography for fermions}},  {\em JHEP} {\bf
  0411} (2004) 058, [\href{http://xxx.lanl.gov/abs/hep-th/0406257}{{\tt
  hep-th/0406257}}].

\bibitem{Batell:2007jv}
B.~Batell and T.~Gherghetta, {\it {Holographic mixing quantified}},  {\em
  Phys.Rev.} {\bf D76} (2007) 045017,
  [\href{http://xxx.lanl.gov/abs/0706.0890}{{\tt arXiv:0706.0890}}].

\bibitem{Batell:2007ez}
B.~Batell and T.~Gherghetta, {\it {Warped phenomenology in the holographic
  basis}},  {\em Phys.Rev.} {\bf D77} (2008) 045002,
  [\href{http://xxx.lanl.gov/abs/0710.1838}{{\tt arXiv:0710.1838}}].

\bibitem{Gherghetta:2010cj}
T.~Gherghetta, {\it {TASI Lectures on a Holographic View of Beyond the Standard
  Model Physics}},  \href{http://xxx.lanl.gov/abs/1008.2570}{{\tt
  arXiv:1008.2570}}.

\bibitem{Goertz:2014qia}
F.~Goertz, {\it {Indirect Handle on the Down-Quark Yukawa Coupling}},  {\em
  Phys.Rev.Lett.} {\bf 113} (2014), no.~26 261803,
  [\href{http://xxx.lanl.gov/abs/1406.0102}{{\tt arXiv:1406.0102}}].

\bibitem{Chatrchyan:2013uxa}
{\bf CMS Collaboration} Collaboration, S.~Chatrchyan et~al., {\it {Inclusive
  search for a vector-like T quark with charge $\frac{2}{3}$ in pp collisions
  at $\sqrt{s}$ = 8 TeV}},  {\em Phys.Lett.} {\bf B729} (2014) 149--171,
  [\href{http://xxx.lanl.gov/abs/1311.7667}{{\tt arXiv:1311.7667}}].

\bibitem{Pappadopulo:2013vca}
D.~Pappadopulo, A.~Thamm, and R.~Torre, {\it {A minimally tuned composite Higgs
  model from an extra dimension}},  {\em JHEP} {\bf 1307} (2013) 058,
  [\href{http://xxx.lanl.gov/abs/1303.3062}{{\tt arXiv:1303.3062}}].

\bibitem{Carena:2009yt}
M.~Carena, A.~D. Medina, N.~R. Shah, and C.~E. Wagner, {\it {Gauge-Higgs
  Unification, Neutrino Masses and Dark Matter in Warped Extra Dimensions}},
  {\em Phys.Rev.} {\bf D79} (2009) 096010,
  [\href{http://xxx.lanl.gov/abs/0901.0609}{{\tt arXiv:0901.0609}}].

\bibitem{Hagedorn:2011un}
C.~Hagedorn and M.~Serone, {\it {Leptons in Holographic Composite Higgs Models
  with Non-Abelian Discrete Symmetries}},  {\em JHEP} {\bf 1110} (2011) 083,
  [\href{http://xxx.lanl.gov/abs/1106.4021}{{\tt arXiv:1106.4021}}].

\bibitem{Hagedorn:2011pw}
C.~Hagedorn and M.~Serone, {\it {General Lepton Mixing in Holographic Composite
  Higgs Models}},  {\em JHEP} {\bf 1202} (2012) 077,
  [\href{http://xxx.lanl.gov/abs/1110.4612}{{\tt arXiv:1110.4612}}].

\bibitem{KerenZur:2012fr}
B.~Keren-Zur, P.~Lodone, M.~Nardecchia, D.~Pappadopulo, R.~Rattazzi, et~al.,
  {\it {On Partial Compositeness and the CP asymmetry in charm decays}},  {\em
  Nucl.Phys.} {\bf B867} (2013) 394--428,
  [\href{http://xxx.lanl.gov/abs/1205.5803}{{\tt arXiv:1205.5803}}].

\bibitem{Hosotani:2005nz}
Y.~Hosotani and M.~Mabe, {\it {Higgs boson mass and electroweak-gravity
  hierarchy from dynamical gauge-Higgs unification in the warped spacetime}},
  {\em Phys.Lett.} {\bf B615} (2005) 257--265,
  [\href{http://xxx.lanl.gov/abs/hep-ph/0503020}{{\tt hep-ph/0503020}}].

\bibitem{Falkowski:2007hz}
A.~Falkowski, {\it {Pseudo-goldstone Higgs production via gluon fusion}},  {\em
  Phys.Rev.} {\bf D77} (2008) 055018,
  [\href{http://xxx.lanl.gov/abs/0711.0828}{{\tt arXiv:0711.0828}}].

\bibitem{Foot:1988aq}
R.~Foot, H.~Lew, X.~He, and G.~C. Joshi, {\it {Seesaw Neutrino Masses Induced
  by a Triplet of Leptons}},  {\em Z.Phys.} {\bf C44} (1989) 441.

\bibitem{Dev:2013ff}
P.~Bhupal~Dev, D.~K. Ghosh, N.~Okada, and I.~Saha, {\it {125 GeV Higgs Boson
  and the Type-II Seesaw Model}},  {\em JHEP} {\bf 1303} (2013) 150,
  [\href{http://xxx.lanl.gov/abs/1301.3453}{{\tt arXiv:1301.3453}}].

\bibitem{Oda:2004rm}
K.-y. Oda and A.~Weiler, {\it {Wilson lines in warped space: Dynamical symmetry
  breaking and restoration}},  {\em Phys.Lett.} {\bf B606} (2005) 408--416,
  [\href{http://xxx.lanl.gov/abs/hep-ph/0410061}{{\tt hep-ph/0410061}}].

\bibitem{Baak:2012kk}
M.~Baak, M.~Goebel, J.~Haller, A.~Hoecker, D.~Kennedy, et~al., {\it {The
  Electroweak Fit of the Standard Model after the Discovery of a New Boson at
  the LHC}},  {\em Eur.Phys.J.} {\bf C72} (2012) 2205,
  [\href{http://xxx.lanl.gov/abs/1209.2716}{{\tt arXiv:1209.2716}}].

\bibitem{Barbieri:1987fn}
R.~Barbieri and G.~Giudice, {\it {Upper Bounds on Supersymmetric Particle
  Masses}},  {\em Nucl.Phys.} {\bf B306} (1988) 63.

\bibitem{Gershtein:2013iqa}
Y.~Gershtein, M.~Luty, M.~Narain, L.~T. Wang, D.~Whiteson, et~al., {\it
  {Working Group Report: New Particles, Forces, and Dimensions}},
  \href{http://xxx.lanl.gov/abs/1311.0299}{{\tt arXiv:1311.0299}}.

\bibitem{Agashe:2013hma}
{\bf Top Quark Working Group} Collaboration, K.~Agashe et~al., {\it {Working
  Group Report: Top Quark}},  \href{http://xxx.lanl.gov/abs/1311.2028}{{\tt
  arXiv:1311.2028}}.

\bibitem{next}
A.~Carmona and F.~Goertz {\em in preparation}.

\end{thebibliography}\endgroup

\end{document}